\documentclass[12pt,a4paper]{article} 
\usepackage{amsmath,amssymb,amsthm,amscd,mathtools}
\numberwithin{equation}{section}
\usepackage{xcolor}
\usepackage{hyperref} 
\usepackage{url}
\usepackage{microtype}
\usepackage{subcaption}
\usepackage{empheq}
\usepackage{enumerate}
\usepackage{booktabs}
\usepackage[makeroom]{cancel}
\usepackage[numbers,sort&compress]{natbib}
\usepackage{dsfont} 
\usepackage{setspace}
\usepackage[skins]{tcolorbox}
\usepackage{mathrsfs}

\newcommand{\beq}{\begin{equation}}
\newcommand{\eeq}{\end{equation}}

\newcommand{\E}{{\epsilon}}

\newcommand{\fg}{\mathfrak{g}}
\newcommand{\fgl}{\mathfrak{g l}}
\newcommand{\fB}{\mathfrak{B}}
\newcommand{\fC}{\mathfrak{C}}
\newcommand{\fa}{\mathfrak{a}}

\newcommand{\fh}{\mathfrak{h}}

\newcommand{\ft}{\mathfrak{t}}

\newcommand{\fm}{\mathfrak{m}}

\newcommand{\fn}{\mathfrak{n}}

\newcommand{\fD}{\mathfrak{D}}

\newcommand{\cA}{\mathcal{A}}

\newcommand{\cC}{\mathcal{C}}
\newcommand{\cN}{\mathcal{N}}
\newcommand{\cW}{\mathcal{W}}

\newcommand{\cY}{\mathcal{Y}}
\newcommand{\cD}{\mathcal{D}}

\newcommand{\cV}{\mathcal{V}}
\newcommand{\sh}{\mbox{sh}}
\renewcommand{\r}{\text{rk}(\fg)}
\renewcommand{\L}{\text{rk}(^L\fg)}

\newmuskip\pFqmuskip

\DeclareRobustCommand{\svdots}{
	\vcenter{%
		\offinterlineskip
		\hbox{.}
		\vskip0.25\normalbaselineskip
		\hbox{.}
		\vskip0.25\normalbaselineskip
		\hbox{.}%
	}%
}

\newcommand*\pFq[6][8]{%
	\begingroup 
	\pFqmuskip=#1mu\relax
	\mathcode`\,=\string"8000
	\begingroup\lccode`\~=`\,
	\lowercase{\endgroup\let~}\pFqcomma
	{}_{#2}\mathbb{F}_{#3}{\left[\genfrac..{0pt}{}{#4}{#5};#6\right]}%
	\endgroup
}
\newcommand{\pFqcomma}{\mskip\pFqmuskip}

\newtheoremstyle{fullit}
{\topsep}      
{\topsep}      
{\normalfont}  
{0pt}          
{\itshape}     
{.\ }          
{0pt}          
{\thmname{#1} \thmnumber{#2}}             

\newtheorem{prop}{Proposition}[section]
\newtheorem{thm}{Theorem}[section]

\theoremstyle{definition}

\newtheorem{example}[thm]{Example}
\newtheorem{remark}{Remark}[section]

\DeclareCaptionSubType*[alph]{figure}
\begin{document}
	\baselineskip=18pt  
	\baselineskip 0.713cm

	\begin{titlepage}
		

		\renewcommand{\thefootnote}{\fnsymbol{footnote}}

		
		\begin{center}
			{\LARGE \bf
				A new realization of quantum algebras  
				\vskip 0.1cm
				in gauge theory and Ramification 
				\vskip 0.4cm
				in the Langlands program
			}
			
			\medskip

			\vskip 0.5cm
			
			{\large
			Nathan Haouzi
			}
			\\
			\medskip
			
			\vskip 0.5cm
			
			{\it
				School of Natural Sciences, Institute for Advanced Study\\
				Einstein Drive, Princeton, NJ 08540 USA\\
			}
			
		\end{center}
		
		\vskip 0.0cm
		
		\centerline{{\bf Abstract}}
		\medskip
		\begin{spacing}{1.08} We realize the fundamental representations of quantum algebras via the supersymmetric Higgs mechanism in gauge theories with 8 supercharges on an $\Omega$-background.
		We test our proposal for quantum affine algebras, by probing the Higgs phase of a 5d quiver gauge theory on a circle. 
		We show that our construction implies the existence of tame ramification in the Aganagic-Frenkel-Okounkov formulation of the geometric Langlands program, a correspondence which identifies $q$-conformal blocks of the quantum affine algebra with those of a Langlands dual deformed $\cW$-algebra.
		The new feature of ramified blocks is their definition in terms of Drinfeld polynomials for a set of quantum affine weights.
		In enumerative geometry, the blocks are vertex functions counting quasimaps to quiver varieties describing moduli spaces of vortices.
		Physically, the vortices admit a description as a 3d $\cN=2$ quiver gauge theory on the Higgs branch of the 5d gauge theory, uniquely determined from the Drinfeld polynomial data; the blocks are supersymmetric indices for the vortex theory supported on a 3-manifold with distinguished BPS boundary conditions. 	
		The top-down explanation of our results is found in the 6d $(2,0)$ little string theory, where tame ramification is provided by certain D-branes. When the string mass is taken to be large, we make contact with various physical aspects of the point particle superconformal limit: the Gukov-Witten description of ramification via monodromy defects in 4d Super Yang-Mills (and their S-duality), the Nekrasov-Tsymbaliuk solution to the Knizhnik-Zamolodchikov equations, and the classification of massive deformations of tamely ramified Hitchin systems.
		In a companion paper, we will show that our construction implies a solution to the local Alday-Gaiotto-Tachikawa conjecture. \end{spacing}
		\noindent\end{titlepage}
	\setcounter{page}{1} 
	
\tableofcontents

	\newpage
	
\section{Introduction}
\label{sec:intro}

In Mathematics, the geometric Langlands program aims to explain an equivalence between certain categories of sheaves on two moduli stacks. On one side, one considers the derived category of coherent sheaves of $O$-modules on the moduli stack of $G$-bundles on $\cC$, where $G$ is a complex connected simply-connected simple Lie group and $\cC$ is a complex projective algebraic curve. On the other side, one considers the derived category of sheaves of $D$-modules on the moduli stack of $^L G$-bundles on $\cC$, where $^L G$ is the Langlands dual group to $G$.

The seminal work of Beilinson and Drinfeld \cite{beilinson1991quantization} implies that such a program can be recast in the language of conformal field theory (CFT) on $\cC$
, which for us will be a curve of genus 0 with punctures. In a CFT, the sheaves of modules are realized as sheaves of conformal blocks, so the geometric Langlands correspondence becomes an equivalence between spaces of conformal blocks for certain representations of two algebras. These algebras were identified using an extraordinary isormophism of Poisson algebras proposed by Feigin and Frenkel \cite{Feigin:1991wy}: on the one hand, one considers the classical commutative $\cW_{\infty}(\fg)$-algebra, which is the algebra of functions on the space of (gauge equivalent classes of) connections on the circle\footnote{$\cW$-algebras are not only labeled by a Lie algebra, but also by a nilpotent orbit inside it. In this work, the nilpotent orbit will always be the maximal one, which we omit writing not to overburden the notations.}. On the other hand, one has the center of the chiral (untwisted) affine Kac-Moody algebra $\widehat{^L \fg}$ at critical level $^L \kappa = - ^L h^\vee$, with $^L h^\vee$ a positive integer called the dual Coxeter number of $^L\fg$. Here, $\fg$ and $^L \fg$ are the Lie algebras of $G$ and $^L G$\footnote{We use the Dynkin notation to label the simple Lie algebras: the classical simple Lie algebras are denoted as $A_r=\text{su}(r+1)$, $B_r=\text{so}(2r+1)$, $C_r=\text{sp}(2r)$ and $D_r=\text{so}(2r)$. Langlands duality maps $\fg$ to itself  whenever $\fg=A_r, D_r, E_{6,7,8}, G_2, F_4$, while $\fg=B_r$ gets mapped to $^L \fg=C_r$ and vice-versa.}.\\ 


The $\cW_{\infty}(\fg)$-algebra famously admits a deformation to a ``quantum" $\cW_{\beta}(\fg)$-algebra, familiar to physicists as a higher spin generalization of the Virasoro algebra (which it reduces to when $\fg=A_1$), and as the chiral symmetry algebra of Toda conformal field theory; the parameter $\beta$ encodes the central charge of the conformal algebra. Meanwhile, it is natural to consider $\widehat{^L \fg}$ at a generic level $^L \kappa$ away from criticality. One therefore predicts a quantum geometric Langlands correspondence, as an isomorphism between the spaces of conformal blocks for certain representations of these chiral algebras\footnote{In the case $\fg=A_1$, this isomorphism has been studied via different methods than will be used in this paper: the techniques rely on the introduction of new degrees of freedom and various changes of variables  \cite{Teschner:2010je,Furlan:1991mm,Frenkel:2015rda}. The generalization beyond $\fg=A_1$ is not clear.}
\beq
\label{quantumlanglands}
\cW_{\beta}(\fg) \;\; \longleftrightarrow \;\; \widehat{^L\fg}_{^L \kappa}  \; .
\eeq
The $\cW_{\beta}(\fg)$-algebra is defined via Drinfeld-Sokolov reduction \cite{FEIGIN199075,Feigin:1991wy,deBoer:1993iz}, the input of which is an affine Lie algebra $\widehat{\fg}_{\kappa}$ at level $\kappa$, on which first class constraints are imposed by means of a BRST complex\footnote{The Drinfeld-Sokolov reduction was first performed for the classical $\cW_{\infty}(\fg)$-algebra, see  \cite{Drinfeld:1984qv} and more recently\cite{ 2013CMaPh.323..663D}.}. The generators of the $\cW_{\beta}(\fg)$-algebra, a spin 2 Virasoro stress tensor and $\r-1$ higher spin currents, are in the zeroth-cohomology of this complex. It follows from this reduction that $\beta = \fn_{\fg}\, (\kappa+h^\vee)$, with $\kappa$ the level of $\widehat{\fg}_{\kappa}$  and $\fn_{\fg}$ the lacing number of $\fg$. In this work, we will take an alternate (but equivalent) approach known as the free field formalism and construct the  $\cW_{\beta}(\fg)$-algebra as the centralizer of a set of screening charges; this construction is also known as the Coulomb gas or Dotsenko-Fateev formalism \cite{Dotsenko:1984nm}\footnote{$\cW$-algebras can also be defined via cosets, though we will not explore this approach here. For recent developments, see the work of Arakawa-Creutzig-Linshaw \cite{Arakawa:2018iyk}.}.\\

Remarkably, both algebras admit a ``$q$-deformation" in the sense of quantum groups\footnote{Throughout this paper, we use the expression ``$q$-deformed" liberally, even when the parameter $q$ has no relation to the deformation in question. For instance, we will often call the deformed conformal block of a quantum affine algebra $U_{\hbar}(\widehat{^L\fg})$ a $q$-conformal block. This terminology is standard practice in the quantum group literature.}. This suggests yet a further generalization of the Langlands program, going back to an old conjecture of Frenkel  \cite{Frenkel:1995zp}, and made precise in a recent work of Aganagic-Frenkel-Okounkov \cite{Aganagic:2017smx}. This \emph{quantum $q$-Langlands correspondence} defines an isomorphism of spaces of $q$-conformal blocks for the following algebras:
\beq
\label{quantumqlanglands}
\cW_{q,t}(\fg) \;\; \longleftrightarrow \;\; U_\hbar(\widehat{^L\fg}_{^L \kappa}) \; .
\eeq
On the right-hand side, one finds the quantization of the universal enveloping algebra of the affine Lie algebra $\widehat{^L\fg}$ at level $^L \kappa$. This quantization was defined in the 80's by Jimbo \cite{Jimbo:1985zk,Jimbo:1985vd} and Drinfeld \cite{Drinfeld1985HopfAA,Drinfeld:1986in}, and appeared in Physics through the study of integrable systems \cite{Pasquier:1989kd}. The algebra $\widehat{^L\fg}$ and its universal enveloping algebra $U(\widehat{^L\fg})$ share the same representation theory, but $U(\widehat{^L\fg})$ enters here because it has the additional structure of a Hopf algebra, so in particular is amenable to a well-defined $\hbar$-deformation. In particular, when $\hbar\rightarrow 1$, the algebra $U_\hbar(\widehat{^L\fg})$ reduces to $U(\widehat{^L\fg})$.

On the left-hand side, one finds the less familiar  $\cW_{q,t}(\fg)$-algebra, a deformation of the conformal ${\cW}_\beta({\fg})$-algebra\footnote{Unlike $U_{\hbar}(\widehat{^L\fg})$, the algebra  $\cW_{q,t}(\fg)$ is \emph{not} properly speaking a quantum group, that is to say a deformation of a Hopf algebra. However, after a minor modification consisting of adding a $q$-deformed Heisenberg algebra $\text{Heis}_{q,t}\oplus{\cW}_{q,t}(A_r)$, one \emph{does} obtain a quantum group. In fact, this direct sum is a subalgebra of a quantum toroidal algebra, $U_{q,t}(\Hat{\Hat{\fgl_1}})$, specifically the subalgebra describing Fock representations at level $r+1$.}. The generators of the $\cW_{q,t}(\fg)$-algebra consist of a deformed stress tensor and $\r-1$ additional ``higher spin" currents (spin is not well-defined after deformation), all constructed in free field formalism\footnote{There is evidence that a Drinfeld-Sokolov-type reduction also exists in the $q$-deformed context: one such construction was proposed in \cite{2001math......7215S}, but relying on the existence of ``semi-infinite" Hecke algebras instead of a BRST complex.} as the commutant of canonical $q$-deformed screening charges\footnote{For completeness, we mention there exists yet another presentation of the deformed $\cW_{q,t}$-algebras, in terms of certain quadratic relations obeyed by the generators. This is a $q$-deformed analog of the usual OPE relations arising in conformal field theory.}. 
The algebra $\cW_{q,t}(A_1)$ is sometimes called the $q$-Virasoro algebra \cite{Shiraishi:1995rp}. 
The algebras $\cW_{q,t}(A_r)$ for $r>1$ were defined in \cite{Feigin:1995sf,Awata:1995zk}, and then generalized to any simple Lie algebra $\cW_{q,t}(\fg)$ by Frenkel and Reshetikhin in \cite{Frenkel:1998}\footnote{There are other generalizations that we will not encounter in this work. For instance, there exist W-superalgebras with additional fermionic symmetry \cite{Kojima2019QuadraticRO,Kojima2021QuadraticRO2,Harada:2021xnm}. Moreover, Kimura and Pestun defined a class of $\cW_{q,t}(\Gamma)$-algebras labeled by general quivers $\Gamma$, not necessarily of Dynkin type \cite{Kimura:2015rgi,Kimura:2016dys,Kimura:2017hez}. Only a small subset of such quivers $\Gamma$ arise in Physics; for us, $\Gamma$ will be the quiver of a simple Lie algebra.}.

Each side of the correspondence \eqref{quantumqlanglands} now features two deformation parameters: the deformation parameters $q$ and $t$ on the left, and the level $^L \kappa$ and quantization parameter $\hbar$ on the right. The quantum $q$-Langlands correspondence relates them as
\beq
\label{dictionary}
\hbar=\frac{q^{\fn_{\fg}}}{t}\; ,\qquad q=\hbar^{-^L(\kappa+h^\vee)} \; .
\eeq

\vspace{10mm}

\subsection{The Quantum Field Theory view on Geometric Langlands}

Kapustin and Witten explained that the geometric Langlands correspondence has a natural realization as S-duality in the language of 4-dimensional supersymmetric gauge theory \cite{Kapustin:2006pk}. Namely, one considers maximally supersymmetric $\cN=4$ Yang-Mills theory with gauge group $G$ on a four-manifold $M_4$ that locally is a product of Riemann surfaces,  $M_4={\cal C}\times M_2$. Here, $\cC$ is the Riemann surface on which the geometric Langlands program ``lives," it is a fiber to the normal bundle to $M_2$.
S-duality is a conjectured generalization of electro-magnetic duality in Maxwell theory; it states that $\cN=4$ $G$-Yang-Mills should be the same theory as $\cN=4$ $^L G$-Yang-Mills. Electric and magnetic BPS degrees of freedom get exchanged under the duality, as well as the weak and strong coupling regimes, according to
\begin{align}
\label{Sdualitytauintro}
\text{S}: \;\;\;(\fg,\tau^{4d}) \longleftrightarrow (^L \fg,^L\tau^{4d}) \; ,\qquad   ^L\tau^{4d} = \frac{-1}{\fn_{\fg}\,\tau^{4d}} \; .
\end{align}
Above, $\tau^{4d}= \frac{\theta}{2\pi}+i\frac{4\pi}{g^2_{4d}}$ denotes the gauge coupling of $\fg$-type Yang-Mills, with $g_{4d}$ the coupling constant and  $\theta$ the theta-angle.

It turns out that the previous quantum deformation parameters can be expressed in terms of these gauge couplings: if we set $t=q^{\fn_\fg\, \tau^{4d}}$ and $^L\tau^{4d}=-^L(k + h^\vee)$, then the dictionary \eqref{dictionary} implies
\beq
\label{TSduality}
\tau^{4d}-1=\frac{-1}{\fn_{\fg}\,^L\tau^{4d}} \; .
\eeq
We recognize the action \eqref{Sdualitytauintro} of S duality on the gauge coupling, up to a shift $\tau^{4d}\rightarrow \tau^{4d}-1$. Assuming (as we will) that the gauge group $G$ is simply-connected, this shift is yet another symmetry of Yang-Mills theory, associated to a $-2\pi$-shift of the $\theta$-angle in $\tau^{4d}$:  
\beq
\label{Tdualitytauintro}
\text{T}: \;\;\;(\fg,\tau^{4d}) \longrightarrow (\fg,\tau^{4d}- 1) \; .
\eeq 
This is one of the surprises of the quantum $q$-Langlands correspondence: it corresponds to  TS-duality in Yang-Mills theory rather than simply S-duality. We will see that this is not an accident, but rather a choice of spectral duality frame.

TS-duality also implies a relation between the parameters of the usual quantum Langlands correspondence \eqref{quantumlanglands}, obtained in the conformal limit  $t= q^{\beta}$ with $\beta$ fixed as $q\rightarrow 1$: 
\beq
\beta-\fn_{\fg} = \frac{1}{^L(\kappa+h^\vee)}\; .
\eeq
In order to recover the geometric Langlands correspondence from gauge theory, one further applies a certain twist of Landau-Ginzburg type to Yang-Mills on $M_4$, which turns it into a topological quantum field theory. Only then, S-duality becomes a statement about the equivalence of two categories on the Riemann surface $\cC$, precisely the ones defined by Beilinson and Drinfeld. As a result, one obtains a rich dictionary between 4d gauge theory and the usual formulation of the geometric Langlands program. For instance, the action of S-duality on local operators gives an interpretation of the cohomology of orbits in the affine Grassmannian of $^L G$ in terms of the representation theory of $G$, a prediction of the geometric Satake correspondence. Furthermore, in gauge theory, it is natural to study extended operators. A 1-dimensional example supported on a 1-cycle of $M_2\subset M_4$ is a 't Hooft loop in gauge theory, which realizes a Hecke eigensheaf on the moduli stack of holomorphic $^L G$-bundles, meaning a $D$-module satisfying a certain property determined by the $^L G$-bundle. Under S-duality, it maps to a Wilson loop, a coherent sheaf supported at a point on the moduli stack of flat $G$-bundles.\\

\begin{figure}[h!]
	\emph{}
	\centering
	\includegraphics[trim={0 0 0 0cm},clip,width=0.99\textwidth]{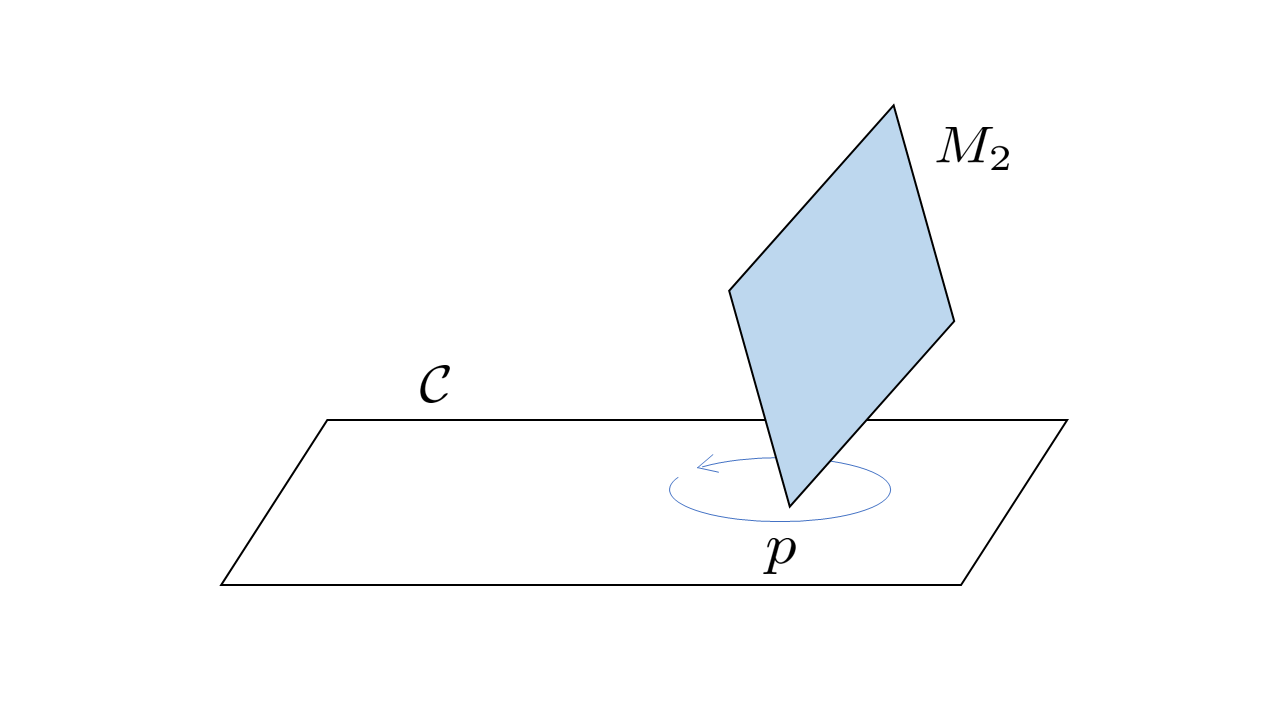}
	\vspace{-10pt}
	\caption{Monodromy defects in 4d Super Yang-Mills wrap $M_2\subset M_4$ and sit at a point $p$ on the Riemann surface $\cC$.} 
	\label{fig:gukovwitten}
\end{figure}

The case of two-dimensional extended operators supported on $M_2\subset M_4$ was later analyzed by Gukov and Witten \cite{Gukov:2006jk,Gukov:2008sn}. In the geometric Langlands program, this corresponds to tame ramification, or specifying additional monodromy and residue data at punctures on the Riemann surface $\cC$. This ramified setting will be the subject of our work\footnote{One could further study operators supported on a 3-manifold inside $M_4$ \cite{Gaiotto:2008sa,Gaiotto:2008ak}. Such operators give a gauge theory realization of the universal kernel in the Langlands program, and the notion of functoriality.}.
For recent gauge theory developments, see also the works \cite{Frenkel:2018dej,Gaiotto:2021tsq,Jeong:2018qpc,Jeong:2023qdr}.

\vspace{10mm}

\subsection{A lift to 6 dimensions, and a two-parameter deformation}

It was further understood that 4-dimensional S-duality has a natural 6d origin, found in a  $\cN=(2,0)$ superconformal field theory (SCFT) ${\cal X}_{\fg_\text{o}}$, labeled by a simply-laced Lie algebra $\fg_\text{o} = ADE$\footnote{The highest spacetime dimension where a quantum field theory can have both conformal and super-Poincar\'{e} symmetries is 6, as was proved by Nahm in the 70's \cite{Nahm:1977tg}.} \cite{Witten:2009at}.

In this picture, 4d Yang-Mills with gauge algebra $\fg_\text{o}$ is recovered by reducing the theory ${\cal X}_{\fg_\text{o}}$ on a 2-torus $T^2=S^1_q\times S^1_t$. Crucially, this reduction depends on the choice of compactification circles: reducing first on $S^1_q$ to 5d and then on $S^1_t$ to 4d, one obtains 4d $\fg_\text{o}$-type Yang-Mills with gauge coupling $\tau^{4d}$. On the other hand, reducing first on $S^1_t$ to 5d and then on $S^1_q$ to 4d, one instead obtains 4d $^L\fg_\text{o}$-type Yang-Mills with gauge coupling $^L\tau^{4d}=-1/\tau^{4d}$. The 4d gauge coupling is therefore identified with the $\tau$-parameter of the torus $T^2$, and S-duality is realized as the S-transformation of the modular group SL$(2,\mathbb{Z})$ acting on the $\tau$-parameter of $T^2$. Because the 6d theory ${\cal X}_{\fg_\text{o}}$ is of $ADE$ type, so is the gauge algebra of the 4d Yang-Mills theory. To realize S-duality for non simply-laced Lie algebras $\fg$, one needs to perform an additional twist by the outer automorphism group of $\fg_\text{o}$ along one of the two circles in the $T^2$ compactification.\\

The theory ${\cal X}_{\fg_\text{o}}$ is the point particle limit of a non-critical theory of strings in 6 dimensions, the $(2,0)$ little string theory, labeled by the same simply-laced Lie algebra $\fg_\text{o}$ \cite{Seiberg:1997zk}. This string theory is not conformal, as it depends on the string mass scale $m_s$; if the natural setting of the quantum geometric Langlands correspondence is the CFT limit $m_s\rightarrow \infty$ as above, then what is the meaning of keeping $m_s$ finite in the Langlands program? 

A first answer was given by Vafa in \cite{Vafa:1997mh}: if we compactify the 6d little string on the torus $T^2=S^1_q\times S^1_t$, then the S-duality in 4d Super Yang-Mills is a consequence of perturbative \emph{T-duality} in 6d little string theory \footnote{Note the T-duality we are using here is \emph{not} T-duality on the worldsheet, which also happens to be related to S-duality in 4d Yang-Mills \cite{Harvey:1995tg,Bershadsky:1995vm,Kapustin:2006pk}.}: which 4d Yang Mills theory one obtains after reducing on $T^2$ (and taking $m_s\rightarrow \infty$) is dictated by which T-dual frame one starts with in little string theory.\\

Recently, Aganagic-Frenkel-Okounkov made a significant breakthrough \cite{Aganagic:2017smx}, by showing that the quantum $q$-Langlands correspondence \eqref{quantumqlanglands} has its origin in the $(2,0)$ little string theory compactified on the genus 0 curve $\cC=\mathbb{C}^\times$\footnote{The choice of curve $\cC$ dictates which quantum algebra enters the story: Yangians for $\mathbb{C}$, quantum affine algebras for $\mathbb{C}^\times$, and quantum elliptic algebras for $T^2$.  Note that unlike in the conformal case, the cylinder $\cC$ is not equivalent to the Riemann sphere with two punctures $\mathbb{CP}^1/\{0,\infty\}$ in the $q$-deformed setting.}. Indeed, the deformation parameters $q$, $t$ and $\hbar$ actually all depend on the scale $m_s$, and the isomorphism of $q$-conformal blocks can be recast as an equality of little string partition functions in the presence of D3 brane defects from type IIB. These branes are self-dual strings from the 6d point of view\footnote{Self-dual strings are BPS objects both in the 6d $(2,0)$ little string and in its SCFT limit ${\cal X}_{\fg_\text{o}}$. They are \emph{not} the same as little strings, which are fundamental strings of type IIB in the zero string coupling limit $g_s\rightarrow 0$.} and it follows from Vafa's argument that T-dualizing them in this background gives a stringy origin to the S-duality of Kapustin-Witten line operators in 4d Yang-Mills.
The various $q$-conformal blocks which enter the Aganagic-Frenkel-Okounkov correspondence are defined as follows:

Following (Igor) Frenkel and Reshetikhin \cite{Frenkel:1991gx}, the $U_{\hbar}(\widehat{^L\fg})$ blocks are defined as solutions of a certain holonomic system of $q$-difference equations. These are called quantum Knizhnik-Zamolodchikov (qKZ) equations, and should be understood as a $q$-deformation of the usual trigonometric Knizhnik-Zamolodchikov equations \cite{Knizhnik:1984nr} that appear in $\widehat{^L\fg}$-type WZW conformal field theory. The solutions are of the form
\beq\label{electriccorrelatorintro}
\left\langle v_{\nu_\infty},\prod_{d=1}^{L} \Phi_{d}(x_d) \; v_{\nu_0} \right\rangle  \; ,  
\eeq
and will be referred to as \emph{electric blocks}.
Each $\Phi_{d}(x_d)$ is a chiral vertex operator, whose details depend on what objects we aim to study in the Langlands program. In the Aganagic-Frenkel-Okounkov correspondence, which we call \emph{unramified}, the operators are labeled by the highest weight $^L w_a$ of a  $U_{\hbar}(\widehat{^L\fg})$-fundamental representation $\widehat{V}_{a}$ for some $a\in\{1,\ldots,\L\}$, and by their position $x_d$ on the cylinder $\cC$. More precisely, such a vertex operator acts as an intertwiner, and the position $x_d$ enters through the use of an evaluation representation $V_a(x_d)$ in the definition of the intertwiner.
The vector $v_{\nu_0}$ is the highest weight vector of an irreducible Verma module $\widehat{V}_{\nu_0}$ over $U_{\hbar}(\widehat{^L\fg})$; in Physics, this vector is often called the Fock vacuum. Likewise, $v_{\nu_\infty}$ is the lowest weight vector of an irreducible Verma module $\widehat{V}^\ast_{\nu_\infty}$\footnote{Mathematically, it would be more accurate to call this setup unramified at $L$ points $x_d$, for $d=1,\ldots,L$, and ramified at the two ends of the cylinder at $0$ and $\infty$, where the Verma modules are.}.

The $q$-conformal blocks do not take values in the full space ${V}_{1}\otimes{V}_{2}\otimes\ldots\otimes{V}_{L}$, but instead in the subspace
\beq
\label{weightsubspaceintro}
\left({V}_{1}\otimes{V}_{2}\otimes\ldots\otimes{V}_{L}\right)_{\nu_0-\nu_\infty}
\eeq
of vectors of weight 
\beq
\nu_0-\nu_\infty = \sum_{a=1}^{\L} m_a\, ^L w_a - \sum_{a=1}^{\L} N_a \, ^L\alpha_a \; .
\eeq 
Here, $^L w_a$ and $^L \alpha_a$ are the $a$-th fundamental weight and positive simple root of $^L\fg$, respectively, and  $N_a$ are $m_a$ are non-negative integers.

These electric blocks admit a well-known integral representation, worked out in the 90's for $\fg=A_r$ \cite{Reshetikhin1992JacksontypeIB,cmp/1104252137,cmp/1104254019,Tarasov:1993vs,1994CMaPh.162..499V,1997q.alg.....3044T}; the generalization to other simple Lie algebras $\fg$, as well as certain quivers beyond the Dynkin classification, was understood recently \cite{Aganagic:2017gsx}.\\

On the $\cW_{q,t}(\fg)$-algebra side, $q$-conformal blocks are also known to admit an integral representation \cite{Shiraishi:1995rp,Feigin:1995sf,Awata:1995zk,Frenkel:1998}:
\beq\label{magneticcorrelatorintro}
\left\langle v_{\mu_\infty},\,\prod_{a=1}^{\r} (Q_a^\vee)^{N_a} \;\prod_{d=1}^{L} {\overline V}^\vee_{d}(x_d)\;   v_{\mu_0} \right\rangle  \, ,
\eeq
and will be referred to as \emph{magnetic blocks}.
All operators are assumed to be normal-ordered in the products. The ${Q}_a^\vee$ are the screening charges of the algebra, and are responsible for turning the blocks into a multi-dimensional complex integral over a set of screening currents: ${Q}_a^\vee=\oint dy\, S_a^\vee(y)$.
The vertex operators ${\overline V}^\vee_{d}(x_d)$ which enter the unramified analysis were introduced by Frenkel and Reshetikhin, and go by the name of \emph{fundamental vertex operators} \cite{Frenkel:1998}; each of them is labeled by the highest weight $^L w_a$ of a  $U_{\hbar}(\widehat{^L\fg})$-fundamental representation $\widehat{V}_{a}$, for some $a\in\{1,\ldots,\L\}$, and by its position $x_d$ on $\cC$. Note that the operators $\overline V^\vee_{d}(x_d)$ are defined with respect to the Langlands dual algebra $^L\fg$. Equivalently, we can always define them in $\fg$ and label $V^\vee_{d}(x_d)$ by the highest \emph{coweight} of a  $U_{\hbar}(\widehat{\fg})$-fundamental representation. 
The vector $v_{\mu_0}$ generates an irreducible Fock space representation of the $\cW_{q,t}(\fg)$-algebra. Up to normalization, the coweights $\mu_0$ and $\mu_\infty$ coincide with the weights $\nu_0$ and $\nu_\infty$ in \eqref{electriccorrelatorintro}.\\

The quantum $q$-Langlands correspondence is the statement that there is an isomorphism between the above integral representations of the electric and magnetic blocks, with two important caveats:

First, on the $\cW_{q,t}(\fg)$-algebra side, the magnetic blocks \eqref{magneticcorrelatorintro} are scalar quantities determined by the choice of contour (equivalently, we can think of them as the components of a vector, where each component is in one-to-one correspondence with a possible contour). We would like to compare these scalars to the electric blocks \eqref{electriccorrelatorintro} on the  $U_{\hbar}(\widehat{^L\fg})$-algebra side, but we face the obstacle that these are instead naturally expressed as vectors, valued in \eqref{weightsubspaceintro}. Then, in order to actually match conformal blocks, each electric block needs to further be paired with a specific covector, or ``Whittaker functional," so that one can compare scalar quantities on both sides of the correspondence. Such covectors can be given a geometric and gauge theory interpretation, which we will come back to in the main text.

Second, the electric blocks are not technically $U_{\hbar}(\widehat{^L\fg})$ $q$-conformal blocks \emph{yet}. This is because an honest conformal block needs to be analytic in \emph{some} chamber of parameters $\{x_d\}$. Now, in writing the electric block \eqref{electriccorrelatorintro}, we have assumed an ordering of the operators $\Phi_{d}(x_d)$, and therefore implicitly specified a chamber, say
\beq\label{chamberintro}
\fC_H: \qquad \; |x_1|<|x_2|<\ldots<|x_L| \; .
\eeq
Unfortunately, the known integral form of electric blocks is not analytic in chamber $\fC_H$; instead, it is a convergent power series in a different  ``K\"{a}hler" parameter $z$ of the qKZ equations, valued in the Cartan subalgebra of ${^L\fg}$, in a chamber
\beq\label{Kahlerdefintro}
\fC_C: \qquad \; |z_a|<1 \; , \qquad \; \; a=1,\ldots,\r \; ,
\eeq
where $z_a = \hbar^{\langle \nu_0, ^L\alpha_a \rangle}$ encodes the Fock vacuum eigenvalues $\nu_0$, with $\alpha_a$ the $a$-th positive simple root of $^L\fg$. Crucially, it is not possible for a $q$-conformal block to be analytic \emph{simultaneously} in chambers $\fC_C$ and $\fC_H$. This is a specific feature of $q$-difference equations which is not shared by ordinary differential equations. Indeed, in the conformal limit, it is possible for the blocks to be analytic in both chambers $\fC_C$ and $\fC_H$ at the same time, following a general theorem of Deligne \cite{deligne1970equations}.
In order to obtain blocks that are analytic in chamber $\fC_H$ instead, we need to perform a linear change of basis, implemented using a finite size matrix of $q$-periodic pseudo-constants whose entries goes by the name of elliptic stable envelopes \cite{Aganagic:2016jmx,Aganagic:2017smx}. At the level of the block integral, this is achieved by modifying the integrand via insertions of products of odd genus 1 theta functions. This change of basis modifies the pole structure of the integrand, but because the modification is exclusively due to $q$-periodic pseudo-constants, the resulting integral is guaranteed to still solve the qKZ equations. 

The same considerations apply to the magnetic blocks, which are naturally analytic in the parameters $z_a=q^{\langle \mu_0, \alpha_a \rangle}$, but not in the parameters $x_d$. Using the same elliptic stable envelopes as for the electric blocks, one builds $\cW_{q,t}(\fg)$ $q$-conformal blocks analytic in chamber $\fC_H$ instead of $\fC_C$.

When they are analytic in a chamber $\fC_C$, we will refer to the electric blocks \eqref{electriccorrelatorintro} and magnetic blocks \eqref{magneticcorrelatorintro} as $z$-solutions. In contrast, when the blocks  are analytic in a chamber $\fC_H$, we will refer to them as $x$-solutions. When the context is unambiguous, we will abuse terminology and refer to both $z$-solutions and $x$-solutions simply as $q$-conformal blocks.\\

The insertion of the operator $\Phi_{d}(x_d)$ (on the $U_{\hbar}(\widehat{^L\fg})$ side) or of the operator $\overline{V}^\vee_{d}(x_d)$ (on the $\cW_{q,t}(\fg)$ side) inside the $q$-conformal blocks has an interpretation in 4d $\fg$-type Super Yang-Mills, after taking the conformal limit: it corresponds to placing respectively a 1/2-BPS Wilson  or 't Hooft line along $\{x_d\}\times M_2$, inside $M_4={\cal C}\times M_2$ \cite{Aganagic:2017smx}. A lot remains to be understood here, such as the precise mathematical description of the Hecke eigensheaves and coherent sheaves in the $q$-deformed context.\\

We will take a different route in this paper: our first goal  will be to further develop the dictionary of the correspondence, by giving a precise definition of what \emph{tame ramification} is.
We will be able to provide evidence for the following claims:\\

\begin{prop}\label{propr1}
\underline{Tamely Ramified quantum $q$-Langlands correspondence:}
There exists a two-parameter deformation of the tamely ramified geometric Langlands correspondence. It is an isomorphism between the space of $q$-conformal blocks of two theories on $\cC$; on one side, we find the blocks of a $U_\hbar(\widehat{^L\fg})$-algebra, with insertions of ``chiral" vertex operators labeling Verma modules. On the other side, we find $q$-conformal blocks of a $\cW_{q,t}(\fg)$-algebra, with insertions of (possibly degenerate) $q$-primary vertex operators, which we define explicitly.\\
\end{prop}

\noindent
Just as in the unramified case, the correspondence can be established by finding an equivalence between the integral form of $U_\hbar(\widehat{^L\fg})$ electric and $\cW_{q,t}(\fg)$ magnetic blocks. With ramification, the definition of the blocks needs to be modified. 

On the $U_\hbar(\widehat{^L\fg})$ side, we propose the following electric blocks:
\beq\label{electric2correlatorintro}
\left\langle v_{\nu_\infty},\,\prod_{d=1}^{L} \Phi_{\nu_d}(\tilde{x}_d) \; v_{\nu_0} \right\rangle  \; . 
\eeq
We claim that the operator $\Phi_{\nu_d}(\tilde{x}_d)$ are labeled by a Verma module ${V}_{\nu_d}$  of highest weight $\nu_d$ over $U_{\hbar}(^L\fg)$ (and the position $\tilde{x}_d$ on $\cC$). The vertex operator acts as an intertwiner of Verma module representations, and the position $\tilde{x}_d$ enters via the use of an evaluation representation in the definition of the intertwiner. The vectors $v_{\nu_0}$  and $v_{\nu_\infty}$ are as in the unramified case. The block is now valued in a space of weight
\beq\label{weightspaceram}
\nu_0-\nu_\infty = \sum_{d=1}^{L} \nu_d \, - \sum_{a=1}^{\L} N_a \, ^L\alpha_a \; .
\eeq
 
\vspace{2mm}

On the $\cW_{q,t}(\fg)$-algebra side, we propose the following magnetic blocks:
\beq\label{magnetic2correlatorintro}
\left\langle v_{\mu_\infty}\;, \, \prod_{a=1}^{\r} (Q_a^\vee)^{N_a} \;\prod_{d=1}^{L} \cV_{\{\lambda\}_d}(\tilde{x}_d) \; v_{\mu_0} \right\rangle  \, .
\eeq
The novelty here is a vertex operator $\cV_{\{\lambda\}_d}(\tilde{x}_d)$, which needs a precise mathematical definition. We call it a deformed primary or $q$-primary, where the terminology comes from the conformal limit: when $t=q^{\beta}$, $q\rightarrow 1$, the Alday-Gaiotto-Tachikawa correspondence predicts the insertion of $\cW_{\beta}(\fg)$-algebra primary vertex operators (or various degenerations thereof) in ramified magnetic-type correlators \cite{Alday:2009aq,Kanno:2009ga}; these are operators of the form
\beq
\label{primaries}
e^{\langle \widetilde{\sigma} , \phi(\tilde{x}_d)\rangle} \; ,
\eeq
with $\phi(\tilde{x}_d)$ a free boson on $\mathbb{C}$ and $\widetilde{\sigma}$ a $^L\fg$-weight encoding the momentum of the corresponding primary state.
The operator $\cV_{\{\lambda\}_d}$ should qualitatively be thought of as a quantum analog of such a primary operator. In particular, a deformed primary should depend on $q$ and $t$ in such a way that it reduces to the ordinary ${\cW}_{\beta}({\fg})$ primary \eqref{primaries} in the conformal limit. Moreover, since the momentum $\widetilde{\sigma}$ is valued in the weight lattice of $^L\fg$, it would be natural to conjecture the ``momentum" of a deformed primary to be  defined as a $q$-exponent, via the representation theory of $U_q(^L\fg)$. However, this candid guess turns out to be too naive: instead, we will argue that a correct definition of a $q$-primary should be given in terms of a set of weights $\{{\lambda}\}_d$ in the quantum affine algebra $U_\hbar(\widehat{^L\fg})$ (or more precisely the quantum loop algebra $U_{\hbar}(L(^L\fg))$, which is the quantum affine algebra without central extensions), with $\hbar = q^{\fn_\fg}/t$.
In hindsight, the fact that the \emph{electric} algebra $U_\hbar(\widehat{^L\fg})$ makes a remarkable appearance in the definition of the \emph{magnetic} blocks should be understood as yet another avatar of the quantum $q$-Langlands correspondence. 
In their original paper, Frenkel and Reshetikhin had already pointed out a deep relation between ${\cW}_{q,t}({\fg})$-algebras and quantum affine algebras \cite{Frenkel:1998,Bouwknegt:1998da}: it was shown there that the $\r$ generators of a ${\cW}_{q,t}({\fg})$-algebra could be reinterpreted as ``characters" of the $\r$ fundamental representations of a quantum affine algebra. Our construction implies a similar relation beyond the generators of ${\cW}_{q,t}({\fg})$.\\

In order for an isomorphism between electric and magnetic block integrals to hold, we find that the form of the $q$-primary  $\cV_{\{\lambda\}_d}(\tilde{x}_d)$ is highly constrained: it is written as a product of more elementary operators $\prod_s\Lambda^{\pm 1}_a({x}_{d,s})$, at  different positions ${x}_{d,s}$ on $\cC$. The details of the product are encoded in the set of quantum affine weights $\{{\lambda}\}_d$.
Based on a fundamental theorem of Chari and Pressley \cite{cmp/1104248585} (see also Drinfeld for the Yangian version \cite{Drinfeld:1987sy}), Frenkel and Reshetikhin \cite{Frenkel:qch} showed that each such weight is in bijection with two sets of $\L$  polynomials valued in $\mathbb{C}[x]$, known as Drinfeld polynomials. 

Our main claim is that the product $\prod_s\Lambda^{\pm 1}_a({x}_{d,s})$ is fixed by the degree of these polynomials, and the argument ${x}_{d,s}$ of the operators is fixed by the roots. The parameter $\tilde{x}_d$ in the actual $q$-primary $\cV_{\{\lambda\}_d}(\tilde{x}_d)$ is a ``center of mass" position with respect to these roots. This gives a definition of the $\cW_{q,t}(\fg)$-algebra primary vertex operator. After specifying the screening charge contours of integration, we therefore obtain a definition of the ramified magnetic block \eqref{magnetic2correlatorintro}. The Langlands isomorphism is established by showing that this magnetic block can be obtained by pairing the electric block with a specific covector, and identifying the (Cartan components of the) Verma module highest weights on both sides:
\beq
q^{\langle\widetilde{\sigma}_d,^L\alpha_a\rangle} = \hbar^{\langle\nu_d,^L\alpha_a\rangle} \; ,\qquad a=1,\ldots,\L \; ,
\eeq
with $^L\alpha_a$ the $a$-th simple root of $^L\fg$.

\vspace{10mm}

\subsection{3d $\cN=2$ Quiver Gauge Theories for Quantum Affine Algebras}

\noindent
Let $\fg_\text{o}$ be a simply-laced Lie algebra. 

\begin{prop}
\label{prop2}
 The data specifying a tamely ramified $q$-conformal block is in one-to-one correspondence with the choice of a 3d $\cN=2$ quiver gauge theory of type $\fg_\text{o}$, on a 3-manifold with distinguished 1/2-BPS boundary conditions; we call such a gauge theory a \emph{Drinfeld quiver}. The block itself is a 3d (half-)index counting the local operators on the boundary. The choice of 3d manifold dictates whether the block is electric or magnetic, while the choice of boundary conditions dictates the analycity, either as a $z$-solution or a $x$-solution.\\
 \end{prop}

In a series of works initiated by the author and Aganagic-Koz\c{c}az-Shakirov \cite{Aganagic:2013tta,Aganagic:2014oia,Aganagic:2015cta}, it was shown that certain counts of BPS states for quiver gauge theories supported on $S^1_{\cC'}\times \mathbb{C}$ coincide with the magnetic $q$-conformal blocks of the $\cW_{q,t}(\fg_\text{o})$-algebra. In particular, the non-compactness of $\mathbb{C}$ must somehow be regularized to obtain a sensible answer. One way to achieve this is to introduce a certain massive deformation of the background, and the associated count of BPS states goes by the name of holomorphic block \cite{Beem:2012mb}. This important result has two major shortcomings.\\

First, the holomorphic block is a quantity only defined in the IR: the 3d theory has massive (isolated) vacua specified ``at infinity" on $\mathbb{C}$, which correspond to a choice of thimble boundary condition\footnote{To our knowledge, the study of such Picard-Lefschetz thimbles in supersymmetric Physics dates back to \cite{Cecotti:1991me,Cecotti:1992rm,Hori:2000ck} in the context of 2d $\cN=(2,2)$ massive sigma and Landau-Ginzburg models, see also more recently \cite{Gaiotto:2015zna,Gaiotto:2015aoa}.}. 

Here, we will reinterpret these boundary conditions from a gauge theory perspective in the UV: we replace $\mathbb{C}$ with a finite disk/hemisphere $D^2$, and trade the choice of a 3d vacuum at infinity for a specific set of 1/2-BPS boundary conditions at finite distance on $T^2 = S^1_{\cC'}\times S^1_{D^2}$, which will flow to a superconformal point in the IR. A priori, either a 2-dimensional $\cN=(1,1)$ or a $\cN=(0,2)$ subalgebra are acceptable, but we will only focus on the latter for geometric Langlands applications. We require the $\cN=(0,2)$ boundary theory to have an unbroken $U(1)_R$ R-symmetry, so that an index can be defined in the first place. Often, this $U(1)_R$ R-symmetry is identified as the 3d bulk one, but not always: we will encounter certain boundary conditions for which $U(1)_R$ is a linear combination of the bulk R-symmetry and other $U(1)$'s in the maximal torus of the UV flavor symmetry group.

Second, no rigorous definition of $q$-primaries was available until now, and in particular previous works have mischaracterized these operators as being labeled by weights of a \emph{classical} Lie algebra, instead of a correct weights \emph{quantum} algebra. As a result, the punctures on $\cC$ were never properly characterized, and the relevance of the analysis to the geometric Langlands program was left obscure. By construction, our definition of $q$-primary operators will remedy these issues.\\

A similar relationship between $q$-conformal blocks and 3d gauge theory indices has been shown to hold in the unramified Langlands correspondence \cite{Aganagic:2017smx}, with the supersymmetry there enhanced to $\cN=4$. Many of the key features of the analysis are common to both the ramified and unramified cases:\\


-- The half-index is defined in the cohomology of the same supercharge, as a trace over the Hilbert space of the theory quantized on $D^2$, where the various fields have twisted boundary conditions along $S^1_{\cC'}$. Such a definition is amenable to a localization computation \cite{Yoshida:2014ssa}. Alternatively, up to boundary mixed 't Hooft anomaly contributions, the BPS count can be done from first principles on a 3-dimensional Minkowski spacetime $\mathbb{R}^{1,1}\times\mathbb{R}_{-}$, and computing a character over the vector space of local operators at the $\cN=(0,2)$ boundary on $\mathbb{R}^{1,1}\times\{0\}$ \cite{Dimofte:2017tpi}.\\


-- The magnetic block of the $\cW_{q,t}(\fg_\text{o})$-algebra is the half-index of a 3d gauge theory supported on the manifold $M_3 = S^1_{\cC'}\times D^2$. For the electric block of the $U_\hbar(\widehat{^L\fg_\text{o}})$-algebra\footnote{Simply-laced Lie algebras are Langlands self-dual, $^L\fg_\text{o} = \fg_\text{o}$, so the notation $^L\fg_\text{o}$ is superfluous here; the distinction will become essential when we discuss $\fg$ non simply-laced.}, the support manifold is $M^\times_3 = S^1_{\cC'}\times (D^2)^{\times}$ instead, with the origin of the disk removed, and 1/2-BPS loop operator insertions at that locus.\\

-- The vector and bifundamental matter multiplet content is the same regardless of ramification, with $\cN=4$ supersymmetry. In particular, the quivers all have $\r$ unitary groups, and therefore $\r$ Fayet-Iliopoulos (F.I.) parameters. These parameters are identified with the K\"{a}hler parameters $z_a$, which are Cartan components of the Fock vacuum $\mu_0$ in the $q$-conformal block. A choice of chamber $\fC_C$ such as \eqref{Kahlerdefintro} therefore translates to a choice of signs for the F.I. parameters, and a $z$-solution is a 3d half-index analytic in such a chamber. In practice, $z$-solutions are constructed in a Higgs vacuum by imposing 1/2-BPS conditions known as exceptional Dirichlet at the $T^2$ boundary.\\ 

\begin{figure}[h!]
	\emph{}
	\centering
	\includegraphics[trim={0 0 0 0cm},clip,width=0.9\textwidth]{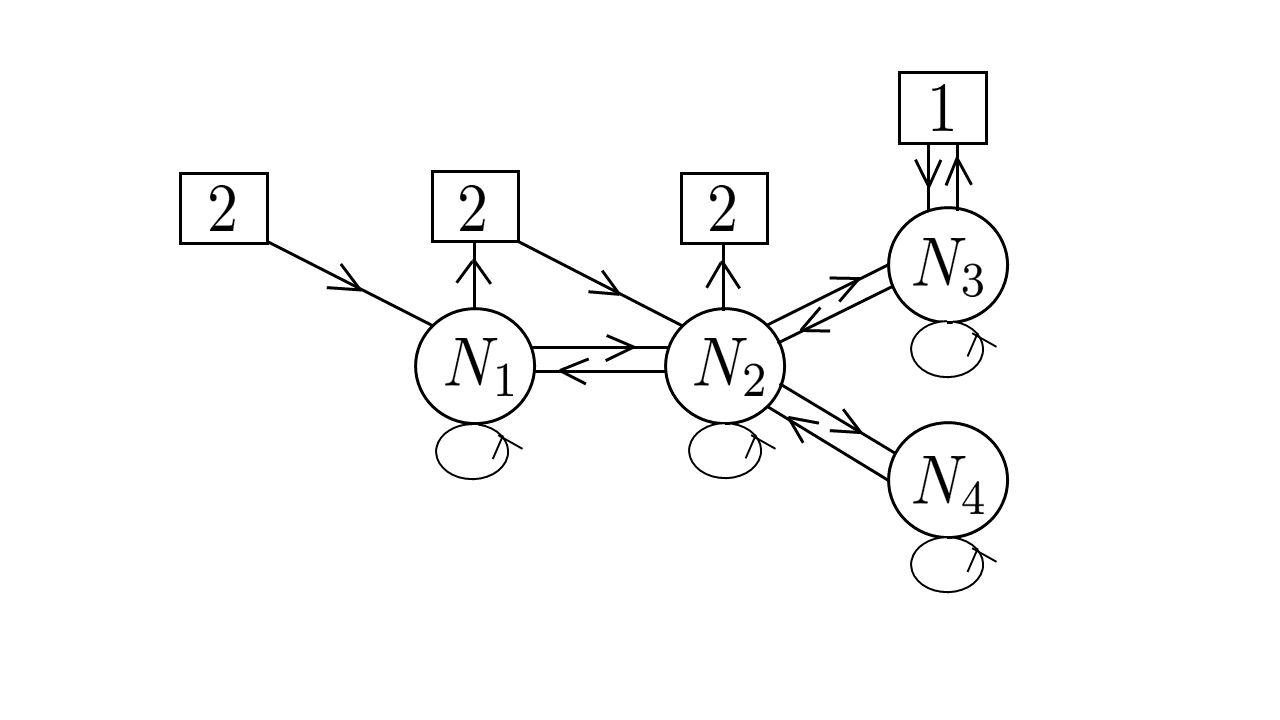}
	\vspace{-1pt}
	\caption{An example of 3d $\cN=2$ Drinfeld quiver for $\fg=D_4$. It is defined via certain weights of a \emph{quantum affine} algebra $U_{\hbar}(\widehat{D_4})$.} 
	\label{fig:drinfeldquiver}
\end{figure}

-- The quivers have nontrivial unitary flavor symmetry groups (framing), for which we turn on generic real masses, complexified by the holonomy around $S^1_{\cC'}$. These masses are identified with the positions of the vertex operators on the Riemann surface.  A choice of chamber $\fC_H$ such as \eqref{chamberintro} is an ordering of the masses, and a $x$-solution is a 3d half-index analytic in such a chamber. In practice, $x$-solutions are constructed in a Higgs vacuum by imposing  1/2-BPS conditions known as enriched Neumann at the $T^2$ boundary.\\

The distinctive features of ramification are encoded in the details of the flavor multiplets: 
In the unramified setup, the contribution of the Frenkel-Reshetikhin fundamental vertex operators ${\overline V}^\vee_{d}$ to the block \eqref{magneticcorrelatorintro} is reproduced in gauge theory by counting BPS states in the presence of 3d $\cN=4$ fundamental hypermultiplets. In the tamely ramified setup,  the contribution of $q$-primary vertex operators $\cV_{\{\lambda\}_d}$ to the block \eqref{magnetic2correlatorintro} is realized instead by explicitly breaking supersymmetry to 3d $\cN=2$, and introducing chiral multiplets.

We call this class of 3d $\cN=2$ gauge theories Drinfeld quivers. The name is chosen to reflect that the flavor content is determined by Drinfeld polynomials in bijection with the quantum affine weights $\{\lambda\}_d$. We propose that the polynomial degrees label the number of fundamental and anti-fundamental chiral multiplets of the theory, while the roots encode the $U(1)_R$ R-charge of the multiplets.\\

In order to engineer $z$-solutions, analytic in chamber $\fC_C$, the boundary conditions we need are exceptional Dirichlet. Their role as $\cN=(2,2)$ boundary conditions in the unramified context of 3d $\cN=4$ gauge theories has recently been an active topic of research in mathematical physics \cite{Bullimore:2016nji,Bullimore:2020jdq,Okazaki:2020lfy,Crew:2020psc,Dedushenko:2021mds}. Such boundary conditions are used in the UV to mimic the presence of an isolated massive vacuum on the Higgs branch in the IR. They depend on a choice of vacuum, and a polarization, meaning a holomorphic Lagrangian splitting of the linear representation $T^*R$ of the gauge group provided by the hypermultiplets. The $\cN=(2,2)$ boundary conditions in question are Dirichlet on vector multiplets, and a mix of Neumann-Dirichlet on the hypermultiplets, depending on the specific vacuum and polarization we choose. In the tamely ramified setup, the 3d $\cN=2$ Higgs branch is only K\"{a}hler, so such a definition of polarization cannot exist. Nevertheless, we will show that the chiral multiplet content of Drinfeld quivers allows for a slightly \emph{weaker} notion of polarization to be well-defined, and with it a certain $\cN=(0,2)$ analog of exceptional boundary conditions.

In order to engineer $x$-solutions, analytic in chamber $\fC_H$, the boundary conditions we need are Neumann, enriched by boundary chiral multiplets. In the unramified setup, these $\cN=(2,2)$ boundary conditions depend on a choice of vacuum and chamber $\fC_H$; they are supported on so-called elliptic stable envelopes, which in gauge theory is a certain canonical basis for the elliptic equivariant cohomology of the Higgs branch  \cite{Aganagic:2016jmx,Aganagic:2017smx}. With tame ramification, we will argue that elliptic stable envelopes are likewise well-defined for the class of Drinfeld quivers.\\

\vspace{10mm}

\subsection{A higher-dimensional Gauge Theory origin}

The relevance of quantum algebras in the study of $D=$ 4, 5, and 6 dimensional supersymmetric gauge theory is well-established by now. The most famous example is the Yangian in $D=$ 4 dimensions: at large $n$, the Yangian of $psu(2,2|4)$ is a symmetry of 4d $\cN=4$ $SU(n)$ Super Yang-Mills \cite{Drummond:2009fd}. There, the quantum algebra is realized as a symmetry of the gauge group. It can also be realized as a  symmetry of the algebra of operators, such as in the holomorphic twist of 4d $\cN=1$ $SU(n)$ Super Yang-Mills \cite{Costello:2013zra,Costello:2017dso}. For us, the Yangian has yet another meaning, in the sense of Nekrasov-Shatashvili \cite{Nekrasov:2009rc}, as a non-perturbative ``symmetry" of a 4d $\cN=2$ \emph{quiver} gauge theory in an $\Omega$-background\footnote{From this quiver point of view, a single $SU(n)$ gauge group would therefore have the Yangian of $A_1 = \text{su}(2)$ as a symmetry.} \cite{2012arXiv1211.1287M,Nekrasov:2012xe,Nekrasov:2013xda}.\\

In this paper, we uncover another avatar of this quantum algebra symmetry for $D$-dimensional theories with 8 supercharges, in their Higgs phase. This is the part of the moduli space of vacua parameterized by hypermultiplet scalars\footnote{The term ``supersymmetric Higgs mechanism"  used in the abstract refers to the breaking of the gauge group by vacuum expectation values of these hypermultiplet scalars.}.
Under some mild conditions\footnote{The conditions are that the gauge theories should have enough matter hypermultiplets to be totally Higgsable, and furthermore be ``asymptotically conformal," in the sense of \cite{Nekrasov:2012xe,Nekrasov:2013xda}. When $D=$ 4, the latter condition simply means that we will restrict ourselves to gauge theories which flow to a superconformal point as the hypermultiplets become massless. When $D=$ 6, it is a condition on the theory being anomaly-free. To our knowledge, it has no direct meaning when $D=$ 5.}, we point out that the Higgs vacua of a $D$-dimensional gauge theory can be labeled by weights in fundamental representations of a Lie algebra $\fg_\text{o}$, where $\fg_\text{o}$ labels the quiver. After turning on an $\Omega$-background, we argue that the Higgs vacua should instead be labeled by weights in fundamental representations of a \emph{quantum} algebra associated to $\fg_\text{o}$. One surprise here is that the quantization parameter of this algebra is \emph{not} the usual one appearing in the Nekrasov-Shatashvili correspondence.

In order to properly identify the quantum algebra, we resort to a dual description of the Higgs branch, in $(D-2)$ dimensions, known as gauge-vortex duality. An early formulation of this duality dates back to the late 90's and is due to Dorey \cite{Dorey:1998yh,Dorey:1999zk}, who noted that in $D=$ 4, the BPS spectra of the following two theories agree: on the one hand, 4-dimensional $\cN=2$ $SU(n)$ Super Yang-Mills with massive hypermultiplets (SQCD) on $\mathbb{C}_q\times\mathbb{C}_t$, and on the other hand, a 2-dimensional  ${\cN}=(2,2)$ $\mathbb{CP}^{n-1}$ gauged sigma-model with chiral multiplets and nonzero twisted masses on $\mathbb{C}_q$ (at the moment, $q$ and $t$ are just labels). The agreement of BPS spectra can be justified by viewing the 2d theory as the worldvolume theory of a codimension-2 Abrikosov-Nielsen–Olesen vortex string solution \cite{Abrikosov:1956sx,Nielsen:1973cs} at the root of the Higgs branch of the 4d theory, the point in moduli space where Coulomb and Higgs branches meet \cite{Shifman:2004dr,Hanany:2004ea}\footnote{On general ground, the duality only holds for F-terms: the vortices arise after F.I. terms are turned on in 4d, which modifies the value of the bare gauge couplings in 2d, but these are D-terms, and in particular do not affect the F-terms.}. The Higgs branch of the 4d theory is probed by turning on vortex flux along $\mathbb{C}_t$, while the dual 2d theory is the theory on the vortices themselves, with worldvolume $\mathbb{C}_q$.

The Bethe-Gauge correspondence of Nekrasov-Shatashvili \cite{Nekrasov:2009ui,Nekrasov:2009rc} suggests a natural extension of the duality, by introducing a nontrivial 2d $\Omega$-background along $\mathbb{C}_t$ with $U(1)$ equivariant parameter $\epsilon_t$ in the 4d theory \cite{Nekrasov:2002qd,Losev:2003py,Nekrasov:2010ka}\footnote{We use the principal branch of the logarithm throughout.}. From the point of view of the 2d vortex theory, the only effect of this modification is to give a twisted mass $\epsilon_t$ to an adjoint chiral multiplet. The duality dictionary becomes more extensive in this context: the chiral rings of the two theories are isomorphic, with matching supersymmetric vacua and corresponding (twisted) superpotentials\footnote{On the 4d side, turning on the $\Omega$-background on $\mathbb{C}_t$ preserves a codimension-2 ${\cN}=(2,2)$ supersymmetry along $\mathbb{C}_q$, and what we mean by twisted superpotential is the superpotential of this transverse 2d theory. The matching of the superpotentials for the two ${\cN}=(2,2)$ theories is a priori highly nontrivial, since the superpotential originating from 4d theory receives an infinite number of non-perturbative corrections due to instantons, while the 2d vortex theory has a 1-loop exact superpotential \cite{Witten:1993yc}.} \cite{Dorey:2011pa,Chen:2011sj}. Below the scale set by the gauge couplings, and the mass scale $\epsilon_t$, the two theories are expected to flow to one and the same theory in the IR, and this is what is meant by gauge-vortex duality \cite{Aganagic:2014kja}.\\

The duality admits a straightforward generalization to quiver gauge theories, and in higher dimensions: a $D$-dimensional theory with 8 supercharges supported on  ${\bf{M}}\times \mathbb{C}_q\times \mathbb{C}_t$ is gauge-vortex dual to a $(D-2)$-dimensional theory with 4 supercharges supported on ${\bf{M}}\times \mathbb{C}_q$, with ${\bf{M}}=\varnothing,\,S^1_{\cC'},\,T^2$ whenever $D=4,5,6$, respectively. From now on, we specialize to $D=5$, since this is the dimension directly relevant to the quantum $q$-Langlands program. Then, for us gauge-vortex duality is a 5d/3d duality of quiver gauge theories compactified on a common circle $S^1_{\cC'}$. 

Concretely, we show that every 3d $\cN=2$ Drinfeld quiver as defined in Proposition 2 is gauge-vortex dual to a 5d $\cN=1$ quiver theory. In particular:\\

\begin{prop}\label{prop3}
The Higgs branch of a 3d $\cN=2$ Drinfeld quiver is a moduli space of 1/2-BPS vortices in a 5d $\cN=1$ quiver gauge theory.
\end{prop} 

Because every weight inside a quantum affine fundamental representation is in one-to-one correspondence with the chiral multiplet content of a 3d Drinfeld quiver, the Proposition implies that each such weight is also realized in the Higgs phase of some 5d gauge theory\footnote{We will not attempt to prove the converse of Proposition 1.3, namely that \emph{every} 1/2-BPS vortex solution in 5 dimensions has a dual description as a Drinfeld quiver. We leave this important question to future work.}.\\ 

\begin{figure}[h!]
	\emph{}
	\centering
	\includegraphics[trim={0 0 0 0cm},clip,width=0.9\textwidth]{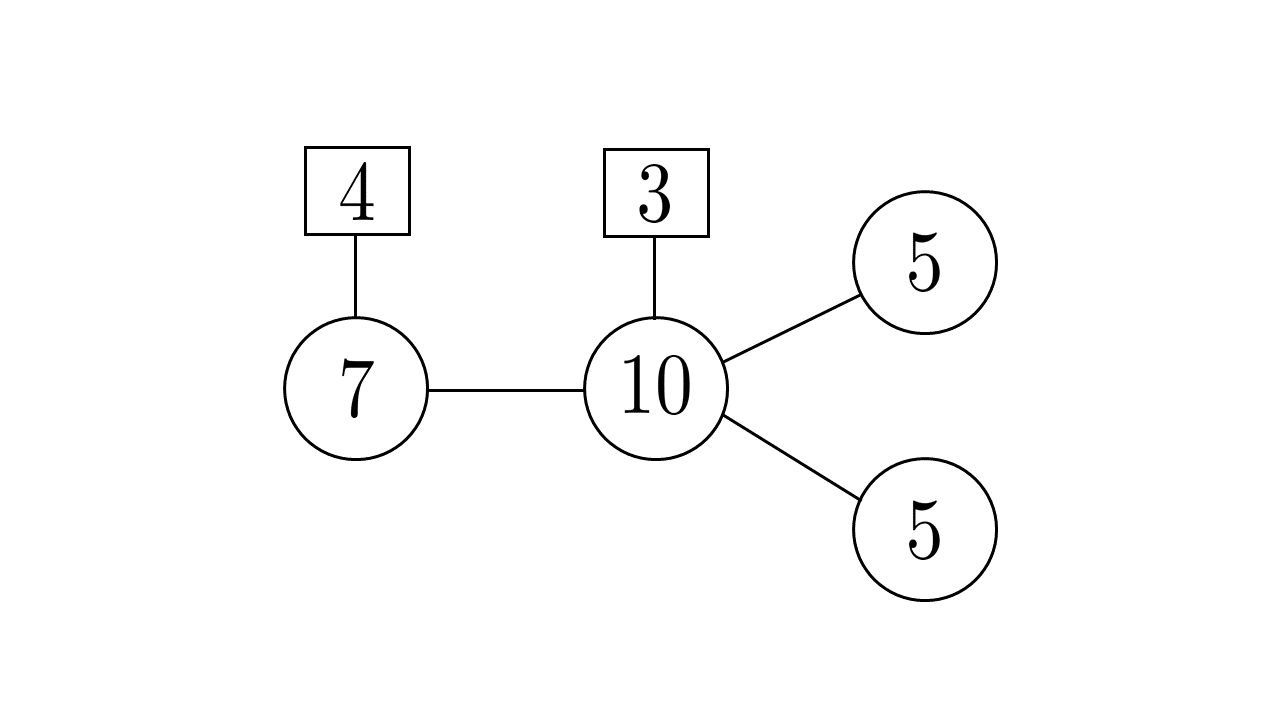}
	\vspace{-1pt}
	\caption{A 5d $\cN=1$ quiver gauge theory. It is defined via certain weights of the \emph{classical} algebra $\fg=D_4$. The 3d Drinfeld quiver from Figure \ref{fig:drinfeldquiver} describes the worldvolume theory of 1/2-BPS vortices on the Higgs branch of this 5d theory, in an $\Omega$-background.} 
	\label{fig:5dquiver}
\end{figure}

Crucially, both the $D$ and $(D-2)$-dimensional theories admit a further $\Omega$-deformation along $\mathbb{C}_q$, with $U(1)$ equivariant parameter $\epsilon_q$. In 5d, this results in a full $\Omega$-background with two parameters on an equal footing, $q=e^{R_{\cC'}\, \epsilon_q}$ and $t=e^{-R_{\cC'}\, \epsilon_t}$. In 3d, $q$ is the only $\Omega$-background parameter, while $t$ remains a mass as before \cite{Aganagic:2013tta,Aganagic:2014oia,Aganagic:2014kja}.

This refinement to $q\neq 1$ is important here for two reasons: 
First, we will identify the quantum algebra on the Higgs branch as $U_{\hbar}(\widehat{^L\fg_\text{o}})$, with $\hbar = q/t$\footnote{In spacetime dimension $D=4$, where gauge-vortex duality is a 4d/2d duality, the relevant quantum algebra will be the Yangian of $^L\fg_\text{o}$, with quantization parameter $\log(q/t)/R_{\cC'}$. For $D=6$, where gauge vortex duality is a 6d/4d duality, the relevant quantum algebra is elliptic.}. This is precisely the algebra that appears in the definition of the dual 3d Drinfeld quiver.
In particular, the quantization parameter is not simply $t$ as in the statement of the Nekrasov-Shatashvili correspondence. 

Second, a background with  $q\neq 1$ is amenable to exact computations, and remarkably, we will show that the quantum algebra symmetry is explicitly manifest at the level of supersymmetric indices. In 3d, this is a holomorphic block on $S^1_{\cC'}\times \mathbb{C}_q$ in the IR, or of a half-index on $S^1_{\cC'}\times D^2$ with distinguished boundary conditions in the UV, counting 1/2-BPS point-like vortex particles on $S^1_{\cC'}\times\{0\}$. In 5d, it is the Nekrasov instanton parition function on $S^1_{\cC'}\times \mathbb{C}_q\times \mathbb{C}_t$, counting 1/2-BPS point-like instanton particles on $S^1_{\cC'}\times\{0\}\times\{0\}$ \cite{Nekrasov:2002qd,Losev:2003py}. It is not clear a priori why one would want to count instantons, which are non-perturbative corrections on the Coulomb branch, when the 3d vortices under study are Higgs branch solutions of the 5d theory;  the justification comes from the work of Nekrasov and Witten, who showed that in the presence of an $\Omega$-background along $\mathbb{C}_t$, turning on $N$ units of $U(1)$ vortex charge in 5d is indistinguishable from having no vortices at all, and instead shifting a corresponding Coulomb modulus by an amount $t^N$ \cite{Nekrasov:2010ka}; that is, in an $\Omega$-background, if we start at the root of the Higgs branch, we can probe the vortices inside the Higgs branch by equivalently probing the Coulomb branch instead, on an integer lattice centered at the hypermultiplet masses, with spacing the positive powers of $t$.

We implement this idea and show that the half-index of a Drinfeld quiver on $S^1_{\cC'}\times D^2$ with exceptional Dirichlet boundary coincides with the Nekrasov instanton partition function of a 5d theory on this discretized Coulomb branch. The 5d instanton partition function is presented as an infinite series expansion in the 5d gauge couplings, which are identified as the (inverse) F.I. couplings in 3d. By construction, this series is analytic in some chamber $\fC_C$ of the 5d gauge/3d F.I. couplings, so it will coincide with a $z$-solution $q$-conformal block in that chamber.

\begin{prop}\label{prop3plus}
For fundamental representations of the quantum affine algebra $U_\hbar(\widehat{^L\fg_{\text{o}}})$, all weight spaces are realized on the Higgs branch of a 5d $\cN=1$ quiver gauge theory of type $\fg_{\text{o}}$.
\end{prop} 
The proposition can be extended to $\fg$ non-simply laced  via a folding operation, by restricting to invariants under outer automorphism group action of a simply-laced algebra.  Our construction of quantum affine weights from 5 dimensions relies on an algorithm which is very similar to the Frenkel-Mukhin algorithm found in the $q$-character literature \cite{Frenkel:1999ky}, in the sense that its input is a dominant weight (in fact, a fundamental weight), along with a prescription to lower the weight\footnote{The lowering algorithm is sometimes called an ``i-Weyl" reflection in the gauge theory literature \cite{Nekrasov:2012xe,Nekrasov:2013xda,Nekrasov:2015wsu}.}. However, the lowering prescription which arises via Higgsing in 5 dimensions \emph{differs} from the Frenkel-Mukhin algorithm in subtle ways, so our algorithm appears to be new.


\begin{remark}
Since the work of Nekrasov and Shatashvili, the quantum symmetry of $D$-dimensional quiver gauge theories with 8 supercharges in an $\Omega$-background has been thoroughly investigated, at a generic point on the Coulomb branch: for instance, in $D=$5, the (quantized) Seiberg-Witten geometry of the theories is captured by characters for the fundamental representations of the quantum affine algebra $U_t(\widehat{\fg_\text{o}})$ \cite{Nekrasov:2012xe,Nekrasov:2013xda}. More generally, this is a prediction of the BPS/CFT correspondence \cite{Nek2008}.
In this paper, we find evidence for another symmetry algebra $U_{\hbar}(\widehat{^L\fg_\text{o}})$, manifest on the Higgs branch and at discrete points on the Coulomb branch.\\
\end{remark}

\begin{remark}
In a setting with 8 supercharges, Nakajima showed that the (classical) equivariant K-theory of the Higgs branch can be described as a certain weight subspace inside the quantum affine algebra $U_\hbar(\widehat{^L\fg})$ (in fact, the quantum loop algebra $U_\hbar(L(^L\fg))$)  \cite{1999math.....12158N}. Here, we find that weight subspaces of the form  \eqref{weightspaceram} can be characterized via the \emph{representation} of a Drinfeld quiver, with 4 supercharges.
\end{remark}

\vspace{10mm}

\subsection{Enumerative Geometry}

The 3d half-index is an example of a vertex function, introduced by Okounkov in the context of K-theoretic counts of quasimaps $\mathbb{P}^1 \rightarrow X$ \cite{Okounkov:2015spn,2011arXiv1106.3724C}, where $X$ is the Higgs branch of the quiver gauge theory. Quasimaps arise here because they are solutions to the vortex equations subjected to certain regularity conditions on the gauge and matter fields, and K-theory enters because we are studying 3d theories compactified on $S^1_{\cC'}$ instead of purely 2d ones; for recent physical applications, see \cite{Bullimore:2018jlp,Dedushenko:2021mds,Dedushenko:2023qjq,Crew:2023tky,Ishtiaque:2023acr}.

The vertex function is a generating function of quasimaps of all degrees to $X$, where the K\"{a}hler/F.I. parameter $z$ keeps track of the map degree. This formalism has been instrumental in establishing the quantum $q$-Langlands program on firm mathematical ground: in the unramified case, Aganagic-Frenkel-Okounkov proved that such vertex functions coincide with $z$-solution $q$-conformal blocks, analytic in a chamber $\fC_C$ \cite{Aganagic:2017smx}. In the quasimap terminology, the magnetic $\cW_{q,t}(\fg)$ blocks are sometimes called scalar vertex functions, while the electric $U_{\hbar}(\widehat{^L \fg})$ blocks are better thought of as vector vertex functions, with descendant insertions at $\{0\}\in\mathbb{P}^1$. These insertions are certain classes in equivariant K-theory which form a stable basis of $X$, the (K-theoretic) stable envelopes.
In this unramified setup, the 3d gauge theories have $\cN=4$ supersymmetry, and their Higgs branch is a Nakajima quiver variety, meaning a holomorphic symplectic quotient $X = T^*\text{Rep}({\cal Q})/\!\!/\!\!/\!\!/G_{V}$.\\  

A scalar vertex function will depend on a choice of vacuum and a polarization. In geometry, a vacuum is a fixed point locus on $X$ under the equivariant tori of the flavor symmetry. Those include the tori of the flavor group, whose equivariant parameters are the masses, and the $\mathbb{C}^\times_{\hbar}$ action rescaling the cotangent direction of $X$, whose equivariant parameter is the quantum affine parameter $\hbar$. A choice of polarization is a choice of holomorphic Lagrangian splitting of the tangent bundle of $X$; it is well-known that such a splitting exists for Nakajima quiver varieties.\\

In fact, the vertex functions solve two sets of $q$-difference equations: the  $z$-solutions we just described are analytic in a K\"{a}hler parameter chamber $\fC_C$, but there are also $x$-solutions, analytic in an equivariant parameter chamber $\fC_H$. In enumerative geometry, these are obtained via insertion of elliptic stable envelopes at $\{\infty\}\in \mathbb{P}^1$, which implements a linear change of basis between $z$- and $x$- solutions\footnote{This change of basis is also a realization of 3d mirror symmetry in enumerative geometry, which has attracted a lot of attention in the Mathematics community in recent years. Computations of vertex functions and explicit tests of mirror symmetry in the $\cN=4$ context have been performed for $X$ of various types: examples include hypertoric varieties, cotangent bundle of the Grassmannian $T^* \text{Gr}(k,N)$, cotangent bundle of flag varieties, bow varieties \cite{2020arXiv200805516D,2020arXiv201108603D,Rimanyi:2019zyi,2023arXiv230807300B,Pushkar:2016qvw,Koroteev:2017nab}.}.\\

In this paper, we extend the quasimap analysis to include tame ramification. The immediate challenge is that Drinfeld quivers only have 3d $\cN=2$ supersymmetry, so in particular their Higgs branch $X$ is only a K\"{a}hler GIT quotient $X = \text{Rep}({\cal Q})/\!\!/G_{V}$, and no longer a holomorphic symplectic quotient. The vacua are still labeled by equivariant masses, but a polarization of the tangent bundle is no longer defined as above, so the meaning of ``vertex function" is no longer clear a priori. Moreover, the geometric interpretation of $\mathbb{C}^\times_{\hbar}$ as an action rescaling the cotangent direction of a bundle on $X$ is lost from the onset. 

But $X$ is not just any K\"{a}hler quotient: it happens to have another interpretation as a moduli space of quasimaps into \emph{another} variety $X'$, where $X'$ itself \emph{is} a Nakajima quiver variety. This is the mathematical formulation of gauge-vortex duality, where $X'$ should be interpreted as the Higgs branch of a 5d $\cN=1$ quiver gauge theory. In particular, we recover the geometric meaning of the parameter $\mathbb{C}^\times_{\hbar}$ as scaling the cotangent direction of $X'$. For the varieties $X$, we show that there exists a canonical notion of polarization in the ramified case, based on the Drinfeld polynomial data, which boils down to the existence of a certain attractive line bundle for $X$ \cite{okounkov2021inductive}. This is the same data that enters the representation of the 3d quiver gauge theory.\\

\begin{prop}
The $q$-conformal block of a $\cW_{q,t}(\fg_o)$-algebra with $q$-primary vertex operator insertions is a generating function of quasimaps $\mathbb{CP}^1\rightarrow X$ of all degrees in equivariant quantum K-theory, with $X$ the Higgs branch of a 3d $\cN=2$ Drinfeld quiver. These are $z$-solutions, analytic in the K\"{a}hler parameter $z$, are vertex functions.  The representation of the quiver is specified by certain Drinfeld polynomials, for which we define an analog of polarization on the tangent bundle of $X$. Elliptic stable envelopes are well defined for $X$, and therefore the existence of $x$-solutions, analytic in the equivariant mass parameters. The $q$-conformal block of a $U_{\hbar}(\widehat{^L\fg_\text{o}})$-algebra is likewise a vector vertex function, with descendant insertions at $\{0\}\in\mathbb{P}^1$.\\ 
\end{prop}

\vspace{10mm}

\subsection{Tame Ramification in Little String Theory}

The string theory engineering of Gukov-Witten surface defects \cite{Gukov:2006jk} in 4d $\cN=4$ Super Yang-Mills is well-known. Perhaps the most straightforward approach is to realize 4d Yang-Mills as the worldvolume theory of D3 branes, and the 1/2-BPS surface defects as another set of partially transverse D3 branes \cite{Constable:2002xt}. For related holographic dual approaches, see \cite{Lin:2004nb,Lin:2005nh,Gomis:2007fi,Chen:2023lzq}. Another realization, when $\fg=A_r$, and close in spirit to the original Gukov-Witten construction, is to consider only one set of D3 branes (still supporting the 4d Yang-Mills theory), and realize the monodromy defects via an orbifold geometry \cite{Alday:2010vg,Kanno:2011fw,Nekrasov:2017rqy,Ashok:2020ekv,Ashok:2020jgb}.

\begin{figure}[h!]
	\emph{}
	\centering
	\includegraphics[trim={0 0 0 0cm},clip,width=1.0\textwidth]{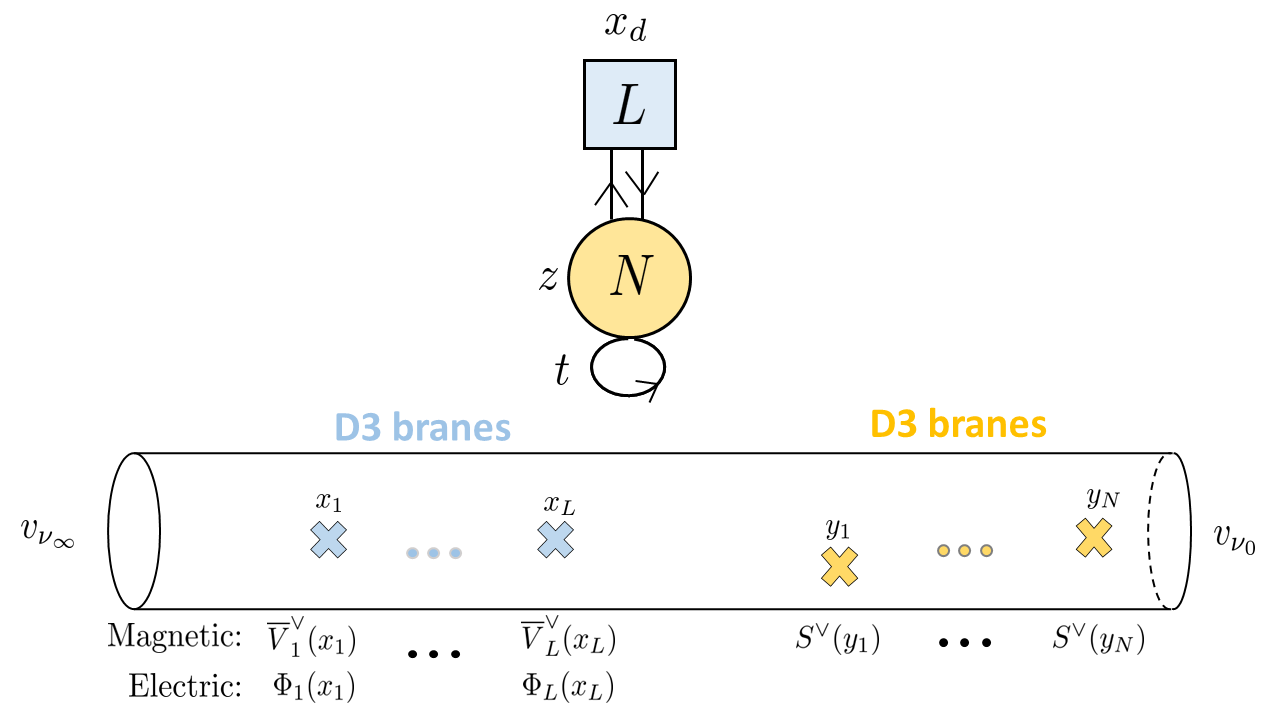}
	\vspace{-10pt}
	\caption{An example of unramified $A_1$ $q$-conformal block on $\mathbb{C}^\times$. In the magnetic block realization, one inserts Frenkel-Reshetikhin $\cW_{q,t}(A_1)$ vertex operators (blue) and screening currents (yellow), which are integrated over. In the electric block interpretation, one inserts intertwiners $\Phi_d(\tilde{x}_d)$ labeled by highest weights of $U_{\hbar}(\widehat{A_1})$ fundamental representations. The block computes an index of the 3d $\cN=4$ gauge theory pictured on top (drawn in terms of $\cN=2$ multiplets).}
	\label{fig:intro}
\end{figure}

\begin{figure}[h!]
	\emph{}
	\centering
	\includegraphics[trim={0 0 0 0cm},clip,width=1.0\textwidth]{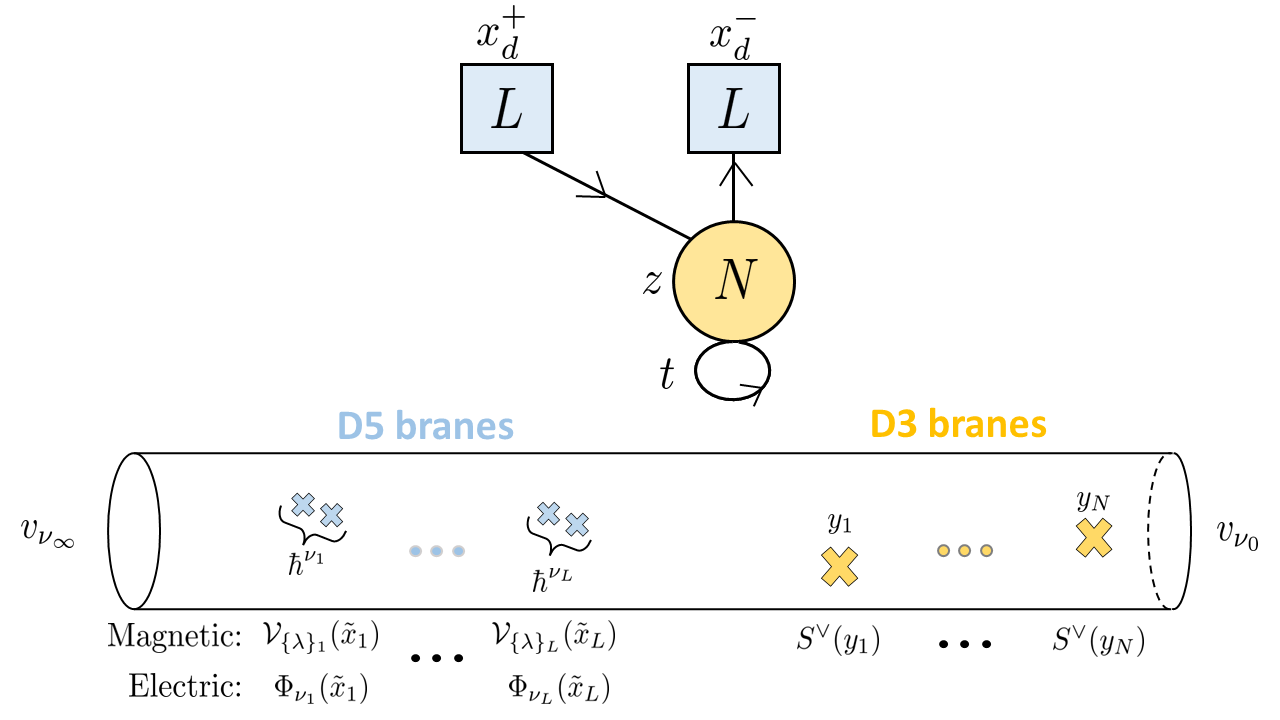}
	\vspace{-10pt}
	\caption{An example of ramified $A_1$ $q$-conformal block on $\mathbb{C}^\times$. In the magnetic block realization, one inserts the $\cW_{q,t}(A_1)$ $q$-primary vertex operators (blue) and screening currents (yellow), which are integrated over. The positions $\tilde{x}_d$ are center of mass locations, since each $q$-primary is a product of more elementary operators defined via Drinfeld polynomial data. In the electric block interpretation, one inserts intertwiners $\Phi_{\nu_d}(\tilde{x}_d)$ labeled by highest weights of $U_{\hbar}(A_1)$ Verma modules. The block computes an index of the 3d $\cN=2$ Drinfeld gauge theory pictured on top.}
	\label{fig:intro2}
\end{figure}

Our approach is in a sense the complete opposite: for us, it is 4d Super Yang-Mills theory which arises from an orbifold geometry, and the surface defects which arise from D-branes \cite{Vafa:1997mh,Witten:1995zh,Strominger:1995ac}. 
In the end, all approaches are related by T and S string dualities, so it is not surprising that they ultimately describe the same low energy physics. 
The perspective we take has the advantage of providing a weakly coupled gauge theory description of the defects, in a mass-deformed setting.\\

Our setup was first introduced in \cite{Aganagic:2015cta}: we consider the $\fg_o$-type $(2,0)$ little string theory, obtained as the zero string coupling limit $g_s\rightarrow 0$ of type IIB on the background
\beq\label{backbasicintro}
{\mathbb C}^2_\Gamma\times \mathbb{C}_q \times \mathbb{C}_t \times \cC  \; .
\eeq
The space ${\mathbb C}^2_\Gamma$ is a minimal resolution  by blowup of a $\fg_o=ADE$ singularity, and $\cC$ is the cylinder of the geometric Langlands program. Gukov-Witten surface defects originate as codimension-2 defects in the little string. These are D5 branes at points on $\cC$ (in the $g_s\rightarrow 0$ limit), and are supported on $\mathbb{C}_q \times \mathbb{C}_t$ and various 2-cycles of ${\mathbb C}^2_\Gamma$. The 5d $\cN=1$ quiver gauge theory from proposition \ref{prop3} is the low energy effective theory on the D5 branes \cite{Douglas:1996sw}. Gauge-vortex duality to 3 dimensions is realized on the Higgs branch after turning on vortex flux on the D5 branes along $\mathbb{C}_t$: this corresponds to introducing  D3 branes ending on D5 branes \cite{Hanany:2003hp}; the vortex D3 branes are also points on $\cC$, and are supported on $\mathbb{C}_q$  and compact 2-cycles of ${\mathbb C}^2_\Gamma$. The 3d $\cN=2$ Drinfeld quiver gauge theory from proposition \ref{prop2} is the low energy effective theory on the D3 branes.
In the language of the $\cW_{q,t}(\fg_o)$ magnetic blocks, the positions of the D3 branes on $\cC$ are the locations of the screening current insertions, while the positions of the D5 branes on $\cC$ are the locations of the $q$-primary vertex operator insertions. From the point of view of the $U_{\hbar}(\widehat{^L\fg_\text{o}})$ electric blocks, the D3 branes are not visible (the screening currents having been integrated over), while the positions of the D5 branes on $\cC$ label the highest weights of Verma modules in the quantum algebra. As far as the Gukov-Witten defects are concerned, we will prove the following proposition:\\

\begin{prop}
\label{prop5} 
The quantum Langlands correspondence on $\cC$ originates from the 6d $\fg_o$-type $(2,0)$ little string, where tame ramification is provided by type IIB D5-branes at points on $\cC$. After taking the point particle limit $m_s\rightarrow \infty$ to the 6d $(2,0)$ SCFT ${\cal X}_{\fg_\text{o}}$, and reducing the theory on $S^1_q\times S^1_t \subset \mathbb{C}_q \times \mathbb{C}_t$, the little string reduces to 4d $\cN=4$ Super Yang-Mills on $\cC\times M_2$, with gauge Lie algebra $\fg_o$, and the D5-brane defects flow to the monodromoy defects of Gukov and Witten, supported on $M_2$. The action of S-duality on Super-Yang Mills and its monodromy defect data is deduced from the action of  little string T-duality along the torus $S^1_q\times S^1_t$, in the presence of the D5 branes wrapping it.\\
\end{prop}

Gukov-Witten surface defects are labeled by discrete data, i.e. the choice of a Levi subalgebra of $\fg$, and by two sets of continuous parameters: $(\beta,\gamma)$, valued in the Cartan subalgebra of $\fg$, and $(\alpha,\eta)$,  respectively valued in the maximal torus of $G$ and $^L G$. 
We will derive the little string origin of the continuous parameters, for $\fg$ an arbitrary simple Lie algebra, and postpone the analysis of the discrete data to the paper \cite{NathanH}. When $\fg$ is non simply-laced, we only allow for D5-brane configurations invariant under the outer automorphism group action of a simply-laced Lie algebra $\fg_\text{o}$, and we furthermore perform a twist by this group action along one of the circles in $S^1_q\times S^1_t$. We will clarify the meaning of the 5d instanton partition function and 3d half-index in the non simply-laced case, and show that they correctly reproduce the $q$-conformal blocks of the corresponding $\cW_{q,t}(\fg)$ and $U_\hbar(\widehat{^L\fg})$ algebras.\\

\vspace{10mm}

\subsection{Implications for the Local Alday-Gaiotto-Tachikawa Correspondence}

The compactification of a 6d $(2,0)$ SCFT on $\cC$ with regular punctures yields a 4d SCFT of class $S$; for such a theory, the integrable system associated to its Coulomb branch \cite{Martinec:1995by,Donagi:1995cf} is a tamely-ramified Hitchin system \cite{Gaiotto:2009we,Gaiotto:2009hg}. We are interested in the classification of defects, by which we mean the geometric or representation theoretic characterization of the data at the punctures on $\cC$. 
	
For a generic point on the Coulomb branch of the 4d theory, the residue data (or more precisely, conjugacy classes thereof)  at a puncture is famously encoded in a nilpotent orbit of $^L\fg$\footnote{Such an orbit is called ``special" in the nilpotent literature, meaning it is in the image of a certain map due to Spaltenstein\cite{Spaltenstein1982ClassesUE}. It is sometimes called the Hitchin orbit for the defect. Likewise, the Higgs branch data of the theory is encoded in another nilpotent orbit, which lives in $\fg$ and is arbitrary, sometimes called the Nahm orbit. More precisely, the Higgs branch is a specific stratum described by the intersection of the Slodowy slice to the nilpotent orbit and the nilpotent cone. The final piece of data is the specification of the Sommers-Achar group for the defect, which is a certain subgroup of Lusztig’s canonical quotient group.} \cite{Gaiotto:2009we,Chacaltana:2010ks,Chacaltana:2012zy,Balasubramanian:2014jca}. For a given Lie algebra $\fg$, the number of nilpotent orbits is finite, so the classification of tame defects is likewise finite. If one turns on mass deformations, the Coulomb branch geometry is generically smoothed, as the residues change from nilpotent to semi-simple. Then, the residue data is instead specified by sheets, which are irreducible components in unions of complex adjoint orbits in $\fg$; this data is equivalent to specifying a Levi subalgebra of $^L\fg$, along with a certain nilpotent orbit inside it \cite{Balasubramanian:2018pbp,Gukov:2006jk,Gukov:2008sn}.\\
	
At the same time, the class $S$ construction is also the $m_s\rightarrow\infty$ limit of the 6d $(2,0)$ little string theory compactified on $\cC$, with  regular punctures provided by the D5 branes of type IIB at points on $\cC$. One could then reasonably expect to be able to ``derive" the classification of SCFT defects as a limit of the little string picture. And indeed, it was shown in \cite{Haouzi:2016ohr,Haouzi:2016yyg} that a large class of mass-deformed tamely ramified Hitchin systems have a natural origin in the little string, namely those theories whose Coulomb branch is fully smooth after mass deformation (``smoothable" SCFTs in the nomenclature of \cite{Balasubramanian:2018pbp}). However, when $\fg\neq A_r$, there are  defects for which a mass deformation does not fully resolve the Coulomb branch (``malleable" SCFTs), or even nontrivial theories for which the flavor symmetry happens to be trivial, implying that no mass deformation is possible in the first place (``rigid" SCFTs)\footnote{These rigid defects are necessarily nilpotent, and known as \emph{strongly parabolic} Higgs bundles in the Mathematics literature \cite{2008arXiv0811.0817L}.}. 
	
In the companion paper \cite{NathanH}, we will explain how the little string construction of this work gives a novel characterization of \emph{all} the above SCFTs, not just the smoothable ones. As a byproduct, we will construct an explicit solution to the so-called ``local" Alday-Gaiotto-Tachikawa (AGT) conjecture, which posits the existence of a map between 6d $(2,0)$ SCFTs defects and specific Verma modules in the $\cW_{\beta}(\fg)$-algebra on $\cC$. This conjecture has been shown to hold for smoothable SCFTs already 15 years ago \cite{Kanno:2009ga}, but to this day remains an open problem for the malleable and rigid SCFTs whenever $\fg\neq A_r$. Our stringy construction will imply that the local AGT conjecture holds for all tame defects in all simple Lie algebras $\fg$, thus providing a new characterization of defects exclusively in terms of $\cW_{\beta}(\fg)$-algebra Verma modules.\\

\vspace{10mm}

\subsection{Implications for the Global Alday-Gaiotto-Tachikawa Correspondence}

The global AGT conjecture states that the Nekrasov instanton partition function of a (asymptotically) conformal 4d $\cN=2$ $SU(r)$ gauge theory on the $\Omega$-background $\mathbb{C}_q\times\mathbb{C}_t$ equals a $\cW_{\beta}(A_r)$ conformal block on $\cC$ \cite{Alday:2009aq,Wyllard:2009hg}. This is another consequence of the class $S$ construction, in the case where the class $S$ theory happens to have a weakly coupled gauge theory description. For a mathematical treatment of the correspondence, see  \cite{2012arXiv1211.1287M,2012arXiv1202.2756S,Braverman:2014xca}.

In our setup, the instanton partition function of a 5d $\cN=1$ $\fg$-type quiver gauge theory on  $S^1_{\cC'}\times\mathbb{C}_q\times\mathbb{C}_t$ equals a 
$\cW_{q,t}(\fg)$ $q$-conformal block on $\cC$  \cite{Aganagic:2013tta,Aganagic:2014oia,Aganagic:2015cta,Haouzi:2017vec}. In order to make contact with the AGT correspondence, it is necessary to consider a 4d limit. However, the conformal limit to $\cW_{\beta}(\fg)$ is strongly-coupled for the gauge theory, as it loses its Lagrangian description in 4d. One could instead reduce the 5d gauge theory on $S^1_{\cC'}$ in such a way that the gauge theory remains weakly coupled with the same Lagrangian in going from 5d to 4d, but this limit does not coincide with the conformal limit to the $\cW_{\beta}(\fg)$ algebra. In conclusion, there is no obvious limit to the setup of the AGT correspondence.

However, it can sometimes happen that 5d theories have more than one inequivalent Lagrangian descriptions to begin with, a phenomenon known as \emph{spectral duality} \cite{Katz:1997eq,Mironov:2012uh,Mironov:2013xva,Mitev:2014jza}. When this is the case, it is typically possible to take the conformal limit to $\cW_{\beta}(\fg)$ while ensuring at the same time that one of the spectral dual 5d Lagrangians remains weakly coupled as it is reduced to 4d. This is a valid limit to the AGT correspondence. Then, our dictionary from gauge theory to $\cW$-algebras is not the one predicted by AGT, but rather it is related to it by spectral duality, whenever such a dual frame exists (the only known examples occur for $\fg=A_r$).\\

In this paper, we briefly revisit this argument when the AGT correspondence is further generalized to include a generic monodromy defect in 4d, meaning one is now counting ramified instantons on $(\mathbb{C}_q/\mathbb{Z}_r)\times\mathbb{C}_t$. This time around, the 4d $\cN=2$ $SU(r)$ Nekrasov instanton partition function is conjectured to compute an affine  $\widehat{A_r}$ conformal block, at a level determined by the rank $r$ and the $\Omega$-background parameters \cite{Alday:2010vg}\footnote{For mathematical background on ramified instanton counting, see \cite{2002math......2208F,Braverman:2004vv,Braverman:2004cr,2008arXiv0812.4656F,Negut:2011aa}. For Physics references, see \cite{Kozcaz:2010yp,Wyllard:2010rp,Wyllard:2010vi,Kanno:2011fw}.}. This conjectured was recently proved as a theorem by Nekrasov and Tsymbaliuk \cite{Nekrasov:2021tik}; see also \cite{Nekrasov:2017gzb,Jeong:2021rll}. 

In our setup, the instanton partition function of a 5d $\cN=1$ $\fg$-type quiver gauge theory on  $S^1_{\cC'}\times\mathbb{C}^{\times}_q\times\mathbb{C}_t$, with the origin of $\mathbb{C}_q$ removed, will be equal to a 
$U_{\hbar}(\widehat{^L\fg})$ $q$-conformal block on $\cC$. In this background, the usual instanton quantum mechanics is modified by additional 1/2-BPS loop insertions along $S^1_{\cC'}$. Though we will not prove it here, we conjecture that these additional insertions should be equivalent to considering the ramified background $\mathbb{C}_q/\mathbb{Z}_r$. 
In the simplest case for which there exists a spectral dual frame, $\fg=A_1$, we find that the conformal limit of our solution yields a $\widehat{A_1}$ conformal block which is spectral dual to the Nekrasov-Tsymbaliuk solution on the Higgs branch. This gives supporting evidence that the ramified quantum $q$-Langlands program as a whole really is formulated in a spectral dual frame compared to the usual ($q$-)AGT program.\\

\vspace{10mm}

\subsection{Outline}

The paper is organized as follows: in Section 2, we review the presentations of the $q$-deformed $\cW$-algebras and quantum affine algebras, along with their conformal blocks. We give a mathematical definition of $q$-deformed primary vertex operators in the $\cW$-algebras and define  tame ramification in the quantum $q$-Langlands correspondence. In Section 3, we define Drinfeld quiver gauge theories with 3d $\cN=2$ supersymmetry, and show that their half-index with appropriate boundary conditions reproduces the various conformal blocks of the quantum algebras. In Section 4, we derive the Drinfeld quivers and their half-index from a 5d $\cN=1$ gauge theory perspective, and show that the fundamental representations of quantum affine algebras can all be constructed in the 5d Higgs phase. In Section 5, we reinterpret the 3d indices as vertex functions in the enumerative geometry of quasimaps. In Section 6, we give the little string theory picture which underlies the results from the previous Sections. In Section 7, we explain how to generalize the simply-laced analysis to the non simply-laced case in gauge and string theory. In Section 8, we comment on various aspects of the conformal limit and make contact with other works in the literature: the Gukov-Witten description of surface defects (and their S-duality), and the description of mass-deformed tamely ramified Hitchin systems, and the Nekrasov-Tsymbaliuk solution to the KZ equations. In Section 9, we work through an example in full detail, for the $\fg=A_1$ algebra. Appendix A reviews standard facts about Lie theory and Langlands duality. Appendix B reviews the various building blocks of the half-index in 3d $\cN=2$ gauge theories.\\

We thank Spencer Tamagni for agreeing to coordinate with us the release of his work on elliptic stable envelopes for moduli spaces of vortices. The existence and computations of $\fg=A_1$ stable envelopes is carried out in detail in that work, for a 3d $\cN=2$ $U(N)$ gauge theory with flavors. In the Examples Section \ref{sec:example}, we compute these elliptic stable envelopes explicitly only for $N=1$.

\vspace{16mm}

\section{Tame Ramification and deformed conformal blocks} 
\label{sec:confblocks}

A review of standard facts about Lie algebras is provided in Appendix \ref{sec:appendixlie}.
Let $\fg$ be a simple Lie algebra and $\cC=\mathbb{R}\times S^1_{\cC}$ be an infinite cylinder. We will assume throughout that $q$ and $t$ are generic complex numbers, not equal to a root of unity.

In this Section, we review the definition of the ${\cW}_\beta({\fg})$ and $U_{\hbar}(\widehat{^L\fg})$-algebras, and proceed to construct the chiral correlators relevant to the tamely ramified Langlands correspondence.

\subsection{The ${\cW}_{q,t}({\fg})$-algebra blocks}

In the work \cite{Frenkel:1998}, Frenkel and Reshetikhin define a canonical deformation of the ${\cW}_\beta({\fg})$-algebras, which they denote as ${\cW}_{q,t}({\fg})$-algebras, with $t=q^\beta$. The cases $\fg=A_1$ and $\fg=A_r$ were previously introduced in \cite{Shiraishi:1995rp} and \cite{Feigin:1995sf,Awata:1995zk}, respectively.  
A $q$-conformal block is a ``chiral" correlator of the form:
\beq\label{correlatordefW}
\left\langle v_{\mu_\infty}\; ,\,\prod_{a=1}^{\r} (Q_a^\vee)^{N_a}\; \prod_{d=1}^{L}\cV_{\{\lambda\}_d}(\tilde{x}_d)\; v_{\mu_0} \right\rangle \, .
\eeq
Our goal in this section is to give a precise definition of such a correlator via an integral representation. In physics terminology, such a representation goes by the name of free field (i.e. Coulomb gas or Dotsenko-Fateev \cite{Dotsenko:1984nm}) formalism. Some of the operators, such as the screening charges $Q_a^\vee$, will be the familiar to the reader, while the vertex operators $\cV_{\{\lambda\}_d}(\tilde{x}_d)$ will be new.\\

Our starting point is a $q$-deformed Heisenberg algebra, written in terms of $\text{rank}(\fg)=r$ generators of ``simple root" type, $\alpha_a[k]$ with $k\in\mathbb{Z}$ and $a=1,\ldots,r$. These satisfy the following commutator relations:
\begin{align}\label{commutatorgenerators}
[\alpha_a[k], \alpha_b[n]] = {1\over k} (q^{k\over 2} - q^{-{k\over 2}})(t^{{k\over 2} }-t^{-{k\over 2} })B_{ab}(q^{k\over 2} , t^{k\over 2} ) \delta_{k, -n} \; ,
\end{align}
along with the zero commutator $[\alpha_a[k], \alpha_b[0]]=0$ for $k\in\mathbb{Z}$. Here, $B_{ab}$ is a certain $q$-deformed symmetrization of the Cartan matrix of $\fg$. Already in the undeformed context, the Cartan matrix 
\beq\label{Cartanlol}
C_{ab} = \langle\alpha_a,\alpha_b^{\vee}\rangle = \frac{2\, \langle\alpha_a,\alpha_b\rangle}{\langle\alpha_b,\alpha_b\rangle}
\eeq 
is not symmetric whenever $\fg$ is non simply-laced. A symmetrization\footnote{We will have more to say about the meaning of symmetrization for the non simply-laced case in Section \ref{sec:nsl}.} is defined as a new matrix
\beq\label{Bmatrix}
B_{ab}=r_a \, C_{ab} \; ,
\eeq 
with 
\beq\label{rlacing}
r_a = \frac{\fn_\fg\,\langle\alpha_a,\alpha_a\rangle}{2} \; .
\eeq
The order of the outer automorphism group of $\fg$ is denoted as $\fn_\fg$, also called the lacing number. In our convention, long roots have length squared 2 under the inner product $\langle\cdot, \cdot\rangle$, which implies
\beq\label{radef}
r_a = \begin{cases}
	\fn_\fg  & \text{if $\alpha_a$ is a long root of $\fg$\ ,} \\
	1 &  \text{if $\alpha_a$ is a short root of $\fg$\ .}
\end{cases}
\eeq
A formal $q$-deformation of the Cartan matrix $C_{ab}$ has canonical definition
\begin{align}\label{qCartanToda}
C_{ab}(q,t)= \left(q^{r_a}t^{-1} +q^{-r_a}t\right) \, \delta_{ab}- [I_{ab}]_q\; ,
\end{align} 
with $I_{ab}= 2 \, \delta_{ab} - C_{ab}$, and the notation $[\ldots]_q$ stands for the quantum number
\begin{align}\label{quantumnumber}
[n]_q = \frac{q^{n}-q^{-n}}{q-q^{-1}} \; .
\end{align}
The symmetrized matrix that appears in \eqref{commutatorgenerators} is simply the $q$-deformation of \eqref{Bmatrix}:
\beq\label{qBmatrix}
B_{ab}(q,t)=\left[r_a\right]_q \, C_{ab}(q,t) \; .
\eeq 
The Fock space representation of this $q$-Heisenberg algebra is denoted as $\pi_{\mu_0}$, and is constructed by acting with the generators on a vacuum state vector $v_{\mu_0}$:
\begin{align}
\alpha_a[0]\, v_{\mu_0} &= \langle\mu_0, \alpha_a\rangle \, v_{\mu_0} \label{eigenvalue}\nonumber\\
\alpha_a[k]\, v_{\mu_0} &= 0\, , \qquad\qquad\;\; \mbox{for} \; k>0\; .
\end{align}
In the free field formalism, an essential ingredient is the introduction of screening current operators. These come in two series, each labeled by their position $y$ on the cylinder $\cC$, as well as one of the fundamental representations of $\fg$\footnote{In \cite{Frenkel:1998}, each of these operators is further multiplied by an overall factor containing a zero mode conjugate to $\alpha_a[0]$. The effect of such a factor is simply to shift the momentum $\mu_0$ of the Fock space $\pi_{\mu_0}$. Then, up to redefinition of the vacuum state $v_{\mu_0}$, we will safely ignore this additional factor in this work.}:
\begin{align}
S_a^\vee(y) &= y^{-\alpha_a[0]/r_a}\,: \exp\left(\sum_{k\neq 0}{ \alpha_a[k] \over q^{k\, r_a\over 2} - q^{-\,{k \, r_a \over 2}}} \, y^k\right): \; , \label{screeningdef}\nonumber\\
S_a(y) &= y^{\alpha_a[0]/\beta}\,: \exp\left(-\sum_{k\neq 0}{ \alpha_a[k] \over t^{k\over 2} - t^{-\,{k  \over 2}}} \, y^k\right): \; .
\end{align}
The ${\cW}_{q,t}({\fg})$-algebra is defined as the associative algebra whose generators are the Fourier modes of operators in the commutant of the screening currents, up to total $q$-difference. Namely, one can construct a basis of $\text{rank}(\fg)=r$ such currents $T_a(z)$, which obey, for all $a,b=1,\ldots,r$,
\begin{align}
&\left[T_a(z),S_b^\vee(y)\right] = {\cD}_{q,y} F(z,y) \; ,
\label{qdif}\nonumber\\
&\left[T_a(z),S_b(y)\right] = {\cD}_{q,y} G(z,y)\; ,
\end{align}
for some functions $F, G$. Above, the $q$-difference operator acting on $F$ (and $G$) is defined as
\beq
{\cD}_{q,y} F(y) =\frac{F(y)-F(q\, y)}{y\, (1-q)} \; .
\eeq
We call screening charge the $(-1)$-st Fourier coefficient of a screening current:
\begin{align}
&Q_a^\vee =\int dy\, S_a^\vee(y) \;\; :\, \pi_0\rightarrow\pi_{-\beta\alpha_a/{r_a}} \; ,
\label{screeningchargedef}\nonumber\\
&{Q_a} =\int dy\, S_a(y) \;\; :\, \pi_0\rightarrow\pi_{\alpha_a} \; .
\end{align}
The commutators \eqref{qdif} can therefore be recast as
\begin{align}
&\left[T_a(z),Q_b^\vee(y)\right] = 0 \; ,
\label{qdif2}\nonumber\\
&\left[T_a(z),Q_b(y)\right] = 0 \; .
\end{align}
For the currents $S_a^\vee$, the contours are defined in such a way that the integrals remain invariant under $q$-shifts, meaning one can multiply the integrand by any $q$-elliptic function without changing the integral,
\beq
\int dy \; S_a^\vee(y) = \int dy\; S_a^\vee(y)\, f(y) \, , \;\;\; \text{where} \;\; f(q\, y) = f(y)\; .
\eeq
Meanwhile, the contours involving the currents $S_a$ give integrals invariant under $t$-shifts instead:
\beq
\int dy \; S_a(y) = \int dy\; S_a(y)\, g(y) \, , \;\;\; \text{where} \;\; g(t\, y) = g(y)\; .
\eeq
This is as much as we will have to say about the currents $S_a$ and charges $Q_{a}$, as they will not enter our correlators. Instead, we will only need the currents $S_a^\vee$ and associated charges $Q_a^\vee$.

For later use, it will be convenient to introduce a set of ``fundamental weight" generators $w_a[k]$, dual to $\alpha_a[k]$ for $a=1,\ldots,r$ and $k\in\mathbb{Z}$, via
\beq\label{etow}
\alpha_a[k] = \sum_{b=1}^n C_{ab}(q^{k\over 2},t^{k \over 2}) w_b[k]\; ,
\eeq
which implies
\begin{align}\label{commutator2}
[\alpha_a[k], w_b[n]] ={1\over k} (q^{k \, r_a\over 2}  - q^{-{k \, r_a\over 2} })(t^{{k \over 2}}-t^{-{k \over 2} })\,\delta_{ab}\,\delta_{k, -n} \, .
\end{align}

\vspace{8mm}

\subsubsection{Dualities and Limits}
\label{sssec:dualities}

The $\cW_{q,t}(\fg)$ algebra enjoy two obvious dualities:  
\begin{align}
& \cW_{q,t}(\fg) \simeq \cW_{q^{-1},t^{-1}}(\fg)\;\; \text{for all simple}\;\; \fg \; , \nonumber\label{isomorphisms}\\
& \cW_{q,t}(\fg) \simeq \cW_{t,q}(\fg)\;\; \text{for simply-laced}\;\; \fg \; .
\end{align}
There is no known generalization of the second isomorphism if $\fg$ is non simply-laced; in particular, $\cW_{q,t}(\fg)$ is not isomorphic to $\cW_{t,q}(^L\fg)$.

The algebra $\cW_{q,t}(\fg)$ has various important limits. Among them, the most well-known one for physics applications is the conformal limit: $t=q^\beta$, $q\rightarrow 1$. In this limit, one obtains the $\cW_{\beta}(\fg)$ algebra, a symmetry of $\fg$-type Toda field theory on the Riemann surface $\cC$, where $\beta$ labels the central charge $c$ of the algebra: 
\beq
c(\beta)=\r+12\left|\beta\, \rho+\frac{1}{\beta}\rho^\vee\right|^2 \; .
\eeq 
Here, $\rho$ the Weyl vector of $\fg$ (half the sum of positive roots), and $\rho^\vee$ the Weyl vector of $^L\fg$. For a comprehensive review, see \cite{Bouwknegt:1992wg}. The central charge formula suggests an isomorphism
\beq\label{isoquantum}
\cW_{\beta}(\fg) \simeq  \cW_{\fn_{\fg}/\beta}(^L\fg)\; ,
\eeq
holds for any simple Lie algebra $\fg$. This isomorphism is not quite what we call the quantum Langlands correspondence; in this paper, the Langlands correspondence stands for isomorphism between conformal blocks of certain representations of the algebras $\cW_{\beta}(\fg)$ and $^L\widehat{\fg}$ at level $^L \kappa$\footnote{More generally, the representations should be viewed as objects in braided tensor categories, at which point the isomorphism can be promoted to an equivalence of categories; see also the related conjecture of Gaitsgory and Lurie \cite{2007arXiv0705.4571G,2016arXiv160105279G}. On the $^L\widehat{\fg}$ side, the relevant category is a certain category of representations whose simple objects are Weyl modules labeled by integral highest weights; such a category is known to be a braided tensor category from the work of Kazhdan and Lusztig \cite{ea33dcad-1399-3228-806e-ca0f28a2822e}. On the $\cW_{\beta}(\fg)$ side, the category is constructed in two steps: one first considers the category of $\cW_{\fn_{\fg}/\beta}(^L\fg)$ modules via Drinfeld-Sokolov reduction on the $^L\widehat{\fg}$ category, and then the isomorphism \eqref{isoquantum} yields a category of $\cW_{\beta}(\fg)$ modules. A good reference is the set of lectures by Frenkel \cite{Frenkel:2005pa}.}.

There exist other interesting limits of the deformed algebra $\cW_{q,t}(\fg)$, which yield other isomorphisms to the quantized universal enveloping algebra of  $^L\widehat{\fg}$. 

The most straightforward limit is $t\rightarrow 1$. The algebra $\cW_{q,1}(\fg)$ becomes commutative, and acquires a Poisson structure. There is evidence that this Poisson algebra $\cW_{q,1}(\fg)$ is isomorphic to the center of the quantum affine algebra $U_{q}(\widehat{\fg})$ at the critical level $k=-h^\vee$ \cite{1996CMaPh.178..237F}. Taking the further limit $q\rightarrow 1$ in $\cW_{q,1}(\fg)$, one obtains an algebra isomorphic to the center of the classical algebra $U(\widehat{\fg})$ at the critical level. It also coincides with the $\beta\rightarrow 0$ limit $\cW_{0}(\fg)$ of the conformal $\cW_{\beta}(\fg)$ algebra \cite{Feigin:1991wy}. 

Starting again in $\cW_{q,t}(\fg)$, the limit $q\rightarrow 1$ happens to be more subtle. First consider $q\rightarrow \epsilon$, with $\epsilon = \exp(2\, \pi\, i/\fn_\fg)$. In the simply-laced case, where $\epsilon=1$, the isomorphism \eqref{isomorphisms} guarantees that $\cW_{1,t}(\fg_\text{o})$ is isomorphic to the algebra $\cW_{t,1}(\fg_\text{o})$ from the previous paragraph. In the non simply-laced case, $\cW_{\epsilon,t}(\fg)$ contains a commutative subalgebra $\cW'_{\epsilon,t}(\fg)$, again endowed with a Poisson structure. The algebra $\cW'_{\epsilon,t}(\fg)$ is isomorphic to the center of the quantum affine algebra $U_{t}(^L\widehat{\fg})$, at the critical level $^L \kappa=-^L h^\vee$.
The further limit $\epsilon = 1$ for $\fg$ non simply-laced is understood algebraically, but its interpretation in representation theory remains an open problem: one finds an algebra $\cW_{1,t}(\fg)$ which is again Poisson, but has no known relation to quantum affine algebras (and in particular no relation to $U_{t}(^L\widehat{\fg})$). The further limit $t\rightarrow 1$ of $\cW_{1,t}(\fg)$ is the $\beta\rightarrow \infty$ limit of the conformal $\cW_{\beta}(\fg)$ algebra, commonly known as ``the classical" $\cW_{\infty}(\fg)$ algebra. Note that $\cW_{\infty}(\fg)$ is isomorphic to $\cW_{0}(^L\fg)$ via \eqref{isoquantum}.

\begin{figure}[h!]
	\emph{}
	\centering
	\includegraphics[trim={0 0 0 0cm},clip,width=0.9\textwidth]{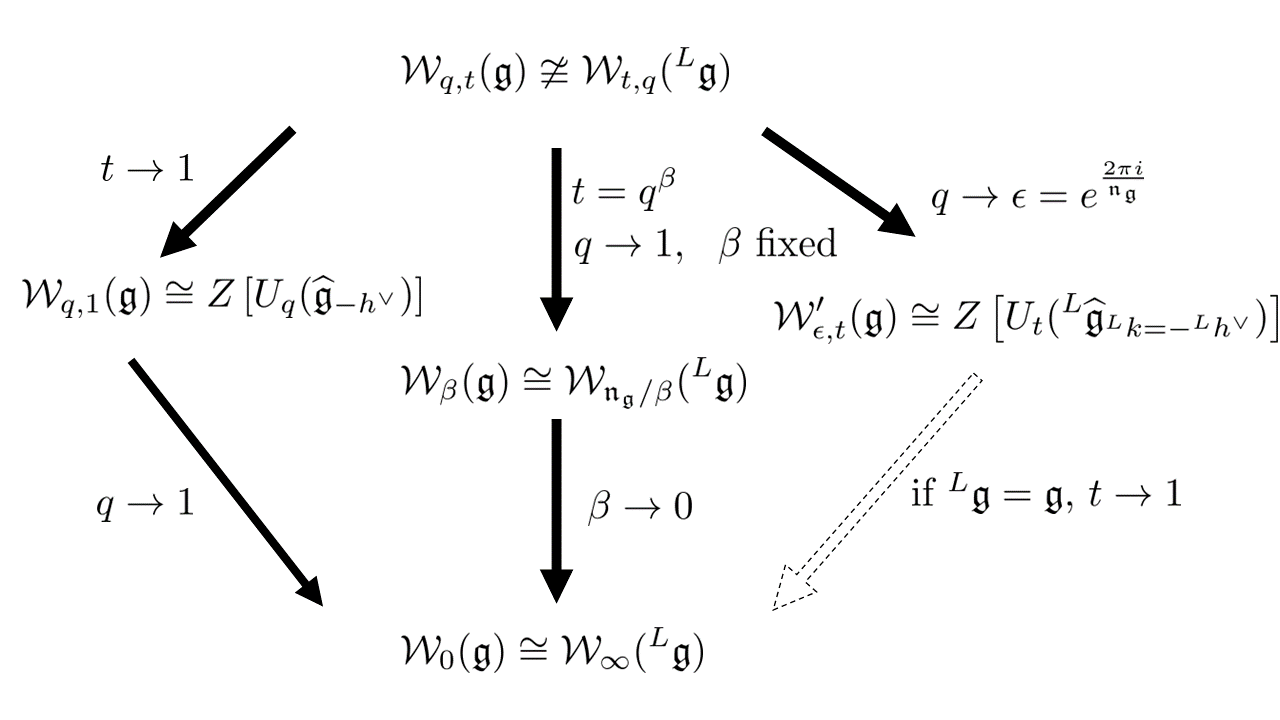}
	\vspace{-1pt}
	\caption{The various limits of the $\cW_{q,t}(\fg)$-algebra described in the main text.} 
	\label{fig:Walgebras}
\end{figure}

Recently, the limit $q\rightarrow 1$ of $\cW_{q,t}(\fg)$ was revisited for $\fg$ non simply-laced \cite{Frenkel:2021bmx,2022arXiv220111918K,Chen:2018ntf}: it is also possible to construct  $\cW_{1,t}(\fg)$ via a ``folding" operation on a simply-laced algebra $\fg_\text{o}$, where a relation to the quantum affine algebra $U_{t}(\widehat{\fg_\text{o}})$ can be made explicit. Here, one should think of $\fg_\text{o}$ as the parent simply-laced Lie algebra from which $\fg$ is constructed as the Lie subalgebra invariant under $\fg_\text{o}$'s outer automorphism group. This interpretation of $\cW_{1,t}(\fg)$ as arising from folding is in fact quite natural in string theory, as we will review in Section \ref{sec:nsl}.\\

\subsubsection{The various vertex operators of ${\cW}_{q,t}({\fg})$-algebras}

The screening currents \eqref{screeningdef} were defined using the generators $\alpha_a[k]$, so it is natural to ask if there also exist operators defined through the dual generators $w_a[k]$. 

The simplest such class of operators was introduced by Frenkel and Reshetikhin in their original paper \cite{Frenkel:1998}; those operators are labeled by their position $z$ on $\cC$ as well as one of the fundamental representations of $\fg$:
\beq\label{YoperatorToda}
{\cY_a}(z)= q^{w_a[0]}\,: \exp\left(\sum_{k\neq 0} w_a[k]\, t^{k \over 2}  \, z^k\right): \; , \qquad\;\;\;\;a=1,\ldots,\r \; .
\eeq
Remarkably, the ${\cW}_{q,t}({\fg})$-algebra generating currents $T_a(z)$ can all be expressed as Laurent polynomials in these ${\cY_a}$ vertex operators. A deep insight of \cite{Frenkel:1998} is that such Laurent polynomials can furthermore be reinterpreted as characters of irreducible representation in quantum affine algebras; this crucial observation will be reviewed in detail.

For example,
The deformed stress tensor of $q$-Virasoro is the generator of ${\cW}_{q,t}({A_1})$; it can be written as a 2-term Laurent polynomial in the $\cY$ operators:
\beq\label{examplestress}
T(z)= \cY(z) + \left[\cY(z\;t/q)\right]^{-1} \; .
\eeq
This is checked explicitly by computing $[T(z),Q_a]=0$. Furthermore, $T(z)$ takes the form of a character for the fundamental representation of $U_{\hbar}(\widehat{A_1})$  (i.e. the spin 1/2 representation in Physics), a point which will be elucidated in \eqref{qtcharacter}. In the conformal limit $t=q^\beta$, $q= e^\epsilon$, $\epsilon\rightarrow 0$, and after rescaling the Heisenberg algebra generators, one finds 
\beq\label{examplestresslimit}
T(z) \;\; \longrightarrow \;\; 2+\epsilon^2\, T_{vir}(z)+ O(\epsilon^3) 
\eeq
The next-to-lowest order term is the holomorphic stress tensor $T_{vir}(z)$ of the usual Virasoro algebra, followed by an infinite number of corrections in the $\epsilon$ expansion. The connection to character theory is lost in the limit.\\

In the same work \cite{Frenkel:1998}, Frenkel and Reshetikhin define a class of vertex operators dual to  the screening currents. Like the screenings, the operators in question come in two series, each labeled by their position $x$ on $\cC$, as well as one of the fundamental representations of $\fg$:
\begin{align}
\overline{V}_a^\vee(x) &= x^{w_a[0]/r_a}\,: \exp\left(-\sum_{k\neq 0}{ w_a[k] \over q^{k\, r_a\over 2} - q^{-\,{k \, r_a \over 2}}} \, x^k\right):  \; , \label{fundvertex}\nonumber\\
\overline{V}_a(x) &= x^{-w_a[0]/\beta}\,: \exp\left(\sum_{k\neq 0}{ w_a[k] \over t^{k\over 2} - t^{-\,{k  \over 2}}} \, x^k\right): \; , \qquad\;\;\;\;a=1,\ldots,\r \; .
\end{align}
These operators are sometimes called ``fundamental" vertex operators. They have a further interpretation as $q$-deformed versions of specific degenerate primary operators. This was in fact the motivation for defining them in the first place \cite{Frenkel:1998}: they are the $q$-deformed generalization to all simple $\fg$ of the Virasoro operators $\Phi_{2,1}(x)$ and $\Phi_{1,2}(x)$, in minimal model notation \cite{Lukyanov:1994re,Lukyanov:1996qs}\footnote{The motivation in those works was very different, and came from Physics: the goal was to realize $q$-Virasoro as a dynamical symmetry of a Restricted Solid-On-Solid integrable system, modeling a crystal-vapor interface as the temperature is varied.}
As explained in the Introduction, these operators were recently shown to be essential ingredients in formulating the \emph{unramified} quantum $q$-Langlands correspondence: after turning off the deformations, they reduce to  Wilson-'t Hooft line operator insertions in the 4d Super Yang-Mills formulation of Kapustin and Witten \cite{Aganagic:2017smx,Kapustin:2006pk}.\\

In this paper, we will need yet another class of operators, again labeled by their position $x$ on $\cC$, and one of the fundamental representations $a$ of $\fg$:
\beq\label{coweightvertexdef}
\Lambda_a(x) =x^{w_a[0]/r_a}\, : \exp\left(\sum_{k\neq 0}{w_a[k] \over (q^{k\, r_a\over 2} - q^{-\,{k \, r_a \over 2}})(t^{k\over 2} - t^{-\,{k \over 2}})} \,t^{k \over 2}\, x^k\right):  \; , \qquad\;\;\;\;a=1,\ldots,\r \; .
\eeq
These operators were not considered by Frenkel and Reshetikhin; to our knowledge, they were first introduced in an Appendix of \cite{Aganagic:2015cta}. In the rest of this section, we will argue that a particular product of $\Lambda_a$ operators provides a good candidate definition of a ${\cW}_{q,t}({\fg})$-algebra primary vertex operator. We will find that such a product has a remarkably simple realization in terms of the representation theory of $U_{\hbar}(\widehat{^L\fg})$, which we now review.

\begin{remark}
The definition of all the above operators can be further motivated from gauge and string theoretic arguments, see Sections \ref{sec:5dgauge} and \ref{sec:littlestring}. Heuristically, in gauge theory, all the above operators represent insertions of various 1/2-BPS defects, of increasing equivariant volume in a 4d $\Omega$-background\footnote{More accurately, these are really defects wrapping a circle $S^1$ in 5d $\cN=1$ Super Yang-Mills on $S^1\times\mathbb{C}_q\times\mathbb{C}_t$. This circle will be important throughout this work.} $\mathbb{C}_q\times\mathbb{C}_t$: each ${\cY_a}$ stands for the insertion of a point defect local operator supported at the origin $\{0\}\times\{0\}\subset\mathbb{C}_q\times\mathbb{C}_t$ \cite{Nekrasov:2015wsu}. The operators $\overline{V}_a^\vee$ and $\overline{V}_a$ are extended defects supported respectively on $\mathbb{C}_q\times\{0\}$ and $\{0\}\times\mathbb{C}_t$. Finally, our operators $\Lambda_a$ and the deformed primary operators, which are products thereof, are supported on the entire $\mathbb{C}_q\times\mathbb{C}_t$. In type IIB string theory, all the above operators are branes wrapping 2-cycles of a transverse resolved $ADE$ singularity (possibly subjected to a nontrivial outer automorphism twist, in the non simply-laced case); the operators ${\cY_a}$ are D1 branes, the operators $\overline{V}_a^\vee$ and $\overline{V}_a$ are D3 branes  supported on $\mathbb{C}_q$ and $\mathbb{C}_t$, respectively, and the operators $\Lambda_a$ are D5 branes supported on $\mathbb{C}_q\times\mathbb{C}_t$ \cite{Haouzi:2019jzk}. D7 branes are likewise expected to play an important role, but their exact interpretation remains to be clarified, see for example \cite{Zenkevich:2020ufs,Kimura:2023bxy} for early results.\\
\end{remark}

\subsubsection{Fundamental representations of quantum affine algebras and characters}

As a warm-up, what does the representation theory of  $U_{q}(^L\fg)$ look like, in the finite case? In many ways, it is analogous to the representation theory of $^L\fg$ itself (assuming as we are that $q$ is not a root of unity). For instance, the construction of finite-dimensional irreducible representations based on the quotient $V_\lambda= M_\lambda/I_\lambda$  of a Verma module $M_\lambda$ by some submodule $I_\lambda$ is essentially the same both for $^L\fg$ and its quantization\footnote{There is a slight subtlety brought by quantization: a finite-dimensional irreducible representation of $U_{q}(^L\fg)$ is actually a tensor product of $V_\lambda$ and a one-dimensional representation. There are $2^r$ such one-dimensional representations: for each $a\in\{1,\ldots,L\}$, the $a$-th one-dimensional representation maps the $a$-th raising and lowering operators to 0, and the $a$-th Cartan element to either $+1$ or $-1$. Without loss of generality, in this work we choose all one-dimensional representations to act as the identity $+1$ on the Cartan elements. The corresponding finite-dimensional irreducible representations are called \emph{type 1}.}. In contrast, the representation theory of $U_{q}(\widehat{^L\fg})$ departs significantly from the representation theory of $\widehat{^L\fg}$, and has been primarily developed in a series of seminal works by Chari and Pressley \cite{cmp/1104248585,Chari:1994pf,Chari:1994kr,Chari:1994pd}\footnote{For the Yangian version of these results, see \cite{Drinfeld:1987sy,Chari1991,KNIGHT1995187}.}:\\

First, we recall the definition of a fundamental representation; let $V$ be a finite-dimensional irreducible representation of $U_{q}(^L\fg)$. An affinization $\widehat{V}$ of $V$ is an irreducible representation of the quantum affine algebra $U_{q}(\widehat{^L\fg})$ which, when viewed as a $U_{q}(^L\fg)$-module, contains $V$ with multiplicity one, and is such that all other irreducible components of $\widehat{V}$ have highest weights strictly smaller than that of $V$. It is a well-known theorem that such affinizations exist. Under this decomposition, one would write $\widehat{V} = V \oplus \ldots$, where the $\ldots$ stand for smaller representations.
In general, affinizations are not unique, and characterizing them is a hard problem, but there are simple instances where they are; in particular, if $V_a$ is a \emph{fundamental} representation of $U_{q}({^L\fg})$, for $a\in\{1,\ldots,\r\}$, Chari and Pressley have proved that the affinization $\widehat{V_a}$ is unique. This affinization $\widehat{V_a}$ is what we call the $a$-th fundamental representation of $U_{q}(\widehat{^L\fg})$.\\

The decomposition $\widehat{V} = V \oplus \ldots$ is economically encoded in a kind of invariant theory known as a ``character". To see how this happens, consider the category $U_{q}(\widehat{^L\fg})-mod$ of finite-dimensional (type 1)\footnote{Among the generators of $U_{q}(\widehat{^L\fg})$, one finds the central element $c^{\pm}$. A representation of $U_{q}(\widehat{^L\fg})$ is said to be of \emph{type 1} if $c$ acts as the identity and if it is of type 1 when restricted to a $U_{q}(^L\fg)$ representation, as defined in the footnote 37. Up to automorphism  twist, it is enough to study the type 1 representation, i.e. the representations of the quantum \emph{loop} algebra $U_q(L(^L\fg)) \cong U_{q}(\widehat{^L\fg})/\langle c -1 \rangle$. This is the quantum version of a well-known fact in classical Lie theory, namely that the finite-dimensional representation theory of $\widehat{^L\fg}$ is that of the loop algebra $L(^L\fg)$, meaning the tensor product $^L\fg\otimes_\mathbb{C} \mathbb{C}[z^\pm]$ of $^L\fg$ with the algebra of Laurent polynomials in $z$. Note in particular that all finite-dimensional representations of quantum affine algebras necessarily have level 0.} representations of $U_{q}(\widehat{^L\fg})$. Note that as a module category, $U_{q}(\widehat{^L\fg})-mod$ is abelian. Furthermore, $U_{q}(\widehat{^L\fg})$ is a Hopf algebra, so the category $U_{q}(\widehat{^L\fg})-mod$ is in fact a monoidal tensor category.   
The monoidal structure guarantees that the associated Grothendieck group $Rep(U_{q}(\widehat{^L\fg}))$ is in fact a ring. This ring $Rep(U_{q}(\widehat{^L\fg}))$, viewed as a free abelian group, is generated by isomorphism classes of irreducible representations. 
Roughly speaking, a character is a morphism from $Rep(U_{q}(\widehat{^L\fg}))$ to a more manageable ring.\\

In the finite-dimensional case $U_{q}(^L\fg)$, the character is defined as the following injective homomorphism: 
\begin{align}
\chi : Rep(U_{q}(^L\fg)) &\rightarrow \mathbb{Z}[y^{\pm}_1, \ldots, y^{\pm}_r]\label{character}\nonumber\\
V\;\;\; &\mapsto \sum_{\lambda=k_1\lambda_1+\ldots+k_r\lambda_r}dim(V_\lambda)\prod_{a=1}^{\L}y_a^{k_a} \; ,
\end{align}
where $V_\lambda=\{v\in V | H.v=\lambda(H) \; \text{for all} \, H\in \fh\}$ is called the weight space for the weight $\lambda$. It is no coincidence that the homomorphism $\chi$ looks like the usual character map from $^L\fg$ Lie algebra theory; this is simply another manifestation of how the representation theory of $U_{q}(^L\fg)$ parallels its classical $^L\fg$ counterpart for generic $q$.\\

Moving on to the infinite-dimensional case, a first guess is to define a character map by making use of the canonical embedding $U_{q}(^L\fg)\hookrightarrow U_{q}(\widehat{^L\fg})$ (in the limit $q\rightarrow 1$, this turns into the familiar embedding $^L\fg \hookrightarrow \widehat{^L\fg}$ where elements of $^L\fg$ are viewed as constant maps). One would hope that by restricting the representations of $U_{q}(\widehat{^L\fg})$ to representations of $U_{q}(^L\fg)$, it would be enough to invoke  the character map \eqref{character} of $U_{q}(^L\fg)$ to classify irreducible representations of $U_{q}(\widehat{^L\fg})$. However, this character is not injective for $U_{q}(\widehat{^L\fg})$, as it does not differentiate between non-isomorphic representations: for instance, when $^L\fg=A_1$, consider two evaluation representations $V_1(x_1)$ and $V_1(x_2)$, where the subscript 1 denotes the fundamental representation of $A_1$, and $x_1\neq x_2$ are two nonzero complex numbers (evaluation representations will be reviewed in Section \ref{ssec:lol}). It follows at once that $\chi(V_1(x_1)) = \chi(V_1(x_2))$, even though the representations are not isomorphic as they have distinct highest weights.

In other words, a more refined invariant is needed after affinization. This can be motivated from Drinfeld's so-called second realization of $U_{q}(\widehat{^L\fg})$ algebras \cite{Drinfeld:1987sy}, where we do not simply have $r$ ``Cartan" generators as in the finite case, but rather an infinite number of them, which we denote as $\psi^{\pm}_{a,k}$, for $a=1,\ldots, \L$ and $k\in\mathbb{Z}^+$. In particular, when considering the embedding $U_{q}(^L\fg)\hookrightarrow U_{q}(\widehat{^L\fg})$, one only uses the fact that a highest weight vector is an eigenvector of $\psi^{\pm}_{a,0}$, but a highest weight vector is also an eigenvector of $\psi^{\pm}_{a,k}$ for all $k > 0$. The highest weights are the corresponding complex eigenvalues $\lambda^{\pm}_{a,k}$. In what follows, only half of these generators $\psi^{+}_{a,k}$ will enter our discussion, so we henceforth ignore the  $\psi^{-}_{a,k}$'s and $\lambda^{-}_{a,k}$'s, and will denote the weights simply as $\lambda^{+}_{a,k}$.\\

Instead of keeping track of each individual  eigenvalue $\lambda^{+}_{a,k}$ separately, it is customary to organize them into an infinite generating series. Remarkably, this series is itself a Laurent expansion around 0 of a ratio of $\L$ polynomials valued in $\mathbb{C}[z]$, commonly known as Drinfeld polynomials. Chari and Pressley have proved that the highest weight space of any finite-dimensional irreducible representation is in bijection with such polynomials \cite{cmp/1104248585}, confirming that the eigenvalues of $\psi_{a,0}$ alone are not enough to tell irreducible representations apart.

Frenkel and Reshetikhin have extended this bijection beyond highest weights, to arbitrary weight spaces inside a finite-dimensional representation $\widehat{V}$ \cite{Frenkel:qch}. Indeed, since $\widehat{V}$ is finite-dimensional, it is always possible to perform a Jordan decomposition into (generalized) eigenspaces of the ``Cartan" generators $\psi^+_{a,k}$:
\beq\label{Jordan}
\widehat{V}=\bigoplus_{\lambda}\, V_{\lambda} \; ,
\eeq
where we denote the weights as $\lambda\equiv \lambda^{+}_{a,k}$ for all $a=1,\ldots, \r$ and $k\geq 0$, and  
\beq\label{Jordan2}
V_{\lambda} =\{v\in\widehat{V} \, |\;\; \exists\, p\geq 1,\, \left(\psi^+_{a,k}-\lambda^{+}_{a,k}\right)^p\cdot v =0 \} \; .
\eeq
In writing the right-hand side, we are implicitly assuming the linear decomposition of $\widehat{V}$ as a $U_{q}(^L\fg)$-module, hence we omit the hats on $V_{\lambda}$.
Then, \emph{any} weight space $V_{\lambda}$ occurring in the above decomposition of $\widehat{V}$ is in bijection with two sets of $\L$ polynomials valued in $\mathbb{C}[z]$:
\begin{align}\nonumber\label{drinfeldpolysl}
& {\cA}^{+}_{\lambda,b}(z)=\prod_{i=1}^{\deg(\cA^{+}_{\lambda,b})}\left(1- q^{\mathfrak{a}^{+}_{\lambda,b,i}}\, z\right) \; ,\\ 
& {\cA}^{-}_{\lambda,b}(z)=\prod_{j=1}^{\deg(\cA^{-}_{\lambda,b})}\left(1- q^{\mathfrak{a}^{-}_{\lambda,b,j}}\, z\right) \; ,\qquad b=1,\ldots,\L \, .
\end{align}
By abuse of notation, in this paper we will also refer to these polynomials as Drinfeld polynomials, regardless of whether they label highest weights or not. Correspondingly, we will further call $q^{-\mathfrak{a}^{\pm}_{\lambda,b,i}}\in\mathbb{C}^\times$  the Drinfeld roots.  The $q$-exponents $\mathfrak{a}^{\pm}_{\lambda,b,i}$ are non-negative half-integers. Note that the polynomials have constant coefficient 1, so they are uniquely fixed by their degree and their roots.  In this paper, we normalize the Drinfeld polynomials such that ${\cA}^{+}_{\lambda_0,b}(z)=(1-z)$ if $\lambda_0$ is the $b$-th fundamental weight (and the other Drinfeld polynomials for $\lambda_0$ are 1).\\

With this data, the character of a finite-dimensional representation $\widehat{V}$ is defined as the following injective homomorphism \cite{Frenkel:qch}:
\begin{align}
\chi_q : Rep(U_{q}(\widehat{^L\fg})) &\rightarrow \mathbb{Z}[Y^{\pm 1}_{1}(z), \ldots, Y^{\pm 1}_{r}(z)]\label{qcharacter}\nonumber\\
\widehat{V}\;\;\; &\mapsto t^{\widehat{V}}(z)=\sum_{\lambda}\dim(V_{\lambda})\prod_{b=1}^{\L}\prod_{i=1}^{\deg(\cA^{+}_{\lambda,b})}Y_b(q^{-\mathfrak{a}^{+}_{\lambda,b,i}}\, z) \prod_{j=1}^{\deg(\cA^{-}_{\lambda,b})}Y^{-1}_b(q^{-\mathfrak{a}^{-}_{\lambda,b,i}}\, z)\, .
\end{align}
The $Y^{\pm}_b$ on the right-hand side are functions of a complex variable $z\in\mathbb{C}^\times$. The sum indexed by $\lambda$ denotes the Jordan decomposition $\widehat{V}=\bigoplus_{\lambda}\, V_{\lambda}$, and is performed over all nonzero eigenspaces $V_{\lambda}\neq 0$.
In the end, the ring of isomorphism classes of finite-dimensional representations of $U_{q}(\widehat{^L\fg})$ is traded for a much simpler Laurent polynomial ring in the $Y^{\pm}_b$, and the map $\chi_q$ is known as a $q$-character.\\

In the same work \cite{Frenkel:qch}, Frenkel and Reshetikhin further introduce the notion of a $(q,t)$-character, as a further one-parameter refinement of the above $q$-character. This was motivated by an intriguing correspondence to the deformed algebra $\cW_{q,t}(^L\fg)$ and its generators \eqref{qdif}  \cite{Frenkel:1998,Bouwknegt:1998da}.

Namely, the $(q,t)$-character of a finite-dimensional representation $\widehat{V}$ is the following injective homomorphism:
\begin{align}
\chi_{q,t} : Rep(U_{q}(\widehat{^L\fg})) &\rightarrow \cW_{q,t}(\fg)\label{qtcharacter}\nonumber\\
\widehat{V}\;\;\; &\mapsto T^{\widehat{V}}(z)=\sum_{\lambda} c^{\widehat{V}}_\lambda(q,t) \;\cY^{\widehat{V}}_{\lambda}(z) \; .
\end{align}
The sum indexed by $\lambda$ once again stands for the decomposition into weight spaces $\widehat{V}=\bigoplus_{\lambda}\, V_{\lambda}$. The coefficients $c^{\widehat{V}}_\lambda(q,t)$ are rational functions of $q$ and $t$ only. 
The key observation is that the $\cY_b$ should now be understood as $\cW_{q,t}(\fg)$ vertex operators, namely those we reviewed in \eqref{YoperatorToda}.  The first term in the character $T^{\widehat{V}}(z)$ is a ``highest weight monomial"
\beq\label{qtfirstterm}
T^{\widehat{V}}(z) = \cY^{\widehat{V}}_{\lambda_0}(z) + \ldots \; ,
\eeq
with $c^{\widehat{V}}_{\lambda_0}(q,t)=1$, and
\beq\label{qthighestweight}
\cY^{\widehat{V}}_{\lambda_0}(z)= \; \prod_{b=1}^{\r} \prod_{i=1}^{\deg(\cA^{+}_{\lambda_0,b})} \cY_b(q^{-\mathfrak{a}^{+}_{\lambda_0,b,i}}\, t^{\widetilde{\mathfrak{a}}^{+}_{\lambda_0,b,i}} \, z) \; .
\eeq 
$\deg(\cA^{+}_{\lambda,b})$ is the degree of the Drinfeld polynomial characterizing the highest weight $\lambda_0$.  The remaining terms $``\ldots"$ in the summand are ``lower weight monomials" determined from a certain $sl(2)$ pasting algorithm  \cite{Frenkel:1998,Bouwknegt:1998da}, which boils down to imposing the vanishing of the commutators between $T^{\widehat{V}}(z)$ and the screening charges \eqref{qdif2}. This is the quantum group analog of constructing all the weights (with multiplicities) of a  $^L \fg$-irreducible representation starting from a highest weight $\lambda_0$, and the property that the classical character $\chi$ is a subring of $\mathbb{Z}[y^{\pm}_1, \ldots, y^{\pm}_r]$ invariant under the Weyl group action of $^L\fg$.
The complex variable $z$ should now be understood as the coordinate of  physical bosonic field on the cylinder $\cC$.\\

The remaining terms in ``$\ldots$" above are normal-ordered products of $\cY^{\pm 1}_b(z)$ operators, of generic form 
\beq\label{qtgenericweight}
c^{\widehat{V}}_\lambda(q,t)\cY^{\widehat{V}}_{\lambda}(z)=  c^{\widehat{V}}_\lambda(q,t)\prod_{b=1}^{\r} \prod_{i=1}^{\deg(\cA^{+}_{\lambda,b})} \cY_b(q^{-\mathfrak{a}^{+,\widehat{V}}_{\lambda,b,i}}\, t^{\widetilde{\mathfrak{a}}^{\widehat{V}}_{\lambda,b,i}} \, z) \prod_{j=1}^{\deg(\cA^{-}_{\lambda,b})} \cY^{-1}_b(q^{-\mathfrak{a}^{-}_{\lambda,b,j}}\, t^{\widetilde{\mathfrak{a}}^{-}_{\lambda,b,j}}\, z)\cdot\bigg[\ldots\bigg]\, ,
\eeq 
Inside the argument of the vertex operators $\cY^{\pm}_b$, the coordinate $z$ is shifted by fugacities $q^{-\mathfrak{a}^{\pm}_{\lambda,b,j}}$ and $t^{\widetilde{\mathfrak{a}}^{\pm}_{\lambda,b,j}}$. The fugacities $q^{-\mathfrak{a}^{\pm}_{\lambda,b,j}}$ are the Drinfeld roots we already encountered in the $q$-character $t^{\widehat{V}}(z)$, while the fugacities $t^{\widetilde{\mathfrak{a}}^{\pm}_{\lambda,b,j}}$ with exponent $\widetilde{\mathfrak{a}}^{\pm}_{\lambda,b,j}\in\frac{1}{2}\mathbb{Z}_+$ are a further refinement by $t$.
By a further abuse of notation,  we will also call the $t$-refined argument of the $\cY^{\pm}_b$ operators  Drinfeld roots. The integers $\deg(\cA^{+}_{\lambda,b})$ and $\deg(\cA^{-}_{\lambda,b})$ are the degrees of the Drinfeld polynomials characterizing the weight $\lambda$; from the $\cW_{q,t}(\fg)$ algebra perspective, these integers are uniquely determined from requiring $T^{\widehat{V}}(z)$ to commute with the screening charges. The expression in bracket $[\ldots]$ equals 1 for almost all weights $\lambda$, except for occasional weights where it equals a $z$-derivative of $\cY^{\pm}_b(z)$ operators\footnote{This happens whenever $sl(2)$ pasting directions lead to coinciding arguments for the $\cY^{\pm}_b$ operators. For example, such a scenario occurs for the "null" weight space  in the second fundamental representation of $U_q(\widehat{D_4})$ (the quantization of what we would call $V_{[0,0,0,0]}$ classically), with corresponding $(q,t)$-character $T^{\widehat{V_2}}(z)$.}. The first term \eqref{qtfirstterm} is a special case of  \eqref{qtgenericweight} for which $\deg(\cA^{-}_{\lambda_0,b})=0$, $c^{\widehat{V}}_{\lambda_0}(q,t)=1$, and $[\ldots]=1$.\\

\begin{remark}
	In the context of the $q$-characters, a simpler $sl(2)$ pasting algorithm exists to construct the entire character starting from the highest weight monomial 
	\beq\label{highestmonom}
	\prod_{b=1}^{\r} \prod_{i=1}^{\deg(\cA^{+}_{\lambda_0,b})} Y_b(q^{-\mathfrak{a}^{+}_{\lambda_0,b,i}} \, z)\; ,
	\eeq
	 known as the Frenkel-Mukhin algorithm \cite{Frenkel:1999ky}.  This algorithm can be applied successfully to find the $q$-character of fundamental representations, which are the ones that arise for us, but it is known to fail when the highest weight monomial \eqref{highestmonom} is not the unique dominant weight monomial in the representation \cite{2008arXiv0801.2239N} (borrowing from the usual Lie algebra terminology, a dominant monomial is defined as a product of $Y_b(z)$'s only in positive powers, i.e. there are no $Y^{-1}_b(z)$'s present)\footnote{In the Lie algebra  $^L\fg$, fundamental representations often contain dominant weights other than their highest weights; for instance, the $G_2$ 14-dimensional  fundamental representation of highest weight $[0,1]$ also contains the dominant weight $[1,0]$. Remarkably, this is \emph{never} the case in quantum affine algebras $U_q(\widehat{^L\fg})$: there are no dominant weight monomials other than highest weight monomials for fundamental representations \cite{Frenkel:1999ky}.}. For an alternate algorithm based on cluster mutations, see the work of Hernandez-Leclerc \cite{2013arXiv1303.0744L} 
\end{remark}

Among the finite-dimensional representations, our construction will only involve the fundamental ones, so we henceforth specialize to those exclusively. For the fundamental representations, the Jordan decomposition reads
\beq 
\widehat{V_a}=\bigoplus_{\lambda}\, V_{a,\lambda} \; ,\qquad a=1,\ldots,\L \, .
\eeq 
Correspondingly, the $a$-th fundamental $q$-character will read
\beq 
t^{\widehat{V_a}}(z)=Y_a(z) + \ldots \;\qquad ,
\eeq 
and the $a$-the fundamental $(q,t)$-character will read
\beq 
T^{\widehat{V_a}}(z)= \cY_a(z) + \ldots \;\qquad .
\eeq 

\begin{remark}
	Historically, the $(q,t)$ character morphism was originally defined as a map  $Rep(U_{q}(\widehat{\fg}))\rightarrow \cW_{q,t}(\fg)$ \cite{Frenkel:1998,Bouwknegt:1998da}. For us, in the context of the Langlands program, the $(q,t)$ character is better understood as a map $Rep(U_{\hbar}(\widehat{^L\fg}))\rightarrow \cW_{q,t}(\fg)$, with a Langlands dual domain and a quantization parameter $\hbar=q^{\fn_{\fg}}/t$ instead of $q$.\\
\end{remark}

\subsubsection{Deformed primary vertex operators: a definition}
\label{sssec:definition}

\underline{Definition}: A $q$-deformed primary vertex operator $\cV_{\{{\lambda}\}}(\tilde{x})$ of $\cW_{q,t}(\fg)$ is labeled by a set of $J$ weights $\{{\lambda}\}=\{\lambda_1,\ldots,\lambda_J\}$ from the quantum affine algebra $U_\hbar(\widehat{^L \fg})$ (where $\hbar=q^{\fn_{\fg}}/t$), as well as $J$ parameters $x_s$ valued on $\cC=\mathbb{C}^\times$ ($s=1,\ldots,J$), subject to the following constraints:
\begin{itemize}
\item[(1)] For all $\lambda_s\in\{{\lambda}\}$, the vector $v_{\lambda_s}$  must belong in a weight space $V_{a,\lambda_s}$ for some $a\in\{1,\ldots,\L\}$, where  $V_{a,\lambda_s}$ appears in the Jordan decomposition of the $a$-th fundamental representation $\widehat{V_a}=\bigoplus_{\lambda}\, V_{a,\lambda}$.
\item[(2)] For all $\lambda_s\in\{{\lambda}\}$, let ${\cA}^{\pm}_{\lambda_s,b}$ denote the Drinfeld polynomials of $\lambda_s$, as in \eqref{drinfeldpolysl}. Then the degrees of the polynomials must satisfy 
\beq\label{constraintpoly}
\sum_{s=1}^{J}\left[ \deg(\cA^{+}_{\lambda_s,b}) - \deg(\cA^{-}_{\lambda_s,b})\right] = 0 \; ,
\eeq
and 
\beq\label{constraintpoly2}
\sum_{s=1}^{J'}\left[ \deg(\cA^{+}_{\lambda_s,b}) - \deg(\cA^{-}_{\lambda_s,b})\right] \neq 0 \; 
\eeq
for all  $b=1,\ldots,\r$, and all proper subsets of $J'< J$ weights.\\
\end{itemize}

\noindent
Explicitly, a $q$-deformed primary operator is a product of $\Lambda^{\pm}_a$ operators \eqref{coweightvertexdef} labeled by the above data, as
\begin{align}
\cV_{\{{\lambda}\}}(\tilde{x}) = \;\; :\prod\limits_{s=1}^{|\{{\lambda}\}|}\prod\limits_{b=1}^{\r}&\prod\limits_{i=1}^{\deg(\cA^{+}_{\lambda_s,b})}\Lambda_b\left(q^{-\mathfrak{a}^{+}_{\lambda_{s},b,i}}\, t^{\widetilde{\mathfrak{a}}^{+}_{\lambda_{s},b,i}}\, x_{s}\right)\times\nonumber\label{qprimarydef}\\ &\times\prod\limits_{j=1}^{\deg(\cA^{-}_{\lambda_s,b})}\Lambda^{-1}_b\left(q^{-\mathfrak{a}^{-}_{\lambda_{s},b,j}}\, t^{\widetilde{\mathfrak{a}}^{-}_{\lambda_{s},b,j}}\, x_s\right): \; ,
\end{align}
where $|\{{\lambda}\}|=J$ and the notations are the same as in the $(q,t)$-character map \eqref{qtgenericweight}.
The complex parameter $\tilde{x}$ in $\cV_{\{{\lambda}\}}(\tilde{x})$ is defined in terms of the right-hand side  parameters $x_s$ as
\beq\label{composition}
x_{s} = \tilde{x} \, q^{\sigma_{s}} \; ,\qquad\;\; s=1, \ldots, J \; .
\eeq
with $\sigma_s\in\mathbb{C}$.\\

To motivate the definition of $\tilde{x}$, note that to each weight space $V_{a,\lambda_s}$, we associated a complex parameter $x_{s}\in\mathbb{C}^\times$ denoting the location of a $q$-puncture $\Lambda^{\pm 1}_b(x_s)$ on $\cC$, $J$ of them in total. This does not mean we have $J$ primary vertex operators, but rather a single $q$-primary vertex operator described by the set $\{{\lambda}\}$. The fine-grained view is that such a $q$-primary always has a presentation as a normal-ordered product of $\Lambda^{\pm 1}_b(x_s)$ operators, each at a different location $x_{s}$.  The coarse-grained view is that a $q$-primary vertex operator is ``located" at a position $\tilde{x}$ on the cylinder, akin to a center of mass defined by \eqref{composition}. In other words, each operator $\Lambda^{\pm 1}_b$ is valued at a point on $\cC$ which is shifted away from $\tilde{x}$ by $q^{-\,\sigma_{s}}$ in \eqref{composition}, and furthermore by  $q^{\mp\mathfrak{a}^{\pm}_{\lambda_{s},b,j}}\, t^{\pm\widetilde{\mathfrak{a}}^{\pm}_{\lambda_{s},b,j}}$ from the Drinfeld roots in \eqref{qprimarydef}.

\begin{remark}
	While the quantum affine weights appear in an additive fashion in the definition of the $\cW_{q,t}(\fg)$ generating currents, namely as a sum of terms in a character, the same weights appear in a multiplicative fashion in the definition of the $\cW_{q,t}(\fg)$ primary operators.
\end{remark}

In the particular case where the Lie algebra is simply-laced, $\fg=\fg_\text{o}$, only half of the exponents in \eqref{qprimarydef} are independent: $\mathfrak{a}^{+}_{\lambda_s,b,i}=\widetilde{\mathfrak{a}}^{+}_{\lambda_s,b,i}$ and $\mathfrak{a}^{-}_{\lambda_s,b,i}=\widetilde{\mathfrak{a}}^{-}_{\lambda_s,b,i}$, for all weights $\lambda_s$. The implication is that in the simply-laced case, the Drinfeld roots do not depend on $q$ and $t$ separately, but only on the ratio $q/t=\hbar$ (when $\fg$ is non simply-laced, $q$ is weighted differently from $t$ according to whether the index $b$ labels a long or a short root). This simple observation implies that the argument of the $\Lambda^{\pm 1}_b$ operators simplifies in the simply-laced case:
\beq\label{qprimarydefslshisha}
\cV_{\{{\lambda}\}}(\tilde{x}) =\; :\prod\limits_{s=1}^{|\{{\lambda}\}|}\prod_{b=1}^{\text{rk}(\fg_\text{o})}\prod_{i=1}^{\deg(\cA^{+}_{\lambda_{s},b})}\Lambda_b\left(\hbar^{-\mathfrak{a}^{+}_{\lambda_s,b,i}}\, x_s\right) \prod_{j=1}^{\deg(\cA^{-}_{\lambda_{s},b})}\Lambda^{-1}_b\left(\hbar^{-\mathfrak{a}^{-}_{\lambda_s,b,j}}\, x_s\right): \; .
\eeq
That is, in the simply-laced case, the relation to the Grothendieck ring $Rep(U_{\hbar}(\widehat{\fg_\text{o}}))$ is direct: ${\cA}^{+}_{\lambda_s,b}(x_s)$ and ${\cA}^{-}_{\lambda_s,b}(x_s)$ are   the literal Drinfeld polynomials \eqref{drinfeldpolysl} (with $q$ there replaced by $\hbar$ here); this is the same weight data which appears in the definition of the $\hbar$-character \eqref{qcharacter} (with $q$ there now written as $\hbar$ here), instead of the more refined  $(q,t)$-character \eqref{qtcharacter} which appears in the non simply-laced case.\\


\begin{remark}
The remarkable way in which the representation theory of $U_\hbar(\widehat{^L \fg})$ enters our definition of $\cW_{q,t}(\fg)$ primaries is a distinctive feature of ramification: in the unramified quantum $q$-Langlands program \cite{Aganagic:2017smx}, the operators which appear inside chiral correlators are the Frenkel-Reshetikhin fundamental vertex operators $\overline{V}_a^\vee$ \eqref{fundvertex}, and the index  ``$a$" labeling a fundamental representation of $U_\hbar(\widehat{^L \fg})$ may just as well label a fundamental representation of $^L\fg$ in the classical sense. In contrast, the labeling of the $q$-primary vertex operators $\cV_{\{{\lambda}\}}$ is intrinsically quantum affine.\\
\end{remark}

\subsubsection{A classification of the primary vertex operators}
\label{sssec:classification}

In the standard terminology of primary vertex operators in conformal field theory, one defines the ``momentum" of the (chiral) vertex operator 
\beq
\label{primaryop}
e^{\langle \widetilde{\sigma} , \phi(\tilde{x})\rangle} \; ,
\eeq
as an element $\widetilde{\sigma}$ of the coweight lattice of $\fg$ (or equivalently the weight lattice of $^L\fg$), and $\phi(\tilde{x})$ is a free $\r$-dimensional boson. In the deformed case, the momentum appears as a $q$-exponent. Ours definitions \eqref{composition} and \eqref{qprimarydef} imply that such an exponent has two components:
\beq\label{totalcommomentum}
\widetilde{\sigma} = \widetilde{\sigma}_{Classical} + \widetilde{\sigma}_{Drinfeld}  \; ,
\eeq
The term $\widetilde{\sigma}_{Classical}$ is the $\fg$-coweight 
\beq\label{commomentum}
\widetilde{\sigma}_{Classical} = \sum_{s=1}^{J} \sigma_{s} \sum_{b=1}^{\r}\left[\deg(\cA^{+}_{\lambda_s,b}) - \deg(\cA^{-}_{\lambda_s,b})\right] w^\vee_b  \; ,
\eeq
where $w^\vee_b$ is the $b$-th fundamental coweight of $\fg$ and $\sigma_{s}\in\mathbb{C}$ is as in \eqref{composition}. This expression can be simplified to 
\beq\label{commomentum2}
\widetilde{\sigma}_{Classical} = \sum_{s=1}^{J} \sigma_{s}\, \underline{\lambda_s}  \; ,
\eeq
with $\underline{\lambda_s}=2\log_{\hbar}\lambda^{+}_{a,0}$ the ``classical" $\fg$-coweight for the quantum affine coweight $\lambda_s=\lambda^{+}_{a,k}$; this is the normalized exponent of $\hbar$ in the 0-th mode eigenvalue $\lambda^{+}_{a,0}$ in Drinfeld's second realization. The equality is established by recognizing that the inner sum in \eqref{commomentum} is the expansion of the $\fg$-coweight $\underline{\lambda_s}$ in Dynkin basis, with $\left[\deg(\cA^{+}_{\lambda_s,b}) - \deg(\cA^{-}_{\lambda_s,b})\right]$ as the $b$-th Dynkin label.\\

The second contribution to the momentum  $\widetilde{\sigma}_{Drinfeld}$ is fully fixed by the representation theory of the quantum affine algebra, namely the Drinfeld roots which shift the positions $x_s$ of the operators \eqref{qprimarydef}. Writing $t=q^{\beta}$, this momentum contribution is the $\fg$-coweight
\beq\label{commomentum3}
\widetilde{\sigma}_{Drinfeld} = \sum_{s=1}^{J} \sum_{b=1}^{\r} \Upsilon_{s,b}\, w^\vee_b \; ,
\eeq
where  
\beq
\Upsilon_{s,b} =\; -\sum_{i=1}^{\deg(\cA^{+}_{\lambda_s,b})}\left(\mathfrak{a}^{+}_{\lambda_{s},b,i} - \beta\; \widetilde{\mathfrak{a}}^{+}_{\lambda_{s},b,i}\right)+ \sum_{j=1}^{\deg(\cA^{-}_{\lambda_s,b})}\left(\mathfrak{a}^{-}_{\lambda_{s},b,j}  -\beta\;\widetilde{\mathfrak{a}}^{-}_{\lambda_{s},b,j} \right) \; .
\eeq
The decomposition $\widetilde{\sigma} = \widetilde{\sigma}_{Classical} + \widetilde{\sigma}_{Drinfeld}$ suggests a canonical classification of $q$-primary vertex operators. Suppose the $J$ coweights $\underline{\lambda_s}$ span the coweight lattice of $\fg$. Without loss of generality, we can take the span to make up a basis of this lattice, meaning we restrict ourselves to only $\r$ linearly independent coweights. This translates to $\{\vec{\underline{\lambda}}\}$ having cardinality $J=\r+1$, with the constraint  $\sum_{s=1}^{\r+1} \underline{\lambda_s} = 0$. A  $q$-primary labeled by such a set $\{\vec{\underline{\lambda}}\}$ will be called \emph{generic}. In the generic case, it is always possible to rescale the $\Lambda^{\pm}_b$ operator positions $x_s$ in order to set $\widetilde{\sigma}_{Drinfeld}=0$, in which case $\widetilde{\sigma} = \widetilde{\sigma}_{Classical}$.

Whenever $\widetilde{\sigma}_{Drinfeld}=0$ but $1< J < \r+1$, the set $\{\vec{\underline{\lambda}}\}$ can no longer span the coweight lattice of $\fg$, and we call the corresponding $q$-primary  \emph{semi-degenerate}, borrowing the conformal terminology of \cite{Kanno:2009ga}.\\

The other extreme case happens whenever $J=1$; there, the center of mass position $\tilde{x}$ can simply be identified with the $\Lambda^{\pm}_b$ operator position $x$: it is always possible to set $\widetilde{\sigma}_{Classical}=0$ by rescaling the relative momentum $\sigma$ away in \eqref{composition}, which simplifies to $\tilde{x}=x$. This does not mean the momentum is entirely trivial, as we still have $\widetilde{\sigma} = \widetilde{\sigma}_{Drinfeld}$. Such $q$-primary operators are therefore fully fixed by representation theory (the Drinfeld roots), and we call them \emph{rigid}.\\

Finally, when $1< J < \r+1$, there also exist $q$-primary operators for which neither $\widetilde{\sigma}_{Classical}$ nor $\widetilde{\sigma}_{Drinfeld}$ can be set to 0, no matter the rescaling of the positions $x_s$. In that case, the momentum is of the form $\widetilde{\sigma} = \widetilde{\sigma}_{Classical} + \widetilde{\sigma}_{Drinfeld}$. We call such $q$-primaries \emph{mixed}.
We summarize the classification in the following table:\\

\begin{tabular}{|c|c|c|}
	\cline{2-3}
	\multicolumn{1}{c|}{}&
	\multicolumn{1}{c|}{$\widetilde{\sigma}_{Classical}$}&
	\multicolumn{1}{c|}{$\widetilde{\sigma}_{Drinfeld}$}
	\\ \hline
	Generic (and semi-degen.) Primary  &  $\neq 0$ & $= 0$ \\
	Rigid Primary  &  $=0$ &  $\neq 0$ \\
	Mixed Primary  &  $\neq 0$ & $\neq 0$ \\    \hline
\end{tabular}

\vspace{8mm}

In the conformal limit \eqref{primaryop}, our definition of generic $q$-primary coincides with the usual definition of a generic primary vertex operator. By the state operator correspondence, this describes an irreducible Verma module state. Meanwhile, the semi-degenerate, rigid and mixed $q$-primaries all turn into primaries with various degeneracies. These are reducible Verma module states that contain null states at various levels. Rigid and mixed operators only arise for $\fg\neq A_r$.
In the companion paper \cite{NathanH}, we will revisit this classification and argue that it yields a solution to the problem of the \emph{local} Alday-Gaiotto-Tachikawa conjecture, see the Introduction for details.\\

\subsubsection{Examples}

Let us illustrate our construction of deformed primary operators in various examples. For ease of notation, we  label the various coweight spaces not by quantum affine coweights $\lambda_s=\lambda^{+}_{a,k}$, but instead by the corresponding ``$Y$ monomials" appearing in the character.\\

\begin{example}
\emph{The generic $q$-primary of $\cW_{q,t}(A_1)$}:\\

Consider the two weight spaces $V_{\lambda_1}=V_Y$ and $V_{\lambda_2}=V_{Y^{-1}}$ in $U_\hbar(\widehat{A_1})$. As vector spaces, $V_{Y}=\mathbb{C} v_{Y}$ with $v_{Y}$ the highest weight vector of the fundamental representation, while $V_{Y^{-1}}=\mathbb{C} v_{Y^{-1}}$, with $v_{Y^{-1}}$ the lowest weight vector of the fundamental representation. Let us check that the set $\{{\lambda}\}=\{Y,Y^{-1}\}$ satisfies the two constraints needed to define a primary. The first constraint states that each weight space should appear in the Jordan decomposition of the fundamental representation $\widehat{V}$ in $U_\hbar(\widehat{A_1})$; and indeed, as $U_\hbar({A_1})$-modules, one has
\beq\label{A1eigenspacewow}
\widehat{V} = V_{Y}  \oplus  V_{Y^{-1}} \; .
\eeq
Second, the (normalized) Drinfeld polynomials for each weight are as follows:
\begin{align}
&\lambda_1 = Y \; : \qquad  {\cA}^{+}_{\lambda_1}(z) = (1- z)\; ,\qquad  {\cA}^{-}_{\lambda_1}(z) = 1 \; ,\nonumber\label{drinfeldex1}\\
&\lambda_2 = Y^{-1} \; : \qquad {\cA}^{+}_{\lambda_2}(z) = 1\; ,\qquad  {\cA}^{-}_{\lambda_2}(z) = (1- \hbar\, z) \; .
\end{align} 
The construction of these polynomials is not familiar to physicists, so we derive them in detail in Section \ref{sec:example}. It follows that the second constraint on the polynomial degrees is satisfied:
\beq
\sum_{s=1}^{2}\left[ \deg(\cA^{+}_{\lambda_s}) - \deg(\cA^{-}_{\lambda_s})\right]=\left[1-0\right]+ \left[0-1\right] =0 \; .
\eeq
The deformed primary is therefore
\beq\label{operatorA11}
\cV_{\{{\lambda}\}}(\tilde{x}) =\; :\Lambda\left(x_{1}\right) \Lambda^{-1}\left(\hbar^{-1}\, x_{2}\right): \; ,
\eeq
where the center of mass position $\tilde{x}$ is defined through $x_{s} = \tilde{x} \, q^{\sigma_{s}}$ for $s=1,2$, and the shift by $\hbar^{-1}$ in the second operator is the Drinfeld root of ${\cA}^{-}_{\lambda_2}(z)$. In our nomenclature, we  call this a generic $q$-primary operator. In the ramified $A_1$ Langlands program, this is the only ``type" of operator one can insert at a puncture on $\cC=\mathbb{C}^\times$. Specializing the momenta $\sigma_{s}$ to specific values will turn $\cV_{\{{\lambda}\}}(\tilde{x})$ into a degenerate operator. For instance, the Frenkel-Reshetikhin fundamental vertex operator \eqref{fundvertex} is such a degenerate specialization.

In the conformal limit, and after rescaling the Heisenberg generators, this is the  $\cW_{\beta}(A_1)$ primary vertex operator
\beq
\label{operatorA11limit}
e^{\langle \widetilde{\sigma}_{Classical}, \phi(\tilde{x})\rangle} \; ,
\eeq
with $\phi(\tilde{x})$ a free boson on $\cC=\mathbb{C}^\times$ and $\widetilde{\sigma}_{Classical}=[\sigma_1-\sigma_2]$ is the (co)weight of $\fg_o=A_1$ denoting the momentum of the primary state. We did not include the momentum due to $\widetilde{\sigma}_{Drinfeld}$, as it can be set to 0 by simply rescaling $x_2$.\\
\end{example}

\begin{example}
\emph{The generic $q$-primary of $\cW_{q,t}(A_r)$}:\\

Consider the $r+1$ weight spaces $V_{\lambda_1}=V_{Y_1}$,  $V_{\lambda_2}=V_{Y_2Y^{-1}_1}$,  $V_{\lambda_3}=V_{Y_3Y^{-1}_2}$, $\ldots$, $V_{\lambda_{r+1}}=V_{Y^{-1}_{r+1}}$  in $U_\hbar(\widehat{A_r})$. Let us check that the set $\{{\lambda}\}=\{Y_1,Y_2Y^{-1}_1,\ldots,Y^{-1}_{r+1}\}$ satisfies the two constraints needed to define a primary. All weight spaces appear in the Jordan decomposition of $\widehat{V_1}$ as $U_\hbar({A_r})$-modules,
\beq
\widehat{V_1} = V_{Y_1}  \oplus  V_{Y_2Y^{-1}_1} \oplus \ldots \oplus V_{Y^{-1}_{r+1}} \;,
\eeq
so the first constraint is satisfied. The nontrivial Drinfeld polynomials (the ones not equal to 1) are
\begin{align}
&\lambda_1 = Y_1 \; : \qquad \qquad {\cA}^{+}_{\lambda_1,1}(z) = (1- z)\; ,\qquad\qquad\qquad\;\;\;\;\;\;\times \; , \nonumber\label{drinfeldex2}\\
&\lambda_2 = Y_2Y^{-1}_1 \; : \qquad {\cA}^{+}_{\lambda_2,2}(z) = (1- \hbar^{1/2}\, z) \; ,\qquad  {\cA}^{-}_{\lambda_2,1}(z) = (1- \hbar\, z)\; , \nonumber\\
&\lambda_3 = Y_3Y^{-1}_2 \; : \qquad {\cA}^{+}_{\lambda_3,3}(z) = (1- \hbar\, z)\; ,\qquad \;\;\;\; {\cA}^{-}_{\lambda_3,2}(z) = (1- \hbar^{3/2}\, z)  \; ,\nonumber\\
&\;\;\;\;\vdots \qquad\qquad\qquad\qquad\qquad\vdots \qquad\qquad\qquad\qquad\qquad\qquad\qquad\vdots \\
&\lambda_r = Y_rY^{-1}_{r-1} \; : \qquad {\cA}^{+}_{\lambda_r,r}(z) = (1- \hbar^{(r-1)/2}\, z)\; ,\,\;\; {\cA}^{-}_{\lambda_r,r-1}(z) = (1- \hbar^{r/2}\, z)  \; ,\nonumber\\
&\lambda_{r+1} = Y^{-1}_{r} \;: \qquad\qquad\;\;\;\; \times \qquad\qquad\qquad \;\;\;\;\;\;\;\;\;\; {\cA}^{-}_{\lambda_{r+1},r}(z) = (1- \hbar^{(r+1)/2}\, z)  \; ,
\end{align} 
so the constraint on the degrees is satisfied. The deformed primary is therefore 
\beq\label{operatorAr1}
\cV_{\{{\lambda}\}}(\tilde{x}) =\; :\Lambda_1\left(x_{1}\right) \left[\Lambda_2\left(\hbar^{-1/2}\,x_2\right) \Lambda^{-1}_1\left({\hbar^{-1}}\,x_2\right)\right]\ldots \left[\Lambda^{-1}_r\left(\hbar^{-(r+1)/2}\,x_{r+1}\right)\right]: \; ,
\eeq
where the center of mass position $\tilde{x}$ is defined through $x_{s} = \tilde{x} \, q^{\sigma_{s}}$ for $s=1,2,\ldots,r+1$. In our nomenclature, we call this a generic $q$-primary operator.
In the conformal limit, this is a  $\cW_{\beta}(A_r)$ primary vertex operator where $\widetilde{\sigma}_{Classical}=[\sigma_1-\sigma_2,\sigma_2-\sigma_3,\ldots,\sigma_r-\sigma_{r+1}]$ is the (co)weight of $\fg_o=A_1$ denoting the momentum of the primary state. The momentum contribution due to $\widetilde{\sigma}_{Drinfeld}$ was set to 0 by rescaling the $x_s$ variables.\\
\end{example}

\begin{example}
\emph{The ``maximally" semi-degenerate $q$-primary of $\cW_{q,t}(A_r)$}:\\
	
Consider the $2$ weight spaces $V_{\lambda_1}=V_{Y_1}$ and  $V_{\lambda_2}=V_{Y^{-1}_1}$ in $U_\hbar(\widehat{A_r})$.  Each weight space appears in the decomposition of some fundamental representation: as $U_\hbar({A_r})$-modules, 
\begin{align}
&\widehat{V_1} = V_{Y_1}  \oplus  \ldots \qquad\;\;\;\;\; , \nonumber\label{decomp}\\
&\widehat{V_r} = V_{Y_r}  \oplus  \ldots \oplus V_{Y^{-1}_1} \; , 
\end{align}
so the first constraint is satisfied. The nontrivial Drinfeld polynomials (the ones not equal to 1) are
\begin{align}
&\lambda_1 = Y_1 \; : \qquad \;\;\; {\cA}^{+}_{\lambda_1,1}(z) = (1- z)\; , \nonumber\label{drinfeldex3}\\
&\lambda_2 = Y^{-1}_1 \; : \qquad {\cA}^{-}_{\lambda_2,1}(z) = (1- \hbar^{(r+1)/2}\, z)\; , 
\end{align} 
so the constraint on the degrees is satisfied. The deformed primary is therefore
\beq\label{operatorAr2}
\cV_{\{{\lambda}\}}(\tilde{x}) =\; :\Lambda_1\left(x_{1}\right)  \Lambda^{-1}_1\left(\hbar^{-(r+1)/2}\,x_{2}\right): \; ,
\eeq
where the center of mass position $\tilde{x}$ is defined through $x_{s} = \tilde{x} \, q^{\sigma_{s}}$ for $s=1,2$. In our nomenclature, we call this a semi-degenerate $q$-primary operator.
In the conformal limit,  this is a  $\cW_{\beta}(A_r)$ degenerate primary vertex operator, where  $\widetilde{\sigma}_{Classical}=[\sigma_1-\sigma_2,0,\ldots,0]$ denotes the $A_r$ (co)weight denoting the momentum of the primary state. Again, $\widetilde{\sigma}_{Drinfeld}$ can be set to 0 by rescaling $x_2$.\\
\end{example}

\begin{example}
	\emph{The rigid $q$-primaries of $\cW_{q,t}(D_4)$}:\\
	
	Consider the single weight space $V_{\lambda}=V_{Y_1Y^{-1}_1}$ in $U_\hbar(\widehat{D_4})$.  This weight space appears in the decomposition of the second fundamental representation: as $U_\hbar({D_4})$-modules, 
	\begin{align}
	\widehat{V_2} = V_{Y_2}  \oplus  \ldots \oplus V_{Y_1Y^{-1}_1} \oplus V_{Y_3Y^{-1}_3} \oplus V_{Y_4Y^{-1}_4} \oplus 2\, V_{Y_2Y^{-1}_2} \oplus  \ldots  \oplus V_{Y^{-1}_2} \label{decompD4} 
	\end{align}
	so the first constraint is satisfied. The nontrivial Drinfeld polynomials are
	\begin{align}
	\lambda = Y_1Y^{-1}_1 \; : \qquad {\cA}^{+}_{\lambda,1}(z) = (1- \hbar^{1/2}\, z) \; ,\qquad  {\cA}^{-}_{\lambda,1}(z) = (1- \hbar^{5/2}\, z)\; , 
	\end{align} 
	so the constraint on the degrees is satisfied. The deformed primary is therefore
	\beq\label{operatorD41}
	\cV_{\{{\lambda}\}}(\tilde{x}) =\; :\Lambda_1\left(\hbar^{-1/2}\,x\right)  \Lambda^{-1}_1\left(\hbar^{-5/2}\,x\right): \; ,
	\eeq
	where the center of mass position $\tilde{x}$ is identified with the $\Lambda^{\pm}_1$ operator position $x$. In our nomenclature, we call this a rigid $q$-primary operator. The term ``rigid" means that one cannot rescale away the relative factor of $\hbar^{-1/2}/\hbar^{-5/2}=\hbar^{2}$ in the argument of the $\Lambda^{\pm}_1$ operators by redefining $x$, in contrast to the generic and semi-degenerate cases. In other words, this operator is determined by the ratio of Drinfeld roots.	In the conformal limit, this is a  $\cW_{\beta}(D_4)$ degenerate primary vertex operator
	\beq
	\label{operatorD41limit}
	e^{\langle \widetilde{\sigma}_{Drinfeld} , \phi(\tilde{x})\rangle} \; ,
	\eeq
	where $\widetilde{\sigma}_{Drinfeld}=[2\beta-2,0,0,0]$ is the $D_4$-(co)weight denoting the momentum of the primary state.\\ 
	
	In fact, there are a total of five rigid $q$-primaries in $\cW_{q,t}(D_4)$. The other four are likewise constructed using single weight spaces found in the decomposition \eqref{decompD4} of the second fundamental representation $\widehat{V_2}$ in $U_\hbar(\widehat{D_4})$:
		\begin{align}
	&\lambda' = Y_3Y^{-1}_3 \; : \qquad {\cA}^{+}_{\lambda,3}(z) = (1- \hbar^{1/2}\, z) \; ,\qquad  {\cA}^{-}_{\lambda,3}(z) =  (1- \hbar^{5/2}\, z)\; , \nonumber\label{D4exmore}\\
		&\lambda'' = Y_4Y^{-1}_4 \; : \qquad {\cA}^{+}_{\lambda,4}(z) = (1- \hbar^{1/2}\, z) \; ,\qquad  {\cA}^{-}_{\lambda,4}(z) =  (1- \hbar^{5/2}\, z)\; , \nonumber\\
			&\lambda''' = Y_2Y^{-1}_2 \; : \qquad {\cA}^{+}_{\lambda,2}(z) = (1- \hbar\, z) \; ,\qquad  {\cA}^{-}_{\lambda,2}(z) =  (1- \hbar^{2}\, z)\; .
	\end{align} 
The weight $\lambda'''$ appears with multiplicity 2 in the decomposition \eqref{decompD4}, so all five weights are accounted for\footnote{This multiplicity 2 for $\lambda'''$ is a feature of affinization: there is no 28-dimensional irreducible representation in $U_\hbar(\widehat{D_4})$; instead, the fundamental representation $\widehat{V_2}$ is 29-dimensional, with decomposition $\widehat{V_2} = V_2\oplus \mathbb{C}$, with $V_2$ the 28-dimensional irreducible representation of $U_{\hbar}(D_4)$. Here, $\mathbb{C}$ stands for one of the two $V_{\lambda'''}$ weight spaces.}. 	
Correspondingly, the vertex operators are
\begin{align}
&\cV_{\{{\lambda'}\}}(\tilde{x}) =\; :\Lambda_3\left(\hbar^{-1/2}\,x\right)  \Lambda^{-1}_3\left(\hbar^{-5/2}\,x\right): \; ,\label{operatorD42}\nonumber\\
&\cV_{\{{\lambda''}\}}(\tilde{x}) =\; :\Lambda_4\left(\hbar^{-1/2}\,x\right)  \Lambda^{-1}_4\left(\hbar^{-5/2}\,x\right): \; ,\nonumber\\
&\cV_{\{{\lambda'''}\}}(\tilde{x}) =\; :\Lambda_2\left(\hbar^{-1}\,x\right)  \Lambda^{-1}_2\left(\hbar^{-2}\,x\right): \; .
\end{align}
In the conformal limit, we obtain degenerate primary operators of momentum $\widetilde{\sigma}'_{Drinfeld}=[0,0,2\beta-2,0]$ and $\widetilde{\sigma}''_{Drinfeld}=[0,0,0,2\beta-2]$, while $\widetilde{\sigma}'''_{Drinfeld}=[0,\beta-1,0,0]$ with multiplicity 2. All these coweights should be thought of as refinements of the null coweight $[0,0,0,0]$ by the ``quantization" parameter $\beta$. Their meaning as highest weight states of reducible Verma modules will be elucidated in \cite{NathanH}.\\
\end{example}

\begin{example}
\emph{A mixed $q$-primary of $\cW_{q,t}(G_2)$}:\\

We choose $\alpha_1$ to be the short root of $G_2$  with  $\langle\alpha_1,\alpha_1\rangle = 2/3$, and $\alpha_2$ to be the long root with $\langle\alpha_2,\alpha_2\rangle = 2$. This sets the integers $r_1=1$ and $r_2=3$ in the definition of the $\cW_{q,t}(G_2)$ algebra \eqref{rlacing}. For convenience, we choose to work with $G_2$ instead of the Langlands dual algebra $^L G_2$ in what follows. Then, consider the two coweight spaces $V^\vee_{\lambda^\vee_1}=V^\vee_{Y^2_2Y^{-1}_2}$ and  $V^\vee_{\lambda^\vee_2}=V^\vee_{Y^{-1}_2}$ in $U_\hbar(\widehat{G_2})$.  Each coweight space appears in the decomposition of a fundamental representation: as $U_\hbar({G_2})$-modules, 
\begin{align}
&\widehat{V^\vee_1} = V^\vee_{Y_1}  \oplus \ldots \oplus V^\vee_{Y^2_2Y^{-1}_2} \ldots \oplus V^\vee_{Y^{-1}_1}\qquad\;\;\;\;\; , \nonumber\label{decompG2}\\
&\widehat{V^\vee_2} = V^\vee_{Y_2}  \oplus  \ldots \oplus V^\vee_{Y^{-1}_2}\; , 
\end{align}
so we find  $V^\vee_{\lambda^\vee_1}$ inside the 15-dimensional representation  $\widehat{V^\vee_1}$ and $V^\vee_{\lambda^\vee_1}$ inside the 7-dimensional representation $\widehat{V^\vee_2}$, and the first constraint is satisfied. Since $G_2$ is non simply-laced, the  Drinfeld polynomials can be identified from the fundamental $(q,t)$-characters; in particular, the Drinfeld roots now depend on $q$ and $t$ separately instead of the ratio $q/t$. The characters are constructed explicitly as the commutant of the screening charges, and the relevant terms are
\begin{align}
&T^{\widehat{V^\vee_1}}(z) = \cY_1(z)  + \ldots +c^{\widehat{V^\vee_1}}_{\lambda^\vee_1}(q,t)\; \cY_2(t^{1/2}q^{-3/2}\,z)\cY_2(t^{1/2}q^{-1/2}\,z)\cY^{-1}_2(t^{3/2}\,q^{-7/2}\,z)+ \ldots \qquad\;\;\;\;\; , \nonumber\label{decompG2char}\\
&T^{\widehat{V^\vee_2}}(z) = \cY_2(z)  +  \ldots + c^{\widehat{V^\vee_2}}_{\lambda^\vee_2}(q,t)\;\cY^{-1}_2(t^3\, q^{-6})\; , 
\end{align}
with $c^{\widehat{V^\vee_1}}_{\lambda^\vee_1}(q,t)= \frac{(q^{3/2}-q^{-3/2})(q^{1/2}t^{-1/2}-q^{-1/2}t^{1/2})}{(q^{1/2}-q^{-1/2})(q^{3/2}t^{-1/2}-q^{-3/2}t^{1/2})}$ and $c^{\widehat{V^\vee_2}}_{\lambda^\vee_2}(q,t)=1$. We readily identify the  nontrivial (refined) Drinfeld polynomials as
\begin{align}
&\lambda^\vee_1 = Y^2_2Y^{-1}_2 \; : \qquad \;\; {\cA}^{\widehat{V^\vee_1},+}_{\lambda^\vee_1,2}(z) = (1-q^{3/2}t^{-1/2} z)\, (1-q^{1/2}t^{-1/2} z)\; , \nonumber\label{drinfeldex4}\\
&\qquad\qquad\qquad\qquad\;\;\;\; {\cA}^{\widehat{V^\vee_1},-}_{\lambda^\vee_1,2}(z) = (1-q^{7/2}t^{-3/2} z)\; ,\nonumber\\
&\lambda^\vee_2 = Y^{-1}_2 \; : \qquad\;\;\;\;\;\; {\cA}^{\widehat{V^\vee_2},-}_{\lambda^\vee_2,2}(z) = (1- q^6 t^{-3}z)\; . 
\end{align} 
The constraint on the degrees is satisfied. The deformed primary is therefore
\beq\label{operatorG2}
\cV_{\{{\lambda^\vee}\}}(\tilde{x}) =\; :\left[\Lambda_2\left(t^{1/2} q^{-3/2}\, x_{1}\right)\Lambda_2\left(t^{1/2} q^{-1/2}\, x_{1}\right)\Lambda^{-1}_2\left(t^{3/2} q^{-7/2}\, x_{1}\right)\right]  \Lambda^{-1}_2\left(t^{3}q^{-6}x_{2}\right): \; ,
\eeq
where the center of mass position $\tilde{x}$ is defined through $x_{s} = \tilde{x} \, q^{\sigma_{s}}$ for $s=1,2$. In our nomenclature, we call this a mixed $q$-primary operator. The term ``mixed" indicates that one can partially rescale away the relative factor of $q$ and $t$ in the argument of the $\Lambda^{\pm}_2$ operators after redefining $x_1$ and $x_2$. Not all Drinfeld roots in $\lambda^\vee_1$ can be rescaled away, however, a feature we already saw in the rigid case. In the case at hand, we will rescale away the Drinfeld root $t^{1/2} q^{-3/2}$ in the first operator $\Lambda_2\left(t^{1/2} q^{-3/2}\, x_{1}\right)$, but the relative factor $\left(t^{1/2} q^{-1/2}\right)/\left(t^{3/2} q^{-7/2}\right)=t^{-1}q^{-3}$ from $\Lambda_2\left(t^{1/2} q^{-1/2}\, x_{1}\right)\Lambda^{-1}_2\left(t^{3/2} q^{-7/2}\, x_{1}\right)$ cannot be rescaled away.
In the conformal limit, we get a pair of  $\cW_{\beta}(G_2)$ degenerate primary operators at $\tilde{x}$, as the momentum coweight splits into two contributions: $\widetilde{\sigma}=\widetilde{\sigma}_{Classical}+\widetilde{\sigma}_{Drinfeld}$, with  $\widetilde{\sigma}_{Classical}=[0,\sigma_1-\sigma_2]$, and $\widetilde{\sigma}_{Drinfeld}=[0,\beta-3]$. The contribution $\widetilde{\sigma}_{Classical}$ is semi-degenerate and originates from the first and last $\Lambda^{\pm}_2$ operators in \eqref{operatorG2}, while $\widetilde{\sigma}_{Drinfeld}$ is rigid and originates from the two middle $\Lambda^{\pm}_2$ operators.\\
\end{example}

\subsubsection{Multiple primary operators}
\label{sssec:multiple}

So far, we have only considered a single  $q$-primary vertex operator insertion on the cylinder $\cC$, labeled by a set of $J$ weights in $U_\hbar(\widehat{^L \fg})$. When we relax the constraint \eqref{constraintpoly2}, this set can be divided into $L\geq 2$ subsets, each of which will obey the condition \eqref{constraintpoly}. This is what we mean by a collection of $L$ deformed primary vertex operators. 

Explicitly, we write the $d$-th vertex operator among them as $\cV_{\{{\lambda}\}_d}(\tilde{x}_d)$, with $d\in\{1,\ldots,L\}$. This operator is labeled by a set of $J_d$ weights $\{{\lambda}\}_d=\{\lambda_{d,1},\ldots,\lambda_{d,J_d}\}$ in $U_\hbar(\widehat{^L \fg})$, as well as $J_d$ complex numbers $x_{d,s}\in\mathbb{C}^\times$, with the same constraints as before: $\sum_{s=1}^{J_d} \lambda_{d,s} = 0$, and each $\lambda_{d,s}\in\{{\lambda}\}_d$ must belong in a weight space $V_{a,\lambda_{d,s}}$ appearing in the decomposition of a fundamental representation $\widehat{V_a}=\bigoplus_{\lambda}\, V_{a,\lambda}$.

Then, inside a chiral correlator, the insertion of $L$ deformed primaries has the form
\begin{align}
\prod_{d=1}^{L}\cV_{\{{\lambda}\}_d}(\tilde{x}_d) =  \prod_{d=1}^{L} :\prod\limits_{s=1}^{|\{{\lambda}\}_d|}\prod\limits_{b=1}^{\L}&\prod\limits_{i=1}^{\deg(\cA^{+}_{\lambda_{d,s},b})}\Lambda_b\left(q^{-\mathfrak{a}^{+}_{\lambda_{d,s},b,i}}\, t^{\widetilde{\mathfrak{a}}^{+}_{\lambda_{d,s},b,i}}\, x_{s}\right)\times\nonumber\label{qprimarydefmany}\\ &\times\prod\limits_{j=1}^{\deg(\cA^{-}_{\lambda_{d,s},b})}\Lambda^{-1}_b\left(q^{-\mathfrak{a}^{-}_{\lambda_{d,s},b,j}}\, t^{\widetilde{\mathfrak{a}}^{-}_{\lambda_{d,s},b,j}}\, x_s\right): \; ,
\end{align}
or in the simply-laced case $\fg=\fg_\text{o}$, 
\beq\label{qprimarydefslshishamany}
\resizebox{0.98\hsize}{!}{$\prod\limits_{d=1}^{L}\cV_{\{{\lambda}\}_d}(\tilde{x}_d) = \prod\limits_{d=1}^{L} :\prod\limits_{s=1}^{|\{{\lambda}\}_d|}\prod\limits_{b=1}^{\text{rk}(\fg_\text{o})}\prod\limits_{i=1}^{\deg(\cA^{+}_{\lambda_{d,s},b})}\Lambda_b\left(\hbar^{-\mathfrak{a}^{+}_{\lambda_{d,s},b,i}}\, x_{d,s}\right) \prod\limits_{j=1}^{\deg(\cA^{-}_{\lambda_{d,s},b})}\Lambda^{-1}_b\left(\hbar^{-\mathfrak{a}^{-}_{\lambda_{d,s},b,i}}\, x_{d,s}\right):$}.
\eeq
There will be $L$ center of mass positions $\tilde{x}_d$, one for each $\cV_{\{{\lambda}\}_d}(\tilde{x}_d)$ operator. Each such position is expressed in terms of the positions $x_{d,s}$ in the argument of the $\Lambda^{\pm 1}_b(x_s)$ operators as
\beq\label{compositionmany}
x_{d,s} = \tilde{x}_d \, q^{\sigma_{d,s}} \; ,\qquad\;\; s=1, \ldots, J_d \; .
\eeq
The ``momentum" coweight of the $d$-th deformed primary will be denoted as  
\beq\label{commomentummany}
\widetilde{\sigma}_d = \widetilde{\sigma}_{d,Classical} + \widetilde{\sigma}_{d,Drinfeld} \; ,
\eeq
with $\widetilde{\sigma}_{d,Classical}$ and $\widetilde{\sigma}_{d,Drinfeld}$ given by \eqref{commomentum2} and \eqref{commomentum3}, respectively. Throughout this work, the positions $\tilde{x}_d$ are always be chosen to be generic. As we saw, the parameters $\sigma_{d,s}$ do not have to be: they are responsible for the rich classification of generic/semi-degenerate/rigid/mixed $q$-primary operators of the $\cW_{q,t}(\fg)$-algebra.\\

\subsubsection{Some comments on characters}

The $q$-character had a clear interpretation as counting isomorphism classes of finite-dimensional representations of $U_{q}(\widehat{\fg})$, but what is the meaning of the $t$-refinement \eqref{qtcharacter}?
Some hints are provided by taking various limits of $\cW_{q,t}(\fg)$, see Section \ref{sssec:dualities}.

First, let $t\rightarrow 1$. In that limit, the $(q,t)$-character morphism \eqref{qtcharacter} reduces to the $q$-character morphism \eqref{qcharacter}, so $t^{\widehat{V}}(z)$ has a natural interpretation as a generating series of central elements in $U_{q}(\widehat{\fg})$ at the critical level \cite{Reshetikhin:1990sq,Ding:1994my}. The further limit $q\rightarrow 1$  in $\cW_{q,1}(\fg)$ reduces the $q$-character of $\widehat{V}$ to the ordinary character \eqref{character}.

Next, consider the limit $q\rightarrow \epsilon$, with $\epsilon = \exp(2\, \pi\, i/\fn_\fg)$. In the simply-laced case, where $\epsilon = 1$, the $(q,t)$-character coincides with the $(t,q)$-character, which therefore reduces to the $t$-character from the previous paragraph in this case. In the non simply-laced case, the limit can be imposed formally in the $(q,t)$-character to produce a $t$-character of $U_{t}(^L\widehat{\fg})$, but the understanding of this limit in representation theory remains elusive.

Alternatively, following the recent works \cite{Frenkel:2021bmx,2022arXiv220111918K,Chen:2018ntf}, the limit $q\rightarrow 1$ in $\cW_{q,t}(\fg)$ for $\fg$ non simply-laced provides a canonical definition for a ring of ``folded $t$-characters" for finite-dimensional representations of a different quantum affine algebra, $U_{t}(\widehat{\fg_\text{o}})$; here, $\fg_\text{o}$ is the simply-laced Lie algebra from which $\fg$ arises as the subalgebra invariant under its outer automorphisms. This folding construction has the advantage of making the connection to representation theory manifest, and we will have more to say about it in Section \ref{sec:nsl}.\\

From these considerations, it is natural to think of the $\cW_{q,t}(\fg)$-algebra generating currents $T^{\widehat{V}}(z)$ (the $(q,t)$-characters \eqref{qtcharacter}) as interpolating between the center of the algebras $U_{q}(\widehat{\fg})$ and $U_{t}(^L\widehat{\fg})$, both at the critical level. This interpolating feature was the preferred interpretation of the $(q,t)$-characters when they were initially constructed \cite{Frenkel:1998}. We stress that in the non simply-laced case, $\cW_{1,t}(\fg)$  is \emph{not} isomorphic to the ring $Rep(U_{t}(^L\widehat{\fg}))$, so the interpolation is by construction quite subtle.\\

Recently, the $(q,t)$-characters have been constructed by Nekrasov from a localization computation in supersymmetric 5-dimensional field theory on $S^1(R') \times \mathbb{C}_q \times \mathbb{C}_t$  \cite{Nekrasov:2015wsu}, a crisp illustration of the BPS/CFT correspondence \cite{Nek2008}\footnote{To clarify the terminology, the term  ``$(q,t)$-character" originally used by Frenkel and Reshetikhin in the 90's \cite{Frenkel:qch} has been largely supplanted by the term ``$qq$-character" since the work of Nekrasov.}. The field theory in question is a $\fg_\text{o}$-type quiver gauge theory with unitary gauge groups; see also the work of Kim \cite{Kim:2016qqs} for the case $\fg_\text{o}=A_1$. Since then, the study of $(q,t)$-characters has undergone a renaissance and remains an active topic of research.

Among the many applications, $(q,t)$-characters have been understood as a generalization of resolvents in matrix models \cite{Nekrasov:2015wsu,Mironov:2016yue}, a certain quantization of Seiberg-Witten geometry \cite{Bourgine:2015szm,Jeong:2019fgx,Haouzi:2020yxy,Haouzi:2020zls}, a generating function of 1/2-BPS Wilson loops in five dimensions \cite{Kim:2016qqs,Assel:2018rcw,Chang:2016iji,Haouzi:2019jzk}, a  quantization of $G$-opers \cite{Frenkel:2020iqq}, the trace of a transfer matrix for certain TQ-integrable systems  \cite{Frenkel:1998,Kimura:2015rgi,Kimura:2017hez},  a count of BPS states in type II string theory with various D-brane defects \cite{Nekrasov:2015wsu,Nekrasov:2016ydq,Assel:2018rcw,Haouzi:2019jzk,Haouzi:2020bso}, and similarly in (refined) topological string theory \cite{Kimura:2017auj}.

In representation theory, the $(q,t)$-characters enter in a more general class of algebras than we have considered so far, the so-called quantum toroidal algebras. For instance, when  $\fg=A_r$, the toroidal algebra in question is  $U_{q,t}(\Hat{\Hat{\fgl_1}})$. This algebra, which is a quantization of the double affinization of $\fgl_1$, goes by many names in the literature: the Ding-Iohara algebra \cite{Ding:1996}, Miki's $(q,t)$-deformed $\cW_{1+\infty}$ algebra \cite{Miki:2007}, the Drinfeld double of the shuffle algebra \cite{2013arXiv1302.6202N}, the elliptic Hall algebra \cite{2005math......5148B,2010arXiv1004.2575S}, the quantum continuous $\fgl(\infty)$ or the spherical part of the $n\rightarrow\infty$ limit in Cherednik's double affine Hecke algebra of type $\fgl_n$ \cite{2010arXiv1002.3100F,FFJMM}\footnote{Technically the definitions of these algebras differ slightly. For instance, some central elements are trivial in the quantum continuous $\fgl(\infty)$ algebra of \cite{2010arXiv1002.3100F,FFJMM}, and the Serre relations are not imposed in the Ding-Iohara algebra of \cite{Ding:1996}.}.  It has two independent complex parameters, which coincide with the parameters $q$ and $t$ of the $\cW_{q,t}(A_r)$ algebras. Explicitly, the usual presentation of the algebra is done in a symmetric way, with three complex parameters $(q_1, q_2, q_3)$ obeying a constraint $q_1\, q_2\, q_3=1$. In the notations of our paper, these parameters are  $(q_1, q_2, q_3) = (q,t^{-1},\hbar^{-1})$. The $(q,t)$-characters are elegantly realized within the algebra $U_{q,t}(\Hat{\Hat{\fgl_1}})$ as the image of a certain Drinfeld current\footnote{One way to visualize this action is to recall that plane partitions (3D Young diagrams) constitute basis elements for representations of $U_{q,t}(\Hat{\Hat{\fgl_1}})$. The Drinfeld current in question acts by simply adding a box to a plane partition to produce another plane partition.} acting on (the tensor product of) Fock representations at level $r$ \cite{2015arXiv151208779B,Bourgine:2016vsq,2020arXiv200304234F}.

The algebra $U_{q,t}(\Hat{\Hat{\fgl_1}})$ enters here because it is known to act on the equivariant K-theory of Hilbert schemes of points on $\mathbb{C}^2$ \cite{2009arXiv0904.1679F,2009arXiv0905.2555S}. In physical terms, this is an action on the instanton moduli space of 5-dimensional supersymmetric gauge theories supported on $S^1\times\mathbb{C}_q\times\mathbb{C}_t$, precisely the ones that we will discuss in Section \ref{sec:5dgauge}. Though we will not pursue the quantum toroidal route in this work, the representation theory of the algebra and its physical applications are currently under active investigation, and are likely to shed new light on the quantum $q$-Langlands program. For a wide range of examples and recent developments, see \cite{2009JMP....50i5215F,Bourgine:2017jsi,Bourgine:2019phm,Bourgine:2021yba,Bourgine:2022scz,Awata:2011ce,Mironov:2016yue,Zenkevich:2020ufs,2022arXiv221214808Z,Gaiotto:2017euk,Liu2022ARA,Negut:2020npc,Awata:2018svb,Harada:2021xnm,Bayindirli:2023byn,Kimura:2023bxy}.\\

\begin{remark}
For completeness, we mention here that the term ``$(q,t)$-character" sometimes refers to a separate construction of $t$-analogs for $q$-characters due to Nakajima \cite{Nakajima:tanalog,Nakajima2001QuiverVA,Nakajima2002tanalogsOQ,Nakajima2010tAnalogsOQ}, which are maps intimately related to our $(q,t)$-characters. Indeed, as was noted by Nekrasov \cite{Nekrasov:2015wsu}, the $(q,t)$-characters have yet another interpretation as difference operators acting on the $\cY^{\pm 1}_b(z)$ fields \eqref{qthighestweight}, and Nakajima's construction conjecturally coincides with the symbol of these difference operators, namely the Poincar\'{e} polynomial of some graded quiver variety. It would be important to make this remark precise. Finally, the space of  ``interpolating $(q,t)$-characters" introduced by Frenkel and Hernandez  \cite{Frenkel:2010wd} also refers to a simpler commutative version of $\cW_{q,t}(\fg)$-algebras closely related to Nakajima's construction.\\
\end{remark}

\subsubsection{Evaluation of the correlators}

Having described the various operators in the free field correlator 
\beq\label{correlatordefW2}
\left\langle v_{\mu_\infty}\; , \prod_{a=1}^{\r} (Q_a^\vee)^{N_a}\; \prod_{d=1}^{L}\cV_{\{\lambda\}_d}(\tilde{x}_d) \; v_{\mu_0} \right\rangle \, ,
\eeq
we proceed to evaluate it via Wick contractions. The results can be expressed exclusively in terms of the $q$-Pochhammer symbol 
\beq
(x \,; q)_\infty = \prod_{k=0}^\infty \left(1-q^k x\right) \; ,
\eeq
for $|x|<1$, and the genus 1 odd theta function
\beq
\theta_{q}(x)= (x \,; q)_\infty\,(q/x \,; q)_\infty \; . 
\eeq
Whenever $|x|>1$ inside the argument of a $q$-Pochhammer symbol, it should instead be replaced by $ (x \,;\, q)_\infty/\theta_{q}(x)=(q/x \,;\, q)^{-1}_\infty$.\\

Explicitly, after writing the screening charges as integrals over positions of screening currents and normal-ordering the operators, the correlator \eqref{correlatordefW2} becomes 
\begin{align}\label{conf1def}
&\prod_{a=1}^{\r}\frac{1}{N_a!} \prod_{i=1}^{N_a}\oint_C \frac{dy_{a,i}}{y_{a,i}}\, y_{a,i}^{\langle\mu_0, \alpha_a\rangle}\cdot\prod_{j\neq i}^{N_a} \left\langle S_a^\vee(y_{a,i})\; S_b^\vee(y_{a,j}) \right\rangle\cdot\nonumber\\
&\qquad\qquad\cdot\prod_{b>a}\;\prod_{j=1}^{N_b}\left\langle S_a^\vee(y_{a,i})\; S_b^\vee(y_{b,j})\right\rangle \cdot \prod_{d=1}^{L}\left\langle  S_a^\vee(y_{a,i})\, \cV_{\{{\lambda}\}_d}(\tilde{x}_d)\right\rangle \; . 
\end{align}
Note we are treating $\prod_{d'\neq d}^{L}\left\langle \cV_{\{{\lambda}\}_d}(\tilde{x}_d)\, \cV_{\{{\lambda}\}_{d'}}(\tilde{x}_{d'})\right\rangle $ as an overall factor outside of the integral, and will ignore it in this work.

The various factors of the integrand are:
\beq\label{zdef}
\prod_{i=1}^{N_a} y_{a,i}^{\langle\mu_0, \alpha_a\rangle} = \prod_{i=1}^{N_a} y_{a,i}^{\frac{\log(z_a)}{\log(q)}} \; 
\eeq
where $\langle \mu_0, \alpha_a\rangle$ is the eigenvalue for the weight vector $v_{\mu_0}$, as defined in \eqref{eigenvalue}.  Physically, $\mu_0$ appears here because the in-state of the correlator is $v_{\mu_0}$ instead of the trivial vacuum state $v_{0}$. In what follows, the various two-point functions will be written with the notation $\langle \mathellipsis\rangle$ to denote the vacuum expectation value  $\langle v_{0}\,\mathellipsis, v_{0} \rangle$. On the right-hand side, we have explicitly introduced parameters $z_a = q^{\langle \mu_0, \alpha_a\rangle}$; when we refer to a $z$-solution, we mean a block analytic in some chamber of these parameters $z_a$.

For screening currents, one finds by direct computation
\beq\label{screena}
\prod_{1\leq i< j\leq N_a}\left\langle S_a^\vee(y_{a,i})\, S_a^\vee(y_{a,j}) \right\rangle = \prod_{1\leq i\neq j\leq N_a}\frac{\left(y_{a,i}/y_{a,j};q^{r_a}\right)_{\infty}}{\left(t\, y_{a,i}/y_{a,j};q^{r_a}\right)_{\infty}}\;\prod_{1\leq i<j\leq N_a} \frac{\Theta\left(t\,y_{a,j}/y_{a,i};q^{r_a}\right)}{\Theta\left(y_{a,j}/y_{a,i};q^{r_a}\right)}
\eeq
for $a=1,\ldots, \r$, and
\beq\label{screenab}
\prod_{1\leq i \leq N_a}\prod_{1\leq j \leq N_b}\left\langle S_a^\vee(y_{a,i})\; S_b^\vee(y_{b,j}) \right\rangle = \prod_{1\leq i \leq N_a}\prod_{1\leq j \leq N_b}\left [ \frac{(\sqrt{q^{r_{ab}}\,t}\, y_{a,i}/y_{b,j};q^{r_{ab}})_{\infty}}{(\sqrt{q^{r_{ab}}/t} \, y_{a,i}/y_{b,j};q^{r_{ab}})_{\infty}}\right]^{\Delta_{a b}}
\eeq
for $b\neq a$. Here,
$\Delta_{ab}=1$ if there is a link connecting nodes $a$ and $b$ in the Dynkin diagram of $\fg$, and 0 otherwise. Note we are taking the product over all $b=2,\ldots, \r$ strictly greater than $a$. We also made use of the non simply-laced notation $r_a=\fn_{\fg}$ when $a$ labels a long root, and $r_a=1$ when $a$ labels a short root, as in \eqref{radef}. Furthermore, $r_{ab}=\text{gcd}(r_a, r_b)$, the greatest common divisor of $r_a$ and $r_b$, meaning
\beq 
r_{ab} = \begin{cases}
\fn_\fg  & \text{when nodes $a$ and $b$ both label long roots in the $\fg$-quiver\ ,} \\
1 &  \text{when either node $a$ or $b$ labels a short root in the $\fg$-quiver\ .}
\end{cases}
\eeq
Each primary vertex operator  $\cV_{\{{\lambda}\}_d}$ is a product of  operators $\prod_{b,s}\Lambda^{\pm}_b(x_{d,s})$, so in order to compute  $\left\langle S_a^\vee(y_{a,i})\, \cV_{\{{\lambda}\}_d}(\tilde{x}_d)\right\rangle $, it is sufficient to evaluate the two-point functions 
\beq\label{vertexscreening}
\left\langle S_a^\vee(y_{a,i})\, \Lambda^{\pm 1}_b(x_{d,s})\right\rangle = \left[(\sqrt{q^{r_a}/t}\; y_{a,i}/x_{d,s}\, ;q^{r_a})_{\infty}\right]^{\pm \delta_{ab}}\; , 
\eeq 
inside the expression \eqref{qprimarydefmany}
The constraint \eqref{constraintpoly} that $\cV_{\{{\lambda}\}_d}$ should have Drinfeld polynomial data of matching degrees implies that its two-point function with screenings can always be expressed as a ratio of $q$-Pochhammer symbols, with the same number of symbols on the numerator and denominator:
\beq\label{qprimarydeflshishatotal}
\left\langle  S_a^\vee(y_{a,i})\, \cV_{\{{\lambda}\}_d}(\tilde{x}_d)\right\rangle =\prod_{b=1}^{\L}\prod_{s=1}^{J_d}\frac{\prod_{k'=1}^{\deg(\cA^{+}_{\lambda_{d,s},b})}\left(q^{\frac{r_a}{2}-\mathfrak{a}^{+}_{\lambda_{d,s},b,k'}}\, t^{\widetilde{\mathfrak{a}}^{+}_{\lambda_{d,s},b,k'}-\frac{1}{2}} y_{a,i}/x_{d,s}\, ;q^{r_a} \right)_{\infty}} {\prod_{k=1}^{\deg(\cA^{-}_{\lambda_{d,s},b})}\left(q^{\frac{r_a}{2}-\mathfrak{a}^{-}_{\lambda_{d,s},b,k}}\, t^{\widetilde{\mathfrak{a}}^{-}_{\lambda_{d,s},b,k}-\frac{1}{2}} y_{a,i}/x_{d,s}\, ;q^{r_a} \right)_{\infty}} \; .
\eeq
In order to fully specify the correlator integral, one also needs to make a choice of contours. We postpone this discussion to Section \ref{sec:3dgauge}.

\begin{remark}
For completeness, we mention in passing that the two-point functions of screening currents with the operators $\cY_b$ from \eqref{YoperatorToda} evaluate to
\beq\label{Yopscreeningex}
\left\langle S_a^\vee(y)\, \cY^{\pm 1}_b(z) \right\rangle = \left[\frac{1-t\, y/z}{1- y/z}\right]^{\pm\delta_{ab}}\; .
\eeq
These are the building blocks needed to derive Ward identities for the $\cW_{q,t}(\fg)$-algebras, computed via insertion of generators $T_a(z)$ in the chiral correlators $\left\langle \ldots\, T_a(z) \right\rangle$, where ``$\ldots$" stands for products of screenings and vertex operators. The Ward identities translate to regularity constraints in the fugacity $z$ which must be obeyed by the correlator on-shell. Since the generators $T_a(z)$ are $(q,t)$-characters, these translate to constraints on the regularity of Laurent polynomials in the monomials $\prod_b\cY^{\pm 1}_b$'s.

In the context of the unramified Langlands correspondence \cite{Aganagic:2017smx}, the relevant vertex operators to insert in a chiral correlator are the fundamental vertex operators $\overline{V}_b^\vee$ from \eqref{fundvertex}; in that case, the two-point functions with screenings give
\beq\label{vertexscreeningex}
\left\langle  S_a^\vee(y)\, \overline{V}_b^\vee(x)\right\rangle =  \left[\frac{\big(\sqrt{q^{r_a}\,t}\; y/x;q^{r_a}\big)_{\infty}}{\big(\sqrt{q^{r_a}/t}\; y/x;q^{r_a}\big)_{\infty}}\right]^{\delta^{ab}}\; .
\eeq
In fact, this expression manifestly coincides with $\left\langle S_a^\vee(y)\,: \Lambda_b(t^{-1} x)\Lambda^{-1}_b(x):\right\rangle$, so each fundamental vertex operator can be written as a product of two ramified vertex operators $\Lambda^{\pm 1}_b$ differing by a $t$-shift, see the discussion under \eqref{fundvertex}. From the perspective of this paper, such a product is a very specific degenerate $q$-primary vertex operator insertion on $\cC$. Likewise, each operator $\cY_b$ can be written as a product of two $\overline{V}^{\pm 1}_b$ operators differing by a $q$-shift, or equivalently as a product of four $\Lambda^{\pm 1}_b$ operators.\\
\end{remark}

\subsubsection{The conformal limit}

The conformal limit from the deformed ${\cW}_{q,t}({\fg})$-algebra to the ${\cW}_{\beta}({\fg})$-algebra is
\beq
t= q^{\beta}\;, \qquad q\rightarrow 1 \; , \;\;   \beta  \;\;\; \text{fixed} \; .
\eeq
Integral blocks in this limit are well-known; for physics applications, see for instance \cite{Dijkgraaf:2009pc,Itoyama:2009sc,Mironov:2010zs,Morozov:2010cq,Maruyoshi:2014eja}.

For the Heisenberg algebra to still make sense in the limit, the generators should be rescaled as
\beq
\alpha_a[k] \rightarrow \frac{\alpha_a[k]}{\log(q)}\; ,\qquad\;\;\; w_a[k] \rightarrow \frac{w_a[k]}{\log(q)} \; .
\eeq
The two-points of screening currents reduce to
\beq\label{screenablimit}
\left\langle S_a^\vee(y_{a,i})\; S_b^\vee(y_{b,j}) \right\rangle \; \rightarrow \;  \left(y_{a,i}-y_{b,j}\right)^{\frac{\beta}{\fn_{\fg}}\langle\alpha^\vee_a,\alpha^\vee_b\rangle}\; ,
\eeq
where $\alpha^\vee_a$ is the $a$-th simple coroot. 

The individual $\Lambda^{\pm 1}_b$ operators do not have a good conformal limit, as is apparent from the two-point \eqref{vertexscreening}. However, their product $\cV_{\{{\lambda}\}_d}$ \emph{does} have a good conformal limit to the primary vertex operator 
\beq
e^{\langle \widetilde{\sigma} , \phi(\tilde{x})\rangle} \; , 
\eeq
where $\widetilde{\sigma}$ is the coweight defined in \eqref{totalcommomentum}. The two-points with a screening current reduces to
\beq
\left\langle S_a^\vee(y_{a,i})\; \cV_{\{{\lambda}\}_d}(\tilde{x}_d) \right\rangle \; \rightarrow \;  \left(y_{a,i}-\tilde{x}_d\right)^{\frac{-\beta}{\fn_{\fg}}\langle\alpha^\vee_a,\widetilde{\sigma}\rangle} \; .
\eeq

\vspace{8mm}

\subsection{The $U_{\hbar}(\widehat{^L\fg})$-algebra blocks}  
\label{ssec:lol}

In the work \cite{Frenkel:1991gx}, I. Frenkel and Reshetikhin define a $\hbar$-deformation of WZW $\widehat{^L\fg}$-type conformal field theories on $\cC$, based on the existence of the quantum affine algebra $U_{\hbar}(\widehat{^L\fg})$. The conformal blocks of the deformed theory are of the form
\beq\label{electriccorrelator}
{\bf F}(\tilde{x}_1, \tilde{x}_2, \ldots, \tilde{x}_L) = \left\langle v_{\nu_\infty},\, \prod_{d=1}^{L} \Phi_{\nu_d}(\tilde{x}_d)  \; v_{\nu_0} \right\rangle 
\eeq
The vectors $v_{\nu_0}$ and $v_{\nu_\infty}$ 
are respectively the highest weight vector of a Verma module $\widehat{V}_{\nu_0}$ and the lowest weight vector of a Verma module $\widehat{V}^\ast_{\mu_\infty}$ over $U_{\hbar}(\widehat{^L\fg})$, at a generic level $^L \kappa\in\mathbb{C}$.\\

In order to give a precise definition of the operators $\Phi_{\nu_d}(\tilde{x}_d)$, one first needs a notion of evaluation representation for quantum affine algebras.

In the more familiar case from classical Lie theory, one considers the composition of an arbitrary representation $V : U(^L \fg)\rightarrow \text{End}(V)$ with a surjective ``evaluation homomorphism",
\begin{align}
\text{ev}_x : & U(\widehat{^L\fg})\rightarrow U(^L \fg) \label{evrep}\nonumber\\
&\text{ev}_x (a\otimes t^n)=x^n\, a \, , \;\;\;\; a\in ^L\fg , \;\; x\in\mathbb{C}^\times \; ,\nonumber\\
&\text{ev}_x (c)=0 \; ,
\end{align}
where $c$ is the central element of $\widehat{^L\fg}$. The composition ${V}\circ\text{ev}_x$ is called an evaluation representation.\\ 

A quantum analog of this construction naively fails because the evaluation homomorphism in no longer well-defined, as the elements $a\otimes t^n$ only make sense classically. Instead, one can define a shift automorphism $D_x: U_{\hbar}(\widehat{^L\fg})\rightarrow U_{\hbar}(\widehat{^L\fg})$, for $x\in\mathbb{C}^\times$, which acts on the Jimbo-Drinfeld generators $e_0$, $f_0$, as $D_x(e^+_0) = x\, e^+_0$, $D_x(e^-_0) = x^{-1}\, e^-_0$, and acts as the identity on all other generators.

In the case of $^L\fg=A_r$, Jimbo proved \cite{Jimbo:1985vd} the existence of a homomorphism of algebras $\text{ev}: U_{\hbar}(\widehat{^L\fg}) \rightarrow U_{\hbar}(^L\fg)$, such that $\text{ev}\circ D_x$ can be thought of as a quantum analogue of the evaluation homomorphism $\text{ev}_x$ we wrote above.

Whenever $^L\fg\neq A_r$, such a homomorphism $\text{ev}$ does not exist, but we can still make use of the shift automorphism, as follows. For an arbitrary simple Lie algebra $^L\fg$, call $V_{\nu}: U_{\hbar}(\widehat{^L\fg})\rightarrow \text{End}(V)$ an evaluation representation of  $U_{\hbar}(\widehat{^L\fg})$ if it has finite length over $U_{\hbar}(^L\fg)$, and if all $U_{\hbar}(^L\fg)$-irreducible subfactors are highest-weight representations of $U_{\hbar}(^L\fg)$.
Then, one defines a family of such representations by composing this map with the shift automorphism $D_x$, for all $x\in\mathbb{C}^\times$. We denote this family by  ${\cal V}_{\nu}(x)=V_{\nu}\circ D_x$.\\

Going back to the correlator \eqref{electriccorrelator}, a ``chiral" vertex operator  $\Phi_{\nu_d}(\tilde{x}_d)$ is labeled by a Verma module ${V}_{\nu_d}$ over $U_{\hbar}({^L\fg})$, of highest weight $\nu_d\in\mathbb{C}$, and by its position $\tilde{x}_d$ on  $\cC=\mathbb{C}^\times$. More precisely, such a vertex operator acts as an intertwiner,
\beq\label{qintertwiner}
\Phi_{\nu_d}(\tilde{x}_d) :\;\; \widehat{V}_{\nu_i} \rightarrow \widehat{V}_{\nu_j} \otimes {V}_{\nu_d}(\tilde{x}_d)\, \tilde{x}_d^{\Delta(\nu_i)-\Delta(\nu_j)}\; .
\eeq
$\widehat{V}_{\nu_{i}}$ and $\widehat{V}_{\nu_j}$ are $U_{\hbar}(\widehat{^L\fg})$ Verma modules of highest weight $\nu_i$ and $\nu_j$, respectively, at generic level $^L \kappa$, and ${V}_{\nu_d}(\tilde{x}_d)$ is an evaluation representation of $U_{\hbar}(\widehat{^L\fg})$ as defined in the previous paragraph. The quantity\footnote{Note that because here the underlying Lie algebra is $^L\fg$, the inner product on the numerator is rescaled compared to the $\fg$ inner product:  $\langle , \rangle_{^L\fg}=\frac{1}{\fn_\fg}\, \langle , \rangle_{\fg}$.}
\beq
\Delta(\nu)=\frac{\langle \nu, \nu+2\,^L\rho\rangle}{2\;^L(\kappa+h^\vee)}
\eeq
is the conformal weight of $v_{\nu}$. As usual, we denoted the Weyl vector and dual Coxeter number of $^L\fg$ by $^L\rho$ and $^L h^\vee$, respectively. The corresponding space of intertwiners is 
\beq\label{Hspace}
H^{\nu_j,\nu}_{\nu_{i}}=\text{Hom}_{U_\hbar(^L\fg)}\left({V}_{\nu_{i}},{V}_{\nu_{j}}\otimes {V}_{\nu}\right) \; ,
\eeq
over the finite quantum group $U_\hbar(^L\fg)$.
The correlator ${\bf F}$ does not take values in the full weight space ${V}_{\nu_1}\otimes{V}_{\nu_2}\otimes\ldots\otimes{V}_{\nu_L}$, but instead in the subspace
\beq
{\bf F}\in \left({V}_{\nu_1}\otimes{V}_{\nu_2}\otimes\ldots\otimes{V}_{\nu_L}\right)_{\nu_0-\nu_\infty}
\eeq
of vectors of weight $\nu_0-\nu_\infty$ in $U_{\hbar}({^L\fg})$, see \eqref{weightspaceram}.\\

Frenkel and Reshetikhin showed  that ${\bf F}$ solves a certain holonomic system of $q$-difference equations called the quantum Knizhnik-Zamolodchikov (qKZ) equations \cite{Frenkel:1991gx}. These equations should be understood as a $q$-deformation of the trigonometric Knizhnik-Zamolodchikov equations \cite{Knizhnik:1984nr}. More precisely, a correlator ${\bf F}$ is fully determined by the qKZ equation, as well as an ordering of the operators $\Phi_{\nu_d}(\tilde{x}_d)$. Put differently, one also needs to specify a chamber such as $|\tilde{x}_1|<|\tilde{x}_2|<\ldots<|\tilde{x}_{L}|$. The equations read, for $s=1,\ldots,L$,
\begin{align}
{\bf F}(\tilde{x}_1,\ldots,q\, \tilde{x}_s,\ldots, \tilde{x}_{L}) &=  R_{s, \,{s-1}}(q\,\tilde{x}_s/\tilde{x}_{s-1})\ldots R_{s, \,1}(q\,\tilde{x}_s/\tilde{x}_1)\left(\hbar^{^L\rho+\frac{\nu_0+\nu_\infty}{2}}\right)_s\times\label{qKZ}\nonumber\\
&\times R_{s, \,{L}}(\tilde{x}_s/\tilde{x}_{L}) \ldots R_{s, \,{s+1}}(\tilde{x}_s/\tilde{x}_{s+1}) \,{\bf F}(\tilde{x}_1,\ldots, \tilde{x}_s,\ldots, \tilde{x}_{L}) \; .
\end{align}
The matrix $R_{i, \, j}(x)$ is an endomorphism  $\text{End}({V}_{\nu_i}(x)\otimes{V}_{\nu_j}(1))$ called the quantum $R$-matrix for the algebra $U_{\hbar}(\widehat{^L\fg})$. 
The notation $(\hbar^{\lambda})_s$, for a given a weight $\lambda$, stands for the action of $\hbar^{\lambda}$ on the $s$-th component ${V}_{\nu_s}$ of the tensor product ${V}_{\nu_1}\otimes{V}_{\nu_2}\otimes\ldots\otimes{V}_{\nu_L}$. For example, 
$\hbar^{\lambda}(v_{\nu_s}) = \hbar^{\langle\lambda , \nu_s\rangle}\, v_{\nu_s}$. The parameter $q$ responsible for the ``$q$-difference" of the equations is related to the quantization parameter $\hbar$ as
\beq
q=\hbar^{-^L(\kappa+h^\vee)} \; .
\eeq
Thinking of ${\bf F}$ as a product of intertwiners \eqref{qintertwiner}, it follows that the solutions to the qKZ equations are valued in 
\beq
\bigoplus_{\lambda_1,\ldots,\lambda_{L-1}} H^{\nu_0,\nu_1}_{\lambda_{1}}\otimes H^{\lambda_1,\nu_2}_{\lambda_{2}}\otimes\ldots\otimes H^{\lambda_{L -1},\nu_{L}}_{\nu_{\infty}} \; , 
\eeq
with $H^{\lambda_i,\nu_s}_{\lambda_{j}}$ as in \eqref{Hspace}.

\vspace{8mm}

\subsubsection{The conformal limit}

In the limit where $\hbar\rightarrow 1$ and $q=\hbar^{-^L(\kappa+h^\vee)}\rightarrow 1$, with $^L(\kappa+h^\vee)$ fixed, the equations \eqref{qKZ} become the trigonometric
Knizhnik-Zamolodchikov equations \cite{Knizhnik:1984nr}\footnote{The reason we recover the trigonometric KZ equations instead of the rational ones is that our Riemann surface is $\mathbb{C}^\times$ instead of $\mathbb{C}$.},
\beq\label{KZ}
^L(\kappa+h^\vee)\, \tilde{x}_s \, \frac{\partial {\bf F}}{\partial \tilde{x}_s} = \left(r_{s0}+r_{s\infty}+ \sum_{\substack{j=1 \\ j\neq k}}^{L} r_{sj}(\tilde{x}_s/\tilde{x}_j)\right) {\bf F}
\eeq
The matrix $r_{ij}(x)$ is called the classical $R$-matrix for the $^L\fg$-type WZW model on $\cC$. It is written explicitly as
\beq
r_{i j}(\tilde{x}_i/\tilde{x}_j)= \frac{\Omega_{i j}\, \tilde{x}_i + \Omega_{j i}\, \tilde{x}_j}{\tilde{x}_i - \tilde{x}_j} \, ,
\eeq
where  
\beq\label{KZdef}
\Omega = \sum_{\alpha>0} e^+_\alpha\otimes e^-_{\alpha} + \frac{1}{2}\sum_{p=1}^r h_p\otimes h_p \; .
\eeq
The first sum is over all positive roots $\alpha$. For each such $\alpha$, we pick a preferred basis of $^L\fg$ with $e^+_\alpha\in{^L\fg}^\alpha$ and $e^-_\alpha\in{^L\fg}^{-\alpha}$, such that $\langle e^+_\alpha , e^-_\alpha\rangle =1$. The vectors $h_p$, for $p=1,\dots,\L$, are chosen 
to make up an orthonormal basis in the Cartan subalgebra of $^L\fg$.


\vspace{10mm}

\section{Tame ramification and 3d gauge theory}
\label{sec:3dgauge}

Throughout this Section, $\fg$ will be a simply-laced Lie algebra $\fg_\text{o}$ of rank $r$.

\subsection{3d $\cN=2$ Drinfeld quiver gauge theories: a definition}
\label{ssec:Drinfelddef}

Our starting point is a 3d quiver gauge theory with $\cN=4$ supersymmetry, on the manifold $S^1_{\cC'}(R_{\cC'})\times M_2$. The quiver is the Dynkin diagram of the Lie algebra $\fg_\text{o} = ADE$. Then, the gauge group is a product of $r=\r$ unitary groups
\beq\label{gaugegroup3d}
G_{3d}=\prod_{a=1}^{r} U(N_a)\; ,
\eeq
where the $N_a$ are all positive integers.
We add bifundamental matter, which consists of $\cN=4$ hypermultiplets in the representation $\oplus_{b>a}\, \Delta_{ab}\,(N_a, \overline{N_b})$ of the group $\prod_{a,b} U(N_a)\times U(N_b)$, where $\Delta_{ab}$ is the incidence matrix of $\fg_\text{o}$, whose entries are $1$ if there is a link connecting nodes $a$ and $b$ in the Dynkin diagram, and $0$ otherwise. The quiver contains a total of $r-1$ such bifundamental hypermultiplets. 
There is an additional global symmetry
\beq\label{topological3d}
F_{C}=\prod_{a=1}^{r} U(1)_{{\cal J}_a} = U(1)^{r}\; ,
\eeq  
called topological symmetry. We denote its maximal torus by $T_C$, and its Cartan subalgebra by $\ft_C$. For all $a=1,\ldots,r$, the symmetry arises from the conserved current ${\cal J}_a=\frac{1}{2\pi}* \text{Tr}\, F_a$, and the associated charge is called the vortex number $k_a$.  Coupling the $a$-th current to an abelian $U(1)$ factor inside the $a$-th gauge group of the quiver results in a Fayet-Iliopoulos (F.I.) term in the action, with corresponding real parameter $\xi_a\in \ft_C$. It is complexified to the parameter $z_a$ by the holonomy along $S^1_{\cC'}$.

The charges of the various fields are constrained by $\cN=4$ supersymmetry, as encoded in a cubic superpotential. In $\cN=2$ notation, this is a superpotential which couples the bifundamental chiral fields to the adjoint chiral field inside the $a$-th vector multiplet, for each quiver node $a$.\\

We explicitly break supersymmetry to $\cN=2$, in two ways: First, let the operators $H$ and $C$ respectively stand for the Cartan generators $J^H_3$ and $J^C_3$ of $U(1)_H\times U(1)_C \subset SU(2)_H\times SU(2)_C$. From a 3d $\cN=2$ standpoint, the off-diagonal Cartan combination $\dfrac{1}{2}(H - C)$ generates a $U(1)_t$ symmetry, which is a flavor symmetry.  By weakly gauging $U(1)_t$, we turn on a real mass $\log(|t|)$ for this symmetry; this is a mass for the adjoint chiral multiplets inside the $\cN=4$ vector multiplets, which breaks supersymmetry to 3d $\cN=2^*$. The mass is complexified to $t$ by the holonomy along $S^1_{\cC'}$.

Second, we introduce additional $\cN=2$ chiral matter\footnote{Even though $\fg_\text{o}$ is simply-laced in this Section, we nevertheless write $^L\fg_\text{o}$ explicitly to keep the notation uniform for all simple Lie algebras; the non simply-laced case will be discussed in Section \ref{sec:nsl}.}:
\begin{align}\nonumber\label{flavorgroup3dgroup}
&\fg = A_r\; : \; \ \,\, \,\qquad F_{H}= U(N^{F,+}_1)\times \left[\prod_{a=2}^{r-1} U(N^{F,\pm}_a)\right]\times U(N^{F,-}_r) \; ,\\
&\fg = D_r\;, E_r\; : \; \;\;\;F_{H}= U(N^{F,+}_1)\times \left[\prod_{a=2}^{r-2} U(N^{F,\pm}_a)\right]\times U(N^{F,-}_{r-1}) \times U(N^{F,-}_r) \; ,
\end{align} 
where the ranks are non-negative integers.
The product indexed by $a$ is a product over all internal nodes of the Dynkin diagram of $\fg$. 
Whenever $a$ and $b>a$ label adjacent nodes in the Dynkin diagram of $\fg$, both $U(N^{F,+}_a)$ and $U(N^{F,-}_b)$ stand for one and the same group $U(N^{F,\pm}_a)$ in our notation \eqref{flavorgroup3dgroup}.

The flavor symmetry $F_{H}$ induces a second type of cubic superpotential term: it couples together chiral fields transforming in bifundamental representations of any two groups among the trio $U(N^{F,\pm}_a)$, $U(N_a)$, $U(N_{a+1})$, for all $a$. Such configurations of chiral multiplets will create ``closed loops" in the quiver diagram. From the explicit form of $F_{H}$ above, this superpotential will only fix \emph{some} of the chiral multiplet charges (for instance, it leaves the charges of chiral multiplets at the ends of a quiver unconstrained).
Not all flavor ranks $N^{F,\pm}_a$  will arise in this work, only a subset relevant to the tamely ramified Langlands program. We propose to define this subset, along with the associated chiral multiplet content of the theory, solely from the definition of the $q$-conformal blocks, as follows:\\

\begin{prop}\label{prop3d}
The data labeling a product of $q$-primary operators $\prod_{d=1}^{L} \cV_{\{\lambda\}_d}(\tilde{x}_d)$ in a $\cW_{q,t}(\fg)$-algebra correlator is in bijection with the chiral multiplet data of a 3d $\cN=2$ quiver gauge theory with flavor symmetry \eqref{flavorgroup3dgroup}. 
\end{prop}

To prove this Proposition, first recall that each vertex operator $\cV_{\{{\lambda}\}_d}$ is labeled by a set of $J_d$ weights $\{{\lambda}\}_d$ taken among the fundamental representations of $U_\hbar(\widehat{^L \fg_\text{o}})$; in particular, any weight $\lambda_{d,s}\in\{{\lambda}\}_d$ is in one-to-one correspondence with the Drinfeld polynomials
\begin{align}\nonumber\label{drinfeldpolygauge}
& {\cA}^{+}_{\lambda_{d,s},b}(z)=\prod_{i=1}^{\deg({\cA}^{+}_{\lambda_{d,s},b})}\left(1- (q/t)^{\mathfrak{a}^{+}_{\lambda_{d,s},b,i}}\, z\right) \; ,\\ 
& {\cA}^{-}_{\lambda_{d,s},b}(z)=\prod_{j=1}^{\deg({\cA}^{-}_{\lambda_{d,s},b})}\left(1- (q/t)^{\mathfrak{a}^{-}_{\lambda_{d,s},b,j}}\, z\right) \; ,\;\; b=1,\ldots,r\; , \;\;\;\; d=1,\ldots,L\, .
\end{align} 
Let
\beq\label{kronecker}
\delta[x] = \begin{cases}
1   & \text{if} \;\; x=0 \\
0   & \text{if} \;\; x\neq 0
\end{cases}
\eeq
be the Kronecker delta function. For all $a=1,\ldots,r$, we will grade the flavor group rank $N^{F,\pm}_a$ by the integer $d\in\{1,\ldots,L\}$ labeling the $d$-th $q$-primary operator, and define 
\beq\label{rankwow}
N^{F,\pm}_a = \sum_{d=1}^{L} N^{F,\pm}_{d,a}\; , \;\text{where}\;\; N^{F,\pm}_{d,a} = \sum_{s=1}^{J_d} \left(1-\delta[\deg({\cA}^{\pm}_{\lambda_{d,s},a})]\right) \; .
\eeq
We further introduce a total of $L^{\pm}_{a}$ chiral multiplets, also graded by $d$, as
\beq\label{chiralnumberwow}
L^{\pm}_{a}=\sum_{d=1}^{L} L^{\pm}_{d,a}\; , \;\text{where}\;\; L^{\pm}_{d,a}=\sum_{s=1}^{J_d}\deg(\cA^{\pm}_{\lambda_{d,s},a}) \; .
\eeq
Here, $L^-_{d,a}$ will stand for the number of \emph{fundamental} chiral multiplets in representation  $(N_a,\overline{N^{F,-}_{d,a}})$  of $U(N_a)\times U(N^{F,-}_{d,a})$, and $L^+_{d,a}$ will stand for the number of \emph{anti-fundamental} chiral multiplets in representation  $(\overline{N_a},N^{F,+}_{d,a})$ of $U(N_a)\times U(N^{F,+}_{d,a})$. The constraint \eqref{constraintpoly} on matching polynomial degrees implies 
\beq\label{equalnumber}
L^+_{d,a} = L^-_{d,a} \; , \qquad\text{for all}\;\; a=1,\ldots,r\; , \;\;\;\; d=1,\ldots,L \; .  
\eeq
The charges of these multiplets under the $U(1)_R$ R-symmetry and $U(1)_t$ flavor symmetry (of mass $t$) will be labeled by the Drinfeld root exponents  $\mathfrak{a}^{\pm}_{\lambda_{d,s},b,j}$  and $-\mathfrak{a}^{\pm}_{\lambda_{d,s},b,j}$, respectively\footnote{On a 3d manifold with boundary, the precise identification actually depends on a further choice of 2d $\cN=(0,2)$ boundary conditions for the chiral multiplets. The exact map between chiral multiplet charges and Drinfeld roots will be given in \eqref{fundmult} for Neumann boundary conditions and in \eqref{antifundmult} for Dirichlet ones.}.\\

Finally, we turn on masses for the chiral multiplets, as follows: denote the maximal torus of the flavor symmetry $F_{H}$ by $T_H$, and the Cartan subalgebra by $\ft_H$. By weakly gauging $F_{H}$, we turn on generic real masses $m_{d,s}\in\ft_H$ with labels $d\in\{1,\ldots,L\}$ and $s\in\{1,\ldots,J_d\}$. That is, to each weight $\lambda_{d,s}$, we associate a unique mass $m_{d,s}$. Once complexified to $x_{d,s}$ by the holonomy around $S^1_{\cC'}$, we obtain as many complex numbers as there are positions on $\cC=\mathbb{C}^\times$ of elementary $\Lambda^{\pm 1}_a(x_{d,s})$ vertex operators inside $\prod_{d=1}^{L} \cV_{\{\lambda\}_d}$. This completes the proof of the proposition.\\

Then, given a magnetic block,
\beq\label{correlatorexample1new}
\left\langle v_{\mu_\infty}\; , \prod_{a=1}^{r} (Q_a^\vee)^{N_a}\; \prod_{d=1}^{L}\cV_{\{\lambda\}_d}(\tilde{x}_d) \; v_{\mu_0} \right\rangle \, ,
\eeq
let the ranks $N_a$ of the 3d gauge groups \eqref{gaugegroup3d} be the above number of screening charges, and the $r$ F.I. parameters \eqref{gaugegroup3d} be the Cartan components of the highest weight $\mu_0$. The bifundamental and adjoint chiral multiplets are fixed by $\cN=4$ supersymmetry, and the $\cN=2$ fundamental/anti-fundamental chiral multiplets are determined from Proposition \ref{prop3d}. The bare Chern-Simons levels are fixed to be 0. We call such a 3d $\cN=2$ quiver gauge theory a {\bf{Drinfeld quiver}}, and denote it as $T^{3d}_{\fg_\text{o}}$.\\

So far, all we have shown is that given the data of a magnetic block \eqref{correlatorexample1new}, there exists a 3d $\cN=2$ gauge theory  \emph{labeled} by this data. 
In Sections \ref{sec:5dgauge} and \ref{sec:littlestring}, we will show that Drinfeld quivers are precisely the ones that arise in the tamely ramified Langlands program, and furthermore all such quivers can be \emph{derived} from 5-dimensional gauge theory and string theory.\\ 

\begin{remark}
By the Langlands correspondence, we could have equally well defined a Drinfeld quiver from the data of an electric block
\beq\label{electriccorrelatornew}
\left\langle v_{\nu_\infty},\, \prod_{d=1}^{L} \Phi_{\nu_d}(\tilde{x}_d)  \; v_{\nu_0} \right\rangle \; .
\eeq
The F.I. parameters are now fixed by the Cartan components of the weight $\nu_0$, the $\cN=2$ chiral flavor content is encoded in the Verma highest weights $\nu_d$, and the gauge group ranks $N_a$ are determined from the requirement that the block is valued in the weight subspace $\nu_0-\nu_\infty$, see \eqref{weightspaceram}.
\end{remark}

We denote the maximal torus of the global symmetry $F_{H}\times F_C$ by $T_H\times T_C$, and the Cartan subalgebra by $\ft_H\oplus\ft_C$. We will keep the parameters $(m_{d,s},\xi_a)\in \ft_H\oplus\ft_C$ generic, and further require  $T^{3d}_{\fg_\text{o}}$ to have isolated massive vacua $\{{\bf{A}}\}$ given those parameters. The global symmetry parameter space is divided into chambers: $m_{d,s}\in\fC_H\in\ft_H$ for the masses and  $\xi_a\in\fC_C\in\ft_C$ for the F.I. parameters. Generically, the vacua are separated by domain walls whose tension is controlled by a mixed Chern-Simons term between $T_H$ and $T_C$.\\ 

All global symmetry parameters are complexified by the holonomy around $S^1_{\cC'}(R_{\cC'})$, with notation 
\begin{align}\label{holonomy}
t=e^{-R_{\cC'}\left(\epsilon_t+i\,A^{\theta}_t\right)}\; , \;\;\;\qquad z_a = e^{-R_{\cC'}\left(\xi_a+i\,A^{\theta}_z\right)} \; ,&\;\;\;\qquad  x_{d,s}=e^{-R_{\cC'}\left(m_{d,s}+i\,A^{\theta}_x\right)} \; .
\end{align}
In this work, we take all F.I. parameters $\xi_a$ generic and positive, and fix the chamber 
\beq\label{Cchamber}
\fC_C= \{\xi_a>0\} \; .
\eeq
We likewise fix an ordering of the chiral multiplet masses, which has a finer structure. First, there is an ordering of masses in a fixed defect: for the $d$-th defect among the $L$ of them, we set $m_{d,1} > m_{d,2} > \ldots > m_{d,J_d}$. Second, there is an ordering of masses between different defects: we set $m_{d,s} > m_{d+1,s'}$ for all $s$ and $s'$.  
It will be convenient to fix a chamber in the center of mass coordinates \eqref{compositionmany}, as
\beq\label{Hchamber}
\fC_H=\{|\tilde{x}_{1}|\ll|\tilde{x}_{2}|\ll\ldots\ll|\tilde{x}_{L}|\}\, ,
\eeq
where we assume that the relative masses  $q^{\sigma_{d,j}}$ are ordered as
\beq
1<|q^{\sigma_{d,1}}|<\ldots<|q^{\sigma_{d,J_d}}|
\eeq
for all $d=1,\ldots, L$, with the understanding that $|q|<1$.

\vspace{8mm}

\subsection{Examples of Drinfeld quivers}

\begin{figure}[h!]
	\emph{}
	\centering
	\includegraphics[trim={4 0 0 0cm},clip,width=1.0\textwidth]{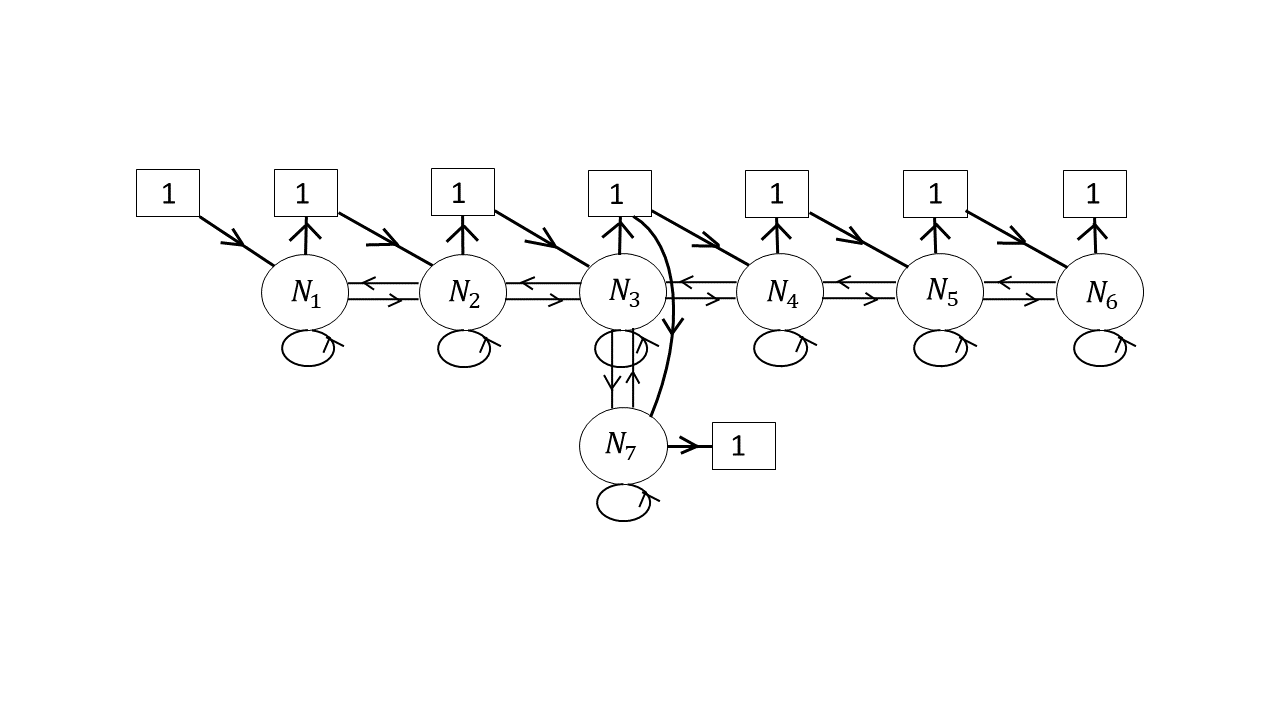}
	\vspace{-2cm}
	\caption{An example of 3d $\cN=2$ Drinfeld quiver for $\fg=E_7$. This particular case is an instance of a handsaw quiver.} 
	\label{fig:drinfeldquiver2}
\end{figure}

Examples of Drinfeld quiver gauge theory are given in Figures \ref{fig:drinfeldquiver} and \ref{fig:drinfeldquiver2}. 

In the example \ref{fig:drinfeldquiver}, the 3d $\cN=2$ quiver gauge theory is of $\fg=D_4$ type, and is defined via the correlator
\beq\label{correlatorexample1}
\left\langle v_{\mu_\infty}\; , \prod_{a=1}^{4} (Q_a^\vee)^{N_a}\; \prod_{d=1}^{3}\cV_{\{\lambda\}_d}(\tilde{x}_d) \; v_{\mu_0} \right\rangle \, ,
\eeq
The specific $q$-primary operators which appear here are labeled by the quantum affine weights
\begin{align}
&\{\lambda\}_1 = \{Y_1, Y_2\, Y^{-1}_1, Y^{-1}_2\}\; ,\label{weightsex}\nonumber\\
&\{\lambda\}_2 = \{Y_1, Y_2\, Y^{-1}_1, Y^{-1}_2\}\; ,\nonumber\\
&\{\lambda\}_3 = \{Y_3\, Y^{-1}_3\}\; ,
\end{align}
where the last weight was already featured in the example \eqref{D4exmore}, and called $\lambda'$ there.\\

The second example \ref{fig:drinfeldquiver2} is labeled by only one operator $\cV_{\{\lambda\}}(\tilde{x})$, and is part of a class of quivers known as \emph{handsaw} quivers. These were first introduced by mathematicians for $\fg_\text{o}=A_r$ \cite{2010arXiv1009.0676F,Nakajima:2011yq}, and later reinterpreted physically in \cite{Aganagic:2014oia,Aganagic:2015cta}. For all $\fg_\text{o}=ADE$, the generic $\fg_\text{o}$-type handsaw quiver is engineered using a single $q$-primary insertion defined by a set of $r+1$ weights $\{{\lambda}\}$ of $U_\hbar(\widehat{^L \fg_\text{o}})$, see the Table \ref{table:handsaws}. It is an instructive exercise to show that each set $\{{\lambda}\}$ below defines a valid generic $q$-primary vertex operator $\cV_{\{{\lambda}\}}$ of $\cW_{q,t}(\fg_\text{o})$, according to our definition \eqref{sssec:definition}. The quiver in the figure represents the $E_7$ column in the table.\\

\begin{tabular}{|l|l|l|l|l|}
	\cline{1-5}
	\multicolumn{1}{|c|}{$\fg=A_r$}&
	\multicolumn{1}{c|}{$\fg=D_r$}& 
	\multicolumn{1}{c|}{$\fg=E_6$}&
	\multicolumn{1}{c|}{$\fg=E_7$}&
	\multicolumn{1}{c|}{$\fg=E_8$}\\ \hline
	$\lambda_1 = Y_1$ &  $\lambda_1 = Y_1$ & $\lambda_1 = Y_1$ & $\lambda_1 = Y_1$ & $\lambda_1 = Y_1$ \\
		$\lambda_2 = Y_2Y^{-1}_1$ &  $\lambda_2 = Y_2Y^{-1}_1$ & $\lambda_2 = Y_2Y^{-1}_1$ & $\lambda_2 = Y_2Y^{-1}_1$ & $\lambda_2 = Y_2Y^{-1}_1$ \\    
	$\lambda_3 = Y_3Y^{-1}_2$ &  $\lambda_3 = Y_3Y^{-1}_2$ & $\lambda_3 = Y_3Y^{-1}_2$ & $\lambda_3 = Y_3Y^{-1}_2$ & $\lambda_3 = Y_3Y^{-1}_2$\\	
	 \;\;\qquad$\svdots$ &  \;\;\qquad$\svdots$ & $\lambda_4 = Y_4 Y_6 Y^{-1}_3$ & $\lambda_4 = Y_4 Y_7 Y^{-1}_3$ & $\lambda_4 = Y_4 Y_8 Y^{-1}_3$\\
	 $\lambda_{r} = Y_{r}Y^{-1}_{r-1}$ & $\lambda_{r-2} = Y_{r-2}Y^{-1}_{r-3}$  & $\lambda_5 = Y_5 Y^{-1}_4$ &  $\lambda_5 = Y_5 Y^{-1}_4$ & $\lambda_5 = Y_5 Y^{-1}_4$\\
	  	  $\lambda_{r+1} = Y^{-1}_{r}$ & $\lambda_{r-1} =Y_{r-1}Y_{r} Y^{-1}_{r-2}$  & $\lambda_6 =  Y^{-1}_5$ &  $\lambda_6 = Y_6 Y^{-1}_5$ & $\lambda_6 = Y_6 Y^{-1}_5$\\
	  	   &  $\lambda_{r} = Y^{-1}_{r-1}$ & $\lambda_7 =  Y^{-1}_6$ &  $\lambda_7 =  Y^{-1}_6$ & $\lambda_7 = Y_7 Y^{-1}_6$\\
	  	   &  $\lambda_{r+1} = Y^{-1}_{r}$ &  &  $\lambda_8 =  Y^{-1}_7$ & $\lambda_8 =  Y^{-1}_7$\\
	  	    &   &  &   & $\lambda_9 =  Y^{-1}_8$\label{table:handsaws}\\ \hline
\end{tabular}

\vspace{8mm}

\begin{remark} 
Handsaw quivers, and more precisely their affine counterparts known as chainsaw quivers, have appeared in a related context: chainsaw varieties physically describe the moduli space of \emph{ramified instantons} in $U(N)$ Super Yang-Mills on $\mathbb{C}_q\times\mathbb{C}_t$, that is the moduli space of instantons in the presence of an additional Gukov-Witten monodromy surface operator supported on  $\mathbb{C}_q\times\{0\}$ \cite{Kanno:2011fw,Alday:2010vg,Nekrasov:2017rqy}. Mathematically, this moduli space is known to be an affine Laumon space of a specific type dictated by the monodromy, that is to say a moduli space of framed parabolic torsion-free sheaves on $\mathbb{P}^1\times\mathbb{P}^1$, where the parabolic structure is encoded in a flag of subsheaves  along  $\mathbb{P}^1\times\{0\}$ \cite{2008arXiv0812.4656F,2010arXiv1009.0676F,Negut:2011aa}. The connection to  quivers arises from the ADHM construction, which realizes such a Laumon space as the moduli space of cyclic (framed) representations of a chainsaw quiver. For recent developments, see \cite{2017arXiv170801795F,Negut:2019agq,Creutzig:2022cch}.
Our construction here is also related to Gukov-Witten monodromy defects, but in a different fashion: the parabolic Laumon space is the Higgs branch of a 3d handsaw quiver gauge theory $T^{3d}_{\fg_\text{o}}$, which further coincides with the moduli space of 1/2-BPS vortices of charge $(N_1,N_2,\ldots,N_r)$ on the Higgs branch of a 5d quiver gauge theory $T^{5d}_{\fg_\text{o}}$ \cite{Nakajima:2011yq,Aganagic:2014oia}. For 3d Drinfeld quivers which are not handsaw, the description of the Higgs branch as a moduli space of codimension-2 vortices still holds, as we will see in see in Section \ref{sec:5dgauge}, though it is no longer parabolic Laumon.
\end{remark}

\vspace{8mm}

\subsection{The half-index of the 3d $\cN=2$ theory}
\label{ssec:3dindex}

In this Section, we will show that a certain supersymmetric (half-)index of $T^{3d}_{\fg_\text{o}}$ coincides with the magnetic and electric $q$-conformal blocks of $\cW_{q,t}(\fg_\text{o})$ and $U_\hbar(\widehat{^L \fg_\text{o}})$ algebras.\\

Perhaps the simplest definition of an index in our context is as a partition function on the 3-manifold $M_3 = S^1_{\cC'}(R_{\cC'})\times\mathbb{C}$, with time along $S^1_{\cC'}$. In this case, a regularization prescription is needed to cure the non-compactness of $\mathbb{C}$. One such prescription is to introduce a massive deformation of the background, where one views $M_3$ as a $\mathbb{C}$ bundle over $S^1_{\cC'}$: let $(X_{\cC'},X_1)$ be coordinates on $S^1_{\cC'}\times\mathbb{C}$, and identify
\begin{align}\label{omega3d}
(0, X_1) \sim  (2\pi R_{\cC'}, q\, X_1) \; .
\end{align}
We will denote this background as $S^1_{\cC'}\times\mathbb{C}_q$, sometimes called a Melvin cigar, or 3d $\Omega$-background with equivariant parameter $q$, meaning the fugacity $\log(q)$ is conjugate to the Cartan generator $J$ of the $U(1)_J$ action rotating $\mathbb{C}$. A count of BPS states in this background is known as a holomorphic block \cite{Beem:2012mb}.  It was shown in \cite{Aganagic:2013tta,Aganagic:2014oia,Aganagic:2015cta} that $\cW_{q,t}(\fg)$-algebra $q$-conformal blocks are examples of such holomorphic blocks.\\

The behavior of the various fields has to be specified at the infinity of $\mathbb{C}$, which in turn labels the choice of a vacuum, also known as a thimble boundary condition. Then, a holomorphic block is intrinsically defined in the IR. It would also be desirable to have a UV definition of the partition function, meaning an index which will count protected BPS operators identified straight from the classical fields of $T^{3d}_{\fg_\text{o}}$'s Lagrangian, along with  $\cN=(0,2)$ boundary conditions (and possibly additional degrees of freedom) prescribed at finite distance. 
From this UV perspective, it is more natural to take $M_3=S^1_{\cC'}\times D^2$, where $D^2$ is a formal disk or hemisphere, meaning  $M_3$ has a torus boundary $S^1_{\cC'}\times S^1_{D^2}$. We will assume that $T^{3d}_{\fg_\text{o}}$ retains $\cN=(0,2)$ superconformal symmetry on the $S^1_{\cC'}\times S^1_{D^2}$ boundary\footnote{Since we have Langlands applcications in mind, we will not consider $\cN=(1,1)$ supersymmetry on the boundary.}, and will require the global symmetry of this boundary theory to preserve at least the maximal torus of the bulk global symmetry. We will also allow symmetries arising from additional degrees of freedom localized on the boundary.

Both in the bulk and on the boundary, the R-symmetry is $U(1)$. We denote the boundary symmetry simply as $U(1)_R$, with generator $R$. The bulk $U(1)$ R-symmetry is either preserved at the boundary, in which case we identify it directly as the $U(1)_R$ symmetry, or it can be spontaneously broken at the boundary; when this is the case, it can still happen that a linear combination of R-symmetry and boundary flavor symmetry is preserved at the boundary, as will be the case when we discuss exceptional Dirichlet boundary conditions. For ease of notation, the preserved symmetry in such a scenario will also be denoted as $U(1)_R$.

The partition function on the 3-manifold $S^1_{\cC'}\times D^2$ is sometimes called a half-index\footnote{The more familiar ``full index" would be related instead to partition functions on $S^1_{\cC'}\times S^2$,  $S^3$ \cite{Kim:2009wb,Imamura:2011su,Kapustin:2011jm,Alday:2013lba,Cecotti:2013mba,Fujitsuka:2013fga,Benini:2013yva}, or $S^1_{\cC'}\times \mathbb{RP}^2$, $S^1_{\cC'}\times S^2_b$, which are circle fibrations over the real projective space $\mathbb{RP}^2$ and the squashed sphere $S^2_b$ \cite{Tanaka:2014oda}, respectively.} \cite{Yoshida:2014ssa}, and is most easily computed via equivariant localization, as a trace over the Hilbert space of states on $D^2$. The fugacity $\log(q)$ is still conjugate to the Cartan generator $J$ of $U(1)_J$, now understood as an eigenvalue for azimuthal rotations on the hemisphere. The partition function can be thought of as a generating function of BPS operators which saturate the BPS bound $D-R-J\geq 0$, with $D$ a Cartan generator for the $U(1)$ rotation along $S^1_{\cC'}$. In the specific case where the bulk 3d theory happens to be trivial, the only contributions to the counting will come from the $T^2$ boundary, and the partition function then goes by the name of (flavored) elliptic genus \cite{Schellekens:1986yi,Schellekens:1986xh,Witten:1986bf,Gadde:2013dda,Benini:2013xpa,Benini:2013nda}.\\

The half-index has yet another meaning, as a BPS count of local operators at the boundary of 3d $\cN=2$ Minkowski spacetime $\mathbb{R}^{1,1}\times\mathbb{R}_{-}$, with the  $\cN=(0,2)$ boundary located at $\mathbb{R}^{1,1}\times\{0\}$. This will be our preferred approach to perform explicit computations.
Then, we introduce coordinates $(x^0, x^1)$ on the boundary $\mathbb{R}^{1,1}\times\{0\}$ and a ``bulk" coordinate $x^\bot\leq 0$ on $\mathbb{R}_{-}$, with boundary located at $x^\bot=0$. The 3d $\cN=2$ algebra has four real supercharges, or equivalently two complex spinors $Q_{\alpha}$ and $\overline{Q}_{\alpha}$, where $\alpha=+,-$. The supercharges $(Q_+, Q_-, \overline{Q}_+, \overline{Q}_-)$ have $U(1)_R$ R-charge $(-1,-1,1,1)$, and obey the anticommutator relations
\begin{align}
&\{Q_{\pm},\overline{Q}_{\pm}\}=\mp 2 \, P_{\pm}\; , \nonumber\label{commutators}\\
&\{Q_{\pm},\overline{Q}_{\mp}\}=\mp 2\, i \, (P_{\bot}\mp i \, Z)\; .
\end{align}
Above, $P_\alpha$ denotes the momentum on $\mathbb{R}^{1,1}$ and $Z$ is a real central charge.\\

It follows from \eqref{commutators} that a valid 1/2-BPS $\cN=(0,2)$ subalgebra preserved on the boundary is one generated by the supercharges  $Q_+$ and $\overline{Q}_+$. The half-index is defined in $\overline{Q}_+$-cohomology, as a character over the vector space of local operators at $x^\bot=0$:
\beq
\label{3dhalfindex}
{\mathcal Z}(T^{3d}_{\fg_\text{o}})  = {\rm Tr}\left[(-1)^F\, q^{J+\frac{R}{2}} \; {\bf{x}}^{\Pi} \right]\;\; .
\eeq
The fugacity $\log(q)$ is conjugate to the Cartan combination $J+\frac{R}{2}$, where $J$ generates the $U(1)_J$ action rotating the boundary plane, while $R$ generates the  $U(1)_R$ R-symmetry. The variable $\log({\bf{x}})$ stands collectively for all fugacities conjugate to the global symmetry generator $\Pi$ preserved on the boundary. Both $J+\frac{R}{2}$ and $\Pi$ commute with the supercharges. $F=2 J$ denotes the fermion number operator.

To our knowledge, this index was first studied and evaluated in the works \cite{Gadde:2013wq,Gadde:2013sca}, where the 3d gauge multiplets were all given Neumann boundary conditions. In this case, the gauge symmetry is preserved at the boundary, and is implemented in the index via Gauss' law, by which we mean that the trace becomes a finite-dimensional contour integral over the gauge fugacities to project over gauge-invariant operators. More recently, the index was revisited in \cite{Dimofte:2017tpi}, where it was further generalized to allow for gauge fields with Dirichlet boundary conditions. In that case, the gauge symmetry is broken to a global symmetry at the boundary, and the index takes the form of a sum over non-perturbative boundary monopole operator contributions. Both sets of boundary conditions will be important for us in interpreting the index as a $q$-conformal block, depending on whether we seek $z$-solutions or $x$-solutions to the qKZ equations.\\

The twist of 3d $\cN=2$ theories with respect to our supercharge $\overline{Q}_+$ was studied in \cite{Aganagic:2017tvx}. This twist is simultaneously topological in the $\mathbb{R}_{-}$-direction and holomorphic on the boundary $\mathbb{C}$. In fact, such a twist is defined for any 3-manifold with a transversely holomorphic foliation; see also the supergravity construction of \cite{Closset:2012ru}. In particular, one finds that the boundary operators in the $\overline{Q}_+$-twisted theory have the structure of a chiral algebra, and the index therefore has a further interpretation as the character of a vacuum module for this algebra. The algebra of local operators in this twisted theory with 2d $\cN=(0,2)$ boundary  was studied in detail in \cite{Costello:2020ndc}.\\

In the end, whether we compute the 3d half-index via the count \eqref{3dhalfindex} or via a hemisphere partition function is not fundamentally important in this work, because both expressions agree up to 1-loop contributions of boundary mixed 't Hooft anomalies \cite{Bullimore:2020jdq}. These contributions always factor out and can be safely ignored as far as the qKZ-equations are concerned. By abuse of notation, we will also use the terms 3d partition function, half-index, and index interchangeably in what follows.\\

In the absence of chiral matter, the supersymmetry of the 3d quiver would have been enhanced to $\cN=4$, with bulk R-symmetry  $SU(2)_H \times SU(2)_C$. The maximal torus of that R-symmetry is $U(1)_H \times U(1)_C$, and is denoted as $U(1)_V \times U(1)_A$ on the $\cN=(2,2)$ boundary. The identification of the $U(1)_R$ boundary R-symmetry in $T^{3d}_{\fg_\text{o}}$ from the $\cN=(2,2)$ perspective turns out to be a subtle point: in this paper, the boundary generator $R$ will \emph{not} be the diagonal combination $\frac{1}{2}(V + A)$ of $U(1)_V \times U(1)_A$, but rather by the generator $V$ of $U(1)_V$.
Choosing $R=\frac{1}{2}(V + A)$ would generically turn the index into a Taylor series in the fugacity $q^{1/4}$, which in turn makes the $U(1)_R$ symmetry unnaturally $4\pi$-periodic. One could always choose $R=V+A$ to properly normalize $U(1)_R$ to be $2\pi$-periodic, but this comes at the expense of giving $\overline{Q}_+$ an R-charge of 2\footnote{For the choice $R=\frac{1}{2}(V + A)$, the index has recently been computed in setups with twice as much supersymmetry: in the absence of $\cN=2$ chiral matter, the index counts local operators in 3d $\cN=4$ gauge theories with 2d (0,4) boundary conditions \cite{Okazaki:2019bok}, with 2d (2,2) boundary conditions \cite{Bullimore:2020jdq,Okazaki:2020lfy}, and in 4d $\cN=4$ Yang-Mills with 2d (0,4) defects \cite{Gaiotto:2019jvo}. Our R-symmetry choice $U(1)_R = U(1)_V$ implies that in all of these works, the fugacity $t$ will be related to ours by a redefinition $t_{\text{here}}=q^{1/2}\,t_{\text{there}}$, where the precise definition of $t_{\text{there}}$ varies slightly from paper to paper.}.

Furthermore, , we will find by direct computation that the fugacity $q$ in the half-index a fortiori matches the fugacity $q$ in $\cW_{q,t}$-algebra blocks precisely when the R-symmetry is chosen as $U(1)_R = U(1)_V$. 
We henceforth write the R-symmetry generator as $R=V$. Then, our index is technically the so-called $A$-twisted half-index \cite{Benini:2015noa,Closset:2016arn,Bullimore:2018jlp}. We stress that we do \emph{not} perform the topological twist with respect to our supercharge here, which among other things would redefine $q^{J+\frac{R}{2}}\rightarrow q^J$ and modify the spin statistics on $J$ inside the index.\\

We decompose the global symmetry explicitly as $U(1)_{t}\times F_C\times F_{H}$, so that the index takes the form
\beq
\label{3dhalfindexmore}
{\mathcal Z}(T^{3d}_{\fg_\text{o}})  = {\rm Tr}\left[(-1)^F\, q^{J+\frac{V}{2}}\; t^{\frac{A-V}{2}} \; z^{\ft_C}\; x^{\ft_H} \right]\;\; .
\eeq
In this notation, $J$ has half-integer eigenvalues, while $V$ and $A$ both have integer eigenvalues, meaning the index is indeed a Taylor series in $q^{1/2}$. The operators $\ft_C$ and $\ft_H$  respectively stand for the Cartan generators of the topological symmetry $F_C=\prod_{a=1}^r U(1)_{{\cal J}^a}$ and the flavor group $F_{H}$ in \eqref{flavorgroup3dgroup} for the quiver preserved at the boundary. Lastly, $\frac{1}{2}(A - V)$  generates the $U(1)_{t}$ flavor symmetry. 

\begin{remark}
	Note it is also possible to define the $B$-twisted index with respect to the operator $A$ generating $U(1)_A\subset U(1)_V \times U(1)_A$; the index was recently computed in \cite{Crew:2023tky,Dedushenko:2023qjq}, with the topological twist implemented\footnote{The parameter $\hbar$ in \cite{Dedushenko:2023qjq} is called $t$ in \cite{Crew:2023tky} and in our paper.}. The $B$-index is related to our $A$-index by the exchange 
	\beq
	U(1)_V \longleftrightarrow U(1)_A \; , \qquad t \longleftrightarrow \hbar=q/t \; .
	\eeq
This exchange is a key aspect of 3d mirror symmetry, which is implemented via a further switch of masses with F.I. parameters \cite{Intriligator:1996ex,Aharony:1997bx,2020arXiv200805516D,2020arXiv201108603D}. Later, when we work in 5d, this symmetry will be realized via spectral duality instead, see Section \ref{ssec:spectral}.
\end{remark}

\vspace{8mm}

\subsection{The $\cW_{q,t}(\fg_\text{o})$ blocks in 3d gauge theory}

\subsubsection{Magnetic $z$-solutions}


In its integral form, the $z$-solution block \eqref{correlatordefW2} is known to be a 3d holomorphic block on $S^1_{\cC'}\times\mathbb{C}_q$ in the IR \cite{Aganagic:2013tta,Aganagic:2014oia,Aganagic:2015cta}. This Section is dedicated to reinterpreting this result in the UV.

We begin by imposing ${\bf N}$ boundary conditions for the $\cN=2$ gauge fields. Because this choice preserves gauge symmetry at the boundary, we must ensure that there are no gauge anomalies. One way to achieve this is to turn on and tune bare Chern-Simons couplings, but we decided to keep their levels at 0 throughout. Instead, we cancel the anomaly by giving N boundary conditions to the $\r$ adjoint chiral fields. The remaining matter content in the quiver is chosen not to introduce new gauge anomalies: the 3d $\cN=4$ bifundamental hypermultiplets between gauge nodes decompose into two $\cN=2$ chiral multiplets, which are respectively given N and D boundary conditions in equal number. For the $\cN=2$ chiral multiplets transforming in $F_H$, we impose N boundary conditions on the $\sum_{a=1}^{r} L^+_a$ fundamental chiral multiplets, and D boundary conditions for the remaining $\sum_{a=1}^{r} L^-_a$ anti-fundamental ones. 
In our definition of Drinfeld quivers, the number of fundamental and anti-fundamental chiral multiplets agree \eqref{equalnumber}, $L^+_a=L^-_a$ for all $a=1,\ldots,r$. This ensures the cancellation of the gauge anomaly.\\

Following the Appendix \ref{sec:appendixindex}, the partition function takes the form
\begin{align}\label{part3d}
{\mathcal Z}(T^{3d}_{{\fg_\text{o}},{\bf N}})= \oint_C  \frac{dy}{y}\, I^{3d}(y,x_{d,s},z) \; .
\end{align}
This is a multi-dimensional complex integral, where the integration variables $y$ collectively stand for the 3d Coulomb parameters, valued in the maximal torus of the gauge group $G_{3d}=\prod_{a=1}^{r} U(N_a)$: 
\beq\label{yvariablesdef}
``\oint_C \frac{dy}{y}\; "\equiv\frac{1}{|W_{G^{3d}}|}\prod_{a=1}^{r} \prod_{i=1}^{N_a}\oint_C \frac{dy_{a,i}}{y_{a,i}} \; .
\eeq
The normalization by the order of the Weyl group $W_{G^{3d}}=\prod_{a=1}^{r} N_a!$ is needed to prevent overcounting contributions to the integral.
The integrand $I^{3d}(y)$ can be read off directly from the 3d $\cN=2$ quiver Lagrangian of the theory. It has the following universal form: 
\begin{align}\label{bulk3d}
I^{3d}(y,x_{d,s},z) = I_a^{3d,F.I.}\cdot I_a^{3d, vec}\cdot I_a^{3d, flavor}\cdot\prod_{b>a}I_{a,b}^{3d, bif}\; .
\end{align}
There will be various 't Hooft anomalies dynamically generated at one-loop, which will further modify the partition function. In particular, with our choice of $I_a^{3d, vec}$, $I_a^{3d, flavor}$ and $I_{a,b}^{3d, bif}$, we should expect a mixed topological/gauge anomaly between $T_C$ and the gauge symmetry maximal torus; this results in a breaking of the topological symmetry $T_C$ at the boundary. In the index, the anomaly manifests itself as the factor
\beq\label{anomalyFI}
I_a^{3d, F.I.} = \prod_{i=1}^{N_a} e^{\frac{\ln(z^\#_a) \ln(y_{a,i})}{\ln(q)}} \; .
\eeq
This notation makes it clear that the corresponding boundary anomaly polynomial is $2 \,f_{a,i}\, f_{z_a}$, where $f_{a,i}$ is the curvature of the corresponding dynamical $U(1)$ gauge symmetry at the boundary, and $f_{z_a}$ is the curvature of the $a$-th $U(1)$ topological symmetry. The parameters 
\begin{align}\label{FI3d}
z_a=q^{\langle \mu_0, \alpha_a \rangle} 
\end{align}
are the $r$ complexified F.I. parameters introduced in the last section, and $z^\#_a$ is a rescaling of $z_a$ by a certain power of $(q/t)$, to be discussed below.  In the $q$-exponent of $z_a$, we recognize the eigenvalues of the $q$-deformed Heisenberg algebra zero modes, obtained after acting on the vacuum state vector $v_{\mu_0}$; that is, $\alpha_a[0]\, v_{\mu_0} = \langle\mu_0, \alpha_a\rangle \, v_{\mu_0}$. We recognize the contribution \eqref{zdef} to the $q$-conformal block.\\

The factor
\begin{align}\label{vec3d}
I_a^{3d, vec}(y_{a,i}/y_{a,j}) =\prod_{1\leq j\neq i\leq N_a} \frac{\left(y_{a,i}/y_{a,j};q\right)_{\infty}}{\left(t\, y_{a,i}/y_{a,j};q\right)_{\infty}}
\end{align}
is the contribution of the $\cN=4$ vector multiplet on node $a$. In $\cN=2$ notation, the numerator is the contribution of the vector multiplet with ${\bf N}$ boundary conditions, along with a $q$-Vandermonde factor to correctly project to gauge invariants. Inside this multiplet, the gaugino and its derivatives contribute to the index with charge
\beq
(\text{vector}_a,{\bf N}) : \qquad\left(\text{adj},n+\dfrac{1}{2},1,0\right)\;\;\; \text{under}\;\;\; U(N_a)\times U(1)_J\times U(1)_R\times U(1)_t\; .
\eeq 
The denominator is the contribution of the adjoint chiral multiplet, also with N boundary conditions. Inside this multiplet, the scalar and its derivatives contribute to the index with charge 
\beq
(\text{adj chiral}_a,\text{N}) : \qquad \left(\text{adj},n,0,1\right)\;\;\; \text{under}\;\;\; U(N_a)\times U(1)_J\times U(1)_R\times U(1)_t\; .
\eeq 
In the $q$-conformal block, we recognize the two-point function 
\beq
I_a^{3d, vec}(y_{a,i}/y_{a,j})=\left\langle S_a^\vee(y_{a,i})\, S_a^\vee(y_{a,j}) \right\rangle
\eeq 
of two screening currents on node $a$. The equality holds up to ratios of theta functions, as in \eqref{screena}. These ratios are a simple constant $(q/t)^{\#}$ representing additional mixed anomalies. This factor is inconsequential from the point of view of solving $q$-difference equations, which is good enough for our purposes, so we simply reabsorb this factor in the definition of the F.I. parameters $z_a \rightarrow z^\#_a$.
The factor
\begin{align}
\label{bif3d}
I_{a,b}^{bif}(y_{a,i}, y_{b,j}) =\prod_{1\leq i \leq N_a}\prod_{1\leq j \leq N_b}\left [ \frac{(\sqrt{q\, t}\, y_{a,i}/y_{b,j};q)_{\infty}}{(\sqrt{q/t} \, y_{a,i}/y_{b,j};q)_{\infty}}\right]^{\Delta_{a b}}
\end{align}
is the contribution of the $\cN=4$ bifundamental hypermultiplets. $\Delta_{ab}$ is the incidence matrix of $\fg_\text{o}$. In $\cN=2$ notation, the numerator is the contribution of bifundamental  chiral multiplets with {D} boundary conditions. Inside these multiplets, the fermion and its derivatives contribute to the index with charge
\beq
(\text{bif chiral}_{ab},\text{D}) : \;\;\; \left(\overline{N_a},N_b,n+\frac{1}{2},1,\frac{-1}{2}\right)\;\; \text{under}\;\; U(N_a)\times U(N_b)\times U(1)_J\times U(1)_R\times U(1)_t .
\eeq 
The denominator is the contribution of bifundamental chiral multiplets with {N} boundary conditions. Inside these multiplets, the scalars and their derivatives contribute to the index with charge
\beq
(\text{bif chiral}_{ab},\text{N}) : \;\;\; \left(N_a,\overline{N_b},n,1,\frac{-1}{2}\right)\;\; \text{under}\;\; U(N_a)\times U(N_b)\times U(1)_J\times U(1)_R\times U(1)_t .
\eeq 
Note that our charge assignment ensures that the $\cN=4$ superpotential term coupling the adjoint scalar to the bifundamentals at each node of the quiver is invariant under $U(1)_t$ and has the correct charge under $U(1)_R$.
In the $q$-conformal block, we recognize the two-point function 
\beq
I_{a,b}^{bif}(y_{a,i}, y_{b,j})=\left\langle S_a^\vee(y_{a,i})\, S_b^\vee(y_{b,j}) \right\rangle
\eeq 
of two screening currents on adjacent nodes $a$ and $b$, see \eqref{screenab}.

So far, all the above contributions are the same as in the unramified quantum $q$-Langlands correspondence \cite{Aganagic:2017gsx}. This is because they all originate from 3d $\cN=4$ multiplets. What distinguishes our 3d Drinfeld quiver theories $T^{3d}_{\fg_\text{o}}$ is the presence of chiral multiplets explicitly breaking the supersymmetry to $\cN=2$.
Their contribution to the index is a factor
\begin{align}\label{matter3d}
I_a^{3d, flavor}(y_{a,i}, x_{d,s}) \; ,
\end{align}
inside the integrand \eqref{bulk3d}, where $x_{d,s}$ are the complexified masses of the chiral multiplets, valued in the maximal torus $T_H$ of the flavor group $F_{H}$. According to Proposition \ref{prop3d}, half of these are fundamental chiral multiplets, with boundary conditions {N}. If the scalars and their derivatives have charge
\beq
(\text{fund chiral}_{a},\text{N}) : \;\;\; \left(N_a,\overline{N^{F,-}_a},n,\rho,\frac{-\rho}{2}\right)\;\; \text{under}\;\; U(N_a)\times U(N^{F,-}_a)\times U(1)_J\times U(1)_R\times U(1)_t ,
\eeq 
their contribution to the index is 
\beq\label{fundmult}
\left(\left(q/t\right)^{\frac{\rho}{2}}  y_{a,i}/x_{d,s}\, ; q\right)^{-1}_\infty \; .
\eeq
In the $q$-conformal block, we recognize the two-point function $\left\langle S_a^\vee(y_{a,i})\, \Lambda^{-1}_a\left((t/q)^{\frac{\rho-1}{2}}\,x_{d,s}\right) \right\rangle$ of a screening current with our vertex operators $\Lambda^{\pm 1}_a$ \eqref{coweightvertexdef}.

The other half are anti-fundamental chiral multiplets, with boundary conditions {D}. If the fermions and their derivatives have charge
\beq
(\overline{\text{fund}}\; \text{chiral}_{a},\text{D}) : \; \left(\overline{N_a},N^{F,+}_a,n+\frac{1}{2},\rho',1-\frac{\rho'}{2}\right)\;\; \text{under}\;\; U(N_a)\times U(N^{F,+}_a)\times U(1)_J\times U(1)_R\times U(1)_t ,
\eeq 
their contribution to the index is  
\beq\label{antifundmult}
\left(\left(q/t\right)^{1-\frac{\rho'}{2}} y_{a,i}/x_{d,s}\, ; q\right)_\infty \; . 
\eeq
In the $q$-conformal block, we recognize the two-point function $\left\langle S_a^\vee(y_{a,i})\, \Lambda_a\left((t/q)^{\frac{1-\rho'}{2}}\,x_{d,s}\right) \right\rangle$.  It follows at once that the total chiral matter contribution  to the index is captured by the two-point function
\beq\label{matter3dwow}
I_a^{3d, flavor}(y_{a,i}, x_{d,s}) =  \left\langle S_a^\vee(y_{a,i})\, :\prod_{d=1}^{L} \cV_{\{\lambda\}_d}(\tilde{x}_d): \right\rangle \; . 
\eeq
Our definition of Drinfeld quivers in Proposition \ref{prop3d} fixes the R-charges and $U(1)_t$ charges labeled by $\rho$ and $\rho'$ in terms of the Drinfeld roots as 
\beq\label{roots}
\mathfrak{a}^{+}_{\lambda_{d,s},b,j} = \frac{1-\rho'}{2} \; , \qquad \;\;\; \mathfrak{a}^{-}_{\lambda_{d,s},b,j} = \frac{\rho-1}{2}
\eeq
Since our Drinfeld polynomials are normalized such that  $\mathfrak{a}^{\pm}_{\lambda_0,b,j}=0$ if $\lambda_0$ is a fundamental weight, this fixes the charge for the associated anti-fundamental chiral multiplet to be $\rho'=1$.\\

For a generic Drinfeld quiver $T^{3d}_{\fg_\text{o}}$, fixing the values of $\rho$ and $\rho'$ for the $U(1)_R\times U(1)_t$ charges of all chiral multiplets straight from gauge theory is a delicate matter: for instance, these values are only \emph{partially} constrained by the superpotential terms. These are the cubic terms which couple fundamental and anti-fundamental chiral fields located at adjacent gauge nodes with a bifundamental chiral field between them. But there will always be chiral multiplets left unconstrained, with no such superpotential term. The simplest illustration is already found at rank 1, see Section \ref{sec:example}. In order to fix all $\rho$ and $\rho'$, one could in principle carry out a $F$-maximization procedure to the IR SCFT \cite{Jafferis:2010un}; we leave this important open question to future work.

In this paper, we will bypass this issue by appealing instead to a dual description of the 3d theory: $T^{3d}_{\fg_\text{o}}$ also arises as the worldvolume theory of 1/2-BPS vortices in a different (3+2)-dimensional gauge theory. This is the route we will take in Section \ref{sec:5dgauge}, where we will see that it is possible to fix all $\rho$ and $\rho'$ straight from 5 dimensions, through a Higgsing procedure. In doing so, we will derive the relation \eqref{roots}, meaning that $\rho$ and $\rho'$ are indeed given by the Drinfeld roots for appropriate quantum affine weights.\\

Then, the integrands of the 3d half-index with boundary  and of the $\cW_{q,t}(\fg_\text{o})$-algebra $q$-conformal block agree. It remains to define the contours. A choice of contour is a choice of vacuum  $\bf{A}$ for the theory $T^{3d}_{\fg_\text{o}}$. Here, the contours are defined to enclose the point at ``$\infty$" on $\cC$ and all poles arising from the $q$-Pochhammer symbols denominators which are compatible with our choice of chamber. Such contours are in one-to-one correspondence with a choice of boundary conditions known as exceptional Dirichlet in the 3d $\cN=4$ context. We will now argue that an analog of these boundary conditions exists for 3d $\cN=2$ Drinfeld quivers.

\vspace{8mm}

\subsubsection{Exceptional Dirichlet for 3d $\cN=4$ theories}

We now describe a distinguished set of boundary conditions  known as exceptional Dirichlet. These boundary conditions have been thoroughly investigated in the 3d $\cN=4$ context \cite{Bullimore:2016nji,Bullimore:2020jdq,Okazaki:2020lfy,Crew:2020psc,Dedushenko:2021mds}. 
There, one considers an isolated vacuum $\{{\bf{A}}\}$, with F.I. and mass parameter deformations turned on. The  boundary condition ${\bf D_{EX}}$ emulates the presence of a massive isolated vacuum $\bf{A}$ at infinity, in a way that is compatible simultaneously with F.I. and mass deformations, meaning it should be compatible with the choice of chambers $\fC_C$ (\eqref{Cchamber}) and $\fC_H$ (\eqref{Hchamber}). 
In particular, this choice of boundary condition breaks the gauge symmetry, but preserves the flavor symmetry $T=U(1)_t\times T_H$ on the Higgs branch, as well as the topological symmetry $T_C$. 

Then, $\cN=(2,2)$ exceptional Dirichlet boundary conditions depend on the choice of vacuum $\{{\bf{A}}\}$ and a certain holomorphic Lagrangian splitting of the (quaternionic) linear representation $T^*R$ of the gauge group, also called a polarization. Equivalently, one performs a decomposition of the tangent bundle of the Higgs branch $X$ into two halves, 
\beq
T_{\{{\bf{A}}\}} X = T^{1/2}_{\{{\bf{A}}\}} X + t\otimes\left({T^{1/2}_{\{{\bf{A}}\}}} X\right)^\vee \; ,
\eeq
with $(\ldots)^\vee$ denoting the dual of $(\ldots)$.
Physically, the tangent bundle is the cohomology of a certain complex describing the fluctuations of the hypermultiplets  satisfying the complex moment map condition (the F-terms) modulo the gauge transformations.\\

The above decomposition of the tangent bundle is not unique. To illustrate this, consider the example of 3d $\cN=4$ SQED: this is an abelian gauge group $U(1)$ and $N^F$ hypermultiplets, meaning $R=\mathbb{C}^{N^F}$ in that case. We turn on masses $x_d$ for the hypermultiplets, and a mass $t$ for the $U(1)_t$ flavor symmetry. If we take $\{{\bf{A}}\}$ to label the $k$-th massive vacuum, then the character of (the fibre on) the tangent bundle of $X$ at that vacuum reads
\beq
T_{k} X = \sum_{d\neq k}^{N^F}\left(\frac{x_d}{x_k} + t\, \frac{x_k}{x_d}\right)  \; .
\eeq
One canonical decomposition of this tangent space is as a sum of ``positive" and ``negative" weight spaces, with positive weight space
\beq\label{decomp1}
T^{1/2}_{k} X  = \sum_{d< k}^{N^F} \frac{x_d}{x_k} + t\,\sum_{d> k}^{N^F} \frac{x_k}{x_d} \; .
\eeq 
This splitting reflects that in a regime where the 3d gauge theory flows to a massive sigma model on $X$,
the BPS equations for the supercharges  $Q_+$ and $\overline{Q}_+$ are inverse gradient (Morse) flow equations for the flavor symmetry real moment map in the $x^\bot$-direction, with $x^\bot\leq 0$. With the convention that the moment map decreases as $x^\bot\rightarrow -\infty$, the boundary condition at $x^\bot= 0$ will have support on an attracting set $X^+_{\{{\bf{A}}\}}\subset X$; this set is the locus of positive gradient flow from the isolated vacuum $\bf{A}$, given a fixed mass chamber $\fC_H=\{|x_1| < |x_2| < \ldots < |x_{N^F}| \}$\footnote{We are assuming $|\epsilon_t|\ll |m_d-m_{d'}|$ throughout.}. Exceptional Dirichlet boundary conditions can then defined to be supported precisely on (the closure of) this attracting set, with corresponding  tangent bundle $T^{1/2}_{k} X$ as in \eqref{decomp1}. This decomposition of the tangent space is also the natural one which appears in the definition of $x$-solutions in the next Section. 

For another canonical decomposition of the tangent space, we could have chosen
\beq\label{decomp2}
T^{1/2}_{k} X = \sum_{d\neq k}^{N^F}\frac{x_d}{x_k} \; .
\eeq 
This splitting reflects the structure of the magnetic $z$-solution block as we have presented it so far following the $\cW_{q,t}(\fg)$-algebra literature. In Section \ref{sec:quasim}, we will use this decomposition to reinterpret the magnetic block as an example of ``vertex function" analytic in the K\"{a}hler parameters $z_a$, in the context of the enumerative geometry of $X$. 
Ultimately, the different choices of polarizations are inconsequential, since they only differ by $q$-shifts of the parameters $z_a$, as far as $q$-difference equations are concerned.\\

At the level of multiplets, the $\cN=(2,2)$ exceptional Dirichlet boundary conditions are implemented as follows:\\

-- The 3d $\cN=4$ vector multiplet is given Dirichlet (${\bf D}$) boundary conditions. In 3d $\cN=2$ notation, this translates to ${\bf D}$ boundary conditions for the vector multiplet and Dirichlet (D) boundary conditions for the adjoint chiral multiplet\footnote{We are assuming all complex masses are 0; otherwise, the adjoint chiral scalar would need deformed Dirichlet ${\text{D}_c}$ boundary conditions with value set by the complex mass in vacuum $\{{\bf{A}}\}$ to preserve supersymmetry.}.\\

-- The 3d $\cN=4$ hypermultiplets are given Dirichlet-Neumann type boundary conditions, depending on the choice of polarization for the vacuum $\{{\bf{A}}\}$.
In 3d $\cN=2$ notation, each hypermultiplet is decomposed in chiral multiplets $(X_d, Y_d)$, and the polarization is labeled by a choice of sign vector $\epsilon=\pm$, where $\epsilon_d=-$ gives D boundary conditions to $X_d$ (and Neumann (N) boundary conditions to $Y_d$), while $\epsilon_d=+$ gives D boundary conditions to $Y_d$ (in which case $X_d$ is assigned N boundary conditions). 
In the vacuum $\{{\bf{A}}\}$, some of the multiplets will have zero mass, and we set their value on the boundary to be equal  to the vev they develop in that vacuum; this is called a deformed Dirichlet boundary condition (${\text{D}_c}$).
For a vacuum where, say, the $k$-th chiral multiplet is a modulus, the exceptional Dirichlet boundary condition has the form:
\beq\label{polarsplitN4}
{\bf D_{EX}}_{\epsilon,k} = \begin{cases}
D_{\bot}Y_{d{|\partial}}=0\;,\qquad\;	X_{d{|\partial}}=c\,\delta_{dk}   & \text{if} \;\; \epsilon_d=- \\
D_{\bot}X_{d{|\partial}}=0\;,\qquad\	Y_{d{|\partial}}=c\,\delta_{dk}   & \text{if} \;\; \epsilon_d=+
\end{cases}
\eeq
For a given 3d $\cN=4$ quiver gauge theory written in terms of $\cN=2$ multiplets, the polarization $T^{1/2}X$ is therefore an assignment $\epsilon=\pm$ for each pair of arrows in the Dynkin diagram of $\fg_{\text{o}}$; ``half" the arrows will label chiral multiplets with D, and the remaining half will label chiral multiplets with N at the boundary.

In the SQED example, there are $N^F$ fundamental hypermultiplets  decomposed into chiral multiplets $(X_d, Y_d)$ of charge $(+1,-1)$, for $d=1,\ldots,N^F$, so the polarization is a sign vector $\{\epsilon\}^{N^F}$ of size $N^F$, with $\epsilon=\pm$. In the F.I. chamber $\fC_C = \{|z|<1\}$ and mass chamber $\fC_H = \{|x_1|<|x_2|<\ldots<|x_{N^F}|\}$, the polarization \eqref{decomp1} is the choice of sign vector $\{+,\ldots,+,-,\ldots,-\}$, where there are $k$ signs ``$+$". This is achieved by giving D to $Y_1, \ldots, Y_k, X_{k+1}, \ldots, X_{N^F}$ (in fact, $Y_k$ gets ${\text{D}_c}$ boundary conditions). The remaining chiral multiplets are given N, and are exponentially suppressed under the gradient flow.

The polarization \eqref{decomp2} is the choice of sign vector $\{+,\ldots,+\}$ instead. This gives D to $Y_1, \ldots, Y_{N^F}$ (actually ${\text{D}_c}$ for $Y_k$), and N to all the $X_d$ chiral multiplets.\\

\vspace{8mm}

\subsubsection{Exceptional Dirichlet for 3d $\cN=2$ Drinfeld quivers}
\label{ssec:excep}


We move on to defining analogs of exceptional Dirichlet boundary conditions for 3d $\cN=2$ Drinfeld quivers, which we will still denote as ${\bf D_{EX}}$. The $\cN=(0,2)$ boundary conditions we need are in many ways analogous to the $\cN=(2,2)$ ones we just reviewed\footnote{In particular, our boundary conditions are \emph{not} the $\cN=(0,2)$ ones used in \cite{Bullimore:2021rnr}.}. This is because the field content of the gauge theory  $T^{3d}_{\fg_\text{o}}$  \emph{almost} preserves a full $\cN=4$ supersymmetry. The only multiplets breaking the supersymmetry to $\cN=2$ are the $L^+_a$ fundamental and $L^-_a$ anti-fundamental chiral multiplets, but they do so in a very specific way: for all $a=1,\ldots,r$, these multiplets always appear in pairs, $L^+_a = L^-_a$, much like in $\cN=4$ hypermultiplets.\\ 

The ${\bf D_{EX}}$ boundary conditions for the quiver are implemented as follows:\\

-- The vector, adjoint chiral, and bifundamental chiral multiplets transforming in $U(N_a)\times U(N_b)$ are given the same boundary conditions as in the $\cN=(2,2)$ setting, see last Section.\\

-- The fundamental and anti-fundamental chiral multiplets transforming in $U(N_a)\times U(N^{F,\pm}_a)$ are given Dirichlet-Neumann type boundary conditions, depending on a choice of ``polarization" of this matter content in the vacuum $\{{\bf{A}}\}$.
Recall that $q$-vertex operators are labeled by an integer $d\in\{1,\ldots,L\}$. The chiral matter on the gauge node $a$ of the quiver is in fact graded by $d$, meaning a fundamental chiral field can be labeled as $\Phi^-_{d,a}$, and its anti-fundamental chiral field partner as $\Phi^+_{d,a}$, see \eqref{equalnumber}. The term ``polarization" here refers to an analog of the polarization in the $\cN=(2,2)$ sense: we assign $\epsilon_{d,a}=\pm$ to each pair $(\Phi^+_{d,a},\Phi^-_{d,a})$, such that $\epsilon_{d,a}=-$ gives D boundary conditions to the field $\Phi^-_{d,a}$ (and N boundary conditions to $\Phi^+_{d,a}$), while $\epsilon_{d,a}=+$ gives D boundary conditions to $\Phi^+_{d,a}$ (and N boundary conditions to $\Phi^-_{d,a}$). In this way, no matter the choice of polarization, ``half" the arrows in the Dynkin diagram of $\fg_{\text{o}}$ will label chiral multiplets with D, and the other half will label chiral multiplets with N. Note the existence of this polarization crucially relies on the assumption that Drinfeld polynomials have matching degrees in our $q$-conformal block, \eqref{constraintpoly}.\\

For our purposes, we will find it convenient to choose the same polarization $\epsilon_{d,1}=\epsilon_{d,2}=\ldots=\epsilon_{d,r}$ for all pairs of chiral multiplets making up the $d$-th defect, and write the corresponding polarization as $\epsilon_{d}=\pm$. 
That is, we assign a polarization $\epsilon_d=-$ (respectively $\epsilon_d=+$) to a defect located at center of mass $\tilde{x}_d$, by imposing D to \emph{all} the fundamental  chiral multiplets for \emph{that} defect (respectively all the anti-fundamental chiral multiplets).\\ 

Just as in the $\cN=(2,2)$ case, in the vacuum $\{{\bf{A}}\}$, some of the chiral multiplets will have zero mass, and we set their value $\Phi^{\pm}_{|\partial}= c$ on the boundary to be equal to the vev they develop in that vacuum, meaning their boundary conditions are of deformed Dirichlet type ${\text{D}_c}$. The vevs are chosen precisely so that the flavor symmetry maximal torus $U(1)_t\times T_H$ is preserved.\\


At the level of the index, the exceptional Dirichlet boundary condition ${\bf D_{EX}}$ is implemented by first writing down the index for usual Dirichlet ${\bf D}$, with $c=0$, and then redefining the symmetries at the boundary to turn the boundary condition into deformed Dirichlet ${\text{D}_c}$ without braking supersymmetry, with $c\neq 0$.

Consider first the rank 1 case of an abelian $U(1)$ gauge theory $T^{3d}_{\fg_\text{o}}$ which preserves $U(1)_V$ at the boundary. Unlike the boundary condition ${\bf N}$, imposing  ${\bf D}$ breaks the bulk gauge symmetry to a global symmetry at the boundary, which we denote as  $U(1)_{\partial}$. We introduce a fugacity $u$ and a generator $\ft_\partial$ for this $U(1)_{\partial}$ symmetry. The index then reads
\beq
\label{3dhalfindexmoreflow1}
{\mathcal Z}(T^{3d}_{{\fg_\text{o}},{\bf D}})  = {\rm Tr}\left[(-1)^F\, q^{J+\frac{V}{2}}\; t^{\frac{A-V}{2}} \; u^{\ft_\partial} \; z^{\ft_C}\; x^{\ft_H}  \right]\;\; .
\eeq
The idea is to generate a flow along the boundary to a new boundary condition by giving one of the chiral fields a vev $c\neq 0$. Suppose the chiral field has charge $(1,1,\sigma_H)$ under $U(1)_V\times U(1)_{\partial}\times T_H$; then the linear combinations $V'=V-\ft_\partial$ and $\ft_H' = \ft_H - \sigma_H\,\ft_\partial$ are preserved along this flow; $V'$ and $\ft_H'$ are the generators of the redefined vector R-symmetry and flavor symmetry, respectively. Since $c$ is an exact deformation of the action, the index trace can safely be taken over the Hilbert space of states counting local operators on the new boundary. In the index, the weight of the corresponding field must be set to 1 to preserve supersymmetry; for our choice of F.I. parameter chamber $\fC_C=\{\xi_a>0\}$, the D-term equations in the vacuum $\{{\bf{A}}\}$ imply $u\, x^{\sigma_H} \, t^{1/2} =1$, meaning we simply substitute $u = t^{-1/2}\, x^{-\sigma_H}$ in the index \eqref{3dhalfindexmoreflow1}. 

In the case where the gauge group is not simply $U(1)$, we consider instead the maximal torus  $T_{G_3d}$ of the gauge group $G_{3d}=\prod_{a=1}^{r} U(N_a)$, and treat each $U(1)_{\partial}\subset T_{G_3d}$ separately as in the rank 1 case above.\\

The contribution of the vector multiplet with ${\bf D}$ boundary conditions to the index is not a mere product of $q$-Pochhammer symbols. Indeed, there are additional non-perturbative contributions to the index coming from monopole operators at the boundary \cite{Dimofte:2017tpi}, reviewed in our Appendix \ref{sec:appendixindex}. The index takes the form of a sum over all abelian magnetic fluxes $m$ on the hemisphere $D^2$, meaning $m$ is an element of the cocharacter lattice  $m\in\Lambda_{cochar}=\text{Hom}(U(1),T_{G_{3d}})$, the space of maps to the maximal torus of the gauge group $G_{3d}$. Because $G_{3d}=\prod_{a=1}^{r} U(N_a)$, the lattice is explicitly $\Lambda_{cochar}=\mathbb{Z}^N$ , with $N=\sum_{a=1}^{r} N_a$. 
\beq
\label{3dhalfindexmoreDiri}
{\mathcal Z}(T^{3d}_{{\fg_\text{o}},\bf D_{EX}})  = \frac{1}{(q\;; q)^N_\infty}\sum_{m\in\mathbb{Z}^{N}}\frac{ z^m}{\prod_{a=1}^{r}\prod_{\alpha_a\in\text{roots}\,[U(N^{F,\pm}_a)]} \left(q^{1+m\cdot\alpha_a}\, x_{\alpha_a}\, ; q\right)_\infty}\; \ldots \; ,
\eeq
where the dots ``$\ldots$" stand for the contributions of effective Chern-Simons terms, along with the remaining matter content of $T^{3d}_{\fg_\text{o}}$ (and possible additional boundary degrees of freedom). We have performed the substitution $u = t^{-1/2}\, x^{-\sigma_H}$ in the entire summand for each boundary $U(1)_{\partial}$ symmetry.\\ 

The half-index ${\mathcal Z}(T^{3d}_{\fg_\text{o},\bf D_{EX}})$ can be directly derived from the Neumann half-index ${\mathcal Z}(T^{3d}_{\fg_\text{o},{\bf N}})$, for an appropriate choice of contour: indeed, this is done by computing the latter integral by residues. A generic ${\bf D_{EX}}$ boundary condition is expected to be defined for an arbitrary massive vacuum and polarization. Meanwhile, our canonical definition of the Drinfeld quiver $T^{3d}_{\fg_\text{o},{\bf N}}$ specifically gave N boundary conditions to all fundamental chiral multiplets, and D boundary conditions to all anti-fundamental chiral multiplets; then, a systematic way to engineer any desired boundary conditions and reproduce a generic ${\mathcal Z}(T^{3d}_{\fg_\text{o},\bf D_{EX}})$ via contour integration is to couple the 3d bulk theory on $S^1_{\cC'}\times D^2$ to a  2d $\cN=(0,2)$ boundary theory via superpotential terms. 
The index $T^{3d}_{\fg_\text{o},{\bf N}}$  is modified accordingly:
\begin{align}\label{part3dnew}
\oint_{\Gamma}  \frac{dy}{y}\, I^{3d}(y,x_{d,s},z) \cdot {\mathcal F}'(y,x_{d,s}) \; ,
\end{align}
where
\beq\label{ratioellram3d}
{\mathcal F}'(y,x_{d,s}) = \frac{{\mathcal F}(y,x_{d,s})}{\Theta(T^{-}X)} 
\eeq
is the contribution of the elliptic genus of the 2d theory. The tangent bundle $T^{-}X$ is the ramified analog of half of the tangent bundle $T^{1/2}X$ which appears in the unramified context of Nakajima quiver varieties; it will be defined precisely in \eqref{polarramif}. The role of the contribution $\Theta(T^{-}X)$ is to introduce new poles  by flipping the boundary conditions $D \leftrightarrow N$ of the multiplets contributing to the numerators. The factor ${\mathcal F}(y,x_{d,s})$ is defined by the requirement that the ratio ${\mathcal F}'(y,x_{d,s})$ be invariant under $q$-shifts (and nonsingular in the masses). We also modified the contour to $\Gamma$, which is now a sum of cycles enclosing all allowed $x_{d,s}$-poles and $q$-tails. Picking a particular component of ${\mathcal F}(y,x_{d,s})$ supported on a specific vacuum $\{{\bf A}\}$ for the exceptional Dirichlet boundary condition will act as a delta function in the space of contours and reduce $\Gamma$ to a specific contour $C$ \cite{Aganagic:2017smx}.

Performing the integral by residue yields an infinite sum of the form \eqref{3dhalfindexmoreDiri}, but with the sum index $m$ only taking values in the sublattice $(\mathbb{Z}^+)^{N}$ with \emph{positive} integer values, as opposed to the full lattice $\mathbb{Z}^{N}$ as we wrote it above. This discrepancy is resolved after careful examination of the summand ${\mathcal Z}(T^{3d}_{{\fg_\text{o}},\bf D_{EX}})$:  among the factors in ``$\ldots$", one finds  $\left(q^{1+m\cdot\alpha_a}\, ; q\right)_\infty$. This is the contribution of the chiral multiplet which acquires a vev in the vacuum, with deformed Dirichlet boundary condition ${\text{D}_c}$. This factor effectively restricts the sum indexed by $m\in\mathbb{Z}^{N}$ to a sum indexed by $m\in(\mathbb{Z}^+)^{N}$, since the summand vanishes otherwise. This argument establishes the equality of the half-indices $T^{3d}_{\fg_\text{o},{\bf N}}$ and ${\mathcal Z}(T^{3d}_{{\fg_\text{o}},\bf D_{EX}})$; for an explicit example, see Section \ref{sec:example}.\\

The sum has a positive radius of convergence in the F.I. parameters $z_a$; by definition, it is what we have been calling the $z$-solution form of the $\cW_{q,t}(\fg)$-algebra blocks, holomorphic in a punctured neighborhood of $0_{\fC_C}$, the origin of the chamber $\fC_C$.\\

In Section \ref{sec:5dgauge}, we will see that the monopole fluxes take values in a much smaller lattice than $(\mathbb{Z}^+)^{N}$: they are in fact labeled by Young diagrams with no more than $N_{a}$ rows, as was first shown for holomorphic blocks \cite{Aganagic:2013tta}.

\vspace{8mm}

\subsubsection{Magnetic $x$-solutions}
\label{sssec:ellipticstab}

A proper $\cW_{q,t}(\fg)$-algebra block should admit an expression analytic in the mass parameters $x_{d,s}$, given the choice of a chamber $\fC_H$. In other terms, it should be holomorphic in a punctured neighborhood of $0_{\fC_H}$, the origin of the chamber $\fC_H$. Such a block is an $x$-solution, related to the previous $z$-solution  by a linear change of basis.\\

The change of basis is implemented by coupling the 3d bulk theory to yet another 2d $\cN=(0,2)$ theory on the boundary, via additional superpotential terms. In the $\cN=(2,2)$ literature, such boundary conditions go by the name of enriched Neumann, and the added matter is sometimes called $\mathbb{C}^*$-valued chiral multiplets; see for instance \cite{Bullimore:2016nji,Bullimore:2021rnr,Dedushenko:2021mds}.
Here, we will also call the $\cN=(0,2)$ analog of these boundary conditions enriched Neumann.

Given the choice of a mass parameter chamber $\fC_H$,
we saw that it was possible to support exceptional Dirichlet boundary conditions on an attracting manifold $X^+_{{\{\bf A\}}}$ under gradient flow for the (real) moment map on $X$; this is possible for Drinfeld quivers because chiral multiplets always come in pairs. 

The support of enriched Neumann boundary conditions is defined so that it contains the attracting manifold  $X^+_{{\{\bf A\}}}$, as well as additional components $\bigcup_{\{{\bf{B}}\}} X^+_{{\{\bf B\}}}$. Here, each attracting manifold $X^+_{{\{\bf B\}}}$ is defined by the gradient flow from a vacuum ${\{\bf B\}}$, and is part of the support only when the image of the moment map is strictly greater than the image of the moment map for the vacuum $\{{\bf{A}}\}$. In the $\cN=(2,2)$ context, this support is known as the stable envelope for the vacuum $\{{\bf{A}}\}$  \cite{2012arXiv1211.1287M}.
This definition makes it clear that enriched Neumann boundary conditions depend on the choice of chamber $\fC_H$; we denote them as ${\bf N_{EN}}_{,\fC_H}$.\\

Concretely, we impose  ${\bf N_{EN}}_{,\fC_H}$ by first giving the vector and adjoint chiral multiplets regular Neumann boundary conditions ${\bf{N}}$, while the chiral matter fields are given N or D boundary conditions, depending on whether they respectively correspond to repelling or attracting directions under the gradient flow, as determined by the chamber $\fC_H$.
This assignment of boundary conditions generically introduces new anomalies, which need to be canceled via additional 2d $\cN=(0,2)$ insertions.\\

At the level of the index, the support of ${\bf N_{EN}}_{,\fC_H}$ modifies the integrand and its pole structure via the insertion of a finite-size matrix, whose size is dictated by the number of massive vacua: 
\begin{align}\label{part3dnew2}
{\mathcal Z}(T^{3d}_{{\fg_\text{o}},{\bf N_{EN}}_{,\fC_H}}) = \oint_{C'}  \frac{dy}{y}\, I^{3d}(y,x_{d,s},z) \cdot  \fB_{\fC_H}(y,x_{d,s},z) \; ,
\end{align}
%
%
The matrix $\fB_{\fC_H}$ is the partition function of $T^{3d}_{\fg_\text{o}}$ on  $S^1_{\cC'} \times S^1_{D^2} \times I$, with $I$ a finite interval. One imposes enriched Neumann ${\bf N_{EN}}_{,\fC_H}$  at one end of the interval and  Dirichlet ${\bf{D}}$ at the other end. The partition function does not depend on the length of the interval; after taking its length to 0, it computes the pole-subtraction matrix of Aganagic-Okounkov \cite{Aganagic:2016jmx,Aganagic:2017smx,Sugiyama:2020uqh,Bullimore:2021rnr,Dedushenko:2021mds}. The integration contour $C'$ is deformed from the $z$-solution one $C$, so as to separate the poles in the new integrand: we now only enclose poles supported on ${\bf {N}_{EN}}$  and leave the ones supported on repelling submanifolds outside the contour. Because the 2d boundary is defined on a torus $S^1_{\cC'} \times S^1_{D^2}$, the support is called the elliptic stable envelope $\text{Stab}^{Ell}_{\fC_H}(y,x_{d,s},z)$. 
To obtain specific entries of the matrix $\fB_{\fC_H}$, the Dirichlet ${\bf{D}}$ at the end of the interval should be specialized to Exceptional Dirichlet ${\bf D_{EX,\{{\bf{A}}\}}}$, and the overlap with ${\bf N_{EN}}_{,\fC_H}$ is the elliptic genus of the corresponding boundary 2d $\cN=(0,2)$ theory on  $S^1_{\cC'} \times S^1_{D^2}$; this specialization amounts to evaluating the fugacities $y_{a,i}$ in vacuum $\{{\bf{A}}\}$ inside the matrix $\fB_{\fC_H,\{{\bf{A}}\}}$. Correspondingly, the support is the elliptic stable envelope in vacuum $\{{\bf{A}}\}$, $\text{Stab}^{Ell}_{\fC_H,\{{\bf{A}}\}}(x_{d,s},z)$.\\

In practice, at least for abelian examples, the insertion $\fB_{\fC_H}$ is best constructed as a two-step process: first, we introduce 2d chiral and Fermi multiplets supported on ${\bf {N}_{EN}}$ as dictated by the choice of chamber $\fC_H$. Second, the remaining boundary degrees of freedom are dictated by anomaly cancellation, which can be readily computed via the anomaly polynomial. This ensures that the entries of the matrix $\fB_\fC$ are $q$-periodic pseudo-constants, and $T^{3d}_{{\fg_\text{o}},{\bf N_{EN}}_{,\fC_H}}$ is anomaly-free on the boundary. We will carry out this procedure explicitly in Section \eqref{sec:example}, and show that it yields an $x$-solution analytic in chamber $\fB_{\fC_H}$, completely analogously to the 3d $\cN=4$ case.\\

We end this section with two comments.
First, on a practical level, determining the support of enriched Neumann boundary conditions  ${\bf {N}_{EN}}$ whenever $T^{3d}_{\fg_\text{o}}$ is a generic Drinfeld quiver gauge theory quickly becomes unmanageable as the ranks of the quiver and gauge groups increase; for details, see the work of Smirnov \cite{Smirnov:2018drm}. For us, all the key features of tame ramification are already present in the abelian setting, so we will treat that case in detail in the Examples Section, for $\fg_\text{o}=A_1$ \ref{sec:example}.

Second, in the $\cN=(2,2)$ context, it has recently been argued that the boundary conditions $\bf D_{EX}$ and ${\bf {N}_{EN}}$ should be exchanged under 3d mirror symmetry \cite{Intriligator:1996ex,Bullimore:2021rnr,2020arXiv200805516D,2020arXiv201108603D,Rimanyi:2019zyi}, along with the chambers $\fC_C$ and $\fC_H$. That is, if $\widetilde{T}^{3d}_{\fg_\text{o}}$ denotes the bulk theory mirror to ${T}^{3d}_{\fg_\text{o}}$, then we should have
\beq
T^{3d}_{\fg_\text{o},\,\bf {D}_{EX}} \longleftrightarrow\; \widetilde{T}^{3d}_{\fg_\text{o},\,\bf {N}_{EN}}
\eeq
after coupling to the relevant boundary. From our presentation, we expect a version of mirror symmetry to likewise hold for 3d $\cN=2$ Drinfeld quiver gauge theories setup \cite{Aharony:1997bx,deBoer:1997ka}, at least for our class of handsaw-type quiver gauge theories, and in the presence of the 2d $\cN=(0,2)$ analog of the $\cN=(2,2)$ $\bf D_{EX}$ and ${\bf {N}_{EN}}$ boundary conditions relevant here.  We leave the detailed investigation of this mirror symmetry to future work.

\begin{remark}
It was initially suggested instead in \cite{Bullimore:2016nji,Okazaki:2020lfy} that $\bf D_{EX}$ boundary conditions should map to other $\bf D_{EX}$ boundary conditions under mirror symmetry, but this is not true in general. The ``accidental" instances where the statement holds happen whenever the support of an enriched Neumann  boundary condition coincides with that of the exceptional Dirichlet one $X^+_{\{{\bf{A}}\}}$, and contains no additional attracting sets: $\bigcup_{\{{\bf{B}}\}} X^+_{\{{\bf{B}}\}}=\varnothing$\footnote{Some counterexamples with nonempty  $X^+_{\{{\bf{B}}\}}$ are provided in \cite{Okazaki:2020lfy}; there, the various indices but should not be interpreted as mapping $\bf D_{EX}$ to a linear combination of multiple $\bf D_{EX}$ in a mirror theory: instead, this linear combination is precisely the support of a single ${\bf {N}_{EN}}$ boundary condition; see also the appendix B of \cite{Bullimore:2016nji}.}.
\end{remark}

%

\vspace{8mm}

\subsection{The $U_{\hbar}(\widehat{^L \fg_\text{o}})$ blocks in 3d gauge theory}

The $U_\hbar(\widehat{^L \fg_\text{o}})$-algebra $q$-conformal blocks are the ``electric" solutions to the $\fg_\text{o}$-type qKZ equations. In the gauge theory, $T^{3d}_{\fg_\text{o}}$ one defines the support manifold to be $M^\times_3= S^1_{\cC'}\times \mathbb{C}^\times_q$, which differs from the previous manifold $M_3$ by removing the origin of the complex line. 

In their integral form, electric $z$- and $x$-solutions now both feature additional insertions $\text{Stab}^K_{\{{\bf A'}\}}(y,x_{d,s})$ of rational functions in the integrand \cite{cmp/1104252137,cmp/1104254019,Tarasov:1993vs,1994CMaPh.162..499V}, labeled by a choice of massive vacuum ${\{{\bf A'}\}}$. The electric $z$-solutions take the form
\beq\label{qKZellstab1}
{\bf F}^{\{{\bf A'}\}}_{\{{\bf A}\}}= \oint_{C}  \frac{dy}{y}\, \text{Stab}^K_{\{{\bf A'}\}}(y,x_{d,s})\cdot I^{3d}(y,x_{d,s},z)\; ,
\eeq
where ${\{{\bf A}\}}$ on the left-hand side labels a choice of contour $C$ on the right-hand side.

The electric $x$-solutions further depend on the choice of chamber $\fC_H$, and take the form
\beq\label{qKZellstab2}
{\bf F}^{\{{\bf A'}\}}_{\fC_H,\{{\bf A}\}} = \oint_{C'}  \frac{dy}{y}\, \text{Stab}^K_{\{{\bf A'}\}}(y,x_{d,s})\cdot I^{3d}(y,x_{d,s},z) \cdot  \fB_{\fC_H}(y,x_{d,s},z)\; ,
\eeq
The contour $C'$ is defined as in the last Section, and in particular will not enclose the possible new poles from  $\text{Stab}^K_{\{{\bf A'}\}}$.\\

\begin{figure}[h!]
	\emph{}
	\centering
	\includegraphics[trim={0 0 0 0cm},clip,width=0.95\textwidth]{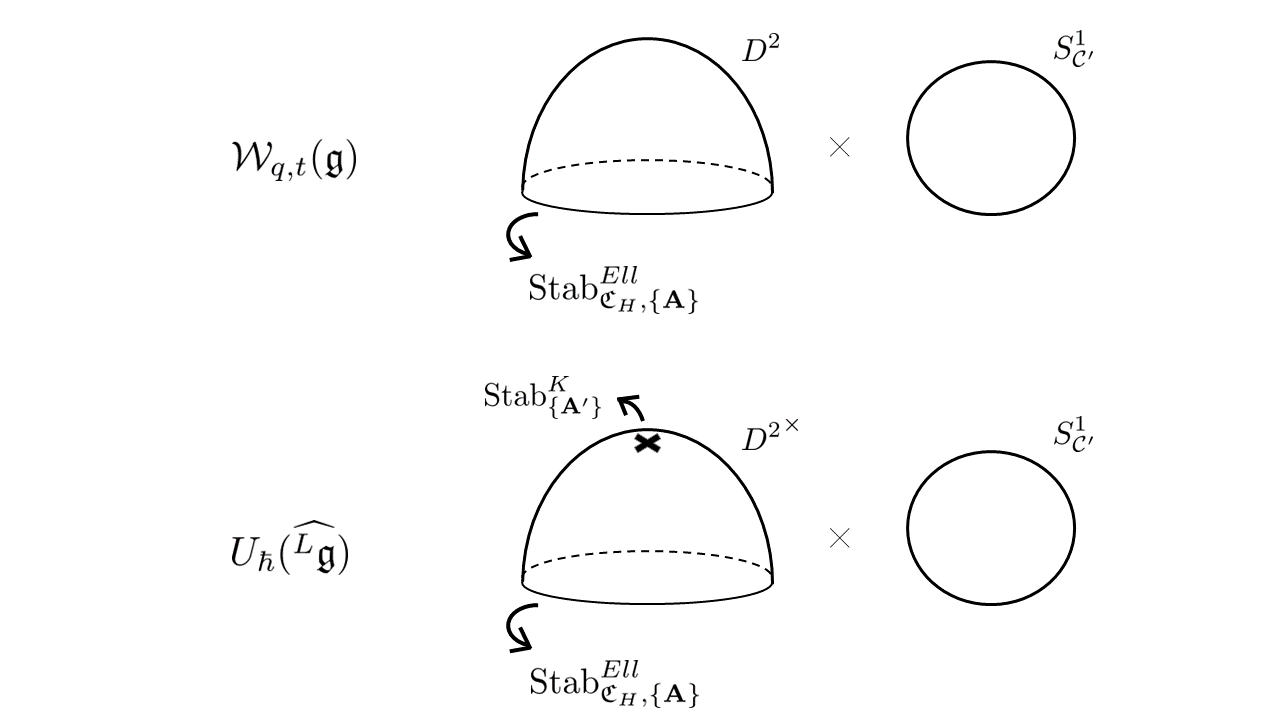}
	\vspace{-1pt}
	\caption{The $x$ solutions in 3d gauge theory. The needed boundary conditions are enriched Neumann, supported on elliptic stable envelopes. To obtain electric blocks, one further excises the origin of $D^2$ and introduces boundary conditions supported on K-theoretic stable envelopes.} 
	\label{fig:electricvsmagnetic}
\end{figure}

The insertions $\text{Stab}^K_{\{{\bf A'}\}}$  stand for the contribution of a 1d quantum mechanics on the circle: these are  1/2-BPS loop operators supported on $S^1_{\cC'}\times \{0\}$ inside the 3d gauge theory. The dependence of the theory on the various masses implies that this quantum mechanics preserves the supersymmetries we previously wrote as $Q_+$ and $\overline{Q}_+$. We couple the 1d theory on the loops to the bulk 3d theory by gauging the flavor symmetries of the quantum mechanics with 3d $\cN=2$ vector multiplets. This translates into gauging the one-dimensional masses, which turns them into scalars for the associated 3d $\cN=2$ vector multiplets. Whenever the vector multiplet is dynamical, the scalar becomes a modulus $y$, which is to be integrated over in the index. In contrast, in the case of a background vector multiplet, the scalar becomes a frozen mass parameter from the 3d point of view.\\ 

The loop contributions to the index are of the form  $(1-y/x_{d,s})^{\pm 1}$, where ``$+1$" and ``$-1$"  respectively stand for (reductions from 2d to 1d of) Fermi and chiral multiplet contributions to the index. The Fermi multiplets should be interpreted as 1/2-BPS Wilson loops in (tensor products of) antisymmetric representations for unitary groups, see \cite{Gomis:2006sb,Assel:2015oxa} and more recently \cite{Haouzi:2019jzk,Haouzi:2020bso}, where similar 3d/1d indices were studied in a closely related context. Presumably, the chiral multiplets could also be interpreted as Wilson loops in tensor products of symmetric representations \cite{Gomis:2006im}, though this would need to be made precise.\\

In order to determine the correct quantum mechanics which results in electric blocks, one constructs stable envelopes as in the previous Section, but this time around in a neighborhood of  $0\in\mathbb{C}^\times$. This amounts to working on the manifold $S^1_{\cC'}\times {\mathcal A}^2$, with ${\mathcal A}^2$ an annulus. The 2d $\cN=(0,2)$ boundary conditions on the outer $T^2$ boundary of ${\mathcal A}^2$ are as before, they are responsible for deciding whether we construct a $z$- or $x$-solution. The 2d $\cN=(0,2)$ boundary conditions on the inner $T^2$  boundary are responsible for engineering an electric or a magnetic block. Imposing enriched Neumann boundary conditions for a vacuum ${\{{\bf A'}\}}$ on the inner $T^2$ is only useful as an intermediate step, since ultimately we are interested in the 1d quantum mechanics obtained by shrinking the boundary back to a point at the origin; this identifies  $\text{Stab}^K_{\{{\bf A'}\}}$ as a K-theoretic stable envelope \cite{Okounkov:2015spn,Aganagic:2017smx}. This shrinking procedure is achieved by sending $q\rightarrow 0$; as a quick check, this is consistent with the index, since $\lim_{q \to 0} \Theta(x\, ; q)=(1-x)$, so the 2d boundary theory on $T^2$ indeed reduces to a 1d quantum mechanics on $S^1_{\cC'}$.\\

From the above presentation, it is clear that magnetic solutions can be recovered from the electric ones via pairing with a specific covector $W$. For instance, in the case of $z$-solutions, consider a magnetic block with exceptional Dirichlet boundary conditions in vacuum ${\{{\bf A}\}}$, and denote it as $\bold{V}_{\{{\bf A}\}}$; then there exists a covector $W$ such that 
\beq\label{covector}
\sum_{\{{\bf A'}\}} W_{\{{\bf A'}\}}\, {\bf F}^{\{{\bf A'}\}}_{\{{\bf A}\}} = \bold{V}_{\{{\bf A}\}} \; ,
\eeq 
with ${\bf F}^{\{{\bf A'}\}}_{\{{\bf A}\}}$ the electric $z$ solution from \eqref{qKZellstab1}. For details on the existence of $W$, we refer the reader to the geometric arguments invoked in the unramified context  \cite{Aganagic:2017smx,Okounkov:2015spn}. The same arguments apply here, without any major modification; see also Section \ref{sec:quasim}.

\vspace{8mm}

\section{Tame ramification from 5d gauge theory}
\label{sec:5dgauge}

The 3-dimensional Drinfeld quiver gauge theory $T^{3d}_{\fg_\text{o}}$ has a 5-dimensional origin, as we now explain.\\ 

Let $T^{5d}_{\fg_\text{o}}$ be a 5d quiver gauge theory with $\cN=1$ supersymmetry, supported on the manifold $M_5=S^1_{\cC'}(R_{\cC'})\times\mathbb{C}^2$. It admits a $SU(2)_R$ R-symmetry. The quiver shape is once again the Dynkin diagram of the simply-laced Lie algebra $\fg_\text{o} = ADE$ of rank $r=\r$, and the gauge group is a product of unitary groups
\beq\label{gaugegroup5d}
G_{5d}=\prod_{a=1}^{r} U(n_a)\; ,
\eeq
where the $n_a$ are all positive integers\footnote{For each $a=1,\ldots,r$ the center $U(1)\subset U(n_a)$  is actually nondynamical, so the gauge group should really be written as $\prod_{a=1}^{r} SU(n_a)$. Nevertheless,  we will later place the theory on the  $\Omega$-background, where the $U(1)$ centers will become physically sensible. For instance, the instanton partition function of the theory in that background depends on $n_a$ equivariant Coulomb parameters, not $n_a-1$, so we will usually keep the gauge group as $G_{5d}$.}.

The bifundamental matter between gauge nodes consists of hypermultiplets in the representation $\oplus_{b>a}\, \Delta_{ab}\,(n_a, \overline{n_b})$ of the group $\prod_{a,b} U(n_a)\times U(n_b)$. The operator $\Delta_{ab}$ is the incidence matrix of $\fg_\text{o}$, with entry $1$ if there is a link connecting nodes $a$ and $b$ in the Dynkin diagram, and  $0$ otherwise. The quiver contains a total of $r-1$ such bifundamental hypermultiplets.

The remaining matter consists of fundamental hypermultiplets, taken in representation  $(n_a,\overline{n^F_a})$ of $U(n_a)\times U(n^F_a)$, with $U(n^F_a)$ the corresponding flavor symmetry group on node $a$:
\beq\label{flavorgroup5d}
G^F_{5d} = \prod_{a=1}^{r} U(n^F_a)\; ,
\eeq  
where $n^F_a$ are non-negative integers. 

There is an additional global symmetry in 5d,
\beq\label{topological5d}
G^{top}_{5d}=\prod_{a=1}^{r} U(1)^{top}_{a} = U(1)^r\; ,
\eeq  
the so-called topological symmetry. For all $a=1,\ldots,r$, the symmetry arises from the conserved current  ${\cal J}_{5d,a}=\frac{1}{8\pi^2}\mbox{Tr}(F_a\wedge F_a)$, and the associated charge is called the instanton number $k_a$. The mass parameters associated to this topological symmetry are instanton counting parameters $z_a$, which in 5d are related to the gauge couplings as
\beq\label{5dinstpar}
z_a=e^{-8\pi^2\,R_{\cC'}/(g^{5d}_{a})^2}\; , \qquad \;\;a=1,\ldots,\r\; .
\eeq
There are various physical constraints to take into account. 

First, as a supersymmetric gauge theory in 5 dimensions, $T^{5d}_{\fg_\text{o}}$ is non-renormalizable, and should be understood as an IR relevant deformation of some UV fixed point. The $r$ gauge coupling $1/(g^{5d}_{a})^2$  have mass dimension 1, and set the scale of the deformation. The requirement of having a UV fixed point imposes constraints on the ranks of the various groups \cite{Intriligator:1997pq}. For instance, in the rank 1 case, meaning a single $SU(n)$ gauge group, $T^{5d}_{A_1}$ has a UV fixed point if $n^F + 2 \,|\kappa_{CS, bare}| \leq 2\, n$, where $\kappa_{CS, bare}$ is a bare Chern-Simons level. Such inequalities exist for all $\fg_\text{o}$-type quivers.

Furthermore, we will see that the 3d theory $T^{3d}_{\fg_\text{o}}$ always arises as a codimension-2 vortex solution on the Higgs branch of $T^{5d}_{\fg_\text{o}}$\footnote{The vortices in question can be described as (non-abelian versions of) semi-local vortices \cite{Achucarro:1999it,Hanany:2003hp}.} In particular, our 5d theories should  be ``fully Higgsable" to begin with. Their existence implies that the ranks of the flavor groups $n^F_a$ have to be ``sufficiently large" integers. In our rank 1 example, assuming $n^F$ is even so we can choose the bare Chern-Simons level to be $\kappa_{CS, bare}=0$, the existence of vortex solutions imposes the constraint $n^F\geq 2\, n$ on the number of fundamental hypermultiplets\footnote{Whenever  $n^F< 2\, n$, full Higgsing is also possible, but only on a baryonic Higgs branch \cite{Argyres:1996eh}. Here, our gauge group is taken to be $U(n)$ instead of $SU(n)$, and baryons are not invariant under the additional gauged $U(1)$. It follows that the baryonic Higgs branch is lifted, and full Higgsing is only possible for $n^F\geq 2\, n$. An identical argument was invoked by Hanany and Witten, but in 3 dimensions instead of 5 \cite{Hanany:1996ie}.}.

Lastly, we will eventually make contact with various quantum field theories in lower dimensions after compactification. Perhaps the most natural reduction is on $S^1_{\cC'}$, down to a 4d $\cN=2$ theory, but even then, this limit is subtle: there are a priori many ways to scale the 5d variables as $R_{\cC'}\rightarrow 0$ which can result in sensible 4d theories. Among them, the most naive one gives a  4d $\cN=2$  $\fg_\text{o}$-type quiver gauge theory with the ``same"  matter content as the 5d one, and gauge couplings which remain finite. We would like these theories to be asymptotically free or conformal in 4d, so the non-positivity of the 1-loop beta function is imposed as a constraint. In our rank 1 $U(n)$ example, this translates to a constraint on the number of flavors\footnote{Gauge theories in 4 dimensions where the number of fundamental hypermultiplets saturates the inequality, such as $n^F = 2\, n$ in the rank 1 case, are sometimes called \emph{asymptotically conformal} \cite{Nekrasov:2012xe}. The expression stands for the theory becoming superconformal as the hypermultiplets become massless.}: $n^F \leq 2\, n$.  Such inequalities exist for all $\fg_\text{o}$-type quiver gauge theories. This compactification is useful to make contact with the representation theory of quantum groups, namely the character of finite-dimensional representations of Yangians \cite{Nekrasov:2012xe,Nekrasov:2013xda}.

Scaling the 5d variables differently, another possible reduction on $S^1_{\cC'}$ will result in a 4d SCFT of class ${\mathcal S}$ \cite{Gaiotto:2009hg,Gaiotto:2009we}. In our rank 1 $SU(n)$ example, the SCFT constraint in the limit would saturate the previous inequality to $n^F = 2\, n$. Generically, and unlike the previous 4d limit, the Lagrangian description is lost in this reduction, since the gauge couplings become infinite as $R_{\cC'}\rightarrow 0$, meaning there is no energy scale at which we can provide a weakly coupled gauge theory description\footnote{Some 5d theories are known to have more than one inequivalent Lagrangian descriptions to begin with, for instance theories which are related by spectral duality \cite{Katz:1997eq}. When this is the case, it is typically possible for one of these Lagrangian descriptions to survive the 4d limit, a point we will revisit in Section \ref{ssec:spectral}.}. This compactification is useful to make contact with the AGT correspondence \cite{Alday:2009aq}.

If we imagine compactifying $\mathbb{C}^2\subset M_5$ on $T^2$ instead, the reduction on this torus yields a 3d $\cN=4$ quiver gauge theory whose Coulomb branch is a slice of the affine Grassmanian, see \cite{Braverman:2016pwk,2017arXiv171203039F,Bullimore:2015lsa} and recently \cite{Bourget:2021siw}; there, constraints similar to the above determine how ``balanced" the theory is, which in turn fixes the behavior in the IR. In our rank 1 example, a theory satisfying the constraint $n^F \leq 2\, n$ would be called ``good" in the terminology of \cite{Gaiotto:2008ak}. Further reducing on $S^1_{\cC'}$ and at low energies, this compactification is useful to make contact with Gukov and Witten's description of surface defects in 4d Super Yang-Mills via 2d sigma models \cite{Gukov:2006jk,Haouzi:2016ohr}. 

In the rest of this work, we decide to impose constraints which are physically sensible in most (and often all) of the above scenarios. First, the flavor content of our theories $T^{5d}_{\fg_\text{o}}$ requires the bare Chern-Simons levels to be integers, which we further specialize to be zero out of convenience, $\kappa_{CS, bare}=0$. There are however nonzero effective Chern-Simons levels $\kappa_{CS}\neq 0$ generated at 1-loop; assuming all masses are positive, these levels are uniquely determined from the ranks of the gauge and flavor groups. Second, we require a sufficiently large number of fundamental hypermultiplets to guarantee a fully Higgsable 5d theory, and the existence of vortex solutions on that Higgs branch. Lastly, in the language of the 4d reductions discussed above, we  only consider theories with vanishing 1-loop beta function. In our rank 1 example, this means $n^F = 2\, n$ already in 5d, so we are considering the gauge group $G_{5d}= U(n)$ and flavor group $G^F_{5d}=  U(2\,n)$. This case will be treated in  detail in Section \ref{sec:example}.\\

For a generic quiver made up of unitary gauge groups, the asymptotic conformality constraint takes the form of $\r$ equations:
\beq\label{constraint5d}
\sum_{b=1}^{\r} C_{ab} \;n_b = n^F_a\; ,\qquad\; a=1,\ldots, \r\; ,
\eeq
where $C_{ab}=\langle\alpha_a,\alpha_b^{\vee}\rangle$ is the Cartan matrix of $\fg_\text{o}$. We will define the quiver theory $T^{5d}_{\fg_\text{o}}$ by specifying its fundamental hypermultiplet content: a vector of $r=\r$ non-negative integers $(n^F_1,n^F_2,\ldots,n^F_r)$, where $n^F_a$ denotes the rank of the $a$-th flavor group $U(n^F_a)$ (and the integers cannot all be zero). This fixes the right-hand side of the linear system \eqref{constraint5d}, which either has no solution or a unique integer-valued solution $(n_1,n_2,\ldots,n_r)\in \left(\mathbb{Z}^+\right)^r$. The solution, whenever it exists, spells out the gauge content of $T^{5d}_{\fg_\text{o}}$, meaning the ranks of the gauge groups are uniquely fixed by the ranks of the flavor groups and the constraint.

We have chosen to set the Chern-Simons levels to 0 on all nodes, $\kappa_{CS, bare,a}=0$.
Given the fundamental and bifundamental matter content of the quiver $T^{5d}_{\fg_\text{o}}$, and fixing the sign of the corresponding masses to be positive, the bare Chern-Simons level on node $a$ is corrected quantum mechanically at 1-loop to give an effective Chern-Simons coupling
\beq\label{effectiveCS0}
\kappa_{CS,a}=\kappa_{CS, bare,a}+\frac{1}{2}\left(n^F_a + \sum_{b<a} \Delta^{ba} \, n_{b} - \sum_{b>a} \Delta_{ab} \, n_{b}\right) \; ,
\eeq
with $\Delta_{ab}=1$ or $0$ depending on whether the gauge nodes $a$ and $b$ are linked or not in the Dynkin diagram of $\fg_\text{o}$. 
After imposing the constraint \eqref{constraint5d} and setting $\kappa_{CS, bare,a}=0$, this simplifies to 
\beq\label{effectiveCS}
\kappa_{CS,a}= n_a - \sum_{b>a} \Delta_{ab}\, n_b \; .
\eeq
for all nodes $a=1,\ldots,\r$.


\vspace{8mm}

\subsection{The instanton partition function}
\label{ssec:5dinstanton}

We are interested in counting self-dual instantons on $\mathbb{C}^2$ inside $M_5=S^1_{\cC'}(R_{\cC'})\times\mathbb{C}^2$. In 5 dimensions and in the moduli space approximation, the dynamics of instantons is captured by a supersymmetric sigma model whose target space is the instanton moduli space.
The partition function of the 5d theory $T^{5d}_{\fg_\text{o}}$ on $S^1_{\cC'}$ factorizes into the contribution of a 1-loop exact perturbative part, counting W-bosons along with their fermionic supersymmetric partners, and an instanton part:
\beq
Z(T^{5d}_{\fg_\text{o}}) = Z_{pert}(T^{5d}_{\fg_\text{o}})\cdot Z_{inst}(T^{5d}_{\fg_\text{o}})\; , \qquad\;\;\;\;  Z_{inst}(T^{5d}_{\fg_\text{o}})=\sum_{\vec{k} =0}^{\infty}\;\prod_{a=1}^{r}  {Z}_{k_a}\;  z_a^{k_a}\; ,
\eeq
where the $a$-th entry of the vector $\vec{k} = (k_1,\ldots,k_r)$ labels an instanton number $k_a$ on node $a$ of the quiver.  We will neglect the perturbative part in what follows, since it does not enter the conformal block physics. The instanton sector is a sum over all instanton numbers, with the counting parameters $z_a$ encoding the 5d gauge couplings.

As is well-known, the instanton moduli space suffers both from IR singularities at infinity of $\mathbb{C}^2$, and from UV singularities due to coincident zero-size instantons, which implies the quantum mechanics description is incomplete. A powerful way to resolve such singularities is to make use of the ADHM construction \cite{Atiyah:1978ri}, which provides a UV completion as a 1d $\cN=4$ gauged quantum mechanics on $S^1_{\cC'}$\footnote{The supersymmetry of the gauged quantum mechanics in question is sometimes called $\cN=4$B. It arises as the 1-dimensional reduction of 2d $\cN=(0,4)$ supersymmetry.}. The instanton dynamics are captured by the Higgs branch of this quantum mechanics, whose gauge group is $\prod_{a=1}^r U(k_a)$. Note that the instanton number $k_a$ now also denotes the rank of the $a$-th gauge group in this UV  quantum mechanics. The instanton partition function takes the form of a Witten index \cite{WITTEN1982253,Alvarez-Gaume:1986ggp}, which can be evaluated via equivariant localization  \cite{Moore:1997dj,Losev:1997hx,Moore:1998et,Nekrasov:2002qd,Hwang:2014uwa}. In the weak coupling regime of the UV quantum mechanics, where $g_{QM,a}\rightarrow 0$ for all $a=1,\ldots, r$, the index reduces to Gaussian integrals around saddle points, meaning it can be evaluated exactly. Meanwhile, the UV incomplete quantum mechanics description is recovered in the low energy regime, after taking the couplings of the gauged quantum mechanics to infinity, $g_{QM,a}\rightarrow \infty$. More precisely, if $E$ is the energy scale at which we wish to study the 5d BPS states, the decoupling of the extra UV degrees of freedom will occur when $g_{QM,a}^{-2/3} \, E \ll 1$\footnote{Technically, we are not varying the couplings $g_{QM,a}$ in those arguments, but rather the radius $R_{\cC'}$ of the circle on which the quantum mechanics is defined. Namely, the BPS index is properly defined for $R^3_{\cC'}\, g_{QM,a}^{2}\gg 1$, and the weak coupling regime where we perform the localization computation is instead defined for  $R^3_{\cC'}\, g_{QM,a}^{2}\ll 1$. This should be interpreted as varying the radius $R_{\cC'}$ continuously from $\infty$ to 0 in the quantity $R^3_{\cC'}\, g_{QM,a}^{2}$, with the couplings $g_{QM,a}^{2}$ kept fixed and large enough ($g_{QM,a}^{2} \gg E^3$) to  decouple the UV degrees of freedom. In the main text, the dependence on $R_{\cC'}$ was kept implicit not to overburden the expressions; it is easily reinstated by dimensional analysis.}.  
The instanton partition of this quantum mechanics once again takes the form of a sum over all instanton numbers,
\beq
{\mathcal Z}_{inst}(T^{5d}_{\fg_\text{o}}) =\sum_{\vec{k} =0}^{\infty}\;\prod_{a=1}^r  {\mathcal Z}_{k_a}\;  \frac{z_a^{k_a}}{k_a!} \; .
\eeq
A subtle but important point is that the UV index ${\mathcal Z}_{inst}$ will sometimes count more states than the ``true" instanton generating function $Z_{inst}$ of the low energy QFT in the decoupling regime, so the two quantities are \emph{not} equal in general. The discrepancy between the two partition functions can be factored out, as it represents the contribution of extra spurious states in the UV. We denote the contribution of these unwanted 5d decoupled bulk states as $Z_{extra}$, and so
\beq
\label{extra}
{\mathcal Z}_{inst} \equiv Z_{inst}\cdot {\mathcal Z}_{extra}
\eeq
The quantity ${\mathcal Z}_{extra}$ has been determined for all classical gauge groups and various types of matter \cite{Nekrasov:2004vw,Shadchin:2005mx}. For some specific gauge groups and matter contents, it can happen that the UV ADHM quantum mechanics admits a realization in string theory  \cite{Witten:1994tz,Douglas:1995bn}; when this is the case, the contribution ${\mathcal Z}_{extra}$ can most easily be understood from moduli associated to light instantonic branes ``escaping to infinity,"  away from the heavier branes on which the 5d gauge theory lives. These non-compact flat directions are moduli coming from the Coulomb branch of the quantum mechanics. Before we say more about these Coulomb moduli, let us discuss the index ${\mathcal Z}_{inst}$.

In $\cN=2$ notation, the index is written with respect to two supercharges $Q_+$ and $\overline{Q}_+$. These are the same supercharges we previously used when we discussed the 3d half-index \eqref{3dhalfindex}. Explicitly, 
\beq
\label{5dhalfindexmore}
{\mathcal Z}_{inst}(T^{5d}_{\fg_\text{o}})  = {\rm Tr}\left[(-1)^F\, q^{J_1+\frac{R}{2}}\; t^{J_2-\frac{R}{2}} \; {\fm}^{\Sigma}\right]\;\; .
\eeq
$F$ is the fermion number operator. $J_1$ and $J_2$ are valued in the Cartan subalgebra of $U(1)\times U(1)$, which acts by rotating $\mathbb{C}\times\mathbb{C}$; they have half-integer eigenvalues.

The variable $\log(\fm)$ collectively stands for all chemical potentials conjugate to the Cartan generator $\Sigma$ of the 5d gauge group $G_{5d}$, 5d flavor group $G^F_{5d}$, and 5d topological symmetry group $G^{top}_{5d}$. These groups are all understood as global symmetries from the point of view of the quantum mechanics. Explicitly, 
\beq
{\fm}^{\Sigma} = \prod_{a=1}^r z_a^{k_a} \prod_{i=1}^{n_a} e_{a,i}^{\ft_{a,i}} \prod_{s=1}^{n^F_a} f_{a,s}^{\ft^{F}_{a,s}} \; ,
\eeq
where each $e_{a,i}$ is a 5d Coulomb parameter with associated Cartan generator $\ft_{a,i}$, and $f_{a,s}$ is a 5d hypermultiplet mass, with associated Cartan generator $\ft^{F}_{a,s}$. As before, $z_a$ encode the 5d gauge couplings, and $k_a$ are the instanton numbers.

The R-symmetry of the $\cN=4$ quantum mechanics is $SU(2)_r\times SU(2)_R$, and the operator $R$ in the index stands for the Cartan generator of $U(1)_R\subset SU(2)_R$. As usual, the presence of the R-symmetry twist is required to preserve supersymmetry. Note that $J_1+\frac{R}{2}$, $J_2-\frac{R}{2}$, and $\Sigma$ all commute with the supercharges.

From the 5-dimensional perspective, we are describing an $\Omega$-background \cite{Nekrasov:2002qd,Losev:2003py,Nekrasov:2010ka}: just as in 3 dimensions, we view $M_5$ as a $\mathbb{C}^2$ bundle over $S^1_{\cC'}(R_{\cC'})$, and let $(X_{\cC'},X_1,X_2)$ be coordinates on $S^1_{\cC'}\times\mathbb{C}^2$, so that the identification
\begin{align}\label{omega}
(0, X_1, X_2) \sim  (2\pi R_{\cC'},\, q\, X_1,\, t^{-1}\, X_2)
\end{align}
holds. That is, the chemical potentials $q$ and $t$  are equivariant parameters for the $U(1)_q\times U(1)_t$ action on $\mathbb{C}\times\mathbb{C}\subset M_5$. We denote this $\Omega$-background as $S^1_{\cC'}\times\mathbb{C}_q\times\mathbb{C}_t$.\\

The index is expected to be well-defined only if the theory is fully gapped. In our quiver quantum mechanics, we will turn on all possible chemical potentials (equivariant parameters) for the global symmetries, but even then, a continuum may remain due to the vector multiplet scalars $\varphi_a$, also called Coulomb branch moduli. Whenever that is the case, we should not expect the index to have a sensible expansion in its parameters with integer coefficients. The scalars $\varphi_a$ are neutral under all global symmetries, meaning such a continuum cannot be lifted by the chemical potentials we turn on. Instead, because our quiver quantum mechanics is exclusively made up of $r$ unitary gauge groups, we can lift the continuum by turning on $r$ Fayet-Iliopoulos (F.I.) parameters $\zeta_a$, one for each gauge group in the quiver quantum mechanics. 

Just like in the 3-dimensional context, the F.I.-parameter space of the quantum mechanics is divided into chambers. In the interior of a chamber, the gauge group is broken to a finite subgroup, while on a boundary between chambers, a continuous unbroken subgroup will emerge, and with it the existence of a Coulomb branch.

For generic values of $\zeta_a$, deep inside a chamber, the F.I. terms generate a mass gap, because the vector multiplet scalars $\varphi_a$ acquire a mass proportional to $g^2_{QM,a}\, \zeta_a$; there, the Witten index is expected to have a sensible integer expansion in its fugacities, and not depend on the parameters $\zeta_a$\footnote{When we evaluate the index by localization, we are taking the limit $g_{QM,a}\rightarrow 0$, which means we are simultaneously taking the limit $g_{QM,a}\, \zeta_a \rightarrow \pm\infty$ in order to produce a mass gap and guarantee a sensible index.}. As the $\zeta_a$ are varied continuously from one chamber to another, the index is expected to jump, a phenomenon known as wall-crossing in quantum mechanics. For instance, in the rank 1 case of a $U(1)$ gauged quantum mechanics, there are two chambers $\zeta>0$ and $\zeta<0$, and a Coulomb branch continuum arises at the boundary $\zeta=0$, at the root of the Higgs branch. The continuum contributes a factor ${\mathcal Z}_{extra}$ to the index formula \eqref{5dhalfindexmore}, which factorizes as \eqref{extra}.

In the rest of this work, we fix all F.I. parameters to lie in a generic chamber, away from the walls, say $\zeta_a$ all chosen generic and positive. In the class of quivers we consider, ${\mathcal Z}_{extra}\neq 1$ for our choice of flavor symmetry and bare Chern-Simons levels. For instance, for a single $SU(n)$ gauge group in 5 dimensions and zero bare Chern-Simons level, the Coulomb branch continuum contributes additional states to the index if the number of fundamental hypermultiplets is twice the rank of the gauge group, $n^F = 2\, n$ \cite{Bergman:2013ala,Bao:2013pwa,Hayashi:2013qwa,Taki:2013vka}. But this is precisely the matter content we consider in this paper, as our constraint \eqref{constraint5d} takes the form $n^F = 2\, n$ in the rank 1 case, see Section \ref{sec:example}. It follows that our index computation of ${\mathcal Z}_{inst}$ will differ from the QFT quantity $Z_{inst}$. Beyond the rank 1 case, it will always be the case that ${\mathcal Z}_{extra}\neq 1$ for our class of quiver gauge theories.

Since we are ultimately interested in comparing gauge theory partition functions to $q$-deformed conformal blocks, it is natural to ask if ${\mathcal Z}_{extra}$ also has a natural interpretation in the chiral algebra context. For $\fg=A_r$ quivers, the  ${\mathcal Z}_{extra}$ contributions have been attributed to the presence of additional free bosons in the $\cW_{q,t}(A_r)$ $q$-conformal block \cite{Alba:2010qc}. In effect, one considers not just the algebra $\cW_{q,t}(A_r)$, but rather the direct sum $\cW_{q,t}(A_r)\oplus u(1)_{q,t}$ with an additional $q$-deformed Heisenberg algebra. The modern perspective on the presence of this additional Heisenberg algebra is that the blocks really are defined via the Fock representation of a quantum toroidal algebra $U_{q,t}(\Hat{\Hat{\fgl_1}})$ \cite{Mironov:2016yue}. 
In our work, we will not worry about singling out ${\mathcal Z}_{extra}$ explicitly. We will simply compute the UV index ${\mathcal Z}_{inst}$ as a whole and the equality to $\cW_{q,t}(\fg)$ $q$-conformal blocks will hold only up to these additional $u(1)_{q,t}$ contributions. We leave the detailed investigation of ${\mathcal Z}_{extra}$ for general quivers to future work, and in particular the interpretation of such factors whenever $\fg\neq A_r$.\\

Going back to the evaluation of the index deep inside a given F.I.-parameter chamber and in the  $g_{QM,a}\rightarrow 0$ limit, one first integrates over
the non-zero modes by performing Gaussian integrals over massive fluctuations, keeping the zero modes fixed. Then, one integrates over these zero modes, which are the 1d $\cN=2$ vector multiplet scalars $\varphi_a$, with $a=1,\ldots,r$ labeling the quiver node. As usual, these scalars get complexified by the holonomy around the circle $S^1_{\cC'}(R_{\cC'})$: we denote the complex eigenvalues as $\phi_{a,I}=\varphi_{a,I}+ i \, A^{\theta}_{a,I}$, \footnote{There are also fermion zero modes coming from the gaugini in the $\cN=2$ vector multiplet. These contribute Fermi multiplets to the index integrand, whose form is that of the Vandermonde determinant for the gauge group of the quantum mechanics, $\prod_{a=1}^r U(k_a)$.} where  $I=1,\ldots,k_a$. For a given instanton configuration $\vec{k}$, there are a total of $|\vec{k}|=\sum_{a=1}^r k_a$ such complex eigenvalues valued on an infinite cylinder. Explicitly, one finds\footnote{We have absorbed a phase $(-1)^{k_a\, \kappa_{CS,a}}$ in the definition of the instanton counting parameter $z^{k_a}_a$; this phase is due to the nonzero effective Chern-Simons levels.}
\begin{align}
\label{5dintegral}
{\mathcal Z}_{k_a}  =\oint  \prod_{I=1}^{k_a}  \left[\frac{d\phi_{a,I}}{2\pi i}\right]{\mathcal Z}^{vec}_a\cdot {\mathcal Z}^{fund}_a\cdot \prod_{b>a}^{r} {\mathcal Z}^{bif}_{a,b} \; .
\end{align}
The factors $Z^{vec}_a$, $Z^{fund}_a$, and $Z^{bif}_{a,b}$ are the 1-loop determinants for the quantum mechanical modes of the vector multiplets, the fundamental hypermultiplets and the bifundamental hypermultiplets of the 5d gauge theory, respectively. The  determinant ${\mathcal Z}^{fund}_a$ is the contribution of 1d $\cN=2$ Fermi multiplets, while ${\mathcal Z}^{vec}_a$ and ${\mathcal Z}^{bif}_{a,b}$ are contributions of Fermi \emph{and} chiral multiplets. 

The chiral multiplets are responsible for providing poles to the integrand. The contours can be specified using an ``$i \E$"  prescription \cite{Nekrasov:2002qd,Nekrasov:2003rj}, or using the Jeffrey-Kirwan residue \cite{Jeffrey:1993}. The latter method was recently used to compute partition functions of 2-dimensional supersymmetric theories \cite{Benini:2013nda,Benini:2013xpa}, and was applied to our quantum mechanical context soon after \cite{Hwang:2014uwa,Cordova:2014oxa,Hori:2014tda}. In particular, the integral should be treated carefully near the zeroes of the chiral multiplets, where the integrand blows up and the assumption that the non-zero modes have a large mass breaks down. Moreover, our integration variables $\phi_{a,I}$ are $\mathbb{C}^\times$-valued, so require regularization with an IR cutoff at the ends of the cylinders\footnote{This last point is intimately related to the identification of the factors ${\mathcal Z}_{extra}$. Namely one should look for $\phi_{a,I}$-poles at $\infty$ in the integrand of the index, signaling the appearance of a Coulomb branch continuum.}; for details, see \cite{Hwang:2014uwa}.\\





Regardless of the residue prescription one uses, in the end one finds that the instanton partition function is evaluated exactly as 
\beq\label{bulk5d}
{\mathcal Z}_{inst}(T^{5d}_{\fg_\text{o}}) =  \sum_{\{\overrightarrow{\boldsymbol{\mu}}\}} \;\prod_{a=1}^r  z_a^{\sum_{i=1}^{n_a}{\left|\boldsymbol{\mu}_{a,i}\right|}}\, {\boldsymbol Z}^{5d,vec}_{a} \cdot {\boldsymbol Z}^{5d,fund}_{a}\cdot {\boldsymbol Z}^{5d,CS}_{a} \cdot \prod^r_{b>a}
{\boldsymbol Z}^{5d,bif}_{a,b}\, .
\eeq
The sum is over a collection of 2d partitions (Young tableaux), one for each $U(1)$ Coulomb parameter in the gauge group $G_{5d}$:
\beq
\{\overrightarrow{\boldsymbol{\mu}}\}=\{\boldsymbol{\mu}_{a,i}\}_{a=1, \ldots, r\, ; \;\; i=1,\ldots,n_a}\; .
\eeq
Every such partition describes a configuration of instantons at the fixed point of the $U(1)_q \times U(1)_t \times U(1)^{\sum_a n_a}$ equivariant action. In particular, the instantons sit at the origin of $\mathbb{C}_q\times\mathbb{C}_t$.\\

Each of the factors in the summand is best expressed in terms of the function
\beq\label{nekrasovN}
{\cN}_{\boldsymbol{\mu}_{a,i}\boldsymbol{\mu}_{b,j}}(Q\, ;q) \equiv \prod\limits_{k,k' = 1}^{\infty} 
\frac{\big( Q \, q^{\boldsymbol{\mu}_{a,i,k}-\boldsymbol{\mu}_{b,j,k'}} \,t^{k' - k + 1}\,;q \big)_{\infty}}{\big( Q\,  q^{\boldsymbol{\mu}_{a,i,k}-\boldsymbol{\mu}_{b,j,k'}}\, t^{k' - k}\, ;q\big)_{\infty}} \,
\frac{\big( Q\,  t^{k' - k}\, ;q \big)_{\infty}}{\big( Q\,  t^{k' - k + 1}\, ;q\big)_{\infty}}\, .
\eeq
The notation $\boldsymbol{\mu}_{a,i,k}$ stands for the length of the $k$-th row in the partition $\boldsymbol{\mu}_{a,i}$.

Then, for a given quiver node $a$, the factor 
\begin{align}\label{5dbulkvec}
{\boldsymbol Z}^{5d,vec}_{a} = \prod_{i,j=1}^{n_a}\left[\cN_{\boldsymbol{\mu}_{a,i}\boldsymbol{\mu}_{a,j}}\left(\frac{e_{a,i}}{e_{a,j}} \, ;q\right)\right]^{-1}
\end{align}
represents the contribution of a 5d vector multiplet, where each $e_{a,i}$ denotes a corresponding Coulomb parameter.

The factor
\begin{align}\label{5dbulkmatter}
{\boldsymbol Z}^{5d,fund}_{a} = \prod_{s=1}^{n^F_a} \prod_{i=1}^{n_a} \cN_{\boldsymbol{\emptyset}\, \boldsymbol{\mu}_{a,i}}\left( \sqrt{\frac{q}{t}}\, \frac{f_{a,s}}{e_{a,i}}\, ; q\right)\; .
\end{align}
represents the contribution of a 5d fundamental hypermultiplet, in representation $(n_a,\overline{n^F_a})$ of $U(n_a)\times U(n^F_a)$, where each $f_{a,s}$ denotes a corresponding mass parameter.

The factor 
\begin{align}\label{5dbulkCS}
{\boldsymbol Z}^{5d,CS}_{a} = \prod\limits_{i=1}^{n_a} \left(T_{\boldsymbol{\mu}_{a,i}}\right)^{\kappa_{CS,a}}\; .
\end{align}
represents the contribution of the effective Chern-Simons term of level $\kappa_{CS,a}$ on node $a$  \eqref{effectiveCS}, where  $T_{\boldsymbol{\mu}_{a,i}} =(-1)^{|\boldsymbol{\mu}_{a,i}|}\; q^{\Arrowvert \boldsymbol{\mu}_{a,i}\Arrowvert^{2}/2}\; t^{-\Arrowvert \boldsymbol{\mu}^t_{a,i}\Arrowvert^{2}/2}$. 

Finally, the factor 
\begin{align}\label{5dbulkbif}
{\boldsymbol Z}^{5d,bif}_{a,b} = \prod_{i=1}^{n_a}\prod_{j=1}^{n_b}\left[\cN_{\boldsymbol{\mu}_{a,i} \boldsymbol{\mu}_{b,j}}\left(f^{bif}_{a,b}\,\frac{e_{a,i}}{e_{b,j}}  \, ; q\right)\right]^{\Delta_{ab}}\; .
\end{align}
represents the contribution of a 5d bifundamental hypermultiplet in representation $\oplus_{b>a}\, \Delta_{ab}\,(n_a, \overline{n_b})$ of $\prod_{a,b} U(n_a)\times U(n_b)$, where $\Delta_{ab}$ is the incidence matrix. We turn on corresponding masses $f^{bif}_{a,b}$, one for each link in the quiver. In order to match to the conventions of the existing $\cW_{q,t}(\fg_\text{o})$-algebra literature\footnote{Our definition of the deformed $\cW_{q,t}(\fg_\text{o})$-algebra relied on the choice of a deformed Cartan matrix of type $\fg_\text{o}$; here, we choose to follow the conventions of \cite{Frenkel:1998,Bouwknegt:1998da} and deform the Cartan matrices as \eqref{qCartanToda}, which fixes all bifundamental masses $f^{bif}_{a,b}$. For a slightly different definition of the deformed $\cW_{q,t}(\fg_\text{o})$-algebra and its associated Cartan matrix, it is also possible to keep all masses arbitrary, see \cite{Kimura:2015rgi,Kimura:2017hez} instead. Ultimately, this is not a crucial distinction, as there is freedom to rescale $f^{bif}_{a,b}$ away by redefining the other parameters. In the quantum affine algebra, this is the freedom to rescale the evaluation parameters.}, we set all bifundamental masses $f^{bif}_{a,b}$ to
\beq\label{bifspecialized}
f^{bif}_{a,b}= \sqrt{q/t} \; .
\eeq

\begin{remark}
	In Section \ref{sec:3dgauge}, we computed the half-index of $T^{3d}_{\fg_\text{o}}$ on $\mathbb{R}^{1,1}\times\mathbb{R}_{-}$ by enumerating all 3d fields in the UV and specifying 1/2-BPS boundary conditions.
	We could have instead considered the adiabatic approximation where the 3d supersymmetric path integral localizes to the supersymmetric quantum mechanics on the moduli space of vortices, analogously to the ADHM computation we did in 5d where the target space of the quantum mechanics was the moduli space of instantons\footnote{In particular, the dependence of the supersymmetric ground states on the various masses is controlled by a Berry connection \cite{Cecotti:1991me,Pedder:2007ff,Sonner:2008fi,deBoer:2008ss,deBoer:2008qe,Cecotti:2013mba,Wong:2015cnt}. For recent developments, see \cite{Dedushenko:2021mds,Bullimore:2021rnr,Dedushenko:2022pem,Galakhov:2023aev}.}.
	
	This approach has some advantages; for example, the equality between 5d instanton and 3d vortex partition functions is manifest when comparing the Witten indices of the two quantum mechanics. Moreover, the electric blocks are more naturally computed in this language: instead of defining a 3d/1d index as we did, one simply introduces the additional chiral and Fermi multiplets directly to the 1d vortex quantum mechanics. For explicit illustrations, see for instance  \cite{Hwang:2017kmk,Haouzi:2019jzk,Crew:2020jyf}. However, this presentation also has drawbacks, which is why we ultimately avoided it: deformed conformal blocks are only known in free field formalism, which \emph{are} the 3d half-indices we computed, and the natural language to best describe their analytic properties is in terms of exceptional Dirichlet and enriched Neumann boundary conditions, which is obscured in the quantum mechanics. Furthermore, the dictionary we found to the representation theory of the quantum affine algebra $U_\hbar(\widehat{^L \fg})$ is only manifest in the 3d picture.
\end{remark}

\vspace{8mm}

\subsection{Gauge-vortex duality on the Higgs branch, and the 5d origin of the blocks}
\label{ssec:duality}

In this section, we analyze the gauge-vortex duality between the 5d gauge theory $T^{5d}_{\fg_\text{o}}$ on its Higgs branch and the 3d Drinfeld gauge theory $T^{3d}_{\fg_\text{o}}$.

Consider first the 5d manifold $S^1_{\cC'}\times\mathbb{C}^2$ in the absence of $\Omega$-deformation. Correspondingly, we set $q, t\rightarrow 1$, which further turns off all bifundamental masses $f^{bif}_{a,b} = \sqrt{q/t}\rightarrow 1$. 
At the root of the Higgs branch, all Coulomb moduli are frozen to (a subset of) the hypermultiplet masses; this process will be described in detail in the next Section. 

We move onto the Higgs branch proper by turning on the real F.I. terms. We are specifically interested in the existence of 1/2-BPS vortices carrying $N_{a,i}$ units of charge:
\beq\label{vortexflux}
N_{a,i} = \frac{1}{2\pi} \int_\mathbb{C} F_{a,i} \; ,
\eeq
where $F_{a,i}$ is the field strength associated to the $i$-th $U(1)$ gauge field, lying in the maximal torus of $U(n_a)$ on quiver node $a$, and the integral is over the complex line transverse to the vortices.

Typically, such a nonzero vortex charge results in additional singular surface operator insertions in spacetime, but in the presence of an $\Omega$-background, the effect is much milder. To see this, let us restore \emph{half} of the $\Omega$-background, meaning we keep $q\rightarrow 1$, but consider an $\Omega$-background on $\mathbb{C}_t$, with $t$ arbitrary. In this background, the theory has an effective description as a 3d $\cN=2$ theory on $S^1_{\cC'}$ with an infinite number of massive modes set by the scale $t$ \cite{Nekrasov:2009rc}. The 5d Coulomb moduli always appear in the combination
\beq\label{omegacoulomb}
e_{a,i} +  t^{X_2\, D_{a,i,X_2}} \; ,
\eeq
where the operator $D_{a,i,X_2}= \partial_{X_2}+ A_{a,i,X_2}$ is the covariant derivative on $\mathbb{C}_t$, with complex coordinate $X_2$ \cite{Nekrasov:2010ka}.

Because of the $\Omega$-background, turning on $N_{a,i}$ units of vortex charge \eqref{vortexflux} will no longer introduce singular surface operators: the only effect will be to shift the values of the 4d Coulomb moduli as
\beq\label{omegacoulombshift}
e_{a,i} \longrightarrow  e_{a,i} \,t^{N_{a,i}} \; .
\eeq
This follows from setting the gauge field to $A_{a,i,X_2}= N_{a,i}/X_2$ in \eqref{omegacoulomb}. In the end, even though we started on the Higgs branch where vortex solutions are well-defined, the $\Omega$-background effectively ``puts us back" onto the Coulomb branch, but only on a lattice of spacing set by positive powers of $t$\footnote{In the large $N$ limit, with $t\rightarrow 1$ and $N_{a,i}\rightarrow\infty$, we can in fact probe the entire continuous Coulomb branch instead of just a lattice. For related perspectives, see \cite{Kimura:2015rgi,Koroteev:2015dja,Koroteev:2018isw,Nieri:2017ntx}.}.

As we have seen, at the root of the Higgs branch, all Coulomb moduli in $T^{5d}_{\fg_\text{o}}$ are frozen to masses according to the coweight decomposition \eqref{weightsdecom}; this remains true in the $\Omega$-background, and we write
\beq\label{5drootHiggs}
e_{a,i} = f_{b,s} \, \sqrt{t/q}^{1+\#_{a,i}} \; ,
\eeq
for some $b\in\{1,\ldots,r\}$ and $s\in\{1,\ldots,n^F_b\}$. The factor $\sqrt{t/q}^{1+\#_{a,i}}$ is a refinement due to the $\Omega$-background. In particular, because $T^{5d}_{\fg_\text{o}}$ is a quiver theory, the bifundamental masses $\sqrt{q/t}$ between adjacent nodes now appear explicitly in the above Higgsing formula: in our conventions, $\#_{a,i}$ is a non-negative integer which records the length of the ``string" from node $a$ where the Coulomb modulus $e_{a,i}$ is located, to node $b$ where the hypermultiplet mass $f_{b,s}$ is located.

After turning on vortices of charge $N_{a,i}$, the Coulomb parameters get further shifted as \eqref{omegacoulombshift}:
\beq\label{5drootHiggsvortex}
e_{a,i} = f_{b,s} \, \sqrt{t/q}^{\;1+\#_{a,i}} \,  t^{N_{a,i}} \; .
\eeq
As we explained, this should be interpreted as the theory being taken away from the root of its Higgs branch and pushed back onto its Coulomb branch, but only on an integer-valued lattice in powers of $t$, due to the transverse $\Omega$-background.

The equality of the 5d instanton partition function and 3d half-index is manifest in the presence of the $\Omega$-background. To see this, one relies on the following property of the function \eqref{nekrasovN}: if $N$ is a positive integer, then 
\beq\label{truncateNekrasov}
\cN_{\boldsymbol{\mu}_{a,i}\boldsymbol{\mu}_{b,j}}\left(\sqrt{q^2/t^2}\; t^{-N}\, ;q\right)=0\;\; \text{unless}\;\; l\left(\boldsymbol{\mu}_{b,j}\right)\leq l\left(\boldsymbol{\mu}_{a,i}\right)+ N \; ,
\eeq
where $l\left(\boldsymbol{\mu}_{b,j}\right)$ is the number of rows in the partition $\boldsymbol{\mu}_{b,j}$, labeling the equivariant Coulomb modulus $e_{b,j}$. 
After specializing all the Coulomb moduli as \eqref{5drootHiggsvortex}, the fundamental matter factor ${\boldsymbol Z}^{5d,fund}_b$ \eqref{5dbulkmatter} in the summand ${\mathcal Z}_{inst}$ will necessarily contain a factor
\beq\label{fundtruncation}
N_{\boldsymbol{\emptyset}\,\boldsymbol{\mu}_{b,j}}\left(\sqrt{q^2/t^2}\;t^{-N_{b,j}} ;q\right)\; ,
\eeq
for some positive integer $N_{b,j}$. It follows that the entire summand vanishes unless the partition $\boldsymbol{\mu}_{b,j}$ has no more than $N_{b,j}$ rows. Note there is no restriction on the number of columns.
All the remaining partitions $\boldsymbol{\mu}_{b',j'}$ are likewise truncated to have a finite number of rows, either directly from ${\boldsymbol Z}^{5d,fund}$ as above, or from previously truncated partitions via the bifundamental factors ${\boldsymbol Z}^{5d,bif}$. 

Whenever two partitions  $\boldsymbol{\mu}_{a,i}$ and $\boldsymbol{\mu}_{b,j}$ have no more than $N_{a,i}$ and $N_{b,j}$ rows respectively, a further factorization occurs,
\begin{align}\nonumber
\cN_{\boldsymbol{\mu}_{a,i}\boldsymbol{\mu}_{b,j}}(Q\, ;q) &= \prod_{k=1}^{N_{a,i}}\prod_{k'=1}^{N_{b,j}} 
\frac{\big( Q \, q^{\boldsymbol{\mu}_{a,i,k}-\boldsymbol{\mu}_{b,j,k'}} \,t^{k' - k + 1}\,;q \big)_{\infty}}{\big( Q\,  q^{\boldsymbol{\mu}_{a,i,k}-\boldsymbol{\mu}_{b,j,k'}}\, t^{k' - k}\, ;q\big)_{\infty}} \,
\frac{\big( Q\,  t^{k' - k}\, ;q \big)_{\infty}}{\big( Q\,  t^{k' - k + 1}\, ;q\big)_{\infty}}\\
&\;\;\times \cN_{\boldsymbol{\mu}_{a,i}\, \boldsymbol{\emptyset}}(Q\, t^{N_{b,j}};q)\;\cN_{\boldsymbol{\emptyset}\boldsymbol{\mu}_{b,j}}(Q\,t^{-{N_{a,i}}};q)\; .\label{NEWnekrasovN}
\end{align}
Since all partitions are truncated, this factorization happens for all functions $\cN_{\boldsymbol{\mu}_{a,i}\boldsymbol{\mu}_{b,j}}$ in the summand ${\mathcal Z}_{inst}$, and it simplifies drastically, to find:

\begin{prop}
The truncated 5d partition function is a 3d half-index with exceptional Dirichlet boundary conditions, for a choice of polarization $\{+,\ldots,+\}$, in a vacuum labeled by truncated Young tableaux.\\
\end{prop}

We can reverse this argument to learn about the combinatorial nature of the contours for the 3d half-index with Neumann boundary conditions. To see how that happens, consider again
\begin{align}\label{part3dagain}
{\mathcal Z}(T^{3d}_{{\fg_\text{o}},{\bf N}})= \frac{1}{|W_{G^{3d}}|}\prod_{a=1}^{r} \prod_{i=1}^{N_a}\oint_C \frac{dy_{a,i}}{y_{a,i}}\, {z_a}^{\frac{\ln(y_{a,i})}{\ln(q)}} \cdot I_a^{3d, vec}\cdot I_a^{3d, flavor}\cdot\prod_{b>a}I_{a,b}^{3d, bif} \; .
\end{align}
In view of the above analysis, we know that the vacua are all labeled by Young diagrams $\{{\boldsymbol{\mu}}_{a,i}\}$ with no more than $N_{a,i}$ rows. Then, let the contours enclose poles that are precisely of this form. Assuming  $|q|, |t| < 1$, these poles are at
\beq\label{newproof6}
y_{a,i,k} =  x_{d,s}\, q^{-\boldsymbol{\mu}_{a,i,k}}\, t^{k-N_{a,i}}\; ,\qquad\;\; k=1, \ldots, N_{a,i}\; ,
\eeq 
with $a=1,\ldots,\r$. The notation $y_{a,i,k}$ for the 3d Coulomb moduli reflects that original integration variables $y_{a,i}$ split after choosing the contours:
\beq\label{ybreaking2}
\prod_{a=1}^{r}\prod_{i=1}^{N_a} y_{a,i} \longrightarrow \prod_{a=1}^{r}\prod_{i=1}^{n_a}\prod_{k=1}^{N_{a,i}}  y_{a,i,k}\; .
\eeq  
Correspondingly, the 3d gauge symmetry group breaks as
\beq\label{breaking2}
G^{3d} =\prod_{a=1}^{r}U(N_a)\qquad\longrightarrow \qquad \prod_{a=1}^{r}\prod_{i=1}^{n_a}U(N_{a,i})\; ,
\eeq
where $N_a=\sum_{i=1}^{n_a}N_{a,i}$.


By gauge-vortex duality, the right-hand side can be reinterpreted purely in 5-dimensional terms: the integers $n_a$ are the ranks of the 5d gauge groups $U(n_a)$, and the ranks of 3d gauge groups $N_{a,i}$ are the vortex charges in the 5d theory. 

The 3d half-index evaluated via residues at \eqref{newproof6} yields
\beq\label{residue3d1}
{\mathcal Z}(T^{3d}_{{\fg_\text{o}},{\bf N}}) =\sum_{\{ \overrightarrow{\boldsymbol{\mu}} \}}\text{res}_{\overrightarrow{\boldsymbol{\mu}}}\left[I^{3d}(y)\right] \; .
\eeq
Namely, both the 3d and 5d partition functions \eqref{bulk5d} are sums over the same set of truncated partitions. 
Both sums are expansions in the parameters $z_a$, which encode the gauge couplings in 5d and the F.I. parameters in 3d. The expressions are analytic in chamber $\fC_C$, by construction.

In order to further compare the summands, the above residue as we have written it here actually yields an infinite answer, so it proves useful to first normalize it by the trivial residue contribution, to obtain a finite answer: 
\beq\label{residue3d2}
{\mathcal Z}(T^{3d}_{{\fg_\text{o}},{\bf N}})=c_{3d}\sum_{\{ \overrightarrow{\boldsymbol{\mu}}  \}}\left[\frac{I^{3d}(y_{\{\overrightarrow{\boldsymbol{\mu}}\}})}{I^{3d}(y_{\{\boldsymbol{\boldsymbol{\emptyset}}\}})}\right].
\eeq
Here, the ``constant" $c_{3d}$ stands for the residue at an empty partition,
\beq\label{c3d}
c_{3d}\equiv \text{res}_{\boldsymbol{\emptyset}}I^{3d}(y)\; .
\eeq

Next, the 5d vector multiplet contributions \eqref{5dbulkvec} become
\begin{align}\label{5dbulkvecto3d}
{\boldsymbol Z}^{5d,vec}_{a} = \frac{I_a^{3d, vec}(y_{\{\boldsymbol{\mu}_a\}})}{I_a^{3d, vec}(y_{\{\boldsymbol{\emptyset}\}})}\cdot V^{vec} \; ,
\end{align}
where the right-hand side is the contribution of the 3d vector multiplets \eqref{vec3d}, and leftover factors we denote as $V^{vec}$.

The 5d bifundamental hypermultiplet factors \eqref{5dbulkbif} become
\begin{align}\label{5dbulkbifto3d}
{\boldsymbol Z}^{5d,bif}_{a,b} = \frac{I^{3d,bif}_{a,b}(y_{\{\boldsymbol{\mu}_a\}}, y_{\{\boldsymbol{\mu}_b\}})}{I^{3d,bif}_{a,b}(y_{\{\boldsymbol{\emptyset}\}}, y_{\{\boldsymbol{\emptyset}\}})}\cdot V^{bif} \; ,
\end{align}
where the right-hand side is the contribution of the 3d bifundamental hypermuliplets \eqref{bif3d}, and leftover factors we denote as $V^{bif}$.

Collecting the above leftover factors, along with the fundamental hypermultiplets and the effective Chern-Simons contributions, we find after some algebra:
\begin{align}\label{5dbulkmatterto3d}
{\boldsymbol Z}^{5d,fund}_{a}\cdot {\boldsymbol Z}^{5d,CS}_{a} \cdot V^{vec} \cdot V^{bif} = \frac{I^{3d,flavor}_{a}(y_{\{\boldsymbol{\mu}_a\}}, \{x_{d,s}\})}{I^{3d,flavor}_{a}(y_{\{\boldsymbol{\emptyset}\}}, \{x_{d,s}\})}\; .
\end{align}
The right-hand side is the contribution of the chiral matter content of the 3d Drinfeld quiver, \eqref{matter3dwow}. By definition, it contains an equal number of fundamental and anti-fundamental chiral multiplets contributions.
In performing this last identification, we implicitly relabeled all 5d hypermultiplet masses as 3d chiral multiplet masses, $f_{a,s}\rightarrow x^{-1}_{d,s}$\footnote{\label{footnotewow}It may not be immediately obvious that such a relabeling is a priori possible, but the decomposition we write in \eqref{weightsdecom} ensures that we have the same number of masses in 5d and in 3d, so the map $f_{a,s}\rightarrow x^{-1}_{d,s}$ is well-defined.}.

After specializing the 5d Coulomb parameters, the sums \eqref{bulk5d} and \eqref{residue3d2} are indexed by the same set of truncated partitions, and the summands match term-by-term, up to the proportionality constant $c_{3d}$; we conclude once again that the indices agree, up to trivial residue contribution\footnote{A slightly more complete treatment would further take into account the perturbative contributions to the partition functions, which we have ignored throughout.}: 
\begin{align}\label{partequality}
{\mathcal Z}(T^{3d}_{\fg_\text{o},\bf D_{EX}}) =c_{3d}\cdot  {\mathcal Z}_{inst}(T^{5d}_{\fg_\text{o}})_{e_{a,i}\propto\; x^{-1}_{d,s}\,t^{N_{a,i}}}\; .
\end{align}
In the $q$-conformal block language, this gives a 5-dimensional description of the magnetic $z$-solutions for the $\cW_{q,t}(\fg)$-algebra blocks.


\subsection{From classical to quantum groups via the Higgs mechanism}
\label{ssec:classtoquant}



Nothing in our treatment of the 5-dimensional gauge theory $T^{5d}_{\fg}$ so far hints at the relevance of a quantum affine symmetry algebra $U_\hbar(\widehat{^L \fg})$. We will now argue that this symmetry arises on the Higgs branch, in the presence of the $\Omega$-background.  

First, consider a trivial spacetime without $\Omega$-background, and define a set containing all hypermultiplet masses $f_{a,s}$ and Coulomb moduli $e_{b,i}$ in $T^{5d}_{\fg}$; at a generic point on the Coulomb branch, there is no natural notion of ordering on this set, but in a Higgs vacuum, it is possible to define a partial order $``\geq"$ on it. Concretely, each hypermultiplet mass will be a maximal element of this ordered set, and the Coulomb moduli are ordered according to how we ``freeze" them to the masses. Namely, given a mass $f_{a,s}$, we freeze a Coulomb modulus by declaring $f_{a,s} \geq e_{a,i}$ as set elements, where the index $a\in\{1,\ldots,\r\}$ is the \emph{same} on both sides. The remaining Coulomb moduli are ordered by further freezing. All in all, this procedure results in $n^F$ partially ordered sets ${\mathcal K}^{\geq}_{a,s}$, each starting with a mass $f_{a,s}$ as a maximal element  (note that the singleton set $\{f_{a',s'}\}^\geq$ is allowed). The superscript notation $\{\ldots\}^\geq$ is here to emphasize that the elements are ordered with respect to $``\geq"$. Different choices of freezing patterns result in different orderings and therefore different sets ${\mathcal K}^{\geq}_{a,s}$, which describe distinct Higgs vacua.

A key observation is that the partial ordering we defined on the set of masses and frozen Coulomb moduli is well-known in the theory of simple Lie algebras: it coincides with the partial ordering on the weight lattice of $^L \fg$. Equivalently, working with the coweight lattice of $\fg$, two coweights $\lambda^\vee$ and $\lambda'^\vee$ are ordered as $\lambda^\vee\geq\lambda'^\vee$ if $\lambda^\vee - \lambda'^\vee$ is a linear combination of positive coroots with non-negative coefficients. This suggests a map from 5d quiver data to Lie algebras, by identifying each mass $f_{a,s}$ as a dominant weight in the $a$-th fundamental representation, and each Coulomb modulus $e_{a,i}$ as the $a$-th positive simple root of $\fg$.\\

So far, all freezing patterns as we have described them are realized on the root of the Higgs branch, where the F.I. terms are zero, but the existence of finite tension BPS vortices imposes further constraints: these solutions appear after the real F.I. terms are turned on, and this typically breaks supersymmetry when the gauge groups are nonabelian. The Coulomb freezing patterns which will preserve supersymmetry are found by solving the F-term and D-term moment map equations for the quiver gauge theory. For example, if $\fg=A_1$, a necessary condition for unbroken supersymmetry with nonzero F.I. terms is that two Coulomb moduli  $e_{1,i}$ and $e_{1,i'\neq i}$  \emph{cannot} be frozen to one and the same mass $f_{1,s}$\footnote{In string theory, the freezing of $e_{1,i}$ and $e_{1,i'\neq i}$ to the same mass $f_{1,s}$ via branes is an example of undesirable ``s-configuration," as defined by Hanany-Witten  \cite{Hanany:1996ie}}. Similar constraints exist for the higher rank Lie algebras, though they take more complicated forms. In particular, we are not aware of a field theory characterization of supersymmetry-preserving patterns for general simple Lie algebras $\fg$. 

We will not aim to classify all supersymmetry-preserving Coulomb freezing patterns in this work. Instead, we will content ourselves with sufficient conditions for their existence, and leave the important question of whether they are also necessary conditions to future work. One such sufficient condition is that the patterns should be labeled by coweights sitting inside fundamental representations\footnote{This condition will also be justified from the string theory engineering of the 5d gauge theory in Section \ref{sec:littlestring}, via the McKay correspondence.}.

Then, we will henceforth label the masses $f_{a,s}$ not just by any dominant coweight, but by the fundamental coweight $w^\vee_a$ (with multiplicity $s$), and all $n_a$ Coulomb parameters in $U(n_a)$ by one and the same positive simple coroot $\alpha^\vee_a$. A frozen Coulomb modulus $f_{a,s}\geq e_{a,i}$ is then labeled by the coweight $w^\vee_a-\alpha^\vee_a$, consistent with the definition of the partial ordering on $\fg$ and the fact that the coweight $w^\vee_a$ is indeed higher than the coweight $w^\vee_a-\alpha^\vee_a$.
A Coulomb parameter $e_{a\pm 1,i'}$ on a neighboring node  can in turn be frozen to the same mass $f_{a,s}$, after first freezing it to $e_{a,i}$; the operation will now be labeled by the coweight $w^\vee_a-\alpha^\vee_a-\alpha^\vee_{a\pm 1}$.
Proceeding in this way for all Coulomb moduli, we end up with a partially ordered set of $\fg$-coweights $\{\vec{\underline{\lambda^\vee}}\}$, all written as a fundamental coweight minus a finite sum of positive simple coroots:
\beq\label{weightsdecom}
\underline{\lambda^\vee_s} = w^\vee_a - \sum_{b=1}^{\r} h_{b,s}\, \alpha^\vee_b \; , \qquad s=1,\ldots,n^F\; ,
\eeq
for some $a\in\{1,\ldots,\r\}$, and $h_{b,s}$ non-negative integers.  Such a decomposition always exists, but is not unique in general. The physical requirement of keeping supersymmetry unbroken after turning on the F.I. terms will put additional constraints on the allowed integers $h_{b}$.  Note that consistency of our notations implies  
\beq\label{consistency}
\sum_{s=1}^{n^F} h_{b,s} = n_b \; ,
\eeq
with $n_b$ the rank of the gauge group $U(n_b)$.
In term of $\fg$-coweights, the asymptotic conformality constraint \eqref{constraint5d} takes an elegant form,
\beq\label{weightsadduptozero}
\sum_{s=1}^{n^F} \underline{\lambda^\vee_s} =0\; ,
\eeq
which follows at once from \eqref{weightsdecom}.\\

%

So far, the relevant quiver algebra has been $\fg$ throughout. We will now see that if the above analysis is repeated in an $\Omega$-background, the relevant representation theory is that of the quantum affine algebra of $\fg$ instead. The reasoning is as before, meaning we can still define a partial ordering on the set of 5d hypermultiplet masses $f_{a,s}$ and Coulomb moduli $e_{a,i}$, where the masses $f_{a,s}$ are once again the maximal elements. However, the freezing of Coulomb moduli now depends explicitly on the $\Omega$-background spacetime equivariant parameters $q$ and $t$, following \eqref{5drootHiggsvortex}. In order for our partial ordering to be compatible with the Higgsing prescription \eqref{5drootHiggsvortex}, it is not hard to show that the following must hold for two consecutive elements in the partially ordered set: 
\begin{align}\label{letsgoset}
&f_{a,s}\geq e_{a,i} & \text{whenever}\;\qquad& e_{a,i}=f_{a,s}\;\hbar^{-1/2}\,t^{N} \; ,&\qquad\nonumber\\
&e_{a,i}\geq e_{a+1,i'} &\text{whenever}\;\qquad& e_{a+1,i'}=e_{a,i}\;\hbar^{-1/2}\,t^{N'}\; ,&\qquad\nonumber\\
&e_{a,i}\geq e_{a-1,i'} & \text{whenever}\;\qquad& e_{a-1,i'}=e_{a,i}\;\hbar^{+1/2}\,t^{-N''}\; ,&\qquad
\end{align}
with $N, N', N''>0$, and $\hbar\equiv q/t$. In particular, the dependence on $\hbar$ between neighboring quiver nodes encodes the contribution of the bifundamental masses $f^{bif}_{a,b} = \hbar^{1/2}$. We are once again only interested in freezing configurations which do not break supersymmetry with nonzero F.I. terms. The frozen configurations do not depend on whether or not the $\Omega$-background is turned on, so there are still $n^F$ sequences ${\mathcal K}^{\geq}_{a,s}$ of ordered elements satisfying \eqref{letsgoset}; each sequence ${\mathcal K}^{\geq}_{a,s}$ starts with a mass $f_{a,s}$ as a maximal element. What the $\Omega$-background does, however, is give a dual description of the fully Higgsed 5d theory as the vortex theory $T^{3d}_{\fg}$. Applying the freezing algorithm to each 5d Coulomb modulus in the 5d instanton partition function results in \eqref{5dbulkmatterto3d}, where the right-hand side is the chiral matter contribution to the 3d half-index, encoding the Drinfeld polynomial data of the quantum affine weights $\{{\lambda}\}$. This immediately implies:

\begin{figure}[h!]
	\emph{}
	\centering
	\includegraphics[trim={0 0 0 0cm},clip,width=0.99\textwidth]{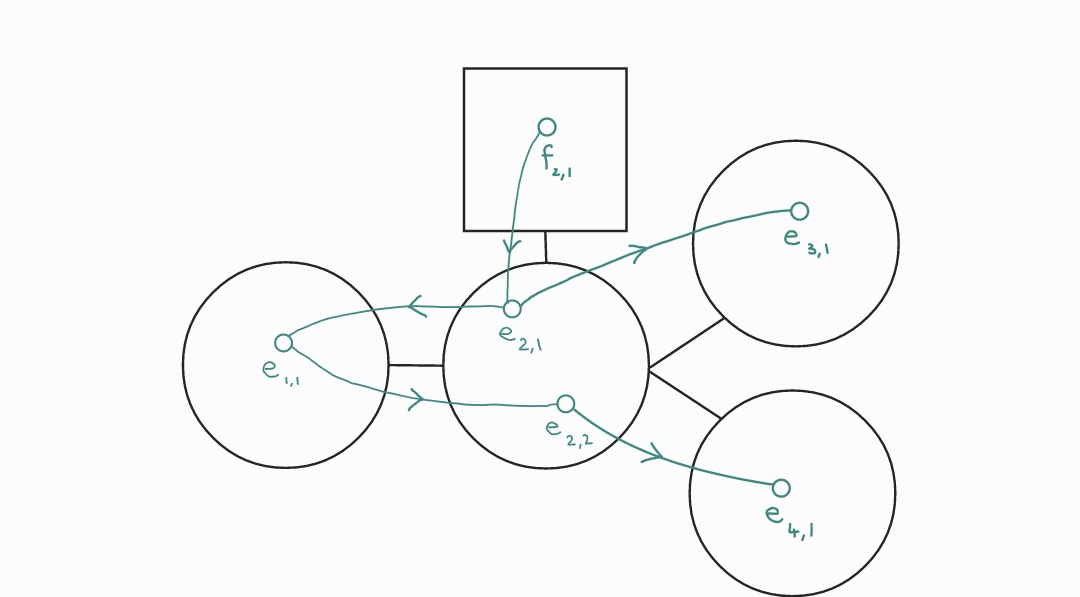}
	\vspace{-10pt}
	\caption{The construction from 5 dimensions of the weight $\lambda' = Y_3Y^{-1}_3$ found in the second fundamental representation  $\widehat{V_2}$ of $U_\hbar(\widehat{D_4})$. The maximal elements are the masses $f_{a,s}$, and the freezing of Coulomb moduli $e_{a,i}$ is featured explicitly, following the Higgsing  rules \eqref{letsgoset}. See \eqref{D4exmore} for the realization of the weight as a $q$-primary insertion, and \eqref{weightsex} for the realization as the chiral matter of a 3d Drinfeld quiver on the Higgs branch of the 5d one.}
	\label{fig:TruncD4}
\end{figure}

\begin{prop}\label{prop5d3d}
Let $V_{a,\lambda_{s}}$ be a weight space appearing in the eigenspace decomposition of a $U_\hbar(\widehat{\fg_{\text{o}}})$ fundamental representation $\widehat{V_a}=\bigoplus_{\lambda_{s}}\, V_{a,\lambda_{s}}$ as $U_\hbar(\fg_{\text{o}})$-modules. Let $\{{\mathcal K}^{\geq}_{a,s}\}$ be the set of all sequences of ordered elements with a 5d mass $f_{a,s}$ as a maximal element. Then the map $\{{\mathcal K}^{\geq}_{a,s}\} \rightarrow V_{a,\lambda_{s}}$ is surjective.
\end{prop}
We checked that this Proposition holds for all weight spaces in all fundamental representations of $U_{\hbar}(\widehat{\fg_{\text{o}}})$, for  $\fg_{\text{o}}=A_{r\leq 4}$, $\;\fg_{\text{o}}=D_{r\leq 6}$, \  and $\fg_{\text{o}}=E_6$, meaning we identified the Drinfeld polynomials of all weights in those cases. For $\fg_{\text{o}}=E_7$ \  and $\fg_{\text{o}}=E_8$, we only performed a partial check for some of the weights. 
As we mentioned, it would be important to clarify whether the map is also injective. The non simply-laced version of this Proposition is written in \ref{prop5d3dnsl}.

\subsection{Comparison with the Frenkel-Mukhin algorithm}
\label{ssec:compare}

We would like to compare our proposal to build weight spaces of quantum affine fundamental representations with the one used in the construction of the Frenkel-Reshetikhin $\hbar$-characters \cite{Frenkel:qch,Frenkel:1999ky}.
By that, we mean a Frenkel-Mukhin type of algorithm alluded to in \eqref{highestmonom} and \eqref{qtfirstterm}, where we start with a ``fundamental weight" operator
\beq\label{coweightvertexdefagain}
\Lambda_a(x) \equiv\; : \exp\left(\, \sum_{k\neq 0}{w_a[k] \over (q^{k\over 2} - q^{-\,{k \over 2}})(t^{k\over 2} - t^{-\,{k \over 2}})} \,t^{-\,{k \over 2}}\, x^k\right):  \; , \qquad\;\;\;\;a\in\{1,\ldots,\r\} \; ,
\eeq
and then proceed to build all the weight spaces in the fundamental representation by successive applications of positive ``simple root operators"\footnote{In the $\hbar$-character literature, the role of the $\Lambda_a$ operators is played by the $\cY_a$-operators \eqref{YoperatorToda}, and the role of the $E_a$ operators is played by operators commonly called $A_a$ \cite{Frenkel:1998,Frenkel:1999ky,Bouwknegt:1998da}.}
\beq\label{rootvertexdef}
E_a(x) \equiv\; : \exp\left(\sum_{k\neq 0}{ \alpha_a[k] \over (q^{k\over 2} - q^{-k \over 2})(t^{k\over 2} - t^{-\,{k \over 2}})}  \,t^{-\,{k \over 2}}\,  x^k\right): \; , \qquad\;\;\;\;a\in\{1,\ldots,\r\} \; .
\eeq

\begin{figure}[h!]
	\emph{}
	\centering
	\includegraphics[trim={0 0 0 0cm},clip,width=0.99\textwidth]{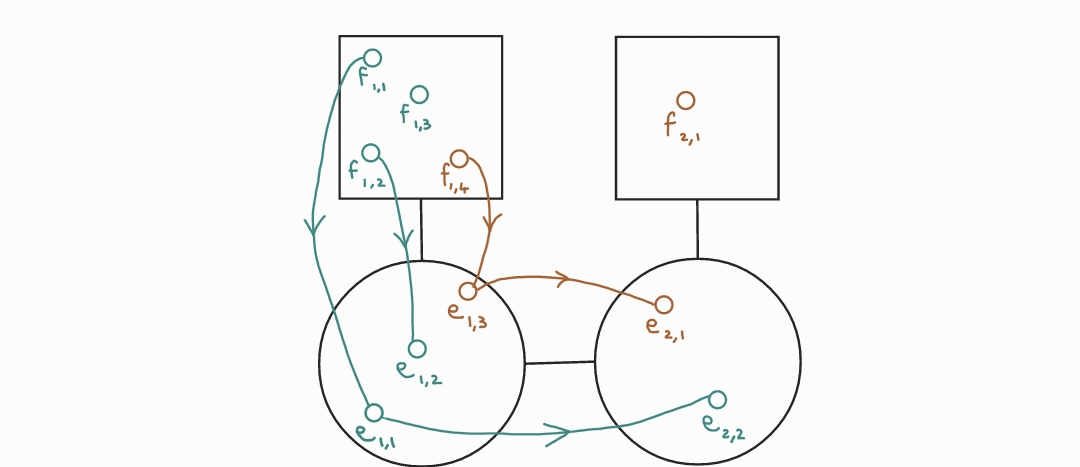}
	\vspace{-10pt}
	\caption{Example of sequences in a 5d quiver gauge theory of type $A_2$. The maximal elements are the masses $f_{a,s}$, and the freezing of Coulomb moduli $e_{a,i}$ is featured explicitly. After Higgsing, the resulting quantum affine weights are grouped by color to denote $q$-primary operator insertions in the $\cW_{q,t}(A_2)$ conformal block. Here, we have a first $q$-primary labeled by a set  of 3 weights $\{\lambda\}_1=\{Y_1,\, Y_2 Y^{-1}_1,\, Y^{-1}_2\}$, and a second $q$-primary labeled by a set of 2 weights $\{\lambda\}_1=\{Y_2, \,Y^{-1}_2\}$. In this example, all the weights happen to be realized as in the Frenkel-Mukhin algorithm, as suggested by the picture.} 
	\label{fig:QuiverA3new}
\end{figure}

\begin{figure}[h!]
	\emph{}
	\centering
	\includegraphics[trim={0 0 0 0cm},clip,width=0.99\textwidth]{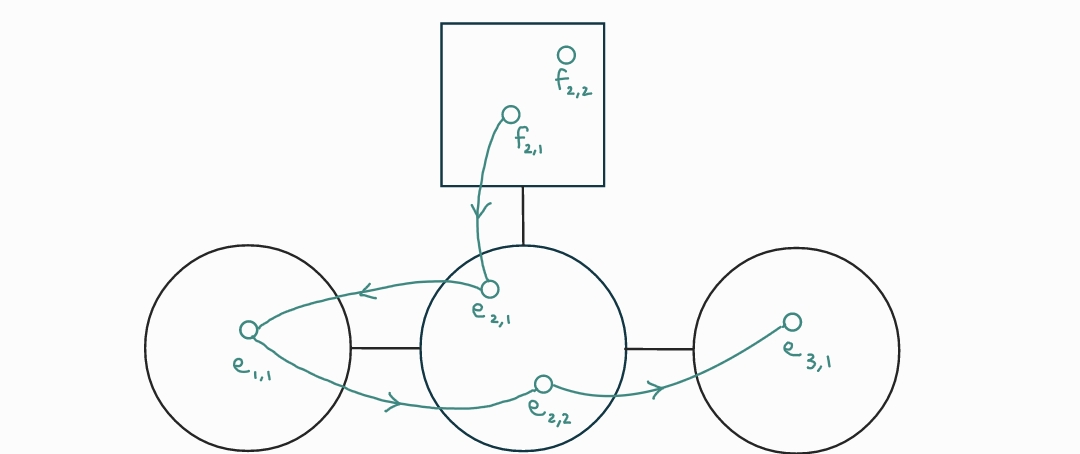}
	\vspace{-10pt}
	\caption{This Higgsing pattern engineers the quantum affine weights $Y_2$ and $Y^{-1}_2$ (and the associated $q$-primary operator of $\cW_{q,t}(A_3)$ is labeled by  $\{\lambda\}=\{Y_2, \,Y^{-1}_2\}$). In this example, the construction of the weight $Y^{-1}_2$ does \emph{not} follow the Frenkel-Mukhin algorithm, as is clear from the truncation path.}
	\label{fig:TruncA3}
\end{figure}

In order to compare to our 5d gauge theory Higgsing construction, consider a sequence  ${\mathcal K}^{\geq}_{a,s}=\{f_{a,s},\ldots\}_\geq$, and identify each mass $f_{a,s}$ as a highest weight operator $\Lambda_a$, and each frozen Coulomb modulus $e_{a,i}$ in the sequence as the operator $E_a$, with some $\hbar$-shifts of the argument. In the Frenkel-Mukhin algorithm, the $a$-th fundamental representation would then be built as 
\beq
:\Lambda_a(x): \;\;\;\rightarrow\;\;\; :\Lambda_a(x)\, E_a(x\,\hbar^{-1/2}): \;\;\;\rightarrow\;\;\; \ldots
\eeq
where the ``$\ldots$" stands for all the remaining weights.
We claim that this procedure disagrees in general with the Higgsing one from 5d gauge theory. For the simplest counterexample, consider the weight $:\Lambda^{-1}_2(\hbar^{-3}x):$ in $\cW_{q,t}(A_3)$, constructed via a sequence
\begin{align}\label{frenk}
:\Lambda_2(x): \;\;\;&\rightarrow\;\;\; :\Lambda_2(x)\, E_2(x\,\hbar^{-1/2}):\nonumber\\
&\rightarrow\;\;\; :\Lambda_2(x)\, E_2(x\,\hbar^{-1/2})\,E_1(x\,\hbar^{-1/2}):\nonumber\\  &\rightarrow\;\;\; :\Lambda_2(x)\, E_2(x\,\hbar^{-1/2})\,E_1(x\,\hbar^{-1/2})\,E_3(x\,\hbar^{-1/2}):\nonumber\\
&\rightarrow\;\;\; :\Lambda_2(x)\, E_2(x\,\hbar^{-1/2})\,E_1(x\,\hbar^{-1/2})\,E_3(x\,\hbar^{-1/2})\,E_2(x\,\hbar^{-1/2}): \; ,
\end{align}
in the Frenkel-Mukhin algorithm. The last line is the weight $:\Lambda^{-1}_2(\hbar^{-3}x):$. By inspection of Figure \ref{fig:TruncA3}, the ``tree" to get to that lowest weight is not the same in gauge theory. This is because the truncation algorithm \eqref{letsgoset} only acts on nearest neighbors $a\pm1$ for a Coulomb modulus on node $a$. It is remarkable that both procedures are distinct and yet yield the correct weight.

\vspace{8mm}

\section{On the enumerative geometry of tame ramification}
\label{sec:quasim}

The aim of the following Section is to recast our results in the language of enumerative geometry and the counting of quasimaps. The standard reference is the set of notes \cite{Okounkov:2015spn} by Okounkov. For recent applications in our context, see \cite{Bullimore:2018jlp,Dedushenko:2021mds,Dedushenko:2023qjq,Crew:2023tky,Ishtiaque:2023acr} in Physics and \cite{Smirnov:2018drm,Rimanyi:2019zyi,2020arXiv200805516D,2020arXiv201108603D,2021arXiv210807202D} in Mathematics.
Throughout this section, the underlying Lie algebra will be simply-laced $\fg=\fg_\text{o}$, of rank $r$. Before we introduce ramification, it will prove worthwhile to briefly recall the analysis in the unramified Langlands correspondence \cite{Aganagic:2017gsx}.

\subsection{Review: Vertex functions for the unramified Langlands correspondence}
\label{ssec:quasim2}

In the unramified setup, the 3d quiver gauge theory of interest has unbroken $\cN=4$ supersymmetry. This implies that its Higgs branch $X$ is a Nakajima quiver variety, meaning a hyper-K\"{a}hler quotient
\beq\label{Nakquot}
X = T^*\text{Rep}({\cal Q})/\!\!/\!\!/\!\!/G_{V} \; ,
\eeq
with
\beq\label{quivquiv}
\text{Rep}({\cal Q})=\oplus_{a\rightarrow b} \text{Hom}(V_a,V_b)\oplus_a \text{Hom}(V_a,W_a) \; ,
\eeq
and ``$\rightarrow$" denotes an arrow in the Dynkin diagram of $\fg_\text{o}$\footnote{In physics terminology, the $\text{Hom}$'s as written represent multiplets for the 3d $\cN=4$ quiver, \emph{not} their decomposition as a $\cN=2$ quiver.}  
\beq
G_V=\prod_{a=1}^{r}GL(V_a) \; , \qquad G_H=\prod_{a=1}^{r}GL(W_a) \; .
\eeq
The vector space $V_a$ has dimension $\text{dim}(V_a) = N_a$, while the vector space $W_a$, also called framing, has dimension $\text{dim}(V_a) = N^F_a$.

We fix maximal tori $T_V$ and  $T_H$ of $G_V$ and $G_H$, respectively. The mass parameters $x_{a,d}$ for $a=1,\ldots,\r$ and $i=1,\ldots,N^F_a$ are $T_H$-equivariant parameters valued in $K_{T_H}(\text{pt})$. We choose a generic chamber $\fC_H$  by choosing generic $x_{a,d}$'s.

The quotient \eqref{Nakquot} is in particular a geometric invariant theory (GIT) quotient, which depends on a choice of stability conditions \cite{10.1093/qmath/45.4.515}. We fulfill these stability conditions by choosing a generic chamber $\fC_C$ (generic positive F.I. terms). Crucially, the quotient is the holomorphic symplectic reduction of a cotangent bundle, so there is a natural action of a group 
\beq
\mathbb{C}^\times_{\hbar}\subset\text{Aut}(X) \; ,
\eeq 
which acts by rescaling the cotangent directions with weight $\hbar^{-1}\in  K_{\mathbb{C}^\times_{\hbar}}(\text{pt})$. We write
\beq\label{Tdef}
T=\mathbb{C}^\times_{\hbar}\times T_H \; .
\eeq
The set of massive isolated vacua we have been calling $\{{\bf{A}}\}$ is the $T$-fixed point locus of $X$, also denoted as $X^{T}$.\\

Mathematically, our moduli spaces of vortices are known as moduli spaces of quasimaps QM$(X)$\footnote{The maps we consider are all stable, meaning they are allowed unstable values in the GIT quotient, but only at finitely many points of the domain $\mathbb{P}^1$.} \cite{2011arXiv1106.3724C,2009arXiv0908.4446C,2010arXiv1005.4125K}, of the form
\beq
f: \mathbb{P}^1 \rightarrow X
\eeq
In practice, the map $f$ induces pull-back bundles $f^*V_a$ and $f^*W_a$ on $\mathbb{P}^1$ (with $V_a$ and $W_a$ as in   \eqref{quivquiv}) and one works with this moduli space of bundles.

On the domain, $\mathbb{P}^1$, we fix two marked points $\{0\}$ and $\{\infty\}$, and consider the subset of quasimaps $QM_{\text{nonsing},\infty}(X)$ for which $f(\{\infty\})$ is stable. Physically, we are asking for the gauge field to be flat at $\{\infty\}$. This space is not compact, because point-like vortices have nontrivial first Chern class, by definition, and so can be placed as close to $\{\infty\}$ as desired without ever being allowed there. This IR divergence was cured in Section \ref{sec:3dgauge} by introducing an $\Omega$-background.
In the quasimap terminology, we let
\beq
\mathbb{C}^\times_q=\text{Aut}(\mathbb{P}^1,\{0\},\{\infty\})  \; 
\eeq
rescale $\mathbb{P}^1$ with weight $q\in K_{\mathbb{C}^\times_q}(\text{pt})$. Quasimaps $QM_{\text{nonsing},\infty}(X)$  fixed under $\mathbb{C}^\times_q$ can only have singularities at the fixed point $\{0\}$ (since $\{\infty\}$ is by definition non-singular), making the moduli space compact. 
Additionally, all matter fields should a approach an isolated vacuum $\{{\bf{A}}\}$ at infinity. Then, we are restricting ourselves to a subset of $QM_{\text{nonsing},\infty}(X)$ which sends $\{\infty\}$ to a fixed point of $X^{T}$.\\

The vertex function ${\bf{V}}$ is a generating function of  equivariant Euler characteristics of the symmetrized virtual structure sheaf over the moduli space of quasimaps, graded by the degree (the vortex number). The counting parameter $z$ is also called K\"{a}hler parameter (the F.I. parameter). For $\r=r$, a degree $k$ quasimap will have counting parameter
\beq\label{quasicount}
z^{deg} = z^{k_1}_1 \, z^{k_2}_2 \, \ldots \, z^{k_r}_r \; ,
\eeq
such that $\sum_{a=1}^r k_r = k$.

In practice, vertex functions are computed via Atiyah-Bott localization using $\mathbb{C}^\times_q$-equivariant K-theory on $QM_{\text{nonsing},\infty}(X)$. Then, given the quiver data \eqref{quivquiv} and the choice of a chamber $\fC_C$ (stability condition), the vertex function can be expressed in some normalization\footnote{These are perturbative contributions, or degree 0 quasimaps, which we have ignored. Back in Section \ref{sec:3dgauge}, we also ignored such contributions in the computation of the 3d half-index; as we noted, they can be recovered by computing instead the 3d hemisphere partition function on $S^1\times D^2$ via localization. For details on the subtleties of the background and the normalization of the vertex function, see for instance \cite{Bullimore:2020jdq,Crew:2023tky,Dedushenko:2023qjq}.} as the following integral:
\beq
{\bf{V}}\; =  \frac{1}{|W_{G_V}|}\prod_{a=1}^{r} \prod_{i=1}^{N_a}\oint_{\Gamma} \frac{dy_{a,i}}{y_{a,i}}\,e^{\frac{\ln(z^\#_a) \ln(y_{a,i})}{\ln(q)}}\, {\mathcal F}'(y_{a,i})\; \left((q-\hbar)\, T^{1/2};q\right)_{\infty}\; .
\eeq
Here, $\left(x;q\right)_{\infty}$ is the usual $q$-Pochhammer symbol, and $z^\#_a$ is a rescaling of the quasimap counting parameter $z_a$ in \eqref{quasicount} by factors of $\hbar$ and $q$.  $T^{1/2}X\in K_T(X)$ stands for a choice of polarization, which picks out half of the tangent bundle $T X$. It is a solution in $K_T(X)$ of the equation
\beq
T^{1/2} X + \hbar^{-1}\otimes \left(T^{1/2} X\right)^{\vee} = T X \; ,
\eeq
where $(\ldots)^{\vee}$ denotes the dual space of $(\ldots)$.

The insertion ${\mathcal F}'(y)$ in the integrand is chosen to be $q$-periodic and nonsingular at the $T_H$-equivariant points $x_{a,d}$. Equivalently, we write the insertion as
\beq\label{ratioell}
{\mathcal F}'(y) = \frac{{\mathcal F}(y)}{\Theta(T^{1/2}X)} \; ,
\eeq
with the requirement that this ratio be invariant under $y\rightarrow q\, y$. This implies ${\mathcal F}(y)$ is a section of a line bundle valued in the $T$-equivariant elliptic cohomology of $X$. The line bundle $\Theta(T^{1/2}X)$ is also called an elliptic Thom line bundle.
The contour $\Gamma$ is a sum of several cycles which enclose the $T_H$-equivariant parameters $x_{a,d}$ and their $q$-shifts in the denominator of the integrand. The insertion  ${\mathcal F}(y)$, when restricted to a class on a particular component in $X^{T}$, reduces the contour to enclose only those equivariant parameters $x_{a,d}$ which pick out the vacuum $\{{\bf{A}}\}$.\\

For the Nakajima variety $X$, a canonical choice of decomposition of the tangent bundle is
\beq
T^{1/2} X = \sum_a V_a\otimes W^*_a + \sum_{a,b}(\Delta_{ab} - \delta_{ab})\, V_a \otimes V^*_b \; ,
\eeq
where we are abusing notations by identifying the vector space $T^{1/2} X$ with its character under the $T_V\times T_H \times  \mathbb{C}^\times_{\hbar}$ torus action.

By inspection of the components of the integral ${\bf{V}}$, we recognize the magnetic blocks, the free field chiral correlators of the $\cW_{q,t}(\fg)$ algebra.
Since many aspects of the proof are identical in the unramified and the ramified cases, we will review it here and highlight the differences. First, the  K\"{a}hler variables $z_a$ are related to the highest weight $\mu_0$ of the deformed Heisenberg algebra Fock space representation as
\beq\label{kahlerparam}
z_a = q^{\langle\mu_0, \alpha_a\rangle} \; .
\eeq
If we choose coordinates 
\beq
V_a = \sum_{i=1}^{N_a} y_{a,i} \, \hbar^{a/2} \; , \qquad\;\; W_a = \sum_{d=1}^{N^F_a} x_{a,d}\, \hbar^{(a-1)/2} \; ,
\eeq
then the contribution from $\text{Hom}(V_a,W_a)$ in \eqref{quivquiv} to $\left((q-\hbar)\, T^{1/2};q\right)_{\infty}$ is
\beq\label{fundamvertex}
\prod_{i=1}^{N_a}\prod_{d=1}^{N^F_a}\frac{\left(q \, y_{a,i}/\hbar^{1/2}\, x_{a,d} ;q\right)_{\infty}}{\left(\hbar \, y_{a,i}/\hbar^{1/2}\, x_{a,d} ;q\right)_{\infty}} \; .
\eeq
Since $\hbar=q/t$, one recognizes the product of 2-point functions 
\beq
\prod_{i=1}^{N_a}\prod_{d=1}^{N^F_a} \left\langle  S_a^\vee(y)\, {\overline V}_b^\vee(x)\right\rangle
\eeq
of a screening current with the $a$-th fundamental vertex operator of Frenkel and Reshetikhin, see \eqref{vertexscreeningex}. Equivalently, this factor is readily identified as the hypermultiplet contribution to the half-index of the corresponding 3d $\cN=4$ gauge theory, with the appropriate $\cN=(2,2)$ boundary conditions.

\noindent
Likewise, the contribution from $\text{Hom}(V_a,V_a)$ to $\left((q-\hbar)\, T^{1/2};q\right)_{\infty}$ is
\beq\label{vecunvertex}
 \prod_{1\leq i\neq j\leq N_a}\frac{\left(\hbar \,y_{a,i}/ y_{a,j} ;q\right)_{\infty}}{\left(q \, y_{a,i}/ y_{a,j} ;q\right)_{\infty}} \; ,
\eeq 
which reproduces the contribution \eqref{screena},
\beq
\prod_{1\leq i< j\leq N_a} \left\langle S_a^\vee(y_{a,i})\, S_a^\vee(y_{a,j}) \right\rangle  \; ,
\eeq
up to ratio of Theta functions. From the point of view of the $q$-difference equations solved by ${\bf{V}}$, the ratio of Theta functions is an inconsequential factor of $\hbar^{\#}$ which can be reabsorbed in the definition of $z_a$ in \eqref{kahlerparam}, hence the notation $z_a \rightarrow z^\#_a$. 

\noindent
Finally, the contribution from $\text{Hom}(V_a,V_b)$ ($b> a$) to $\left((q-\hbar)\, T^{1/2};q\right)_{\infty}$ is
\beq\label{bifunvertex}
\prod_{i=1}^{N_a}\prod_{j=1}^{N_b}\frac{\left(q \,\hbar^{a/2}\, y_{a,i}/\hbar^{b/2}\, y_{b,j} ;q\right)_{\infty}}{\left(\hbar \,\hbar^{a/2}\, y_{a,i}/\hbar^{b/2}\, y_{b,j} ;q\right)_{\infty}} \; ,
\eeq 
which reproduces the contribution \eqref{screenab},
\beq
\prod_{i=1}^{N_a}\prod_{j=1}^{N_b} \left\langle  S_a^\vee(y_{a,i})\, S_b^\vee(y_{b,j})\right\rangle \; .
\eeq
Therefore, the integrands agree. As far as the integration contours are concerned, one notes that a choice of insertion  ${\mathcal F}'(y_{a,i})$ in ${\bf{V}}$, and in particular of a component supported in $X^{T}$, is in one-to-one correspondence with a choice of screening charges for the magnetic $\cW_{q,t}(\fg)$-algebra block. This completes the identification.\\

Electric blocks, which are solutions to the $U_{\hbar}(\widehat{^L \fg})$ qKZ equations, are likewise vertex functions. More precisely, the qKZ equations being matrix equations, it is more accurate to call their solutions vector vertex functions: these are generating functions of quasimaps with descendant insertions at $\{0\}\in\mathbb{P}^1$. These are K-theoretic stable envelope insertions, which are certain classes in $K_T(X)$ which provide the stable basis of $X$ \cite{Okounkov:2015spn}.

The content of the unramified quantum $q$-Langlands isomorphism is that there exists a certain Whittaker covector whose pairing with the vector electric block equals the scalar magnetic $\cW_{q,t}(\fg)$-algebra block. Physically, this is why electric blocks arise as 3d partition functions on $\mathbb{C}^\times_q\times S^1_{\cC'}$ with the origin removed, while magnetic blocks are partition functions on $\mathbb{C}_q\times S^1_{\cC'}$, with the puncture at $\{0\}$ closed up.

Note that both the electric and magnetic blocks so far are $z$-solutions, since they are by construction analytic in the K\"ahler variables $z_a$ in chamber $\fC_C$. To construct the $x$-solutions instead, analytic in the equivariant parameters in chamber $\fC_H$, one further inserts the corresponding elliptic stable envelope at $\{\infty\}\in\mathbb{P}^1$ via evaluation map, that is to say an assignment of an elliptic cohomology class for each $T$-fixed point. The existence and uniqueness of the envelopes for this problem was proved in \cite{Aganagic:2016jmx}.

\vspace{8mm}



\subsection{Enter tame ramification}
\label{ssec:quasim3}

The 3d Drinfeld quiver gauge theory $T^{3d}_{\fg_\text{o}}$ was defined in Section \ref{ssec:Drinfelddef}. In particular, let
\beq
G_V=\prod_{a=1}^{r}GL(V_a) \; , 
\eeq
with $\text{dim}(V_a) = N_a$ as in the unramified case. The framing is
\begin{align}\nonumber\label{flavorgroup3dgroupframing}
&\fg = A_r\; : \; \ \,\, \,\qquad G_{H}= GL(W^+_1)\times \left[\prod_{a=2}^{r-1} GL(W^{\pm}_a)\right]\times GL(W^-_r) \; ,\\
&\fg = D_r\;, E_r\; : \; \;\;\;G_{H}= GL(W^+_1)\times \left[\prod_{a=2}^{r-2} GL(W^{\pm}_a)\right]\times GL(W^-_{r-1}) \times GL(W^-_r) \; ,
\end{align} 
where the dimensions $\text{dim}(W^{\pm}_a) = L^{\pm}_{a}$ are defined in terms of Drinfeld polynomial data, see Proposition \ref{prop3d}. The vector spaces $W^{\pm}_a$ are graded by an integer $d$ labeling subsets of quantum affine weights, as $L^{\pm}_{a} = \sum_{d=1}^L L^{\pm}_{d,a}$ with $L^{\pm}_{d,a}=\sum_{s=1}^{J_d}\deg(\cA^{\pm}_{\lambda_{d,s},a})$.\\

We fix maximal tori $T_V\subset G_V$ and $T_H\subset G_H$. The (complexified) mass parameters $x^{\pm}_{d,a}$ for $d=1,\ldots,L$ and $a\in\{1,\ldots,\L\}$ are the $T_H$-equivariant parameters. Equivalently, we will also use the notation $x_{d,s}$ for $d=1,\ldots,L$ and $s=1,\ldots,J_d$, to emphasize that each equivariant parameter is in one-to-one correspondence with a weight $\lambda_{d,s}$ taken in some fundamental representation of $U_{\hbar}(\widehat{^L \fg})$.\\

The Higgs branch $X$ is now only a GIT quotient
\beq\label{GITquot}
X = \text{Rep}({\cal Q})/\!\!/G_{V} \; ,
\eeq
As far as the stability condition on $X$ is concerned, we work in the same generic chamber $\fC_C$ as before.\\

Because the GIT quotient $X$ is no longer a holomorphic symplectic reduction of a cotangent bundle, at first sight, $\mathbb{C}^\times_{\hbar}$ (which used to rescale the cotangent directions of such a bundle) appears to have lost its geometric meaning in the ramified setting. However, $X$ has another interpretation: it is also the moduli space of degree $(N_1,N_2,\ldots,N_r)$ quasimaps into  
\beq\label{Nakquot2}
X' = T^*\text{Rep}({\cal Q}')/\!\!/\!\!/\!\!/G'_{V} \; ,
\eeq
with $\text{Rep}({\cal Q}')$ defined by the 5d $\cN=1$ quiver diagram of Section \ref{sec:5dgauge}:
\beq\label{quivquiv2}
\text{Rep}({\cal Q}')=\oplus_{a\rightarrow b} \text{Hom}(V'_a,V'_b)\oplus_a \text{Hom}(V'_a,W'_a) \; ,
\eeq
and
\beq
G'_V=\prod_{a=1}^{r}GL(V'_a) \; , \qquad G'_H=\prod_{a=1}^{r}GL(W'_a) \; ,
\eeq
with dimensions $\text{dim}(V'_a) = n_a$ and $\text{dim}(W'_a) = n^F_a$. Crucially, $X'$ \emph{is} a Nakajima quiver variety, so the $\mathbb{C}^\times_{\hbar}$ still retains a geometric meaning in the ramified setting, as
\beq
\mathbb{C}^\times_{\hbar}\subset\text{Aut}(X') \; .
\eeq 
This interpretation of $X$ in terms of quasimaps to $X'$ is the mathematical formulation of gauge-vortex duality from Section \ref{ssec:duality}.\\ 

Therefore, we will denote as before
\beq\label{Tdef2}
T=\mathbb{C}^\times_{\hbar}\times T_H \; ,
\eeq
and  $X^{T}$ will be the $T$-fixed point locus of $X$.

In the definition of $T_H$, there is an equivariant parameter $x_{d,s}$ for every quantum affine weight $\lambda_{d,s}$, with a constraint 
\beq\label{constraintpolyagain}
\sum_{s=1}^{J_d}\left[ \deg(\cA^{+}_{\lambda_{d,s},a}) - \deg(\cA^{-}_{\lambda_{d,s},a})\right] = 0 \; , \qquad a=1,\ldots,r \; , \;\;\;d=1,\ldots,L 
\eeq
on the degrees of their Drinfeld polynomials. This constraint translates to $\text{dim}(W^{-}_{d,a}) - \text{dim}(W^{+}_{d,a})=0$ for the dimensions of the vector spaces. Physically, this means all fundamental chiral multiplets labeled by ``$-$" and anti-fundamental chiral multiplets labeled by ``$+$" will be grouped in pairs, so by analogy with the unramified case, it is still possible to ``erase" half the arrows in the quiver diagram of $T^{3d}_{\fg_\text{o}}$. This intuitive argument was made precise by Okounkov \cite{okounkov2021inductive}, see also the related concept of partial polarization recently introduced in \cite{Ishtiaque:2023acr}.
Explicitly, we define
\beq\label{polarramif} 
T^{\pm}X = T^{\text{vec}}X + \sum_{d}T^{\pm}_{\{\lambda\}_d}X\; ,
\eeq
where
\beq
T^{\text{vec}}X = \sum_{a,b}(\Delta_{ab} - \delta_{ab})\, V_a \otimes V^*_b
\eeq
is the same as in the unramified setting, and the spaces
\beq
T^{\pm}_{\{\lambda\}_d}X = \sum_{a} V_a\otimes W^{\pm *}_{d,a} 
\eeq
are unique to tame ramification. In $K_T(X)$, we will decompose the tangent bundle as
\beq\label{tangentdecomp}
\bigg[T^{\text{vec}} X + \hbar^{-1}\otimes \left(T^{\text{vec}} X\right)^{\vee} \bigg]+ \sum_{d}\bigg[ T^{+}_{\{\lambda\}_d}X + \left(T^{-}_{\{\lambda\}_d}X\right)^{\vee}\bigg] = T X \; 
\eeq
As before, we consider the moduli space of quasimaps $QM_{\text{nonsing},\infty}(X)$ which sends $\{\infty\}$ to a fixed point of $X^{T}$, and its associated vertex function ${\bf{V}}$\footnote{For 3d $\cN=2$ theories and their associated GIT quotients \eqref{GITquot}, it is perhaps more accurate to call ${\bf{V}}$ a ``big K-theoretic $I$ function" \cite{2013arXiv1304.7056C,2011arXiv1106.3724C,2016arXiv160206494T}.}. The degree  $k= \sum_{a=1}^r k_r$ of the quasimaps is called vortex number from the point of view of $X$, or instanton number from the point of view of $X'$. We keep the notation $\mathbb{C}^\times_q$ for the automorphism of the $\mathbb{P}^1$ domain with $\{0\}$ and $\{\infty\}$ as marked points. The vertex function takes the form:
\begin{align}\nonumber\label{vertexnew}
{\bf{V}}\; =  \frac{1}{|W_{G_V}|}\prod_{a=1}^{r} \prod_{i=1}^{N_a}\oint_{\Gamma} &\frac{dy_{a,i}}{y_{a,i}}\,e^{\frac{\ln(z^\#_a) \ln(y_{a,i})}{\ln(q)}}\, {\mathcal F}'(y_{a,i})\cdot\\
&\cdot \left((q-\hbar)\, T^{\text{vec}};q\right)_{\infty}\prod_{d=1}^{L}\left(T^{+}_{\{\lambda\}_d}-T^{-}_{\{\lambda\}_d};q\right)_{\infty}\; .
\end{align}

The insertion ${\mathcal F}'(y)$ in the integrand is chosen to be $q$-periodic and nonsingular at the $T_H$-equivariant points $x_{d,s}$:
\beq\label{ratioellram}
{\mathcal F}'(y) = \frac{{\mathcal F}(y)}{\Theta(T^{+}X)} \; ,
\eeq
with ${\mathcal F}(y)$ the section of a line bundle valued in the $T$-equivariant elliptic cohomology of $X$. 
Here, $\Theta(T^{+}X)$ is the analog of the square root of the line bundle $\Theta(T X)$ one had in the unramified context. More precisely, it is called an attractive line bundle for $X$ in the terminology of Okounkov \cite{okounkov2021inductive}. The existence of this bundle in our context has important consequences: it guarantees the existence of elliptic stable envelopes, following the same proof as in the unramified case \cite{Aganagic:2016jmx}. This is why it sensible to construct $x$-solutions to the qKZ-equations starting from the $z$-solutions {\bf{V}}. Having defined the Vertex function, We want to show:

\begin{prop}
The integral ${\bf{V}}$ is a magnetic block of the $\cW_{q,t}(\fg)$ algebra with $L$ $q$-primary vertex operators insertions $:\prod_{d=1}^{L} \cV_{\{{\lambda}\}_d}(\tilde{x}_d):$.
\end{prop}
To prove this proposition, choose coordinates 
\beq
V_a = \sum_{i=1}^{N_a} y_{a,i} \, \hbar^{a/2} \; , \qquad\;\;\; W^{\pm}_{d,a} = \sum_{s=1}^{J_d}\sum_{k=1}^{\deg(\cA^{\pm}_{\lambda_{d,s},a})} \tilde{x}_{d,s}\, \hbar^{(a-1)/2}\, \hbar^{\mathfrak{a}^{\pm}_{\lambda_{d,s},a,k}}\; ,\label{coordram} 
\eeq
where $\hbar^{-\mathfrak{a}^{\pm}_{\lambda_{d,s},a,k}}$ are the Drinfeld roots of the polynomials $\cA^{\pm}_{\lambda_{d,s},a}$.  
The framing-independent contribution to $\left((q-\hbar)\, T^{\text{vec}};q\right)_{\infty}$ is the same as in the unramified case, so we do not repeat here. The contribution to $\prod_{d=1}^{L}\left(T^{+}_{\{\lambda\}_d}-T^{-}_{\{\lambda\}_d};q\right)_{\infty}$ is specific to tame ramification, and evaluates to
\beq\label{vertexproof}
\prod_{d=1}^{L}\prod_{i=1}^{N_a}\prod_{b=1}^{\r}\prod_{s=1}^{J_d}\frac{\prod_{k'=1}^{\deg(\cA^{+}_{\lambda_{d,s'},b})}\left(y_{a,I}/(\hbar^{-1/2}\,\hbar^{\mathfrak{a}^{+}_{\lambda_{d,s'},b,k'}}\, x_{d,s});q\right)_{\infty}}{\prod_{k=1}^{\deg(\cA^{-}_{\lambda_{d,s},b})}\left(y_{a,I}/(\hbar^{-1/2}\,\hbar^{\mathfrak{a}^{-}_{\lambda_{d,s},b,i}}\, x_{d,s}) ;q\right)_{\infty}} \; .
\eeq
Since $\hbar=q/t$, one recognizes the product of two-point functions 
\beq
\prod_{d=1}^{L}\prod_{i=1}^{N_a}\left\langle  S_a^\vee(y_{a,i})\, \cV_{\{{\lambda}\}_d}(\tilde{x}_d)\right\rangle
\eeq
with the screening currents \eqref{qprimarydeflshishatotal}.\\

It remains to justify the appearance of the line bundle $\Theta(T^{+}X)$ (with $T^+$ defined in \eqref{polarramif}) in the denominator of the ${\mathcal F}'(y)$ insertion. For this purpose, the parametrization \eqref{compositionmany} of the $T_H$-equivariant parameters $x_{d,s}$ is not particularly illuminating.  Instead, we relabel $x_{d,s}$ as $x^{\pm}_{d,a}$, with $d=1,\ldots,L$ and $a=1,\ldots,r$, with a constraint: whenever $a$ and $b>a$ label adjacent nodes in the Dynkin diagram of $\fg$, the $T_H$-equivariant parameters $x^{-}_{d,a}$ and $x^{+}_{d,b}$ will stand for one and the same parameter $x^{\pm}_{d,a}$. It is not hard to show that there is a one-to-one mapping between the labeling of parameters as $x^{\pm}_{d,a}$ and our previous labeling as $x_{d,s}$ with $s$ running over all weights in $\{{\lambda}\}_d$. Then, define center of mass and relative $T_H$-parameters as
\beq\label{compositionmany3}
x^{\pm}_{d,a} = \tilde{x}_d \, \hbar^{{\sigma}^{\pm}_{d,a}} \; ,\qquad\;\; a=1, \ldots, r \;  .
\eeq
We proceed to choose new coordinates
\beq
V_a = \sum_{i=1}^{N_a} y_{a,i} \, \hbar^{a/2} \; , \qquad\;\;\; W^{\pm}_{d,a} = \tilde{x}_{d}\; \hbar^{(a-1)/2}\; \hbar^{\overline{\sigma}^{\pm}_{d,a}}\; \sum_{s=1}^{J_d}\sum_{p=1}^{\deg(\cA^{\pm}_{\lambda_{d,s},a})}\hbar^{\mathfrak{a}^{\pm}_{\lambda_{d,s},a,p}}   \; .
\eeq
This makes it clear that for a given $\tilde{x}_{d}$, the poles in the integrand  \emph{only} depend on parameters labeled by ``$-$", while the numerators are $q$-Pochhammer symbols which depend on ``$+$" parameters instead.
It follows that the line bundle  $\Theta(T^{+}X)$ will remove the numerators, and introduce instead new ``$+$"-dependent poles, in complete analogy with the unramified case. By construction,  since the vector spaces $W^{+}_{d,a}$ and $W^{-}_{d,a}$ always appear in pairs for all $d$ and $a$, the number of new ``$+$" poles will equal the number of ``$-$" poles already present; the contour $\Gamma$ is defined as a sum of cycles enclosing the parameters $\tilde{x}_{d}$ along with all their $q$-shifts. Ultimately, we are only interested in specific components of $X^T$ supported on the vacuum $\{{\bf{A}}\}$, which will reduce the contour $\Gamma$ to enclose only the relevant poles for that vacuum. In \cite{Aganagic:2013tta,Aganagic:2014oia,Aganagic:2015cta}, it was shown that for handsaw quivers, poles are combinatorially labeled by truncated Young diagrams, with no more than $N_{a,i}$ rows. This remains true of general Drinfeld quivers, and the proof is the same as in those papers (see Section \ref{sec:5dgauge}). From the point of view of the GIT quotient $X$, this labeling of fixed points is an unexpected result, since the integer $N_{a,i}$ only labels the number of moduli $y_{a,i}$ in the maximal torus $T_{V_a}\subset GL(V_a)$. Instead, from the point of view of the Nakajima variety $X'$, the integer $N_{a,i}$ has another meaning as the first Chern class of the $i$-th $U(1)$ gauge bundle in the maximal torus $T_{V'_a}\subset GL(V'_a)$, where the combinatorial interpretation in terms of Young diagrams becomes explicit \cite{Aganagic:2013tta}.
This completes the proof of the Proposition.\\

In order to obtain the vector electric blocks, which are solutions to the $U_{\hbar}(\widehat{^L \fg})$ qKZ equations, one proceeds as in the unramified case and considers quasimaps with descendant insertions at $\{0\}\in\mathbb{P}^1$. Equivalently, a scalar magnetic $\cW_{q,t}(\fg)$-algebra block is recovered from a vector electric block by pairing the latter with the appropriate  Whittaker covector functional.

Both the electric and magnetic blocks are $z$-solutions, analytic in the K\"ahler variables $z_a$ for a choice of chamber $\fC_C$. Meanwhile, the existence of $x$-solutions, analytic in the $T_H$-equivariant parameters for a choice of chamber $\fC_H$, is guaranteed from the existence of elliptic stable envelopes in the ramified setting. As we have seen, this follows from the decomposition of the tangent bundle $T X$ as \eqref{tangentdecomp}, and viewing $\Theta(T^{+}X)$ as an attractive line bundle. The elliptic stable envelopes are inserted at $\{\infty\}\in\mathbb{P}^1$.


\vspace{8mm}

\section{Ramification and little string theory}
\label{sec:littlestring}

\subsection{String Theory Construction}
\label{ssec:string5d}

Our starting point is type IIB string theory on the background
\beq\label{backbasic}
{\mathbb C}^2_\Gamma\times \mathbb{C}^3  \; .
\eeq
The space ${\mathbb C}^2_\Gamma$ is the minimal resolution by blowup of a Du Val (or Kleinian) singularity ${\mathbb C}^2/\Gamma$, with $\Gamma$ a discrete subgroup of $SU(2)$, labeled by an $\fg_\text{o}=A D E$ Lie algebra of rank $r$. As is well-known, such a resolution consists of $r$ 2-spheres, which intersect following the Dynkin diagram of $\fg_\text{o}$.

We decouple gravity by taking the string coupling limit $g_s\rightarrow 0$, which does not entirely trivialize the physics \cite{Seiberg:1997zk,Losev:1997hx}. In particular, one obtains a 6-dimensional chiral string theory on $M_6$, the $(2,0)$ little string theory of type $\fg_\text{o}$, with the same $\fg_\text{o}=ADE$ that labeled the resolved singularity ${\mathbb C}^2_\Gamma$ in the full 10-dimensional type IIB. The little string theory is not a local quantum field theory, it contains strings of finite tension equal to the square of the string mass, $m_s^2$, again inherited from the type IIB fundamental strings. 

For a gentle introduction to the elementary properties of little strings, the standard reviews are \cite{Aharony:1999ks,Kutasov:2001uf}. For a precise account of how the Wightman axioms of quantum field theory are modified in little strings, see \cite{Kapustin:1999ci}.

In the regime where the string length is taken to be very small, or equivalently  $m_s\rightarrow\infty$, we lose the one scale of the theory and recover a point particle theory, a $(2,0)$ superconformal field theory (SCFT), again labeled by the Lie algebra $\fg_\text{o}$. For the time being, we will keep $m_s$ finite, and will only take the limit to infinity when we make contact with the (undeformed) quantum geometric Langlands program.

The moduli of the $(2,0)$ little string all have their origin in the parent 10-dimensional type IIB superstring, as periods of various 2-forms along the 2-cycles $S^2_a$ of the resolved singularity ${\mathbb C}^2_\Gamma$. There are $r$ compact moduli, which are periods associated with the R-R 2-form $C^{(2)}$:
\beq\label{RRtwoform}
m_s^2\int_{S^2_a}C^{(2)}\, ,\;\qquad  a=1,\ldots, r\, .
\eeq
The corresponding compact directions all have radius $m_s^2$.
One also finds $4 r$ non-compact moduli, which are the periods associated with the NS-NS B-field $B^{(2)}$ and a triplet of self-dual 2-forms $\omega_{I,J,K}$:
\beq\label{periods}
\frac{m_s^4}{g_s}\int_{S^2_a}\omega_{I,J,K}\, ,\qquad \frac{m_s^2}{g_s}\int_{S^2_a}B^{(2)}\, ,\;\qquad  a=1,\ldots, r\, .
\eeq
The 2-forms $\omega_{I,J,K}$ arise here since ${\mathbb C}^2_\Gamma$ is a hyperk\"ahler manifold. We have chosen a canonical normalization of the moduli such that they all have mass dimension 2, ensuring no dimensional factors in their kinetic terms.

In the little string limit $g_s\rightarrow 0$, one requires that all these periods remain fixed. Therefore, the moduli space of the $(2,0)$ little string is simply
\beq\label{modulispace}
\left(\mathbb{R}^4\times S^1\right)^{r}/W(\fg_\text{o})\; ,
\eeq
where the quotient by Weyl group $W(\fg_\text{o})$ is performed not to overcount moduli.\\

The R-symmetry of the little string theory is $SU(2)_R$, which acts by rotating the $3\, r$ scalars that are the periods of $\omega_J$, $\omega_K$, and the NS-NS 2-form $B^{(2)}$. The operator $R$ in the index stands for the Cartan generator of $U(1)_R\subset SU(2)_R$, which only rotates the periods of $\omega_J$ and $\omega_K$.

Next,  we perform a compactification of the type IIB background on a Riemann surface:
\beq\label{back}
{\mathbb C}^2_\Gamma\times M_6 \; , \;\qquad \text{where}\;\; M_6=\cC\times \mathbb{C} \times \mathbb{C} \; .
\eeq
As in the rest of this work, the Riemann surface $\cC$ is an infinite cylinder\footnote{One immediate reason for this restriction is that the Riemann surface better be flat for the background to be a solution of type IIB string theory. A more technical reason is that the deformed $\cW$-algebras were naturally related to gauge theories \emph{compactified on a circle}, see Sections \ref{sec:3dgauge} and \ref{sec:5dgauge}. Without much additional work, the framework we developed in this paper can be used to further study the complex line $\cC=\mathbb{C}$ by considering an appropriately scaled  $R_{\cC}\rightarrow\infty$ limit, or the torus $\cC=T^2$ by identifying the ends of the cylinder. In those cases, the relevant quantum algebra would be a Yangian and a quantum elliptic algebra, respectively. The finite-dimensional representation theory of these algebras is essentially the same as that of the quantum affine algebra.} of radius $R_{\cC}$,
\beq
\cC= \mathbb{R} \times S^1(R_{\cC}) \; .
\eeq
The $(2,0)$ little string compactified on the circle of radius $R_\cC$ enjoys T-duality, where it becomes equivalent to a non-chiral little string theory compactified on a circle of radius
\beq
R_{\cC'}=\frac{1}{m_s^2\, R_\cC}\; ,
\eeq
the (1,1) little string theory. Equivalently, this (1,1) little string is the $g_s\rightarrow 0$ limit of the 10-dimensional type IIA superstring on our background \eqref{back}\footnote{Note that the existence of T-duality in compactified little string is a strong indication that it is not a local quantum field theory.}. 

From the IIB perspective, the moduli space of vacua of the $(2,0)$ little string compactified on $S^1(R_{\cC})$ is again \eqref{modulispace}, where the periods \eqref{RRtwoform} and \eqref{periods} are simply rescaled by the radius $R_\cC$. From the IIA perspective, this is also the moduli space of the $(1,1)$ little string compactified on $S^1(R_{\cC'})$, where $4\, r$ moduli are as in the uncompactified theory (again up to radius), and $r$ new compact moduli appear as the holonomy $A^{\theta}_{a}$ of gauge fields around $S^1(R_{\cC'})$, related to the R-R 2-form period in type IIB by 
\beq\label{wowgauge}
A^{\theta}_{a} = R_{\cC}\,m^2_s\int_{S^2_a} C^{(2)}\; ,
\eeq 
with periodicity $2\,\pi/R_{\cC'}$.

\vspace{8mm}

\subsection{D5 branes and 5d gauge theory}
\label{ssec:simplylaced}

In order to make contact with the Langlands program and its (tame) ramification, we need to introduce punctures on the Riemann surface $\cC$.
In string theory, a natural way to achieve this is to introduce D-branes at points on $\cC$. In our type IIB background, there are a priori many such branes which fit this requirement. The correct branes to consider here are D5 branes wrapping 2-cycles of the resolved $ADE$ singularity ${\mathbb C}^2_\Gamma$, and wrapping $\mathbb{C}^2\subset M_6=\cC\times\mathbb{C}\times\mathbb{C}$. The relevance of these branes in our context was first noted in \cite{Aganagic:2015cta,Haouzi:2016ohr}.\\ 

In order to be more quantitative, we introduce some notations: by the McKay correspondence, the second homology group $H_2({\mathbb C}^2_\Gamma, \mathbb{Z})$ is identified with the root lattice $\Lambda_{rt}$ of $\fg$. Indeed, the group $H_2({\mathbb C}^2_\Gamma, \mathbb{Z})$ is spanned by $r$ vanishing 2-cycles $\{S^2_{a}\}$. Correspondingly, as is well-known, $\Lambda$ is spanned by $r$ positive simple roots $\alpha_{a}$. Furthermore, the intersection pairing in homology is identified with the Cartan Killing form of $\fg_\text{o}$, up to an overall sign:
\beq\label{pairingCartan}
\# (S^2_a \cap S^2_b)=-C_{ab}\; , \qquad\;\;\;\;a,b=1,\ldots,r \; ,
\eeq
where $C_{ab}$ is the Cartan matrix of $\fg_\text{o}$.

Then, we first introduce $n_a$ D5 branes wrapping the compact 2-cycle $S^2_{a}$, for all $a=1,\ldots,r$, and $\mathbb{C}^2\subset M_6$. This results in a total of $n=\sum_{a=1}^r n_a$ D5 branes, and a net non-zero D5 brane R-R (magnetic) flux, measured by a nontrivial homology class $[S]\in H_2({\mathbb C}^2_\Gamma, \mathbb{Z})$. By abuse of notation, we write $[S]$ in a basis of positive simple roots as
\beq\label{compact}
[S] = - \sum_{a=1}^{r}  \,n_a\,\alpha_a\;\;  \in  \,\Lambda_{rt} \; ,
\eeq
with $n_a$ non-negative integers.\\

We also introduce the second relative homology group $H_2({\mathbb C}^2_\Gamma, \partial \left({\mathbb C}^2_\Gamma\right), \mathbb{Z})$. Its elements are 2-cycles of ${\mathbb C}^2_\Gamma$ which are allowed a nontrivial boundary at infinity $\partial \left({\mathbb C}^2_\Gamma\right)$. The group is spanned by $r$ non-compact 2-cycles $\{{S^2_{a}}^*\}$, each of which is constructed as the fiber of the cotangent bundle ${T^*}{S^2_a}$ over a generic point on the $a$-th 2-sphere $S^2_a$. The McKay correspondence further identifies this relative homology group with the  weight lattice $\Lambda_{wt}$ of $\fg_\text{o}$. Equivalently, the $r$ non-compact 2-cycle $\{{S^2_{a}}^*\}$ are identified with the $r$ fundamental weights $w_{a}$ of $\fg_\text{o}$. It follows that the pairing between positive simple roots and fundamental weights in $\fg_\text{o}$ translates in geometry to 
\beq
\# (S^2_a \cap {S^2_b}^*)=\delta_{ab} \; , \qquad\;\;\;\;a,b=1,\ldots,r \; .
\eeq
Note that the homology group $H_2({\mathbb C}^2_\Gamma, \mathbb{Z})$ is a subgroup of the relative homology group $H_2({\mathbb C}^2_\Gamma, \partial \left({\mathbb C}^2_\Gamma\right), \mathbb{Z})$, since compact 2-cycles are elements of $H_2({\mathbb C}^2_\Gamma, \partial \left({\mathbb C}^2_\Gamma\right), \mathbb{Z})$ with trivial boundary at infinity. This is the homological version of the more familiar statement that the root lattice $\Lambda_{rt}$ is a sublattice of the weight lattice $\Lambda_{wt}$ in $\fg_\text{o}$.

Correspondingly, we  introduce $n^F_a$ D5 branes wrapping the non-compact 2-cycle ${S^2_a}^*$, for all $a=1,\ldots,r$, and $\mathbb{C}^2\subset M_6$. This results in a total of $n^F=\sum_{a=1}^r n^F_a$ D5 branes, and an associated R-R flux measured by the class $[S^*]\in H_2({\mathbb C}^2_\Gamma, \partial \left({\mathbb C}^2_\Gamma\right), \mathbb{Z})$. Again by abuse of notation, we write $[S^*]$ in a basis of fundamental weights as
\beq\label{noncompact}
[S^*] =  \sum_{a=1}^{r} \, n_a^F \, w_a \;\;  \in\, \Lambda_{wt}\; ,
\eeq  
where $n_a^F$ are non-negative integers sometimes called Dynkin labels. In this basis, the fundamental weight $w_a$ is the vector $w_a=[0,\ldots , 0, 1, 0, \ldots, 0]$ of size $r$ with a 1 in the $a$-th entry. The homology class \eqref{noncompact} is written as $[S^*]=[n_1^F, n_2^F, n_3^F, \ldots, n_r^F]$.

All in all, the net D5 brane R-R flux is encoded in the homology class $[S+S^*]$, understood as an element of the lattice $\Lambda_{wt}$.\\


In the absence of branes, the IIB background has 16 supercharges, solely due to the background ${\mathbb C}^2_\Gamma$. In order to preserve supersymmetry after introducing the wrapped D5 branes, one needs to properly align the triplet of self-dual 2-forms $\vec{\omega}=(\omega_I, \omega_J, \omega_K)$: as far as the $n^F$ D5 branes wrapping non-compact 2-cycles are concerned, supersymmetry is preserved only if the vectors $\int_{{S^2_a}^*} \vec{\omega}$ point in the same direction, meaning the central charges are aligned for all $a=1, \ldots, r$. Then, let us explicitly choose  $\vec{\omega}$ to satisfy
\beq
\int_{{S^2_a}^*}\omega_{I}>0\, ,\qquad \int_{{S^2_a}^*}\omega_{J}=0\, ,\qquad \int_{{S^2_a}^*}\omega_{K}=0\,  .
\eeq
With this choice of periods, the $n^F$ D5 branes are 1/2-BPS, meaning 8 supercharges remain. 

In order for the $n$ D5 branes to preserve the same supersymmetry, one considers the periods of the 2-forms through the compact 2-cycles $S^2_a$, corresponding to the choice of a metric on ${\mathbb C}^2_\Gamma$. For all $a=1, \ldots, r$, we will explicitly choose
\beq\label{couplings}
\tau^{6d}_a\equiv \int_{S^2_a}\left(\frac{m^2_s}{g_s}\,\omega_{I}+i\, C^{(2)}\right)\, ,\qquad \int_{S^2_a}\omega_{J}=0\, ,\qquad \int_{S^2_a}\omega_{K}=0\;  ,
\eeq
with $\text{Re}(\tau^{6d}_a)>0$\footnote{The choice of sign in $\text{Re}(\tau^{6d}_a)>0$ will also ensure the convergence of the various $z$-solution blocks we computed in this work.}, along with
\beq\label{NSNS}
\int_{S^2_a}B^{(2)}=0\; .
\eeq
With this choice of periods, the D5 branes wrapping the compact and non-compact 2-cycles break the same half supersymmetry, so the background still has 8 supercharges.

Our discussion so far was phrased in the full type IIB background. In the $(2,0)$ little string, one simply considers the D5 branes as above, but in the $g_s\rightarrow 0$ limit. Inspired by the nomenclature of \cite{Losev:1997hx}, one option would be to call the branes in that limit ``d branes." However, we choose to keep the terminology ``D branes" throughout, and the context will hopefully make it clear whether the string coupling is null or not.\\

We consider an intermediate energy scale $E$ well below the string scale, $E\ll m_s$.
At that scale, and in the limit $g_s\rightarrow 0$, the theory on the D5 branes is a 5-dimensional gauge theory with $\cN=1$ supersymmetry, on the manifold $S^1(R_{\cC'})\times\mathbb{C}^2$ \cite{Aganagic:2015cta}.\\

The theory is not simply 4-dimensional on $\mathbb{C}^2$, due to the presence of the cylinder $\cC=S^1(R_{\cC})\times\mathbb{R}$ in the transverse directions. Indeed, strings will wind around the cylinder circle, so one should expect a Kaluza-Klein tower of states on the T-dual circle of radius
\beq
R_{\cC'}=\frac{1}{m_s^2\, R_{\cC}} \; ,
\eeq
and one should therefore expect that the low-energy physics depends on the scale $R_{\cC'}$.
Phrased differently, the  D5 branes are points on the circle $S^1(R_{\cC})$, but they have an equivalent description as D6 branes wrapping $S^1(R_{\cC'})$ in the (1,1) little string, by T-duality. From this T-dual picture, it is clear that the low energy theory on the D6 branes really is a 5d theory wrapping the circle $S^1(R_{\cC'})$.\\

The characterization of the 5d gauge theory comes from geometry. Since the D5 branes are wrapping 2-cycles in ${\mathbb C}^2_\Gamma$, they are described at energy scale $E$ by a quiver gauge theory \cite{Douglas:1996sw,Johnson:1996py}, of shape the Dynkin diagram of the Lie algebra $\fg_\text{o}$\footnote{In the analysis of Douglas and Moore \cite{Douglas:1996sw}, our wrapped D5 branes would have been treated as (fractional) D3 branes probing the orbifold singularity. The authors show that in the full type IIB background, the effective gauge theory on the branes is described by an \emph{affine} quiver, of rank $r+1$. The affine nature of the quiver is due to $n_0$ (full) D3 branes transverse to the $ADE$ surface, resulting in an additional diagonally embedded $U(n_0)$ gauge group. Its inverse gauge coupling is the sum of inverse gauge couplings for all quiver nodes, but this sum equals $1/g_s$, the string coupling of type IIB string theory. It follows that in the little string limit $g_s\rightarrow 0$, the $U(n_0)$ gauge fields are effectively frozen, explaining why our quiver gauge theories are only of \emph{finite} Dynkin type \cite{Cachazo:2001gh,Cachazo:2001sg}.}.
In 5 dimensions, the inverse gauge couplings  $1/(g^{5d}_a)^2$ have mass dimension 1. Since we would like the gauge theory to be weakly coupled at energy $E$, the inequality $E\, (g^{5d}_a)^2\ll 1$ should hold for all $a=1,\ldots,r$. In terms of the dimensionless couplings $\tau^{6d}_a$, which scale as $1/(m_s\, (g^{5d}_a)^2)$, this is equivalent to requiring  $E/m_s\ll  \tau^{6d}_a$.
The gauge group of the quiver is
\beq\label{D5gaugegroup}
G_{5d}=\prod_{a=1}^r U(n_a)\; ,
\eeq
where the ranks $n_a$ are the number of D5 branes wrapping the compact 2-cycle $S^2_a$ \eqref{compact}.
Each $U(n_a)$ gauge group contains a massive $U(1)$ center by the Green-Schwarz mechanism, so the gauge groups should really be  $SU(n_a)$. Nevertheless, because the 5d theory is eventually placed on an $\Omega$-background $S^1_{\cC'}\times\mathbb{C}_q\times \mathbb{C}_t$, it convenient to simply drop the special unitary condition and treat all $n_a$ equivariant parameters on an equal footing. 
The positions $e_{a,i}$ of the $n=\sum_{a=1}^{r} \, n_a$ D5 branes on $\cC$ are the 5d Coulomb moduli.\\ 

The flavor symmetry of the quiver is
\beq\label{D5flavorgroup}
G^F_{5d}=\prod_{a=1}^r U(n_a^F)\; ,
\eeq 
where the ranks $n_a^F$ are the number of D5 branes wrapping the non-compact 2-cycle ${S^2_a}^*$ \eqref{noncompact}.
These fundamental hypermultiplets are in representation $(n_a,\overline{n^F_a})$ of $U(n_a)\times U(n^F_a)$, they arise from the zero modes of open strings at the intersection of D5 branes wrapping the compact 2-cycle $S^2_a$ and the non-compact 2-cycle ${S^2_a}^*$ on the fiber. The positions $f_{a,s}$ of the $n^F=\sum_{a=1}^{r} \, n_a^F$ D5 branes on $\cC$ are masses for these hypermultiplets.

Finally\footnote{We do not consider the open strings between D5 branes wrapping ${S^2_a}^*$ and ${S^2_b}^*$, the intersection pairing of which is dictated by the quadratic form matrix of $\fg_\text{o}$ (the inverse Cartan matrix): the resulting bifundamental hypermultiplets represent nondynamical degrees of freedom, which we have not kept track of throughout this work.}, there are hypermultiplets in the bifundamental representation $\oplus_{b>a}\, \Delta_{ab}\,(n_a, \overline{n_b})$ of the group $\prod_{a,b} U(n_a)\times U(n_b)$, where $\Delta_{ab}=1$ if there is a link connecting nodes $a$ and $b$, and $\Delta_{ab}=0$ otherwise. They arise from the zero modes of open strings with one end on the D5 branes wrapping $S^2_a$, and the other end on D5 branes wrapping $S^2_b$.\\

The $\tau^{6d}_a$ variables \eqref{couplings} are not moduli of $T^{5d}_{\fg_{\text{o}}}$. Indeed, the 2-forms in the definition of $\tau^{6d}_a$ are valued in all six dimensions of $M_6=\cC\times \mathbb{C}^2$, so they are not dynamical. Instead, they  are parameters: they determine the $r$ gauge couplings of the quiver theory.

The periods of the 2-forms $\omega_J$, $\omega_K$, and of the NS-NS B-field $B^{(2)}$ through the 2-cycles $S^2_a$ are the F.I. parameters of $T^{5d}_{\fg_{\text{o}}}$, set to zero for now as in \eqref{couplings}.

The Coulomb moduli, hypermultiplet masses and F.I. parameters all start out as real variables, but get complexified by the holonomy around the (T-dual) cylinder circle $S^1(R_{\cC'})$.\\

The 5-dimensional quiver gauge theory is the theory $T^{5d}_{\fg_{\text{o}}}$ from Section \ref{sec:5dgauge}. There, we also imposed $r$ additional constraint equations:
\beq\label{constraint5dencore}
\sum_{b=1}^r C_{ab} \;n_b = n^F_a\; ,\qquad\; a=1,\ldots, r\; ,
\eeq
with $C_{ab}$ the Cartan matrix of $\fg_\text{o}$; if we were to reduce on $S^1(R_{\cC'})$, these constraints would imply the vanishing of the beta function. In the string language, the constraints translate to requiring the net D5 brane R-R flux to be zero. From the intersection pairing \eqref{pairingCartan}, our constraint equations are equivalent to requiring the class $[S+S^*]$ to vanish in homology. In practice, given a fixed class $[S^*]$, the class $[S]$ is uniquely determined (if it exists) from solving the constraint
\beq\label{vanishingflux}
[S+S^*]=0 \; .
\eeq
For instance, for $\fg_\text{o}=A_1$, only the even classes $[S^*]=[2\, n]$ yield solutions $[S]$ for $n$ a non-negative integer. This is simply the homological version of the more familiar gauge theory statement that the beta function vanishes in 4 dimensions if and only if the number of fundamental hypermultiplets is twice the rank, $n^F = 2\, n$.\\

We realize the ADHM instanton quantum mechanics of $T^{5d}_{\fg_\text{o}}$ in the little string picture by introducing $k_a$ 1/2-BPS D1 branes which wrapping the compact 2-cycle $S^2_{a}$, for all $a=1,\ldots,r$ \cite{Douglas:1995bn}, and then taking the $g_s\rightarrow 0$ limit. There are a total of $k=\sum_{a=1}^r k_a$ such branes, each located at a point $\phi_{a,I}$ on $\cC$. The effective theory on the D1 branes is a 1-dimensional $\cN=4$ quantum mechanics on $S^1(R_{\cC'})$, for the same reason that the theory on the D5 branes is 5-dimensional on $S^1(R_{\cC'})$.

Equivalently, in the T-dual frame, we are studying the $(1,1)$ little string on $S^1(R_{\cC'})$ with $n+n^F$ D6 branes and $k$ D2 branes wrapping the circle.
The Witten index of the ADHM quantum mechanics \eqref{5dhalfindexmore} is the partition function of the $(1,1)$ little string on $M'_6=\cC'\times\mathbb{C}_q\times\mathbb{C}_t$ in the presence of the branes\footnote{Qualitatively, a localization argument will show that the partition function of the bulk little string with D6 brane defects is the same as the partition function of the brane defects \emph{themselves}: away from the defect branes, the supersymmetry is restored to the 16 supercharges of the bulk little string, where the index trivializes due to the pairwise cancellation of bosons and fermions in the enhanced supersymmetry. The only nonzero contributions to the index come from the defect branes.}:
\beq
\label{5dhalfindexmoreagain}
{\mathcal Z}_{inst}(T^{5d}_{\fg_\text{o}})  = {\rm Tr}\left[(-1)^F\, q^{J_1+\frac{R}{2}}\; t^{J_2-\frac{R}{2}} \; {\fm}^{\Sigma}\right]\;\; .
\eeq
The integral representation of the index is now understood as the contribution of strings stretching between the various branes: the integrand is made up of 1-loop determinants which are the contributions of D2/D2 and D2/D6 strings, while the integration variables are the D2 brane positions on $\cC'$. In the IIA language, these variables are parameterized by $\phi_{a,I}=\varphi_{a,I}+ i \, A^{\theta}_{a,I}$, with $A^{\theta}_{a,I}$ the holonomy around  $S^1(R_{\cC'})$. The integration over the D2 brane positions enforces the gauge invariance of the group $\prod_{a=1}^r U(k_a)$.


\subsection{D3 branes and 3d gauge theory}
\label{ssec:3dgauge}

The 3d gauge theory $T^{3d}_{\fg_\text{o}}$ also has a realization in the $(2,0)$ little string. To see this, we force the D5 brane theory on its Higgs branch by turning on some of the periods: 
\beq\label{periods3d}
\int_{S^2_a}\omega_{J}=0\, ,\qquad \int_{S^2_a}\omega_{K}=0\, ,\qquad \int_{S^2_a}B^{(2)}>0\, .
\eeq
Correspondingly, this turns on $r$ real F.I. parameters (complexified as usual) for $T^{5d}_{\fg_\text{o}}$, while the complex F.I. parameters remain null. With the F.I. parameters aligned as above, a common supersymmetry is preserved,  and the remaining R-symmetry is $U(1)_R\subset SU(2)_R$. In the little string, we bind the $n$ D5 branes wrapping the cycles $\{S^2_{a}\}$ to (a subset of) the $n^F$ D5 branes wrapping the cycles $\{{S^2_{a}}^*\}$, and move their positions $e_{a,i}$ to coincide with the positions $f_{a,s}$ on the cylinder $\cC$. This results in a new configuration of $n^F$ D5 branes wrapping non-compact 2-cycles only. We denote these 2-cycles as $\{{S^2_s}^*\}$, with corresponding homology classes in $H_2(X, \partial X, \mathbb{Z})$ given by
\beq\label{non-compactrecom}
[S_s^*] =  \underline{\lambda_s} \;\;  \in\, \Lambda_{wt}\; ,
\eeq  
The class $[S_s^*]$ can be expanded as
\beq\label{non-compactrecomagain}
[S_s^*] =  [S_a^*]  - \sum_{b=1}^r h_{s,b}\, [S_b] \; ,
\eeq 
where $h_{s,b}$ are non-negative integers and $[S_a^*]= w_a$ and $[S_a]= \alpha_a$ are the classes associated to the $a$-th fundamental weight and $a$-th simple root, respectively. This decomposition is the same as \eqref{weightsdecom} in the gauge theory $T^{5d}_{\fg_\text{o}}$, where we had specified which Coulomb moduli get tuned to which mass. Once all Coulomb moduli are frozen in this way, the hypermultiplets become massless and can acquire a vev, defining the Higgs branch. The $n^F$ classes $[S_s^*]$ are not independent, since we required the net D5 brane R-R flux to vanish \eqref{vanishingflux}. This constraint now reads
\beq\label{weightsadduptozerohom}
\sum_{s=1}^{n^F} [S_s^*] =0\; ,
\eeq
which is the homological version of \eqref{weightsadduptozero}.

In type IIB, the 1/2-BPS semi-local vortex solutions on the Higgs branch of $T^{5d}_{\fg_\text{o}}$ are D3 branes at points on the cylinder $\cC$, wrapping the compact 2-cycles of $\mathbb{C}^2_{\Gamma}$ and one of the two complex lines, say $\mathbb{C}_q$ \cite{Hanany:2003hp}. Overall, we introduce a total of $N =\sum_{a=1}^r N_a$ D3 branes, where $N_a$ of them wrap $S^2_{a}$, with corresponding $H_2(X, \mathbb{Z})$ homology class 
\beq\label{d3compact}
[N] = \sum_{a=1}^r  \,N_a\,\alpha_a\;\;  \in  \,\Lambda \; .
\eeq
From the point of view of the D5 brane theory, the D3 branes are codimension-2 and end on them, which this turns on magnetic flux, consistent with the vortex interpretation \cite{Hanany:1996ie}. Recall that on $\mathbb{C}^2_{\Gamma}$, the D5 branes wrap holomorphic cycles; meanwhile, the D3 branes are Lagrangian, intersecting the D5 branes transversely in $\mathbb{C}^2_{\Gamma}$. The low energy effective theory on the D3 branes is the 3d $\cN=2$ Drinfeld quiver gauge theory $T^{3d}_{\fg_\text{o}}$: the quiver shape is the Dynkin diragram of $\fg_\text{o}$, and the theory is supported on the manifold $S^1(R_{\cC'})\times\mathbb{C}_q$\footnote{Once again, the D3 brane theory is 3-dimensional and not simply 2-dimensional with $\cN=(2,2)$ supersymmetry on $\mathbb{C}_q$, because of the fundamental strings wrapping the cylinder $\cC$, meaning there is a Kaluza Klein tower of states on the T-dual cylinder $\cC'$. Alternatively, $T^{3d}_{\fg_\text{o}}$ is the effective theory on D4 branes wrapping $S^1(R_{\cC'})$ in the $(1,1)$ little string.}. The little string couplings $\tau^{6d}_a$ in \eqref{couplings} are the F.I. parameters. The zero modes of D3/D3 strings on a node $a$ produce the 3d $\cN=4$ supersymmetric vector multiplets, while the D3/D3 strings between neighboring nodes are responsible for the $\cN=4$ bifundamental hypermultiplets. The $n^F$ D5 branes further break half the supersymmetry, so D3/D5 strings provide the 3d $\cN=2$ fundamental and anti-fundamental chiral matter\footnote{Gauge-vortex duality in gauge theory is now realized as a large $N$-duality in the little string  \cite{Aganagic:2013tta,Aganagic:2014kja}. The argument was originally given in the unramified limit $q=t$ by Dijkgraaf and Vafa \cite{Dijkgraaf:2009pc}, see also the early works \cite{Strominger:1995cz,Greene:1996dh}. In fact, in the $\Omega$-background, $N$ does not have to be large at all for the duality to hold, and can even be equal to 1, as we will do in the Examples Section.}.\\

We found from our gauge theory analysis that this chiral matter is naturally labeled by weight spaces of a quantum affine algebra $U_\hbar(\widehat{^L \fg})$, and furthermore that these weight spaces are recorded in the half-index of $T^{3d}_{\fg_\text{o}}$. This is the partition function of the theory on the D3 branes\footnote{Alternatively, we argued that the ADHM instanton partition function of $T^{5d}_{\fg_\text{o}}$ was realized as the Witten index of the gauged quantum mechanics on $k$ D1 branes wrapping the 2-cycles $\{S^2_{a}\}$. We could follow the same route here, and consider the gauged quantum mechanics on the $k$ D1 branes in the presence of D3 and D5 branes; the half-index of $T^{3d}_{\fg_\text{o}}$ coincides with the Witten index of this quantum mechanics.}.
Therefore, these weight spaces can now be reinterpreted as the zero modes of D3/D5 strings in $(2,0)$ little string theory on $M_6=\cC\times\mathbb{C}_q\times\mathbb{C}_t$. Equivalently, they the zero modes of D4/D6 strings in $(1,1)$ little string theory on $M'_6=\cC'\times\mathbb{C}_q\times\mathbb{C}_t$ are labeled by quantum affine weights.\\

Following our dictionary to $q$-conformal blocks, we conclude that $\cW_{t,q}(\fg)$ magnetic blocks are $(2,0)$ little string partition functions on $M_6$ in the presence of D3 and D5 branes. Each such block has a presentation as the half-index on $S^1_{\cC'}\times D^2$ for the effective theory $T^{3d}_{\fg_\text{o}}$ on the D3 branes. The $z$-solutions will be analytic in a given $\tau^{6d}_a$-parameter chamber, and are constructed by imposing exceptional Dirichlet boundary conditions on $S^1_{\cC'}\times S^1_{D^2}$.  The $x$-solutions will be analytic in a given mass-parameter chamber, the positions of the D5 branes on $\cC$, and are constructed by imposing enriched Neumann boundary conditions on $S^1_{\cC'}\times S^1_{D^2}$. 

The $U_\hbar(\widehat{^L \fg})$ electric blocks are constructed in the same way, except one should consider the $(2,0)$ little string on the manifold $M^\times_6= \cC\times\mathbb{C}^\times_q\times\mathbb{C}_t$ in the presence of D3 and D5 branes. $M^\times_6$ differs from $M_6$ only by removing the origin of $\mathbb{C}_q$, so the D3 and D5 branes are now supported on $\mathbb{C}^\times_q\times \{S^2_{a}\}$ and $\mathbb{C}^\times_q\times\mathbb{C}_t\times \{{S^2_s}^*\}$ respectively, with appropriate boundary conditions at the excised origin.

\vspace{8mm}

\subsection{The conformal limit}
\label{sec:conflim}

The $(2,0)$ little string is supported on $M_6 = \cC \times {\bf M}_4$, with ${\bf M}_4=\mathbb{C}_q\times \mathbb{C}_t$. We now take the limit to the $(2,0)$ SCFT:
\beq
m_s\rightarrow\infty \; ,
\eeq 
where we require $M_6$, and  in particular the cylinder $\cC=S^1_{\cC}\times \mathbb{R}$, to remain a fixed background throughout. 
The gauge theory $T^{5d}_{\fg_\text{o}}$ on the D5 branes was supported on $S^1_{\cC'}\times {\bf M}_4$, where $S^1_{\cC'}$ is the T-dual circle of radius
\beq
R_{\cC'} = \frac{1}{m^2_s\, R_{\cC}} \; .
\eeq
It follows that the radius $R_{\cC'}\rightarrow 0$ in the conformal limit, since $R_{\cC}$ is kept fixed. We further keep the $\Omega$-background parameters $\epsilon_q$ and $\epsilon_t$ fixed on ${\bf M}_4$, and so all the quantization parameters 
\beq
q=e^{R_{\cC'}\, \epsilon_q}\, ,\qquad t=e^{-R_{\cC'}\, \epsilon_t} \, ,\qquad \hbar=q/t \, ,
\eeq
trivialize to 1. This is why we lose the quantum algebra deformations in the limit, and recover the usual chiral algebras on $\cC$. 

$T^{5d}_{\fg_\text{o}}$ becomes a 4-dimensional theory with $\cN=2$ supersymmetry supported on $\mathbb{C}_{\epsilon_q}\times \mathbb{C}_{\epsilon_t}$. The 6d little string moduli \eqref{RRtwoform} and \eqref{periods} are kept fixed in the limit and become moduli of the 6d SCFT; $r$ of these  moduli are $m_s^2 \tau^{6d}_a$, so we must have $\tau^{6d}_a\rightarrow 0$. But these are the inverse gauge couplings of the theory on the D5 branes, so the 4-dimensional theory no longer has a description as a weakly coupled gauge theory\footnote{Except in rare ``accidental" instances, see Section \ref{ssec:spectral}.}. The constraint \eqref{constraint5dencore} becomes a conformal condition in 4d, namely the vanishing of the beta function. Since  $T^{5d}_{\fg_\text{o}}$ originated as a description of a codimension-2 defect in the little string, we deduce that it becomes a 4d theory of class $S$ in the limit \cite{Gaiotto:2009we,Gaiotto:2009hg}.\\

Likewise, the Drinfeld gauge theory $T^{3d}_{\fg_\text{o}}$ on the D3 branes is initially supported on $M_3 = S^1_{\cC'}\times \mathbb{C}_q$. After taking the conformal limit, it becomes a 2-dimensional theory with $\cN=(2,2)$ supersymmetry supported on $\mathbb{C}_{\epsilon_q}$ instead. 
The gauge couplings of  $T^{3d}_{\fg_\text{o}}$ are  $m_s^{-4}$ times fixed moduli of the (2,0) SCFT, so the Lagrangian description of the theory is again lost in the limit.\\

In the following Sections, it will prove useful to reinterpret the manifold  $M_6$ as a torus fibration $T^2=S^1_q(R_q)\times S^1_t(R_t)$:
\beq\label{fibration}
T^2\hookrightarrow M_6 \hookrightarrow M_4 = \cC\times\mathbb{R}_+\times\mathbb{R}_+ \; .
\eeq 
This is the perspective that makes the relation to the geometric Langlands program manifest. Indeed, in the conformal limit, and reducing on $T^2$, we obtain 4d Super Yang-Mills on $M_4=\cC\times\mathbb{R}_+\times\mathbb{R}_+$. The D5 branes wrapping $T^2$ become D3 branes wrapping $\mathbb{R}_+\times\mathbb{R}_+\subset M_4$, so they flow to codimension-2 defects at points on $\cC$ in the limit. These are Gukov-Witten monodromy defects \cite{Gukov:2006jk,Gukov:2008sn}, as we will analyze in detail in Section \ref{sec:ym}.\\ 

It would be important to make precise contact with other recent descriptions of this background. For instance, it is known that there exist $Y_{L,M,N}$ chiral algebras found ``at the corner" of $\mathbb{R}_+\times\mathbb{R}_+$ in 4d Super Yang-Mills \cite{Gaiotto:2017euk,Prochazka:2018tlo}. In particular, for $\fg=A_n$, the fact that our D5 branes realize certain Fock modules in the $\cW_{q,t}(A_n)$-algebra is consistent with this picture: in the conformal limit, we have $Y_{n,0,0}=u(1)\oplus\cW_{\beta}(A_n)$, and Gukov-Witten defects should describe Verma modules of the algebra, via a free field realization just as in our construction. This also fits into a larger program where such modules are known to be realized as certain representations of the affine Yangian of $\fgl(1)$ \cite{Gaiotto:2020dsq}, while in the $q$-deformed context they should naturally appear as representation of the quantum toroidal algebra $U_{q,t}(\Hat{\Hat{\fgl_1}})$ \cite{Zenkevich:2020ufs,Harada:2021xnm,NathanH2}.\\

If we consider the starting manifold to be $M^{\times}_6$, with the origin removed, then $M_4 = \cC\times\mathbb{R}\times\mathbb{R}_+$, and the D5 branes flow in the conformal limit to monodromy defects at points on $\cC$ and wrapping $\mathbb{R}\times\mathbb{R}_+$. Gaiotto and Witten explained that imposing Nahm-type boundary conditions at the origin of $\mathbb{R}_+$ naturally leads to a description of the system via $\fg$-opers, see Section 8.4 in \cite{Gaiotto:2011nm}. It would be important to understand how the classical limit of our $\cW(\fg)$-algebra construction reduces to this description.

\vspace{8mm}

\section{The non simply-laced case in gauge and string theory}
\label{sec:nsl}

So far, our realization of the $q$-conformal blocks in gauge and string theory was discussed in the simply-laced case $\fg_\text{o}$. However, the $\cW_{q,t}(\fg)$ and $U_\hbar(\widehat{^L \fg})$ blocks of Section \ref{sec:confblocks} were defined for an arbitrary simple Lie algebra $\fg$. This begs the question of whether or not the cases $\fg=B_r, C_r, G_2, F_4$ could also be understood in gauge and string theory. The purpose of this Section is to explain how this is done, and point out some open questions.\\

\begin{figure}[h!]
	\emph{}
	\centering
	\includegraphics[trim={0 0 0 0cm},clip,width=0.99\textwidth]{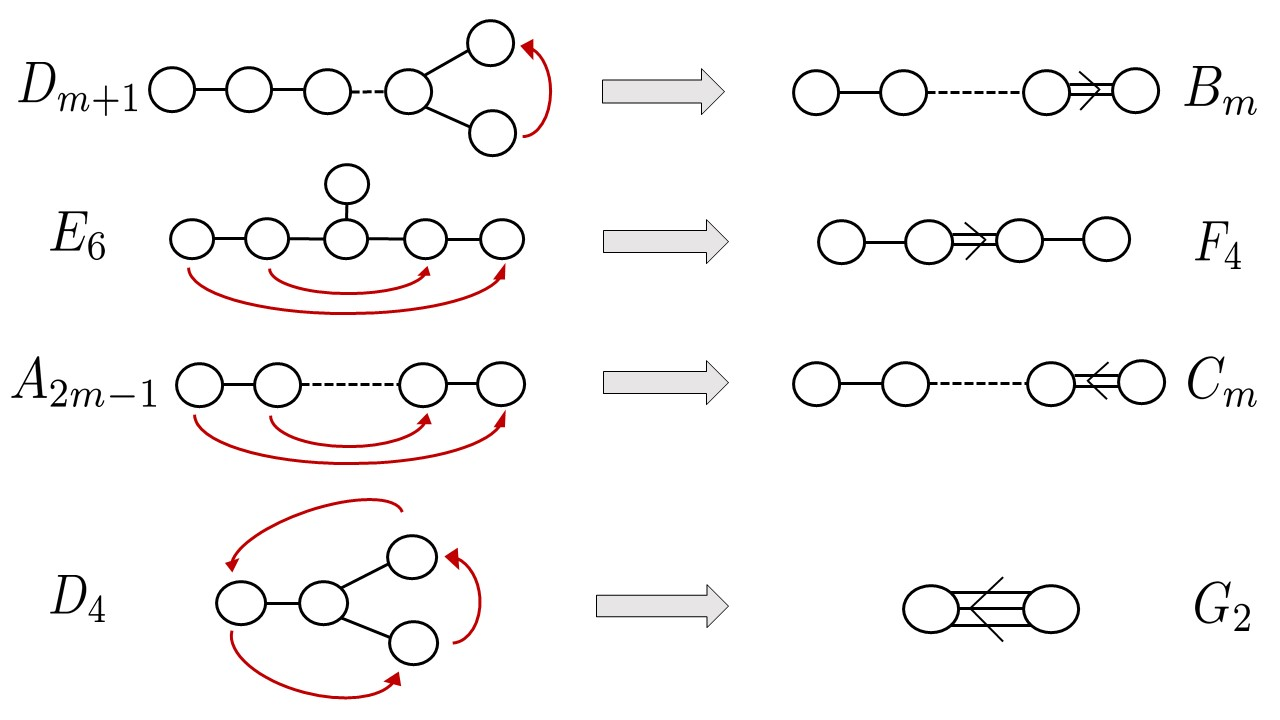}
	\vspace{-10pt}
	\caption{Construction of the non simply-laced Lie algebras starting from the simply-laced ones via folding.} 
	\label{fig:nonsimplylaced}
\end{figure}

The construction of the non simply-laced Lie algebras from the simply-laced ones is well-known: one observes that a non simply-laced Lie algebra $\fg$ is always a subalgebra of a simply-laced Lie algebra $\fg_\text{o}$ invariant under an $A$-action, where $A$ is a non-trivial outer automorphism group of $\fg_\text{o}$.  The group $A$ is abelian\footnote{The outer automorphism group of $D_4$ is in fact nonabelian, as it is the symmetric group $S_3$. For us, only the abelian subgroup $\mathbb{Z}_3$ will be relevant.}, and it is a well-known fact that the outer automorphisms of $\fg_\text{o}$ are in one-to-one correspondence with the automorphisms of its Dynkin diagram. It follows at once that the only possible outer automorphism groups are either $A=\mathbb{Z}_2$ or $A=\mathbb{Z}_3$, see figure \ref{fig:nonsimplylaced}. The lacing number $\fn_\fg$ is the order of the group $A$.

Let $\fa\in A$, and let $\Delta$ be the set of simple roots of $\fg_\text{o}$. As reviewed in Appendix \ref{sec:appendixlie}, the simple roots of $\fg$ split into two sets:
First, we take the long roots of $\fg$ to be the simple roots of $\fg_\text{o}$ invariant under $A$-action,
\begin{align}
\Delta_l = \left\{\alpha \; | \;\alpha\in\Delta,\; \alpha=\fa(\alpha) \right\}\; .
\end{align}
In our convention, long roots have length squared $2$, the same as the simple roots of $\fg_\text{o}$.
Second, the short roots of $\fg$ are the simple roots of $\fg_\text{o}$ which belong in orbits of $A$ of length $\fn_\fg$,
\begin{align}
\Delta_s &= \left\{\frac{1}{\fn_\fg}\left(\alpha+\fa(\alpha)+\ldots+\fa^{\fn_\fg-1}(\alpha)\right)\; | \; \alpha\in\Delta,\; \alpha\neq \fa(\alpha) \right\} \; .
\end{align}
These short roots of $\fg$ have length squared $2/\fn_\fg$, per our convention.

\subsection{String theory realization}

The above construction has a natural realization in string theory \cite{Aspinwall:1996nk,Bershadsky:1996nh} (see also 
the extension of the usual simply-laced McKay correspondence can to the non simply-laced case \cite{Slodowy1980SimpleSA,happel1980binary}). In the simply-laced case, our starting point was the type IIB string theory on $\mathbb{C}^2_{\Gamma}\times M_6$, with $\mathbb{C}^2_{\Gamma}$ a resolved $\fg_\text{o}$ singularity, and the $(2,0)$ little string living on $M_6=\cC\times\mathbb{C}_q\times\mathbb{C}_t$. If we view $M_6$ as the fibration \eqref{fibration}, then it is natural to consider the following twisted background:
\begin{equation}\label{backgroundnsl}
\frac{\mathbb{C}^2_{\Gamma}\times S^1_q(R_q)}{\mathbb{Z}_{\fn_\fg}}\times S^1_t(R_t)\times M_4 \; .
\end{equation}
Namely, as we go around the circle $S^1_q$, the resolved $ADE$ singularity $\mathbb{C}^2_{\Gamma}$ comes back to itself, but only up to the action of a generator $\fa\in \mathbb{Z}_{\fn_\fg}$ of the outer automorphism group.

In particular, this is a non-trivial action on the root lattice of $\fg_\text{o}$, and therefore a non-trivial $\fa$-action  on the homology group $H_2(\mathbb{C}^2_{\Gamma}, \mathbb{Z})$. Once we introduce $n$ D5 branes (and later $N$ D3 branes as vortices) wrapping the compact 2-cycles $\{S^2_{a}\}$ in $H_2(\mathbb{C}^2_{\Gamma}, \mathbb{Z})$, this action will permute the branes, and the configuration of D5 branes (and D3 branes) is henceforth required to be invariant under this action.

The same considerations apply to the weight lattice of $\fg_\text{o}$, and therefore to the relative homology group $H_2({\mathbb C}^2_\Gamma, \partial \left({\mathbb C}^2_\Gamma\right), \mathbb{Z})$: the configuration of $n^F$ flavor D5 branes wrapping the non-compact 2-cycles $\{{S^2_{a}}^*\}$ also has to be invariant under $\fa$-action.

A simple coroot of $\fg$ is a sum of simple roots of $\fg_\text{o}$, where all the roots are in a single $A$-orbit. Correspondingly, the $n$ D5 branes wrapping $\{S^2_{a}\}$ in the fibered geometry have a net magnetic flux measured by an element of $\Lambda_{cort}$, the coroot lattice of $\fg$; the corresponding homology class is denoted as 
\[
[S^\vee] = \sum_{a=1}^r  \,n_a\,\alpha^\vee_a\;\;  \in  \,\Lambda_{cort} \; .
\]	
Likewise, a fundamental coweight $w_a^{\vee}$ of $\fg$ is a sum of fundamental weights of $\fg_\text{o}$ where all the weights are in a single $A$-orbit. We conclude that the $n^F$ D5 branes wrapping $\{{S^2_{a}}^*\}$ in the fibered geometry have a net magnetic flux measured by an element of  $\Lambda_{cowt}$, the coweight lattice of $\fg$; the corresponding relative homology class is denoted as $[{S^*}^\vee]$.
\[
[{S^*}^\vee] = \sum_{a=1}^r  \,n^F_a\,w^\vee_a\;\;  \in  \,\Lambda_{cowt} \; .
\]	

In order to make sense of the class $[{S+S^*}^\vee]=[S^\vee]+[{S^*}^\vee]$, we simply consider $[S^\vee]$ as a class in relative homology with trivial boundary at infinity. Equivalently, the coroot lattice $\Lambda_{cort}$ being a sublattice of the coweight lattice $\Lambda_{cowt}$, the net flux $[{S+S^*}^\vee]$ can always be written as a coweight of $\fg$. 

Just as in the simply-laced case, we impose the asymptotic conformality constraint $[{S+S^*}^\vee]=0$, which once again takes the form of $r$ equations:
\beq\label{constraint5dnsl}
\sum_{b=1}^r C_{ab} \;n_b = n^F_a\; ,\qquad\; a=1,\ldots, r\; ,
\eeq
with $C_{ab}=\langle\alpha_a,\alpha_b^{\vee}\rangle$ the Cartan matrix of $\fg$. 
In the limit $g_s\rightarrow 0$, the background \eqref{backgroundnsl} describes the $ADE$ $(2,0)$ little string on 
\beq\label{littlestringnsl}
\frac{S^1_q(R_q)}{\mathbb{Z}_{\fn_\fg}}\times S^1_t(R_t)\times M_4 \; ,
\eeq
subjected to the $\mathbb{Z}_{\fn_\fg}$-twist. The D5 and D3 branes likewise survive the little string limit and are subjected to the twist\footnote{Though we refrain from doing so in this Section, we comment in passing that if one further takes the limit $m_s\rightarrow\infty$, the compactified  $(2,0)$ $ADE$ little string reduces to a compactified $(2,0)$ $ADE$ SCFT, again subjected to the $\mathbb{Z}_{\fn_\fg}$-twist. This SCFT background has been invoked in various contexts to engineer non simply-laced gauge groups and defects; notable examples can be found in the works \cite{Drukker:2010jp,Tachikawa:2010vg,Witten:2011zz}.}.


\vspace{8mm}

\subsection{On the notion of moduli space of vacua for a non simply-laced quiver}

We start with a basic question: are there non simply-laced gauge theories?

In the simply-laced case, both $T^{5d}_{\fg_\text{o}}$ and $T^{3d}_{\fg_\text{o}}$ arose as the low energy effective gauge theory on the $n$ D5 and $N$ D3 branes in the $(2,0)$ little string, respectively. After the $\mathbb{Z}_{\fn_\fg}$-twist, is it still sensible to define non simply-laced quiver gauge theories $T^{5d}_{\fg}$ and $T^{3d}_{\fg}$ as the effective theories on the D5 and D3 branes in the  background \eqref{littlestringnsl}?

Let us first consider the 3d quiver gauge theory $T^{3d}_{\fg_\text{o}}$ on the $N$ D3 branes, where the $r$ gauge nodes $U(N_a)$ are linked by bifundamental hypermultiplets following the shape of the Dynkin diagram of a simply-laced Lie algebra $\fg_\text{o}$. 
An immediate obstacle to considering $T^{3d}_{\fg}$ with $\fg$ non simply-laced is that there is no well-defined notion of bifundamental matter in that case, and in particular no associated Lagrangian: there are always at least two adjacent nodes $a$ and $b$ in the Dynkin diagram of $\fg$ for which the roots $\alpha_a$ and $\alpha_b$ have different lengths, and the representation theory of the supersymmetry algebra does not allow for a hypermultiplet to exist for these nodes.\\

Not all is lost, however. Most notably, it was conjectured in \cite{Hanany:2012dm,Cremonesi:2014xha} that for any simple Lie group $G$, the moduli space of $G$-instantons on  $\mathbb{C}^2$ could be studied via the Coulomb branch of an auxiliary 3-dimensional $\widehat{\fg}$-shaped quiver (Note the fundamental matter has $\cN=4$ supersymmetry in that setup instead of our $\cN=2$ Drinfeld flavors). One implication is that even though a non simply-laced quiver does not make sense as a gauge theory, the notion of a Coulomb branch as a variety could potentially still be well-defined. The conjecture was rigorously established in \cite{Braverman:2016pwk}: even though $\fg$ is non simply-laced, its Cartan matrix is symmetrizable, and a meaningful $\fg$-type Coulomb branch is defined via a folding action on the Coulomb branch of a $\fg_\text{o}$-theory, whose Dynkin graph is simply-laced. Namely, folding is realized via the $\fa$-action on the $\fg_\text{o}$-type Coulomb branch, whose fixed point gives a definition of a $\fg$-type Coulomb branch variety.
For physical considerations on folding and other subtleties in discrete gauging, see the recent work \cite{Bourget:2020bxh}. 

Physically, let $\psi_a(X_1)$ be a vector multiplet field of the simply-laced theory $T^{3d}_{\fg_\text{o}}$, where the index $a\in\{1,\ldots,r\}$ denotes the node of the $\fg_\text{o}$-type quiver, and $X_1$ is a complex coordinate on $\mathbb{C}_q$, where the D3 branes are supported. We require $\psi_a$ to obey
\beq
\label{foldingfields3d}
\psi_a(e^{2\pi i}X_1) =\fa\cdot \psi_a(X_1)\; , \qquad a\in\{1,\ldots,r\} \; ,
\eeq
where $\fa\in A$ is acting on the field $\psi_a$ on the right-hand side.

In order to define the ``vector multiplet" of a tentative non simply-laced theory $T^{3d}_{\fg}$, we consider a subset of the fields $\psi_a$ in $T^{3d}_{\fg_\text{o}}$ which split into two sets: first, there exist fields $\psi_a$ which are fixed points of the $\fa$-action. We call these fields $\psi_{a, long}(X_1)$, where $a$ now labels a long root of $\fg$. In other words, for such fields,
\begin{align}
\psi_{a, long}(e^{2\pi i}X_1) = \psi_{a, long}(X_1)\; .
\end{align} 
Second, we organize the fields of $T^{3d}_{\fg_\text{o}}$ which belong in $A$-orbits of length $\fn_\fg>1$ into a sum $\psi_a+\fa\cdot \psi_a+\ldots+\fa^{\fn_\fg-1}\cdot \psi_a$. The sum gives a well-defined single-valued field on the $\fn_\fg$-cover of $S^1_q/\mathbb{Z}_{\fn_\fg}\times \mathbb{R}$, with covering map $X_1\mapsto X_1^{\fn_\fg}$; we call such a field $\psi_{a, short}$, with $a$ labeling a short root of $\fg$. Note in particular that the ranks of the gauge groups $U(N_a)$ belonging in the same $A$-orbit must be equal in the original $T^{3d}_{\fg_\text{o}}$ theory. 

The same considerations extend to chiral fields in adjoint and fundamental representations on a given node $a$. For instance, the chiral matter of $T^{3d}_{\fg}$ originates from D3/D5 strings with the $\mathbb{Z}_{\fn_\fg}$-twist.\\

The above considerations suggest a definition of what we mean by a $T^{3d}_{\fg}$ theory: when $a$ labels a long root in $\fg$,  the gauge and matter field content on node $a$ is identical to the one in the original theory $T^{3d}_{\fg_\text{o}}$. When $a$ labels a short root in $\fg$, the gauge and matter field content on node $a$ is derived from $T^{3d}_{\fg_\text{o}}$ by restricting to $A$-orbits of length $\fn_\fg$ as above. In particular, all gauge and flavor groups which belong in the same $\mathbb{Z}_{\fn_\fg}$-orbit will have identical ranks. 
Between two long nodes or two short nodes, the bifundamental matter is well-defined with a Lagrangian origin found in $T^{3d}_{\fg_\text{o}}$. Between a long node and a short node, the ``bifundamental matter" is at best exotic and non-Lagrangian, and at worst physically ill-defined. 
At any rate, the construction makes it clear that the notion of $T^{3d}_{\fg}$ Coulomb branch is sensible as a variety, at least classically.\\

It turns out that given a general symmetrizable Cartan matrix \cite{2014arXiv1410.1403G}, there exist different possible definitions of a Coulomb branch, each with distinct advantages; for instance, a construction\footnote{Think of a simply-laced theory $T^{3d}_{\fg_\text{o}}$ supported on $S^1_{\cC'}(R_\cC')\times\mathbb{C}_q$ as an assignment of various vector bundles over the ring $R=\mathbb{C}[X_1]$, with $X_1$ a coordinate on $\mathbb{C}_q$ \cite{Braverman:2016wma}. Then, define a ``fractional quiver" $T^{3d}_{\fg}$ by assigning a different ring $R_a=\mathbb{C}[X_{1,a}]$ to each node $a=1,\ldots,r$ in the Dynkin graph of $\fg$ \cite{Kimura:2017hez,Nakajima:2019olw}. Our D3 branes in the string theory background \eqref{backgroundnsl} realize a particular instance of this construction.
} proposed by Nakajima-Weekes \cite{Nakajima:2019olw} lends itself to various geometric interpretations, such as the identification of $\fg$-type Coulomb branches as generalized slices of the affine Grassmanian. For the non simply-laced Lie algebras $\fg=B_r, C_r, G_2, F_4$  of interest to us, the different definitions of the Coulomb branch all happen to yield isomorphic varieties, though this need not be true for more general symmetrizable Cartan matrices.

In the Physics literature, $\fg$-type Coulomb branches have been defined for supersymmetric quiver gauge theories from 1 to 5 dimensions, in a similar fashion to what we presented here in 3 dimensions \cite{Cecotti:2012gh,Dey:2016qqp,Aganagic:2017smx,Haouzi:2017vec,Kimura:2017hez,Haouzi:2019jzk}.

For instance, the quiver theory $T^{5d}_{\fg}$ is similarly defined as the effective theory on the $n$ D5 branes in the little string background \eqref{littlestringnsl}. We let $\widetilde\psi_a(X_1,X_2)$ denote a field of the simply-laced theory $T^{5d}_{\fg_\text{o}}$ on node $a$ of the quiver, with $(X_1,X_2)$  coordinates on $\mathbb{C}_q\times\mathbb{C}_t$, where our D5 branes are supported. We submit the background to the $\mathbb{Z}_{\fn_\fg}$-twist as before, meaning there is a monodromy around  $X_1=0$ (but not around $X_2=0$), and we obtain
\beq
\label{foldingfields5d}
\widetilde\psi_a(e^{2\pi i}X_1,X_2) =\fa\cdot \widetilde\psi_a(X_1,X_2)\; , \qquad a\in\{1,\ldots,r\} \; ,
\eeq
as in \eqref{foldingfields3d}. The fields $\widetilde\psi_{a, long}$ and $\widetilde\psi_{a, short}$ are defined in the same way as their 3d counterparts $\psi_{a, long}$ and $\psi_{a, short}$.

In practice, the gauge content of a theory $T^{5d}_{\fg}$ can be defined by first specifying the fundamental hypermultiplet content: a vector of $r$ non-negative integers $(n^F_1,n^F_2,\ldots,n^F_r)$, where $n^F_a$ denotes the rank of the $a$-th flavor group $U(n^F_a)$ (and the integers cannot all be zero). This fixes the right-hand side of the linear system \eqref{constraint5dnsl}, which either has no solution or a unique solution $(n_1,n_2,\ldots,n_r)\in \left(\mathbb{Z}^+\right)^r$. The solution, whenever it exists, defines the gauge content of $T^{5d}_{\fg}$, meaning the ranks of the gauge groups are uniquely fixed by the ranks of the flavor groups, together with the constraint.

\subsection{Supersymmetric indices}

It is clear that a Higgs branch $X_{\fg}$ cannot be sensibly defined, at least not in the usual sense. For instance, the dimension of this tentative branch would necessarily come out a non-integer rational number, due to the problematic bifundamental hypermultiplet. Nevertheless, it follows from the work of Nakajima \cite{Nakajima:2015txa} that a K-theoretic count of quasimaps  $\mathbb{P}^1 \rightarrow X_{\fg}$  \emph{is} well-defined\footnote{Reversing the argument, the existence of those quasimaps suggests that there should exist a more general definition of a Higgs branch, which would reduce to the usual one in the simply-laced case.}. 
For us, $X_{\fg}$ would have been the moduli space of ``vortices" of the non simply-laced quiver theory $T^{5d}_{\fg}$, which is ill-defined, but the half-index ${\mathcal Z}(T^{3d}_{\fg})$ has combinatorial meaning \cite{Aganagic:2017smx,Haouzi:2017vec,Okounkov:2016sya,Haouzi:2019jzk}\footnote{Physically, we would say that Coulomb branch localization is well-defined on $T^{3d}_{\fg}$ and yields the integral form ${\mathcal Z}(T^{3d}_{\fg},{\bf N})$ of the half-index.}. 
By abuse of notation, we will sometimes refer to $T^{3d}_{\fg}$ as the theory on the vortices of $T^{5d}_{\fg}$, keeping  in mind that by ``vortex" on a ``Higgs branch," we formally mean the $\mathbb{Z}_{\fn_\fg}$-twist of the physically sound simply-laced version.

Various enumerative counts of BPS solitons in non-simply laced quiver theories have been computed in this fashion:  instanton partition functions on $S^1_{\cC'}\times \mathbb{C}_q\times \mathbb{C}_t$ from index theorems \cite{Haouzi:2017vec,Kimura:2017hez}  a generalized ADHM quantum mechanics \cite{Haouzi:2019jzk}\footnote{It was argued in that work that a bifundamental hypermultiplet between a short node and a long node should be reproduced by the contribution of $\fn_{\fg}$ $(0,2)$ usual Fermi multiplets in the ADHM quantum mechanics, as many as the number of arrows between the two nodes of the non simply-laced quiver. It would be important to derive this conjecture from first principles.}, the Hilbert series of $G$-type instantons on $\mathbb{C}^2$ \cite{Cremonesi:2014xha}, and various partition functions on $S^3$ or superconformal indices on $S^2\times S^1$, $S^3\times S^1$ and Lens space \cite{Dey:2016qqp}.

The 3d half-index is once again defined as a trace, but only counting BPS states in orbits of the automorphism group $A=\mathbb{Z}_{\fn_\fg}$, whether they are bulk or boundary degrees of freedom. Our twisted background \eqref{backgroundnsl} implies that the index will now be a Taylor series in the fugacity  $q^{1/(2\, \fn_\fg)}$ in our previous conventions \eqref{3dhalfindexmore} (with $\fn_\fg=1,2,$ or $3$). In the simply-laced case, where $\fn_\fg=1$, the index was a Taylor series in $q^{1/2}$. We decide to keep this feature universal across all  quivers, including the non simply-laced ones, by simply rescaling $q\rightarrow q^{\fn_\fg}$:
\beq
\label{3dhalfindexmorensl}
{\mathcal Z}(T^{3d}_{\fg})  = {\rm Tr}\left[(-1)^F\, q^{\fn_\fg(J+\frac{V}{2})}\; t^{\frac{A-V}{2}} \; z^{\ft_C}\; x^{\ft_H} \right]\;\; .
\eeq
The only effect of this rescaling is to give a different $q$-weight to the BPS states according to whether they are associated to long or short roots in the $\fg$-quiver.

The index is evaluated as before in terms of $q$-Pochhammer symbols and Theta-functions, with the following modifications
\beq
\left(x_a \,; q\right)_\infty\; \longrightarrow \;\left(x_a \,; q^{r_a}\right)_\infty \; , \qquad\;\;\; \Theta\left(x_a\,; q\right)\; \longrightarrow \;\Theta\left(x_a\,;q^{r_a}\right)\; .
\eeq
where
\beq\label{radefagain}
r_a = \begin{cases}
	\fn_\fg  & \text{when node $a$ labels a long root in the $\fg$-quiver\ ,} \\
	1 &  \text{when node $a$ labels a short root in the $\fg$-quiver\ .}
\end{cases}
\eeq
Note that by definition of the Theta function, the ``flips" of boundary conditions from Neumann to Dirichlet (and vice-versa) in the index can be extended without trouble to the non simply-laced case. For instance, a 3d chiral field on node $a$ with D boundary conditions is dual to a 3d chiral multiplet with N boundary conditions after introducing a boundary $\cN=(0,2)$ Fermi multiplet:
\beq
\left(q^{r_a}/x_a \,; q^{r_a}\right)_\infty = \frac{\Theta\left(x_a\,;q^{r_a}\right)}{\left(x_a \,; q^{r_a}\right)_\infty} \; .
\eeq
For the bifundamental matter contributions between two gauge nodes $a$ and $b$, we replace 
\beq
\left(x_a/x_b \,; q\right)_\infty\; \longrightarrow \;\left(x_a/x_b \,; q^{r_{ab}}\right)_\infty \; , \qquad\;\;\; \Theta\left(x_a/x_b\,;q\right)\; \longrightarrow \;\Theta\left(x_a/x_b\,;q^{r_{ab}}\right)
\eeq
where $r_{ab}=\text{gcd}(r_a, r_b)$ is the greatest common divisor of $r_a$ and $r_b$.
Explicitly, the bifundamental matter contribution to the index takes the form
\begin{align}
\label{bif3dnsl}
I_{a,b}^{bif}(y_{a,i}, y_{b,j}) =\prod_{1\leq i \leq N_a}\prod_{1\leq j \leq N_b}\left [ \frac{(\sqrt{q^{r_{ab}}\, t}\, y_{a,i}/y_{b,j};q^{r_{ab}})_{\infty}}{(\sqrt{q^{r_{ab}}/t} \, y_{a,i}/y_{b,j};q^{r_{ab}})_{\infty}}\right]^{\Delta_{a b}}
\end{align}
between two gauge nodes $a$ and $b$ in the Dynkin diagram of $T^{3d}_{\fg}$. Note in particular that if the roots $\alpha_a$ and $\alpha_b$ have different lengths, the above contribution is non-Lagrangian by construction.\\

Repeating the derivation of Section \ref{sec:3dgauge}, we find that after folding, the 3d half-index with exceptional Dirichlet boundary conditions ${\mathcal Z}(T^{3d}_{\fg,{\bf D_{EX}}})$ is well-defined, and matches the expression for the $\cW_{q,t}(\fg)$-algebra magnetic block analytic in the chamber $\fC_C$ \cite{Haouzi:2017vec,Haouzi:2019jzk}. It is likewise expected that ${\mathcal Z}(T^{3d}\fg_\text{o},\,\bf {N}_{EN})$ should be well-defined, but the geometric interpretation of elliptic stable envelopes is not well-established in the non simply-laced case, so we will not comment on it further. For analogous results in the unramified case, see \cite{Aganagic:2017smx}.\\

The various blocks and 3d half-indices can be derived from the instanton quantum mechanics of the 5-dimensional theory $T^{5d}_{\fg}$, just as in the simply-laced case. The non simply-laced instanton quantum mechanics is again defined via folding, by simply imposing the $\mathbb{Z}_{\fn_\fg}$-twist on our previous simply-laced ADHM quantum mechanics.

At the level of the Witten index, this translates to rescaling the fugacity $q\rightarrow q^{\fn_\fg}$, as we have done in the 3d index:
\beq
\label{5dhalfindexmorensl}
{\mathcal Z}_{inst}(T^{5d}_{\fg})  = {\rm Tr}\left[(-1)^F\, q^{\fn_\fg(J_1+\frac{R}{2})}\; t^{J_2-\frac{R}{2}} \; {\fm}^{\Sigma}\right]\;\; .
\eeq
The index is evaluated as a sum over $\sum_{a=1}^{r} n_a$ 2-dimensional partitions, one for each $U(1)$ equivariant Coulomb parameter: 
\beq
\{\overrightarrow{\boldsymbol{\mu}}\}=\{\boldsymbol{\mu}_{a,i}\}_{a=1, \ldots, r\, ; \;\; i=1,\ldots,n_a}\; .
\eeq
One computes
\beq\label{bulk5dnsl}
{\mathcal Z}_{inst}(T^{5d}_{\fg}) =  \sum_{\{\overrightarrow{\boldsymbol{\mu}}\}} \;\prod_{a=1}^r  z_a^{\sum_{i=1}^{n_a}{\left|\boldsymbol{\mu}_{a,i}\right|}}\, {\boldsymbol Z}^{5d,vec}_{a} \cdot {\boldsymbol Z}^{5d,fund}_{a}\cdot {\boldsymbol Z}^{5d,CS}_{a} \cdot \prod^n_{b>a}
{\boldsymbol Z}^{5d,bif}_{a,b}\, ,
\eeq
which has the same functional form as the simply-laced index, but the various factors now explicitly depend on root lengths in the $\fg$-quiver.

For example, the factor 
\begin{align}\label{5dbulkbifnsl}
{\boldsymbol Z}^{5d,bif}_{a,b} = \prod_{i=1}^{n_a}\prod_{j=1}^{n_b}\left[\cN_{\boldsymbol{\mu}_{a,i} \boldsymbol{\mu}_{b,j}}\left(f^{bif}_{a,b}\,\frac{e_{a,i}}{e_{b,j}}  \, ; q^{r_{ab}}\right)\right]^{\Delta_{a,b}}\; .
\end{align}
stands for the contribution of a 5d bifundamental hypermultiplet between nodes $a$ and $b$ in the $\fg$-quiver, with corresponding mass $f^{bif}_{a,b}$. Note that the argument of the function $\cN_{\boldsymbol{\mu}_{a,i} \boldsymbol{\mu}_{b,j}}$ now takes into account whether $\alpha_a$ and $\alpha_b$ are long or short roots:
\beq\label{nekrasovNnsl}
{\cN}_{\boldsymbol{\mu}_{a,i}\boldsymbol{\mu}_{b,j}}(Q\, ;q^{r_{ab}}) \equiv \prod\limits_{k,k' = 1}^{\infty} 
\frac{\big( Q \, q^{r_a\boldsymbol{\mu}_{a,i,k}-r_b\boldsymbol{\mu}_{b,j,k'}} \,t^{k' - k + 1}\,;q^{r_{ab}} \big)_{\infty}}{\big( Q\,  q^{r_a\boldsymbol{\mu}_{a,i,k}-r_b\boldsymbol{\mu}_{b,j,k'}}\, t^{k' - k}\, ;q^{r_{ab}}\big)_{\infty}} \,
\frac{\big( Q\,  t^{k' - k}\, ;q^{r_{ab}} \big)_{\infty}}{\big( Q\,  t^{k' - k + 1}\, ;q^{r_{ab}}\big)_{\infty}}\, .
\eeq

Our definition of the $\cW_{q,t}(\fg)$-algebras for $\fg$ non simply-laced in Section \ref{sec:confblocks} follows the conventions of    \cite{Frenkel:1998,Bouwknegt:1998da}; this fixes the bifundamental masses $f^{bif}_{a,b}$ between adjacent nodes $a$ and $b$ to
\beq\label{bifspecializednsl}
f^{bif}_{a,b}= \sqrt{q^{r_{ab}}/t} \; .
\eeq
For example, let $\fg_\text{o}=E_6$ and the outer automorphism group be $A=\mathbb{Z}_2$. The simple roots in $\mathbb{Z}_2$-orbits of length 1 become the long simple roots of $\fg=F_4$, which we label as $\alpha_1$ and $\alpha_2$. in $\fg=F_4$. The remaining $E_6$ simple roots are in $\mathbb{Z}_2$-orbits of length 2, they become the short simple roots  of $\fg=F_4$, which we label as $\alpha_3$ and $\alpha_4$. Then, in our conventions, the bifundamental masses in $T^{5d}_{F_4}$ are fixed to be  
\begin{align}\label{bifspecializedF4}
&f^{bif}_{1,2}= \sqrt{q^2/t} \; ,\nonumber\\
&f^{bif}_{2,3}= \sqrt{q/t} \; ,\nonumber\\
&f^{bif}_{3,4}= \sqrt{q/t} \; .
\end{align}

The equality of 3d and 5d partition functions is a consequence of gauge-vortex duality in the presence of the outer automorphism twist. A proof was presented for handsaw quivers in \cite{Haouzi:2017vec}, and is essentially the same for all Drinfeld quivers, so we will not repeat it here. Instead, we will point out the essential new features to make contact with the quantum algebras.

We start by freezing each 5d Coulomb modulus $e_{a,i}$ to a hypermultiplet mass $x^{-1}_{d,s}$, and moreover turn on $N_{a,i}$ units of vortex flux as
\beq\label{5drootHiggsvortexnsl}
e_{a,i} = x^{-1}_{d,s} \; t^{\frac{\#_{a,i}+1}{2}} q^{\frac{-\widetilde{\#}_{a,i}-1}{2}}\,  t^{N_{a,i}} \; .
\eeq
The factors $t^{\frac{\#_{a,i}+1}{2}}$ and  $q^{\frac{-\widetilde{\#}_{a,i}-1}{2}}$ are refinements due to the $\Omega$-background, which is further corrected by $q^{\frac{-\widetilde{\#}_{a,i}}{2}}$. In our conventions, $\#_{a,i}$ and $\widetilde{\#}_{a,i}$ are non-negative integers;  $\#_{a,i}$  records the length of the ``string" from node $a$ where the Coulomb modulus $e_{a,i}$ is located, to the node where the hypermultiplet mass $x^{-1}_{d,s}$ is located. In the simply-laced case, we had $\widetilde{\#}_{a,i}=\#_{a,i}$, but the presence of the twist as we go around $S^1_q$ implies that $q$ and $t$ now scale differently. 
From the 5d bulk perspective, the locus \eqref{5drootHiggsvortexnsl} describes the Coulomb branch of $T^{5d}_{\fg}$, but only on an integer-valued lattice in powers of $t$, due to the transverse $\Omega$-background on $\mathbb{C}_t$.  

The integers $\#_{a,i}$ and $\widetilde{\#}_{a,i}$ are uniquely determined from the requirement that the instanton partition function of $T^{5d}_{\fg}$ should truncate to the vortex partition function of $T^{3d}_{\fg}$.
Namely, for $N$ a positive integer, 
\beq\label{truncateNekrasovnsl}
\cN_{\boldsymbol{\mu}_{a,i}\boldsymbol{\mu}_{b,j}}\left(q^{r_b}\,t^{-N-1}\, ;q^{r_{ab}}\right)=0\;\; \text{unless}\;\; l\left(\boldsymbol{\mu}_{b,j}\right)\leq l\left(\boldsymbol{\mu}_{a,i}\right)+ N
\eeq
where $l\left(\boldsymbol{\mu}_{b,j}\right)$ is the number of rows in a partition $\boldsymbol{\mu}_{b,j}$, labeling the equivariant Coulomb modulus $e_{b,j}$. 
But after specializing all the Coulomb moduli as \eqref{5drootHiggsvortexnsl}, the fundamental matter factor ${\boldsymbol Z}^{5d,fund}_{b}$ in the summand of ${\mathcal Z}_{inst}(T^{5d}_{\fg})$ will necessarily contain a factor
\beq\label{fundtruncationnsl}
N_{\boldsymbol{\emptyset}\,\boldsymbol{\mu}_{b,j}}\left(\sqrt{q^{2\, r_b}/t^2}\;t^{-N_{b,j}} \,;q^{r_b}\right)\; ,
\eeq
for some positive integer $N_{b,j}$. It follows that the entire summand vanishes unless the partition $\boldsymbol{\mu}_{b,j}$ has no more than $N_{b,j}$ rows. A new subtlety unique to the non simply-laced case is that the definition of the $\cN_{\boldsymbol{\mu}_{a,i}\boldsymbol{\mu}_{b,j}}$ function depends on whether $a$ and $b$ are indices for long or short roots of $\fg$. Making use of the trivial identity 
\beq\label{Nekrasovpropoerties}
\cN_{\boldsymbol{\mu}_{a,i}\boldsymbol{\mu}_{b,j}}(Q\, ;q) =\prod_{k=0}^{r_{ab}-1}N_{\boldsymbol{\mu}_{a,i}\boldsymbol{\mu}_{b,j}}(q^k\, Q\, ;q^{r_{ab}})\; ,
\eeq
one proceeds to show that \eqref{truncateNekrasovnsl} is responsible for the truncation of \emph{all} partitions $\overrightarrow{\boldsymbol{\mu}}$, and find:

\begin{align}\label{partequalitynsl}
{\mathcal Z}(T^{3d}_{\fg,\bf D_{EX}}) = c_{3d}\cdot  {\mathcal Z}_{inst}(T^{5d}_{\fg})_{e_{a,i}\propto\; x^{-1}_{d,s}\,t^{N_{a,i}}}\, .
\end{align}
The normalization constant $c_{3d}$ stands for the empty partition contribution to the partition function. In particular, the Higgsing procedure implies the following conjecture:

\begin{prop}\label{prop5d3dnsl}
	Let $V_{a,\lambda_{s}}$ be a weight space appearing in the eigenspace decomposition of a $U_\hbar(\widehat{^L \fg})$ fundamental representation $\widehat{V_a}=\bigoplus_{\lambda_{s}}\, V_{a,\lambda_{s}}$ as $U_\hbar(^L \fg)$-modules. Let $\{{\mathcal K}^{\geq}_{a,s}\}$ be the set of all sequences of ordered elements with a 5d mass $f_{a,s}$ as a maximal element. Then the map $\{{\mathcal K}^{\geq}_{a,s}\} \rightarrow V_{a,\lambda_{s}}$ is surjective.
\end{prop}
We checked that this Proposition holds for all weight spaces in all fundamental representations of $U_{\hbar}(\widehat{^L\fg})$, for  $\fg=B_{r\leq 3}$, $\;\fg=C_{r\leq 3}$, \  and $\fg=G_2$, meaning we identified the Drinfeld polynomials of all weights in those cases. For $\fg=F_4$, we only performed a partial check for some of the weights. Once again, we found that the weights are typically realized in a different fashion compared to the Frenkel-Mukhin algorithm.

\vspace{8mm}

\section{Surface defects in $\cN=4$ Super Yang-Mills and S-duality}
\label{sec:ym}

Kapustin and Witten showed that the geometric Langlands correspondence can be formulated at the level of supersymmetric gauge theory \cite{Kapustin:2006pk}. Namely, consider maximally supersymmetric $\cN=4$ Yang-Mills theory with a (simple and connected) gauge group $G$ on a four-manifold $M_4$ that locally is a product of Riemann surfaces,  $M_4={\cal C}\times M_2$. Here, $\cC$ is the Riemann surface on which the geometric Langlands program ``lives," it is a fiber to the normal bundle to $M_2$. 
Among the various fields of the theory, one finds a connection $A$ on a $G$-bundle  $E\rightarrow\cC$, and a field $\Phi$ that is a one-form on $\cC$ with values in $\text{ad}(E)$.

More precisely, $\Phi$ arises from a Landau-Ginzburg-type twist on some of the scalar fields of $\cN=4$ super Yang-Mills. One finds that a necessary condition for $A$ and $\Phi$ to preserve supersymmetry is that they should solve Hitchin's equations \cite{Hitchin:1986vp}:
\begin{align}
&F-\phi\wedge\phi =0 \nonumber\label{Hitcheq}\\
&D\Phi=0\nonumber\\
&D\star\Phi=0
\end{align}
Here, we denoted the gauge-covariant exterior derivative by $D=d+[A,\cdot]$, and the Hodge star operator by $\star$. One can think of these equations as a reduction of the self-dual Yang-Mills equations from four to two dimensions. In this picture, $\Phi$ corresponds to the two components of the gauge field lost in the dimensional reduction from four to two dimensions.\\

\vspace{8mm}

\subsection{Tamely ramified Hitchin system: a short review}

For a mathematical treatment, we refer the reader to earlier works by Simpson \cite{MR1159261,Simpson1992}, Konno \cite{10.2969/jmsj/04520253} and Nakajima \cite{MR1397988}; see also the review by Donagi and Pantev \cite{Donagi2008GeometricLA}. For conformal field theory considerations, see Frenkel \cite{Frenkel:2005pa,Frenkel:2006nm}.\\

We are interested in a class of solutions to the above system with a prescribed singularity behavior at punctures on $\cC$, as studied by Gukov and Witten \cite{Gukov:2006jk}. Explicitly, we choose local Euclidean coordinates $(x^2,x^3)\in\mathbb{R}^2$ near a point of ramification on $\cC$, say $x^2=x^3=0$, and further change to polar coordinates $x^2+i\,x^3 = r\, e^{i\theta}$. A surface defect is defined at $x^2=x^3=0$ by specifying the singular profile of the fields $A= A_2+i\, A_3$ and  $\Phi=\Phi_2+i\, \Phi_3$ near there. A natural ansatz is to require $\Phi$ to have a pole of some prescribed order at $r=0$. One speaks of \emph{tame} ramification in the geometric Langlands program when $\Phi$ has at most a \emph{simple} pole (of order 1) at each ramification point on $\cC$. All the defects of our paper are tamely ramified in that sense.
In contrast, when $\Phi$ is defined to have a pole \emph{not} of order 1 at some ramification point, one speaks instead of \emph{wild} ramification \cite{Witten:2007td}. 

Writing down an explicit ansatz is greatly simplified if one further demands the defects to be superconformal and preserve as much supersymmetry as possible. A conformal ansatz will be invariant under scaling and rotations in the  $\mathbb{R}^2$-plane, and will take the form
\begin{align}
&A= a(r)\, d\theta + b(r)\,\frac{dr}{r} +\ldots\nonumber\label{Hitchsol}\\
&\Phi=  c(r)\,  \frac{dr}{r} + d(r)\,d\theta +\ldots \; , 
\end{align}
where the dots ``$\ldots$" stand for regular terms that will not concern us.
One can always apply a gauge transformation to $A$ which sets $d(r)=0$. Moreover, the invariance under scaling further constraints $a(r), b(r),$ and $c(r)$ to be constants. We write these constants as $a(0)=\alpha$, $b(0)=\beta$, and $c(0)=\gamma$. Hitchin's equations imply that  $\alpha, \beta, \gamma$ all commute among themselves, so these constants are naturally valued in the Cartan subalgebra $\fh$ of $\fg$. Near $r=0$, one can write
\begin{align}
&A= \alpha\, d\theta +\ldots\nonumber\label{Hitchsol2}\\
&\Phi=  \beta\,  \frac{dr}{r} - \gamma \,d\theta +\ldots \; . 
\end{align}
The surface defect breaks the gauge group $G$ to a maximal torus $T$, or rather a subgroup $L$ which contains $T$, called a Levi subgroup. Which Levi subgroup we get depends on the type of defect we insert; for instance, the Levi subgroup $L$ coincides with $T$ whenever the defect is ``generic"\footnote{The notion of generic here coincides with the conformal limit of a generic defect from Section \ref{sssec:classification}. This is not a coincidence, but a consequence of the (local) Alday-Gaiotto-Tachikawa correspondence \cite{Kanno:2009ga}.}. The global R-symmetry group likewise breaks as $SO(6)\rightarrow SO(4)\times SO(2)$.\\

The moduli space of solutions to Hitchin's equations \eqref{Hitcheq} is called the Hitchin moduli space. In the absence of singularities, it is a finite-dimensional hyper-K\"{a}hler manifold. In particular, there exists a $\mathbb{C P}^1$ worth  of complex structures on it. In some complex structure we call $I$, a solution to Hitchin's equations on $\cC$ can be viewed as a pair $(E, \varphi)$, with $E$ a holomorphic $G$-bundle and $\varphi$ a holomorphic section of the bundle $K\otimes \text{ad}(E)$, with K the canonical bundle on $\cC$. This data $(E, \varphi)$ goes by the name of Higgs bundle. 

When we further impose that $A$ and $\phi$ have  the singular behavior \eqref{Hitchsol2} at points $\{p_1, \ldots, p_L\}$, the moduli space of solutions is still hyper-K\"{a}hler, and in complex structure $I$ it is known as the moduli space of ramified Higgs bundle.

The holomorphic structure can be justified as follows: First consider the complex-valued connection ${\bold A} = A+i\Phi$, which is flat by Hitchin's equations \eqref{Hitcheq}. The gauge-covariant exterior derivative can be written as $d_{\bold A} = \partial_{\bold A} + \overline{\partial}_{\bold A}$, with $\partial_{\bold A}$ a derivative of (1,0) type, and $\overline{\partial}_{\bold A}$ a derivative of (0,1) type. 
It is this operator $\overline{\partial}_{\bold A}$ which makes the bundle $E$ holomorphic, at least away from the points $\{p_1, \ldots, p_L\}$. For instance, consider a neighborhood of a point $p$ located, say, at $z=0$, where $z=x^1+i\, x^2$; then the operator  $\overline{\partial}_{\bold A}$ simplifies to $d\overline{z}\,\frac{\partial}{\partial \overline{z}}$. We further decompose the Higgs field as $\Phi=\varphi^{(1,0)} + \varphi^{(0,1)}$, with $\varphi\equiv\varphi^{(1,0)}$ of (1,0) type, and $\varphi^{(0,1)}$ of (0,1) type. It follows that $(E, \varphi)$ is a Higgs bundle. In complex structure $I$, we can rewrite the solution \eqref{Hitchsol2} near $z=0$ as 
\begin{align}
&A= \alpha\, d\theta +\ldots\nonumber\label{Hitchsol3}\\
&\varphi^{(1,0)}=  (\beta+i\,\gamma)\,  \frac{dz}{z}  +\ldots \; . 
\end{align}
The complex structure $I$ depends holomorphically on $\beta+i\,\gamma$, but does not depend on $\alpha$. At the same time, in cohomology, the (1,1) K\"{a}hler form $\omega_I$ depends linearly on $\alpha$, but does not depend on $\beta$ or $\gamma$. 

K\"{a}hler moduli are complexified due to supersymmetry, so we expect additional real parameters $\eta$ to be present besides the real K\"{a}hler moduli in $\alpha$. The existence of $\eta$ can be qualitatively motivated by noting that the 2-form field strength $F$ serves a dual purpose in four dimensions: on the one-hand,  $F$ must take the form
\beq
\label{thooftloop}
F=2\pi\alpha\,\delta_{M_2}+\ldots \; ,
\eeq
with $\delta_{M_2}$ is a 2-form delta-function Poincar\'{e} dual to the support of the defect $M_2\subset M_4$.  On the other-hand, we can consider integrating the 2-form $F$ with support on the manifold $M_2$ itself, resulting in a coupling
\beq\label{theta2d}
\text{exp}\left(i\,\eta\int_{M_2} F\right) \; .
\eeq
An analogous phenomenon occurs in the study of line defects: \eqref{thooftloop} is the surface defect analog of a magnetic 't Hooft loop in $M_4$, while \eqref{theta2d} is the analog of an electric Wilson loop.
In the case where the gauge group is $G=U(1)$, we recognize the integrand in \eqref{theta2d} as the first Chern class of the $U(1)$ bundle over $M_2$, whose degree is measured by the integral. Therefore, $\eta$ should be thought of as a two-dimensional theta angle, and the exponential should be included as an additional factor inside the path integral. If $G$ is nonabelian of rank $r$, the parameter $\eta$ instead stands for a vector of $r$ two-dimensional theta-angles, one for each abelian factor of the maximal torus $T\cong U(1)^r$.\\ 

The parameters $\alpha$, $\beta$ and $\gamma$ take values in $\fh$, but a gauge transformation will shift $\alpha$ by an elements of the coroot lattice, meaning $\alpha$ actually takes values in $\fh/\Lambda_{cort} \cong T$, the maximal torus of $G$. The parameter $\eta$ takes values in $\fh^\vee$, the Cartan subalgebra of the Langlands dual algebra $^L \fg$; a gauge transformation will shift $\eta$  by an element of the weight lattice, so it is valued in $\fh/\Lambda_{wt} \cong\, ^L T$, the maximal torus of $^L G$. All parameters are furthermore invariant under Weyl group action, so in the end,
\beq
(\alpha,\beta,\gamma,\eta)\in (T\times \fh \times \fh \times ^L T)/W \; .
\eeq

\vspace{8mm}

\subsubsection{The S-transformation}

S-duality is a symmetry of $\cN=4$ Yang-Mills which exchanges the gauge group $G$ with its Langlands dual $^L G$. Let $\tau^{4d}= \frac{\theta}{2\pi}+i\frac{4\pi}{g^2_{4d}}$ denote the gauge coupling of $G$-type Yang-Mills, where $g_{4d}$ is the coupling constant and  $\theta$ the theta angle. Likewise, let $^L \tau^{4d}$ be the gauge coupling of the  S-dual  $^L G$-type Yang-Mills theory. At the level of their Lie algebras, S-duality acts as
\beq 
\label{Sdualitytau}
\text{S}: \;\;\;(\fg,\tau^{4d}) \longleftrightarrow (^L \fg,^L\tau^{4d}) \; ,\qquad   ^L\tau = \frac{-1}{\fn_{\fg}\,\tau^{4d}} \; .
\eeq
Turning our attention to the surface defect, let $(\alpha,\beta,\gamma,\eta)$ be the set of defect parameters of the $G$-type theory, and $(^L\alpha,^L\beta,^L\gamma,^L\eta)$ be the set of defect parameters of the $^L G$-type theory. The Higgs field residues $(\beta,\gamma)$ transform almost trivially under S-duality: they are simply rescaled as
\beq\label{Sdualitybetagamma}
(^L\beta,^L\gamma) = \text{Im}(\tau)\,(\beta^*,\gamma^*) \; .
\eeq
The notation $(\beta^*,\gamma^*)$ on the right-hand side is necessary since the Higgs fields $\phi$ and $^L \phi$ are not valued in the same space, so they cannot be directly compared. We therefore introduce a linear map\footnote{Our map is rescaled by a factor of $\fn_\fg$ compared to the map introduced in \cite{Gukov:2006jk}, because there the short roots have length squared $2$, whereas for us they have length squared $2/\fn_\fg$. It is perfectly fine to use the same map as \cite{Gukov:2006jk}, at the expense of multiplying the right-hand side of \eqref{Sdualitybetagamma} by an extra factor of $\fn_\fg$, and understanding $\alpha \mapsto \alpha^*$ as a map from the lattice $\Lambda_{rt}$ to the lattice $\frac{1}{\fn_\fg}\cdot\Lambda_{cort}$ instead.} from the dual Cartan subalgebra $\fh^\vee$, where the roots of $\fg$ are valued, to the Cartan subalgebra $\fh$, where the coroots of $\fg$ are valued. The map acts on simple roots as $\alpha \mapsto \alpha^*$, defined such that if $\alpha_l$ is a long root, then  $\alpha_l^*$ is $\fn_\fg$ times the short coroot $\alpha^\vee_s$, and if $\alpha_s$ is a short root, then  $\alpha_s^*$ is the long coroot $\alpha^\vee_l$. This gives a well-defined map from $\Lambda_{rt}$ to $\Lambda_{cort}$. The above map naturally extends to a map of Higgs fields $\phi \mapsto \phi^*$ (and its residues at the pole $(\beta,\gamma)\mapsto(\beta^*,\gamma^*)$), where the $a$-th component $\phi_a$ is mapped to $\phi^*_a$ as above, according to whether $a$ labels a long or a short root of $\fg$. 

Note that when $\fg$ is simply-laced, roots and coroots coincide, and the map $\alpha \mapsto \alpha^*$ is therefore the identity map, which simplifies the discussion.\\

Meanwhile, the definition of the parameters $\alpha$ and $\eta$ reviewed in last Section suggests that S-duality should act on them in a more nontrivial way, as a proper electro-magnetic duality. In the abelian case $G=U(1)$, Gukov and Witten prove \cite{Gukov:2006jk} that this is indeed the case, and conjecture it should still hold when $G$ is nonabelian. Explicitly, the conjecture reads
\beq\label{Sdualityalphaeta}
(^L\alpha,^L\eta) = (\eta,-\alpha)
\eeq

\begin{remark}
The works \cite{Dorey:1996hx,Argyres:2006qr} introduce explicitly a linear transformation $R$ which maps the roots to the coroots of $\fg$. For instance, in the case $\fg=G_2$, one can define $R$ as acting on the root lattice by an angle $\pi/2$ and multiplication by a factor of $\sqrt{3}$ \cite{Argyres:2006qr}; this results in a ``flip" in the Dynkin diagram of $G_2$. The transformation $R$  acts nontrivially on the moduli space of the theory, on the lattice of electric and magnetic charges, and therefore presumably also on the defects parameters $(\alpha,\eta)$. In the case at hand, since our conventions for the root lengths match those of \cite{Argyres:2006qr}, we would then conjecture the S-duality action $\left(^L\alpha,^L\eta\right) = \left(R\,\eta/\fn_\fg,-R\,\alpha\right)$. However, we choose instead to adopt the notational convenience of Appendix A in \cite{Gukov:2006jk} throughout our work, and write our expressions without explicitly referring to $R$.\\
\end{remark}

\subsubsection{The T-transformation}

Besides the non-perturbative S-duality we just reviewed, we also have a classical T-symmetry from $G$ to itself, which acts on the gauge coupling $\tau^{4d}= \frac{\theta}{2\pi} + i\, \frac{4\pi}{g^2_{4d}}$ as
\beq
\label{Tdualitytau}
\text{T}: \;\;\;(\fg,\tau^{4d}) \longrightarrow (\fg,\tau^{4d}-1) \; .
\eeq 
This is a symmetry of $G$-Yang-Mills under a shift of the theta angle $\theta$  by $2\pi$. 
The surface defect parameters  $(\beta,\gamma)$ are invariant under the T-symmetry, while the parameters $(\alpha,\eta)$ transform as
\beq\label{Tdualityalphaeta}
\text{T}: (\alpha,\eta) \longrightarrow (\alpha,\eta-\alpha)
\eeq
A gauge theory proof of this result is essentially the one found in the ``Witten effect" \cite{Witten:1979ey,Gukov:2006jk}. Namely, $(\alpha,\eta)$  behaves just like the charge of a dyon, which transforms precisely as above under the action of T.\\

\subsubsection{The full duality group}

For completeness, we briefly comment here on the symmetry group when both S and T are present. The S and T symmetries generate a discrete infinite subgroup of SL$(2,\mathbb{R})$. If $\fg$ is simply-laced, that group is well-known to be the modular group SL$(2,\mathbb{Z})$.

If $\fg$ is non simply-laced, then S : $\tau\mapsto -1/(\fn_\fg\, \tau)$, where $\fn_\fg=2$ or $\fn_\fg=3$, is no longer a modular transformation. Instead, the transformations T, STS, and the central element S$^2$ (which belongs in the Weyl group) generate a subgroup $\Upsilon_0(\fn_\fg)\subset\;$SL$(2,\mathbb{Z})$. In that case, the full duality group is the $\mathbb{Z}_2$ quotient of the semi-direct product $\left(\Upsilon_0(\fn_\fg)\rtimes\mathbb{Z}_4\right)/\mathbb{Z}_2$, generated by S and T, with relations  \cite{Dorey:1996hx}. Here, the transformation S generates the $\mathbb{Z}_4$ factor, and the reason for the quotient is that the transformation S$^2$ generates a $\mathbb{Z}_2$ which is already in $\Upsilon_0(\fn_\fg)$.

In the cases $\fg=B_r$ and $\fg=C_r$, the full duality group is an equivalence between different theories, since the transformation S maps the algebra $B_r$ to the algebra $C_r$, and vice-versa. It follows that the self-duality group of each of these theories is only $\Upsilon_0(2)$.

In the cases $\fg=G_2$ and $\fg=F_4$, the full duality group we described above is a $\mathbb{Z}_2$ quotient of what is known as a Hecke group \cite{Beardon1995TheGO}. For details on how these Hecke groups act on the moduli space and on the electric and magnetic charges of these theories, a good reference is \cite{Argyres:2006qr}.

\vspace{8mm}

\subsection{4d S-duality from T-duality of little strings}
\label{ssec:sdual}

We will now derive the S-duality of 4d Yang-Mills and its monodromy defects straight from the T-duality of the little string. The argument in the absence of D5 branes defects was first given by Vafa \cite{Vafa:1997mh} (see also the last Section of \cite{Argyres:2006qr}). We will first review his argument and then proceed to add the D5 branes.

\subsubsection{Warm-up: The Unramified Case}

Consider the twisted background from Section \ref{sec:nsl}:
\begin{equation}
\label{IIBVafa}
\frac{\mathbb{C}^2_{\Gamma}\times S^1_q(R_q)}{\mathbb{Z}_{\fn_\fg}}\times S^1_t(R_t)\times M_4 \; ,
\end{equation}
where the outer automorphism twist by $\mathbb{Z}_{\fn_\fg}$ acts by folding the Dynkin diagram of $\fg_\text{o}$ after going once around $S^1_q$. If  $\fg_\text{o}$ doesn't admit a nontrivial outer automorphism twist, meaning whenever $\fg_\text{o}=A_{2p}, E_7, E_8$, then $\mathbb{C}^2_{\Gamma}\times S^1_q(R_q)$ will be understood without the twist by $\mathbb{Z}_{\fn_\fg}$. We further take the $g_s\rightarrow 0$ limit to obtain the $\fg_\text{o}$-type $(2,0)$ little string theory on $S^1_q(R_q)\times S^1_t(R_t)\times M_4$, in the presence of a $\mathbb{Z}_{\fn_\fg}$-twist along $S^1_q$. We now T-dualize the little string, in two different ways.\\

\begin{figure}[h!]
	\emph{}
	\hspace{-7ex}
	\centering
	\vspace{-10pt}
	\includegraphics[width=1.0\textwidth]{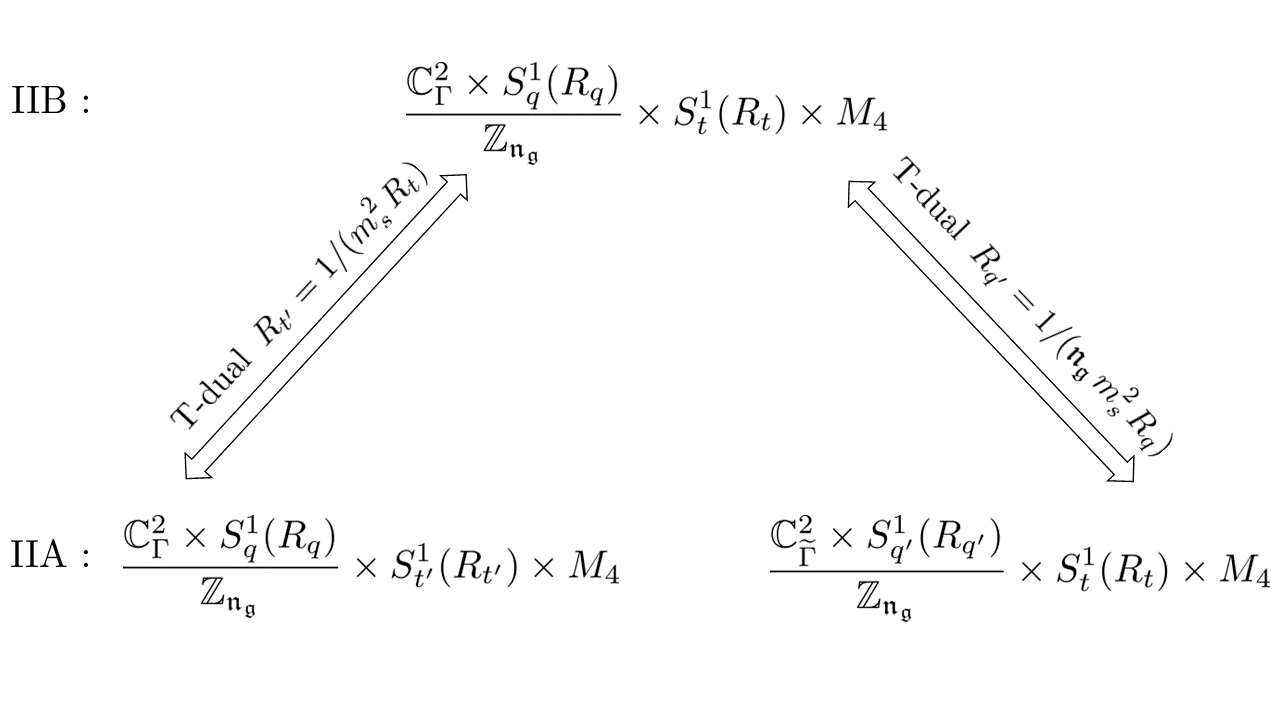}
	\vspace{-0pt}
	\caption{The two T-dualities from which one derives the S-duality of 4d Super Yang-Mills. Adding D5 branes, one further derives the S-duality of the Gukov-Witten monodromy defects.} 
	\label{fig:Vafabranes}
\end{figure}

First, we T-dualize the $(2,0)$ little string theory along $S^1_t$, and obtain the (1,1) little string theory on  $S^1_q(R_q)\times S^1_{t'}(R'_t)\times M_4$, with $\mathbb{Z}_{\fn_\fg}$-twist along $S^1_q$, and where the circle $S^1_{t'}$ has radius $R'_t=1/(m_s^2\, R_t)$. In the full 10-dimensional superstring background, this is a  T-duality to type IIA string theory on
\begin{equation}
\label{IIA1Vafa}
\frac{\mathbb{C}^2_{\Gamma}\times S^1_q(R_q)}{\mathbb{Z}_{\fn_\fg}}\times S^1_{t'}(R'_t)\times M_4 \; ,
\end{equation}
Unlike its $(2,0)$ counterpart, the (1,1) little string theory has a gauge symmetry. The gauge group has an $ADE$ Lie algebra $\fg_\text{o}$, and inverse gauge coupling\footnote{More precisely, the gauge theory description of the (1,1) little string is valid at weak coupling. That description breaks down when the energy scale becomes of order $m_s$, at which point new stringy degrees of freedom need to be added. For a relatively recent quantitative account of these matters, we refer the reader to the work \cite{Chang:2014jta}.}
\beq
\frac{1}{(g^{6d})^2} = m^2_s \; .
\eeq
The $\fh$-valued gauge field has $r$ Cartan components $A_a=\int_{S^2_a} C^{(3)}$, which is the Kaluza-Klein reduction of the type IIA R-R 3-form over the $r$ compact 2-cycles of the resolved $ADE$ singularity.

In order to obtain a 4-dimensional gauge theory, we take the point particle limit $m_s\rightarrow \infty$, and we further reduce the six-dimensional theory on the torus $S^1_q\times S^1_{t'}$ by taking $R_q\rightarrow 0$ and $R'_t\rightarrow 0$. Since $R_t = 1/(m_s^2\, R'_t)$, the radius of the initial $(2,0)$ circle $S^1_t$ stays finite in the limit. Then, the torus volume goes to zero, $i\, R_q\,R'_t\rightarrow 0$. The resulting theory is 4d $\cN=4$ Super Yang-Mills on $M_4$. The gauge group has a Lie algebra of type $\fg$, and the (imaginary part of the) inverse gauge coupling is given by the finite combination   
\beq
\label{4dcoupling1}
\tau^{4d}= m_s^2\, (i\, R_q\, R'_t) = i\,\frac{R_q}{R_t} \; .
\eeq
An alternate way to derive the above gauge coupling is to note that an instanton in 4 dimensions is a fundamental string with Euclidean worldsheet wrapping the torus, with action $2\pi m_s^2\,  R_q\,R'_t$. But the action of a gauge theory instanton is $-2\pi i \tau^{4d}$, from which \eqref{4dcoupling1} follows.\\

Alternatively, we can T-dualize the $(2,0)$ little string theory along $S^1_q$, and obtain the (1,1) little string theory on  $S^1_{q'}(R'_q)\times S^1_{t}(R_t)\times M_4$, with $\mathbb{Z}_{\fn_\fg}$-twist along $S^1_{q'}$, and where this circle $S^1_{q'}$ has radius $R'_q=1/(\fn_\fg\, m_s^2\, R_q)$. In the full 10-dimensional type IIB background \eqref{IIBVafa}, this is a T-duality along  $S^1_{q}$  to type IIA string theory on
\begin{equation}
\label{IIA2Vafa}
\frac{\mathbb{C}^2_{\widetilde{\Gamma}}\times S^1_{q'}(R'_q)}{\mathbb{Z}_{\fn_{\fg}}}\times S^1_{t}(R_t)\times M_4 \; .
\end{equation}
Above, $\mathbb{C}^2_{\widetilde{\Gamma}}$ is a resolved singularity labeled by a $ADE$ Lie algebra $\widetilde{\fg_\text{o}}$ defined as follows: if $\fg_\text{o}=A_{2p-1}$, then $\widetilde{\fg_\text{o}}=D_{p+1}$, and vice versa (if $\fg_\text{o}=D_{p+1}$, then $\widetilde{\fg_\text{o}}=A_{2p-1}$), while $\widetilde{\fg_\text{o}}=\fg_\text{o}$ in all other cases. In particular, $\widetilde{\fg_\text{o}}=\fg_\text{o} = E_6$ admits a twist of order 2, while $\widetilde{\fg_\text{o}}=\fg_\text{o} = D_4$ admits a twist of order 3.  As before, if $\fg_\text{o}=\widetilde{\fg_\text{o}}=A_{2p}, E_7, E_8$,  then $\mathbb{C}^2_{\widetilde{\Gamma}}\times S^1_{q'}(R'_q)$ is  understood without the twist by $\mathbb{Z}_{\fn_\fg}$.
The algebra $\widetilde{\fg_\text{o}}$ has an outer automorphism group $\mathbb{Z}_{\fn_{\fg}}$, and the subalgebra invariant under the $\mathbb{Z}_{\fn_{\fg}}$-action is the Lie algebra $^L \fg$. In order to correctly identify the radius $R'_q$ of $S^1_{q'}$, note that in the type IIB setup \eqref{IIBVafa}, the momentum along the original circle $S^1_{q}$ is quantized to be an integer multiple of $1/(\fn_\fg\, R_q)$, due of the $\fn_{\fg}$-twist. Indeed, the fields are single-valued only on the $\fn_\fg$-fold cover of $S^1_{q}$, meaning one has to go around $S^1_{q}$ a total of $\fn_{\fg}$ times. But then, by T-duality, a string which winds around $S^1_{q'}$ will have an energy quantized in the same units, $m_s^2\, R'_{q} = 1/(\fn_\fg\, R_q)$. We conclude that the radius of the T-dual circle is as claimed, $R'_q =  1/(\fn_\fg\, m_s^2\, R_q)$. 

In the limit $g_s\rightarrow 0$, this describes the (1,1) little string theory on $S^1_{q'}(R'_q)\times S^1_{t}(R_t)\times M_4$, with $\mathbb{Z}_{\fn_{\fg}}$-twist along $S^1_{q'}$. The low energy effective theory is a gauge theory, with $ADE$ Lie algebra $\widetilde{\fg_\text{o}}$, and inverse gauge coupling
\beq
\frac{1}{(\widetilde{g^{6d}})^2} = m^2_s \; .
\eeq
We once again take the point particle limit $m_s\rightarrow \infty$, and further reduce on the torus $S^1_{q'}\times S^1_{t}$ by taking $R'_q\rightarrow 0$, $R_t\rightarrow 0$\footnote{This time around, it is $R_q = 1/(\fn_\fg \, m_s^2\, R'_q)$, the radius of the initial $(2,0)$ circle $S^1_q$, which stays finite in the limit.}. The torus volume $i\, R_{q'}\,R_t\rightarrow 0$ then goes to zero. The resulting theory is 4d $\cN=4$ Super Yang-Mills on $M_4$. The gauge group has a Lie algebra of type $^L\fg$, and the (imaginary part of the) inverse gauge coupling is given by the finite combination
\beq
\label{4dcoupling2}
^L\tau^{4d}= m_s^2\, (i\, R_{q'}\, R_t) = i\,\frac{R_t}{\fn_{\fg}\,R_q}
\eeq
Comparing \eqref{4dcoupling1} and \eqref{4dcoupling2}, we recover the transformation of the 4-dimensional gauge coupling under S-duality \eqref{Sdualitytau},
\beq
\label{Sdualitytau2}
\text{S}: \;\;\;(^L\fg,^L\tau) \Longleftrightarrow (\fg,\tau) \; ,\qquad   ^L\tau = \frac{-1}{\fn_{\fg}\,\tau} \; .
\eeq

\vspace{6mm}

\subsubsection{The Ramified Case}

We consider again the twisted type IIB background \eqref{IIBVafa}, with $M_4=\cC\times M_2$, and add a collection of D5 branes wrapping
\begin{equation}
\label{IIBD5branes}
\frac{S^2_{\{a\}}\times S^1_q(R_q)}{\mathbb{Z}_{\fn_\fg}}\times S^1_t(R_t)\times M_2 \; .
\end{equation}
As before, we take the $(2,0)$ little string limit $g_s\rightarrow 0$, and T-dualize in two different ways.

\vspace{4mm}

\subsubsection{D-branes as monopoles, and the Hitchin integrable system}

First, a T-duality along $S^1_t$ results in a configuration of D4 branes in the (1,1) little string theory, wrapping
\begin{equation}
\label{IIA1D4branes}
\frac{S^2_{\{a\}}\times S^1_q(R_q)}{\mathbb{Z}_{\fn_\fg}}\times M_2 \; .
\end{equation}
In particular, the D4 branes are points on $\cC\times S^1_{t'}(R'_t) = \mathbb{R}\times S^1_{\cC}(R_{\cC})\times S^1_{t'}(R'_t)$, where $R'_t=1/(m_s^2\, R_t)$. They are magnetically charged under the gauge field of the (1,1) $\fg_\text{o}$-little string, whose $r$ Cartan components are $A_a=\int_{S^2_a} C^{(3)}$. As such, they are $\fg_\text{o}$-monopoles, subjected to the $\mathbb{Z}_{\fn_\fg}$-twist along $S^1_q$. It follows that the gauge field strength $F$ obeys the Bogomolny equations on $\mathbb{R}\times S^1_{\cC}\times S^1_{t'}$,
\beq
\label{bogomolny1}
D\psi = *F \; ,
\eeq
where $\psi$ is a real scalar field which approaches a constant value at the end of the cylinder, far away from the D4 brane insertions on $\cC$. In the (1,1) little string, this constant scalar has Cartan components $\psi_a = \lim_{g_s'\rightarrow 0} m_s^3\int_{S^2_a}\frac{\omega_I}{g'_s}$, for $a=1,\ldots,r$ and $g_s'$ the type IIA coupling.\\
\begin{figure}[h!]
	\emph{}
	\hspace{-7ex}
	\centering
	\vspace{-10pt} 
	\includegraphics[width=1.0\textwidth]{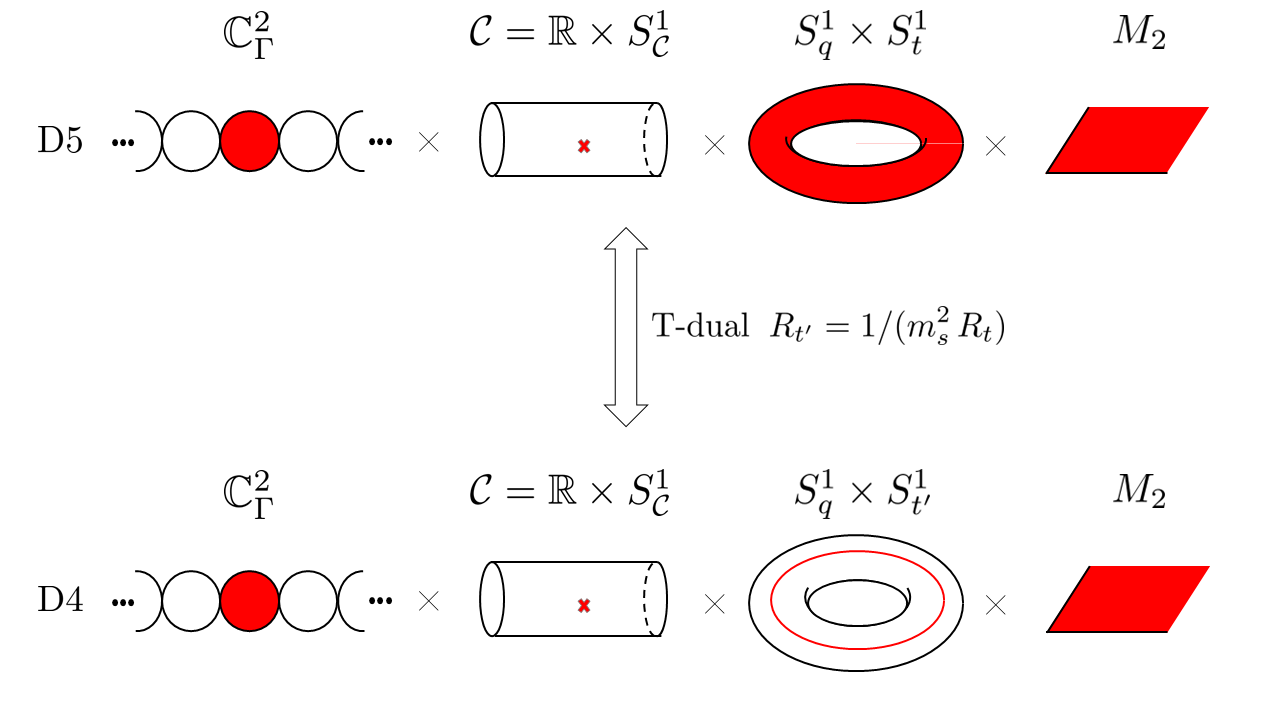}
	\vspace{-0pt}
	\caption{Details of the first T-duality in the presence of D5 branes} 
	\label{fig:Vafabranes1}
\end{figure}

If we instead T-dualize the D5 branes \eqref{IIBD5branes}  along $S^1_q$, we obtain D4 branes in the (1,1) little string, wrapping
\begin{equation}
\label{IIA2D4branes}
\frac{\widetilde{S^2_{\{a\}}}}{\mathbb{Z}_{\fn_\fg}} \times S^1_t(R_t)\times M_2 \; .
\end{equation}
Above, $\widetilde{S^2_{\{a\}}}$ are the compact 2-cycles which span the homology group $H_2({\mathbb C}^2_{\widetilde{\Gamma}}, \mathbb{Z})$, where the resolved singularity ${\mathbb C}^2_{\widetilde{\Gamma}}$ is the one from \eqref{IIA2Vafa}. This time, the D4 branes are points on $\cC\times S^1_{q'}(R'_q) = \mathbb{R}\times S^1_{\cC}(R)\times S^1_{q'}(R'_q)$, where $R'_q =  1/(\fn_\fg\, m_s^2\, R_q)$. These are $\widetilde{\fg_\text{o}}$-monopoles, subjected further to a $\mathbb{Z}_{\fn_\fg}$-twist along $S^1_{q'}$. They are magnetically charged under the gauge field of the (1,1) $\widetilde{\fg_\text{o}}$-little string, and the associated  field strength $^L F$ will obey the Bogomolny equations
\beq
\label{bogomolny2}
D ^L\psi = * ^L F \; ,
\eeq
where $^L\psi$ is again a real scalar field originating from the (1,1) little string.\\

\begin{figure}[h!]
	\emph{}
	\hspace{-7ex}
	\centering
	\vspace{-10pt} 
	\includegraphics[width=1.0\textwidth]{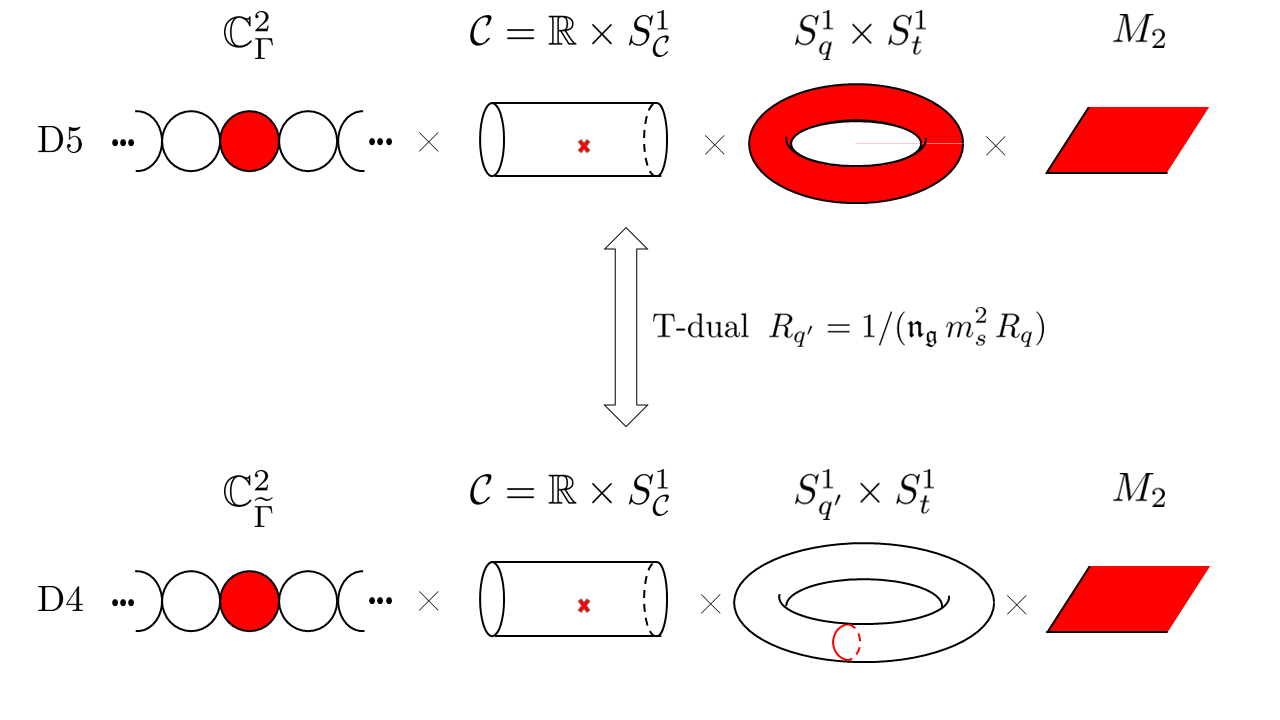}
	\vspace{-0pt}
	\caption{Details of the second T-duality in the presence of D5 branes} 
	\label{fig:Vafabranes2}
\end{figure}

When we say that the D4 branes are monopoles, we should be more specific, since the branes wrap different types of 2-cycles in $\mathbb{C}^2_{\Gamma}$: we have $n=\sum_{a=1}^{r}  \,n_a$ of them wrap the \emph{compact} 2-cycles, meaning their flux is identified with a class $[S] = \sum_{a=1}^{r}  \,n_a\,\alpha_a$ in $H_2({\mathbb C}^2_\Gamma, \mathbb{Z})$, or the root lattice of $\fg_\text{o}$. By Poincar\'{e} duality, this group is identified with the cohomology group $H_{c}^2(\mathbb{C}^2_{\Gamma},\mathbb{Z})$, or the coroot lattice of $\fg_\text{o}$. But elements of the coroot lattice are in one-to-one correspondence with non-abelian monopole charges. We conclude that the $n$ D4 branes are non-abelian monopoles.  

Moreover, we have  $n^F=\sum_{a=1}^{r}  \,n^F_a$ branes wrapping the \emph{non-compact} 2-cycles. The flux of such branes is identified with a class $[S^*] = \sum_{a=1}^{r}  \,n^F_a\,\lambda_a$ in  $H_2({\mathbb C}^2_\Gamma, \partial \left({\mathbb C}^2_\Gamma\right), \mathbb{Z})$, or the weight lattice of $\fg_\text{o}$. By Poincar\'{e} duality, we further identify this group with the cohomology group $H^2(\mathbb{C}^2_{\Gamma},\mathbb{Z})$,  or the coweight lattice of $\fg_\text{o}$. But elements of the coweight lattice are in one-to-one correspondence with Dirac monopole charges. We conclude that the $n^F$ D4 branes are singular abelian Dirac monopoles.
Both the non-abelian and Dirac monopoles are subjected to the $\mathbb{Z}_{\fn_\fg}$-twist.\\



Consider first the T-duality along $S^1_{t}$, with the D4 branes in the type IIA background \eqref{IIA1D4branes}. In order to make contact with surface defects of 4-dimensional $\fg_\text{o}$-type Yang Mills, we should reduce the torus $S^1_{q}\times S^1_{t'}$ to zero size and take the point particle limit $m_s\rightarrow \infty$, just as we did in the unramified case. Note that the circle $S^1_{q}$ is essentially a spectator here, since the D4 branes wrapping $S^1_{q}$ are monopoles on $\cC\times S^1_{t'}$, meaning the limit $R_q\rightarrow 0$ is a ``trivial" dimensional reduction. More interestingly, we should clarify what happens to the branes as the circle $S^1_{t'}$ becomes small. In that regime, the Bogomolny equations lose the dependence on the $S^1_{t'}$-coordinate and we obtain:
\begin{align}
&F- [\varphi,\overline{\varphi}] =0 \nonumber\label{HitcheqMult}\\
&\overline{D}_x\varphi=0\nonumber\\
&D_x\overline{\varphi}=0 \; ,
\end{align}
with $x$ a complex coordinate on $\cC$, related to our previous coordinate in \eqref{Hitchsol3} by $z=e^x$; the associated one-form on $\cC$ is $\varphi(x)\, dx$. This is not quite the Hitchin system \eqref{Hitcheq}, but rather a ``multiplicative" version of it on $\cC$ \cite{Elliott:2018yqm}; the term multiplicative here stands for the imaginary part of $\varphi(x)$ being a periodic scalar, with period $2\pi /R'_t$. This period is large, as we are taking $S^1_{t'}$ to be small. 

In IIA variables, the Cartan components of the complex scalar field should be written as $\varphi_a=\psi_a+i\, A^{\theta}_{a}$, with the imaginary part as the holonomy $A^{\theta}_{a}$ of the (1,1) gauge field around $S^1_{t'}$. Its periodicity is a consequence of invariance under large gauge transformations, $R'_t\,A^{\theta}_{a} \rightarrow R'_t\, A^{\theta}_{a} +2\,\pi$.

In IIB variables, $\varphi_a$ takes a constant value $(R_t\,m^2_s)\, \tau^{6d}_a$ at the end of the cylinder. The holonomy is rewritten as $A^{\theta}_{a} = R_t\,m^2_s  \int_{S^2_a} C^{(2)}$, so that $\tau^{6d}_a$ is a linear combination of $(2,0)$ little string moduli in the bulk six-dimensions \eqref{couplings}, hence nondynamical from the point of view of the compactified theory\footnote{Alternatively, $m^2_s\,\tau^{6d}_a$ are the inverse gauge couplings of the effective theory on the D5 branes, hence nondynamical.}:
\beq\label{varphi}
\varphi_a = \lim_{g_s\rightarrow 0}(R_t\,m^2_s)\int_{S^2_a}\left(\frac{m^2_s}{g_s}\,\omega_{I}+i\, C^{(2)}\right) \; , \;\qquad a=1,\ldots, \r \; .
\eeq 
The $n$ non-abelian monopoles and $n^F$ Dirac monopoles on $\cC\times S^1_{t'}$ brought by the D4 branes will modify the profile of the Higgs field $\varphi$, away from this constant.\\ 

In order to figure out the precise singularity profile of $\varphi$, it proves useful to write the spectral curve of the multiplicative Hitchin system, which can be derived from the Seiberg-Witten curve of the gauge theory on the D-branes.
For our quiver gauge theories $T^{5d}_{\fg_\text{o}}$, the curve at a generic point on the Coulomb branch was derived by Nekrasov and Pestun \cite{Nekrasov:2012xe}. The method consists in writing down the instanton partition function ${\mathcal Z}_{inst}(T^{5d}_{\fg})$ in a full $\Omega$-background $S^1_{\cC'}\times\mathbb{C}_q\times\mathbb{C}_t$ and extracting the curve from a set of saddle point equations in the limit $q,t\rightarrow 1$. Their analysis further implied that the Coulomb branch of a 5d $\cN=1$ $\fg_\text{o}$-type quiver gauge theory was describing the moduli space of singular $\fg_\text{o}$-monopoles on $\cC \times S^1_{t'}$, with $\cC=\mathbb{C}^\times$ serving as the geometric Langlands curve\footnote{Soon after, Nekrasov-Pestun-Shatashvili noted that the saddle point analysis can still be performed in the limit $q\rightarrow 1$ while keeping $t$ arbitrary \cite{Nekrasov:2013xda}. The saddle point equations should now be understood as objects in ``quantum geometry;" physically, the ``classical" Seiberg-Witten curve is upgraded to a Schr\"{o}dinger-type equation, and the curve coordinates $(p,x)$ are now better understood as operators whose failure to commute is proportional to $\log(|t|)$. The moduli space of monopoles likewise becomes quantized.}. In order to make contact with the quantum geometric Langlands program, the saddle point analysis should be performed in a different limit: $t=q^\beta$, $q\rightarrow 1$ with $\beta$ fixed. The end result is a vanishing determinant, evaluated in a representation $V$ of $\fg$:
\beq\label{determ}
0=\text{Det}_V\left(e^{R'_t\, p} - e^{R'_t\, \varphi(x)}\right)\; .
\eeq
Here, the analysis is greatly simplified on the special point where the Coulomb branch meets the Higgs branch\footnote{The curve at that locus is sometimes called the UV curve; it should be understood as a multi-cover of the punctured curve $\cC$ \cite{Gaiotto:2009we}. The Seiberg-Witten at a generic point on the Coulomb branch is recovered from the UV curve by turning on all possible normalizable Coulomb moduli.}. At that locus, the theory abelianizes and all monopoles are Dirac monopoles. If we denote the multiplicative Higgs field there as $\varphi^{\bf{S}}(x)$, then $e^{R'_t\, \varphi^{\bf{S}}(x)}$ is diagonal and the determinant factorizes into $\text{dim}(V)$ factors, with
\beq\label{Higgs1}
e^{R'_t\, \varphi^{\bf{S}}_a(x)} = e^{\tau^{6d}_a}\cdot \prod_{d=1}^L \prod_{s=1}^{|\{{\lambda}\}_d|}\prod_{b=1}^{\L} \left(1-e^{x}  \,e^{-R'_t\, \Upsilon_{d,s,b}} x^{-1}_{d,s}\right)^{\pm\deg(\cA^{\mp}_{\lambda_{d,s},b})} \; .
\eeq
The factor $e^{\tau^{6d}_a}$ is the constant vacuum contribution, in the absence of D-brane monopole insertions. The D4 branes located at $x_{d,s}$ on $\cC$ modify the right-hand side  the product on the right-hand side, where the D4 branes are , and we are following the $q$-primary vertex operator notation of . In particular, $\{{\lambda}\}_d$ is a set of quantum affine weights labeling the $d$-th primary operator, and $\cA^{\mp}_{\lambda_{d,s},b}$ are the associated Drinfeld polynomials, whose roots are encoded in $\Upsilon_{d,s,b}$:
\beq
\Upsilon_{d,s,b} =\; -\sum_{i=1}^{\deg(\cA^{+}_{\lambda_{d,s},b})}\mathfrak{a}^{+}_{\lambda_{d,s},b,i} \left(1-\beta\right)+ \sum_{j=1}^{\deg(\cA^{-}_{\lambda_{d,s},b})}\mathfrak{a}^{-}_{\lambda_{d,s},b,j} \left(1-\beta \right) \; ,
\eeq
as we defined it in \eqref{commomentum3}, with $\beta=\log(t)/\log(q)$ encoding the central charge of the $\cW_{\beta}(\fg)$ algebra.

\begin{remark}
If the saddle point analysis of the instanton partition function is performed in the fully classical limit $q,t\rightarrow 1$ without scaling by $\beta$, one finds the curve \eqref{determ} but with the Higgs field \cite{Nekrasov:2012xe,Aganagic:2015cta}
\beq\label{Higgsclassical}
e^{R'_t\, \varphi^{\bf{S}}_a(x)} = e^{\tau^{6d}_a}\cdot \prod_{d=1}^L \prod_{s=1}^{|\{\vec{\underline{\lambda}}\}_d|} \left(1-e^{x}  \, x^{-1}_{d,s}\right)^{-\underline{\lambda^\vee_{d,s}}} \; .
\eeq
Most notably, the dependence on $\beta$ and on the Drinfeld polynomials of the quantum algebra is lost: each $\underline{\lambda^\vee_{d,s}}=2\log_{\hbar}\lambda^{+}_{a,0}$ is a classical $\fg$-coweight, which only captures the 0-th mode eigenvalue among the infinite number of modes in the quantum affine coweight $\lambda^\vee_{d,s}=\lambda^{+}_{a,k}$, $k\geq 0$. For generic and semi-degenerate defects, the Higgs fields \eqref{Higgsclassical} and \eqref{Higgs1} will coincide, but for rigid and mixed defects, only \eqref{Higgs1} will correctly capture the singular monopole profile; in particular, the Higgs field will explicitly depend on $\beta$ for those punctures, see Section \ref{sssec:classification}.\\
\end{remark}

We are now ready to take the point particle limit, as well as reducing the torus $S^1_{q}\times S^1_{t'}$ to zero size:
\beq\label{limits1}
m_s\rightarrow \infty\; ,\qquad R_q\rightarrow 0 \; ,\qquad R'_t\rightarrow 0 \; .
\eeq 
At the same time, we demand that the T-dual radii $R'_q=1/(\fn_{\fg}\,m_s^2\, R_q)$ and $R_t=1/(m_s^2\, R'_t)$. The geometric Langlands curve $\cC = \mathbb{R}\times S^1_{\cC}(R_{\cC})$ should remain fixed as well. 
As in the unramified case, the (1,1) $\fg_\text{o}$-type little string on $S^1_{q} \times S^1_{t'} \times \cC \times \mathbb{C}$ with $\mathbb{Z}_{\fn_{\fg}}$-twist reduces to $\cN=4$ $\fg$-type Super Yang-Mills theory on $\cC\times\mathbb{C}$ in the limit.\\

The D4 branes of the (1,1) little string wrapping $S^2_{\{a\}}\times S^1_q\times \mathbb{C}$ flow to the monodromy defects \eqref{Hitchsol3} of Gukov and Witten. One way to show this is to carefully take the limit while keeping the  Higgs field $\varphi^{\bf{S}}(x)$ fixed. In particular, as far as the constant contribution is concerned, $\varphi^{\bf{S}}_a= \tau^{6d}_a/R'_t$ must remain fixed, so the $\tau^{6d}_a$'s must go to 0. In the $\cW_{\beta}(\fg)$-algebra, these $\r$ constants are the Cartan components of the Fock vacuum eigenvalue $\mu_0$ \eqref{eigenvalue},
\beq
\tau^{6d}_a/R'_t = \langle\mu_0, \alpha_a\rangle \; .
\eeq
For the D-brane contributions, introduce the usual center of mass coordinates $x_{d,s} = \tilde{x}_d \, e^{R'_t \,\sigma_{d,s}}$ for the $d$-th primary operator insertion, and keep $\tilde{x}_d$ and $\sigma_{d,s}$ fixed as $R'_t\rightarrow 0$. This translates to the D4 branes ``coalescing" to $\tilde{x}_d$ on $\cC$, keeping their relative separation fixed. Note that the imaginary part of the Higgs field $\varphi^{\bf{S}}(x)$ is no longer a periodic scalar in the limit. For instance, the periods $\text{lim}_{m_s\rightarrow\infty} m^2_s\int_{S^2_a} C^{(2)}$ are now non-compact scalars in the moduli space of the $\fg_\text{o}$-type $(2,0)$ SCFT, subjected to the $\mathbb{Z}_{\fn_{\fg}}$-twist. 
All in all, including the brane defects, we obtain in the limit
\beq
\varphi^{\bf{S}}(x) = \langle\mu_0, \alpha_a\rangle + \sum_{d=1}^L \sum_{s=1}^{|\{{\lambda}\}_d|}\sum_{b=1}^{\L} \frac{\pm\left(\Upsilon_{d,s,b}+\sigma_{d,s}\right)\deg(\cA^{\mp}_{\lambda_{d,s},b})}{e^{-x}\, \tilde{x}_d-1} \; .
\eeq
Switching to the coordinate $z=e^x$ on $\cC$, the canonical one-form is $\varphi^{\bf{S}}(x) \, dx = \varphi^{\bf{S}}(z) \, dz/z$, so we get
\beq\label{Higgs}
\varphi^{\bf{S}}(z) = \frac{(\beta_0+i\,\gamma_0)}{z} + \sum_{d=1}^L  \frac{(\beta_d+i\,\gamma_d)}{\tilde{x}_d- z} \; ,
\eeq
with 
\beq\label{residuedata}
\beta_0+i\,\gamma_0 = \langle\mu_0, \alpha_a\rangle \; ,\qquad\;\;\; \beta_d+i\,\gamma_d = \sum_{s=1}^{|\{{\lambda}\}_d|}\sum_{b=1}^{\L} \pm\left(\Upsilon_{d,s,b}+\sigma_{d,s}\right)\deg(\cA^{\mp}_{\lambda_{d,s},b}) \; .
\eeq
This is the Higgs field for a mass-deformed tamely ramified Hitchin system on $\cC$. The data is the residue $\beta_0+i\,\gamma_0$ of the generic puncture at one end of the cylinder at $z=0$, and the residues at $z=\tilde{x}_d$ for $d=1,\ldots,L$. There is another generic puncture  at $z=\infty$, but its data is entirely determined in terms of the previous residues, as well as the number of normalizable deformations of the UV curve (the number of Coulomb moduli). 
In the companion paper \cite{NathanH}, we will classify the residues \eqref{residuedata} via the Drinfeld polynomial data, and show that it leads to a new classification of mass deformations for tamely ramified Hitchin systems. As a byproduct, we will obtain a description of all punctures in terms of Verma modules in the $\cW_{\beta}(\fg)$-algebra, thus giving a solution to the local Alday-Gaiotto-Tachikawa conjecture.




\vspace{8mm}

\subsubsection{S-transformation of $(\beta,\gamma)$}

Recall that the starting point of this analysis was a T-duality along $S^1_{t}$. We now repeat the same arguments as in the previous section, but with the T-duality along $S^1_{q}$ instead, with the D4 branes in the type IIA background \eqref{IIA2D4branes}. In that case, the D4 branes are  monopoles on $\cC\times S^1_{q'}$. This time, it is the circle $S^1_{t}$ which is essentially a spectator, and in the limit of small $R'_q$, the Bogomolny equations \eqref{bogomolny2} reduce to a multiplicative Hitching system, just as before. 

The Higgs field Cartan components are $^L\varphi_a = ^L\psi_a + i\; ^L A^{\theta}_a$ in IIA notation, with periodic imaginary part. They take constant values at the end of the cylinder $\cC$; in IIB notation,
\beq\label{varphinew}
^L \varphi_a= (\fn_{\fg}\,R_q\,m^2_s)\int_{\widetilde{S^2}_a}\left(\frac{m^2_s}{g_s}\,\omega_{I}+i\, C^{(2)}\right) \; , \;\qquad a=1,\ldots, \L \; .
\eeq 
where the compact 2-cycles $\widetilde{S^2}_{\{a\}}$ describe a resolved singularity of type $\widetilde{\fg_\text{o}}$ (recall that $\widetilde{\fg_\text{o}}=D_{p+1}$ if $\fg_\text{o}=A_{2p-1}$, and vice-versa, while $\widetilde{\fg_\text{o}}=\fg_\text{o}$ in all other cases).
Note the presence of the lacing number $\fn_{\fg}$ after rewriting  $R'_q$ in terms of $R_q$. The corresponding multiplicative Hitchin system has spectral curve
\beq\label{determ2}
0=\text{Det}_V\left(e^{R'_q\, p} - e^{R'_q\, ^L\varphi(x)}\right)\; .
\eeq
This time, the point-particle limit is 
\beq\label{limits2}
m_s\rightarrow \infty\; ,\qquad R'_q\rightarrow 0 \; ,\qquad R_t\rightarrow 0 \; ,
\eeq 
and we demand that the T-dual radii $R_q=1/(\fn_{\fg}\,m_s^2\, R'_q)$ and $R'_t=1/(m_s^2\, R_t)$ remain fixed. 
In this limit, the (1,1) $\widetilde{\fg_\text{o}}$-type little string on $S^1_{q'} \times S^1_{t} \times \cC \times \mathbb{C}$ with $\mathbb{Z}_{\fn_{\fg}}$-twist becomes the $\cN=4$ $^L \fg$-type Super Yang-Mills theory on $\cC\times\mathbb{C}$, and the D4 branes wrapping $\widetilde{S^2}_{\{a\}}\times S^1_t\times \mathbb{C}$ become the monodromy surface defects for the $^L \fg$ theory. In particular, the Higgs field of 4-dimensional $^L \fg$-type Yang-Mills is identified as
\beq\label{Higgs2}
^L\varphi^{\bf{S}}(z) = \frac{(^L\beta_0+i\, ^L\gamma_0)}{z} + \sum_{d=1}^L  \frac{(^L\beta_d+i\, ^L\gamma_d)}{\tilde{x}_d- z} \; .
\eeq

Now, unless $\fg$ is simply-laced, we cannot directly compare $({^L\beta}, {^L\gamma})$ and $({\beta}, {\gamma})$, since $\phi(x)$ is valued in the Cartan subalgebra of $\fg$, whereas $^L\phi(x)$ is valued in the Cartan subalgebra of $^L\fg$. We therefore use the notation introduced in \eqref{Sdualitybetagamma}, namely $\alpha \mapsto \alpha^*$, where if $\alpha_l$ is a long root, then  $\alpha_l^*$ is $\fn_\fg$ times the short coroot $\alpha^\vee_s$, and if $\alpha_s$ is a short root, then  $\alpha_s^*$ is the long coroot $\alpha^\vee_l$. We note that the multiplicative Higgs fields enter as $e^{R'_t\, \varphi}$ and $e^{R'_q\, ^L \varphi}$. For instance, for the puncture at $z=0$, this is written explicitly in  \eqref{varphi} and \eqref{varphinew}, respectively. Then, after correctly accounting for the factor of $\fn_{\fg}$, we obtain:
\beq
({^L\beta}, {^L\gamma}) = \frac{R_q}{R_t}\, (\beta^*,\gamma^*) \; .
\eeq
This relation holds for any of the $d$ punctures located at $z=\tilde{x}_d$, with $d=0,1,\ldots,L$ (and $\tilde{x}_0 = 0$).
Because the Yang-Mills gauge coupling is $\tau^{4d}= i\,R_q/R_t$, this also reads
\beq
({^L\beta}, {^L\gamma}) = \text{Im}(\tau^{4d})\, (\beta^*,\gamma^*) \; .
\eeq
This is precisely the 4-dimensional S-duality action on the Higgs field parameters \eqref{Sdualitybetagamma}.

\subsubsection{S-transformation of $(\alpha,\eta)$}

Once again, we start with the type IIB background \eqref{IIBVafa}, in the presence of the $\mathbb{Z}_{\fn_{\fg}}$-twist. Recall that the D-branes are at points on the Riemann surface $\cC$, and these points are the stringy definition of the punctures on $\cC$. Let $S^1_{\{p\}}$ be a small circle on $\cC$ surrounding the location of the $d$-th puncture, located at a point $p$. We define two $\fh$-valued periods, whose  Cartan components read
\beq\label{alphaetaIIB}
\alpha_a^{\text{IIB}} = m_s^4\int_{S^2_a\times S^1_{\{p\}} \times S^1_q} C^{(4)} \; , \qquad\;\; \eta_a^{\text{IIB}} =  m_s^4\int_{S^2_a\times S^1_{\{p\}} \times S^1_t} C^{(4)} \; , \qquad  \;\;a=1,\ldots,\r \; .
\eeq 
Above, $C^{(4)}$ is the R-R 4-form of type IIB. Consider the usual T-duality along $S^1_t$ to type IIA \eqref{IIA1Vafa}.  The above periods get mapped to 
\beq\label{alphaetaIIA1}
\alpha_a^{\text{IIA}} = m_s^5\int_{S^2_a\times S^1_{\{p\}} \times S^1_q \times S^1_{t'}} \widetilde{C}^{(5)} \; , \qquad\;\; \eta_a^{\text{IIA}} =  m_s^3\int_{S^2_a\times S^1_{\{p\}}} C^{(3)} \; , \qquad \;\;a=1,\ldots,\r \; .
\eeq 
Above, $C^{(3)}$ is the R-R 3-form of type IIA, and  $\widetilde{C}^{(5)}$ is the associated 5-form defined through Hodge duality $d\widetilde{C}^{(5)}=\star dC^{(3)}$. 
In the limit $g_s\rightarrow 0$, these are periods of the (1,1) $\fg_\text{o}$-type little string subjected to a $\mathbb{Z}_{\fn_{\fg}}$-twist. In particular, the integral of $C^{(3)}$ along $S^2_a$ is the gauge field $A_a$ of the (1,1) little string, whose holonomy around the puncture $p$ on $\cC$ is precisely $\eta_a^{\text{IIA}}$. We further take the CFT limit and reduce on the torus $S^1_q \times S^1_{t'}$, requiring  the periods $\alpha_a^{\text{IIA}}$ and $\eta_a^{\text{IIA}}$ to remain finite in the limit. We identify them with the surface defect parameters  $(\alpha,\eta)$ in $\fg$-type gauge theory:
\beq\label{alphaeta1}
\lim_{\substack{g_s\rightarrow 0 \\ m_s\rightarrow\infty \\ R_q,R'_t\rightarrow 0}}\alpha_a^{\text{IIA}} = \alpha_a \; , \qquad \lim_{\substack{g_s\rightarrow 0 \\ m_s\rightarrow\infty \\ R_q,R'_t\rightarrow 0}}\eta_a^{\text{IIA}} = \eta_a \; ,\qquad \;\; \;\;a=1,\ldots,\r \; .
\eeq
Alternatively, we perform T-duality from type IIB to type IIA along $S^1_q$. The periods \eqref{alphaetaIIB} get mapped to 
\beq\label{alphaetaIIA2}
^L\alpha_a^{\text{IIA}} = m_s^3\int_{S^2_a\times S^1_{\{p\}}} C^{(3)} \; , \qquad\;\; ^L\eta_a^{\text{IIA}} = m_s^5\int_{S^2_a\times S^1_{\{p\}} \times S^1_t \times S^1_{q'}} \widetilde{C}^{(5)} \; \qquad , \;\;a=1,\ldots,\L \; .
\eeq 
In the limit $g_s\rightarrow 0$, these are periods of the (1,1) $\widetilde{\fg_\text{o}}$-type little string subjected to a $\mathbb{Z}_{\fn_{\fg}}$-twist.  We further take the CFT limit and reduce on the torus $S^1_q \times S^1_{t'}$. Requiring the periods $^L\alpha_a^{\text{IIA}}$ and $^L\eta_a^{\text{IIA}}$ to remain finite in the limit, we identify them with the surface defect parameters  $(^L\alpha_a,^L\eta_a)$ in $^L \fg$-type gauge theory:
\beq\label{alphaeta2}
\lim_{\substack{g_s\rightarrow 0 \\ m_s\rightarrow\infty \\ R_q,R'_t\rightarrow 0}} {^L\alpha_a^{\text{IIA}}} = {^L\alpha_a} \; , \qquad \lim_{\substack{g_s\rightarrow 0 \\ m_s\rightarrow\infty \\ R_q,R'_t\rightarrow 0}} {^L\eta_a^{\text{IIA}}} = {^L\eta_a} \; \qquad , \;\;a=1,\ldots,\L \; .
\eeq
Then, from the IIA definition of the parameters, and taking into account the reversal of orientation between $\alpha_a^{\text{IIA}}$ and $^L\eta_a^{\text{IIA}}$, we obtain in the limit:
\beq
(^L\alpha,^L\eta) = (\eta,-\alpha)  \; .
\eeq
This is precisely the 4-dimensional S-duality relation \eqref{Sdualityalphaeta}.

\vspace{8mm}

\subsubsection{T-transformation of the parameters}

So far, $\tau^{4d}=i\, \frac{4\pi}{g^2_{4d}}$ was purely imaginary. A non-zero $\theta$-angle $\tau^{4d}= \frac{\theta}{2\pi} + i\, \frac{4\pi}{g^2_{4d}}$ can be incorporated by turning on an NS-NS $B^{(2)}$-field flux on the torus:
\beq
\tau^{4d} = m_s^2\, (i\, R_q\, R'_t - B^{(2)}) \; .
\eeq
The S-duality of $\tau^{4d}$ is as before, \eqref{Sdualitytau2}. It is derived once again from the T-duality of little strings, this time in a background with $B^{(2)}$ flux. This follows from a straightforward application of ``Buscher rules" \cite{Buscher:1987sk}, which we omit writing explicitly here not to overburden the presentation; see for example Section 2 of \cite{Amariti:2020lua}.  Note that $B^{(2)}$ must go to zero in the conformal limit, so that $m_s^2\, B^{(2)}$ remains fixed.

There is now a T-symmetry corresponding to shifting $B^{(2)}$ by $m_s^{-2}$, which translates to the $2\pi$ shift of the $\theta$-angle: 
\beq
\label{Tdualitytau2}
\text{T}: \;\;\;(\fg,\tau^{4d}) \Longleftrightarrow (\fg,\tau^{4d}-1) \; .
\eeq 
We now turn to the surface defect parameters: from the stringy definition of $(\beta_a,\gamma_a)$, it is clear that those parameters are invariant under $B^{(2)}$-shifts. 

It follows from \eqref{alphaetaIIA1} that the parameters $\eta_a^{\text{IIA}}$ are likewise invariant, while we find the following term in $\alpha_a^{\text{IIA}}$:
\beq
m_s^5\int_{S^2_a\times S^1_{\{p\}} \times S^1_q \times S^1_{t'}} \widetilde{C}^{(3)}\wedge B^{(2)} \; .
\eeq
Then, a shift of $B^{(2)}$ by $m_s^{-2}$ will map $\alpha_a \rightarrow \alpha_a - \eta_a$. We have therefore recovered the action of the T symmetry on the surface defect parameters:
\beq\label{Tdualityalphaeta2}
\text{T}: (\alpha,\eta) \longrightarrow (\alpha,\eta-\alpha) \; .
\eeq

\vspace{8mm}

\subsection{On TS-duality versus S-duality}

The quantum $q$-Langlands correspondence implies the dictionary
\beq\label{dictionagain}
q=\hbar^{-^L(\kappa+h^\vee)}\, ,\;\;\qquad t = \hbar/q^{\fn_{\fg}}\, ,
\eeq
between the parameters of $\cW_{q,t}(\fg)$ and $U_{\hbar}(\widehat{^L\fg}_{^L \kappa})$. The physical origin of this dictionary was first elucidated in the work \cite{Aganagic:2017smx}, in the unramified context. We will start by briefly reviewing those arguments, which are unaffected by ramification, and then comment on the distinction between TS duality and S duality in gauge theory.\\

First, we have shown that the partition function of the $\fg_\text{o}$-type $(2,0)$ little string theory on the manifold 
\begin{equation}\label{background1}
M^\times_6 =\frac{\mathbb{C}^\times_q}{\mathbb{Z}_{\fn_\fg}}\times \mathbb{C}_t\times \cC 
\end{equation}
is a deformed conformal block of $U_{\hbar}(\widehat{^L\fg_\text{o}})$, subjected to the $\mathbb{Z}_{\fn_{\fg}}$ outer automorphism twist, with tame ramification at points on $\cC$. These are $U_{\hbar}(\widehat{^L\fg})$ blocks, by definition\footnote{As previously mentioned, the construction of the folded qKZ equations in the non simply-laced case is still not completely clear and deserves further study \cite{Frenkel:2021bmx}.}. In writing $\mathbb{C}^\times_q$, we removed the origin of $\mathbb{C}_q$, consistent with the vector-valued nature of the conformal blocks. 

In the point particle limit $m_s\rightarrow \infty$, this is the chiral conformal block of a $\widehat{^L\fg_\text{o}}$-type WZW model.
To make contact with 4d SYM, we consider once again $M^\times_6$ as a $T^2=S^1_q(R_q)\times S^1_t(R_t)$ fibration:
\beq
T^2\hookrightarrow M^\times_6 \hookrightarrow M_4 = \cC\times\mathbb{R}\times\mathbb{R}_+ \; .
\eeq 

As explained by Nekrasov and Witten \cite{Nekrasov:2010ka}, the reduction of the 6d theory on this torus leads to 4d $\cN=4$ SYM on $M_4$ with the topological Landau-Ginzburg twist. Whether the 4d gauge group is of $\fg$-type or $^L\fg$-type depends on the order of the $T^2$ reduction, as we saw throughout this paper. Far away from the origin of $\mathbb{C}_q\times\mathbb{C}_t$ (the fixed point locus of the $U(1)^2$ equivariant action in the $\Omega$-background), the topological twist becomes trivial, and $M^\times_6$ is flat. This makes it possible to identify the circle radii in $T^2=S^1_q(R_q)\times S^1_t(R_t)$ with the $\Omega$-background parameters:
\beq\label{radii}
R_q=\frac{2\pi}{i\,\fn_{\fg}\,\epsilon_q}\; ,\qquad \; R_t=\frac{2\pi}{\epsilon_t} \; .
\eeq

The dictionary relation \eqref{dictionagain} is a consequence of \eqref{radii} and results of Ooguri and Vafa \cite{Ooguri:1999bv}. Namely, consider the 6d $(2,0)$ SCFT on 
\begin{equation}\label{background11}
M^\times_6 =\left(\frac{S^1_q}{\mathbb{Z}_{\fn_\fg}}\times \mathbb{C}\right)\times M_3 \; , \qquad \; M_3 = \mathbb{R}\times\cC \; .
\end{equation}
This is the same manifold $M^\times_6$ as before, but now reinterpreted as a product of two 3-manifolds instead of a product of a 2-manifold and a 4-manifold. The partition function of the 6d theory on $M^\times_6$ computes the partition function of $^L\fg$-type Chern-Simons theory on $M_3$, at level $^L(\kappa+h^\vee)$. The level itself is determined by the $\Omega$-background parameter on ${S^1_q}\times \mathbb{C}$: as we go around ${S^1_q}$, the complex line $\mathbb{C}$ is rotated by $q_{C.S.} = e^{2\,\pi\, i/^L(\kappa+h^\vee)}$, which needs to be compensated by an R-symmetry twist to preserve supersymmetry. In our background, this is a twist not by $t^{-1}$ as one could have naively expected, but instead by $t^{-1}\cdot q^{\fn_{\fg}}=\hbar$;  the appearance of $q$ is due to the topological twist on the cylinder $\mathbb{C}^\times_q$ being trivial, and the exponent $\fn_{\fg}$ accounts for the possible outer automorphism twist. 

In other words, had the origin of $\mathbb{C}_q$ not been removed, the twist would have been by $e^{R_q\, (-\epsilon_{t})}$. Here, the R-symmetry twist as we go around $S^1_q$ is instead $q_{C.S.} = e^{\fn_{\fg}\, R_q\, \epsilon_{\hbar}}$, with $\epsilon_{\hbar}=\fn_{\fg}\, \epsilon_q - \epsilon_t$.

It follows from the first relation in \eqref{radii} that $q_{C.S.}= e^{-2\, \pi \, i \, \epsilon_{\hbar}/\epsilon_{q}}$, from which we deduce 
\beq
\frac{\epsilon_{q}}{\epsilon_{\hbar}} = -^L(\kappa+h^\vee) \; . 
\eeq
After exponentiation, this is precisely the dictionary \eqref{dictionagain}. If we perform the reduction on $T^2$, this ratio of $\Omega$-background parameters is also the gauge coupling of $^L\fg$-type 4d SYM: 
\beq
\frac{\epsilon_{q}}{\epsilon_{\hbar}}=\, ^L\tau^{4d}\; .
\eeq

When we ``fill the puncture" at the origin and replace $\mathbb{C}^\times_q$ by $\mathbb{C}_q$, the partition function of the $\fg_\text{o}$-type $(2,0)$ little string theory on the manifold 
\begin{equation}\label{background2}
M_6 =\frac{\mathbb{C}_q}{\mathbb{Z}_{\fn_\fg}}\times \mathbb{C}_t\times \cC \; .
\end{equation}
is instead a scalar-valued deformed conformal block of $\cW_{q,t}(\fg_\text{o})$, subjected to the $\mathbb{Z}_{\fn_{\fg}}$ outer automorphism twist, with tame ramification at points on $\cC$. These are $\cW_{q,t}(\fg)$ blocks, by definition.  The parameters $q$ and $t$ are the 5d $\Omega$-background parameters $q=e^{R_{\cC'}\,\epsilon_q}$ and $t=e^{-R_{\cC'}\,\epsilon_t}$. After setting $t = q^{\beta}$, this implies
\beq
\beta = \frac{-\epsilon_t}{\epsilon_q} \; .
\eeq 
This relation was first discovered as part of the AGT correspondence \cite{Alday:2009aq,Nekrasov:2010ka}, in the conformal limit $m_s\rightarrow \infty$ to the $\cW_{\beta}(\fg)$-algebra (this is the limit $q\rightarrow 1$ with $\beta$ fixed in the $\cW_{q,t}(\fg)$-algebra).

This time around, the $T^2=S^1_q(R_q)\times S^1_t(R_t)$ reduction yields $\fg$-type 4d SYM with gauge coupling
\beq
\tau^{4d} = \frac{\epsilon_t}{\fn_{\fg}\,\epsilon_q} \; .
\eeq
In a frame where the 4d gauge coupling is $\tau^{4d}= i\,R_q/R_t$, it follows that
\beq
\tau^{4d}-1=\frac{-1}{\fn_{\fg}\,^L\tau^{4d}} \; .
\eeq
Then, the quantum $q$-Langlands correspondence is a TS-duality in gauge theory. As we now argue, in order to realize Langlands duality as S-duality proper, we need to engineer the 5d gauge theory $T^{5d}_{\fg_\text{o}}$ in a spectral frame, or equivalently the 3d Drinfeld gauge theory   $T^{3d}_{\fg_\text{o}}$ in a mirror frame.

\vspace{6mm}

\subsubsection{Remarks on spectral duality}
\label{ssec:spectral}

Sometimes, when $\fg_\text{o} = A_r$, it can happen that the 5d gauge theory on $S^1_{\cC'}$ has more than one Lagrangian description, related by spectral duality. This duality was first described in the context of geometric engineering \cite{Katz:1997eq}, and later discovered in integrable systems \cite{Mironov:2012uh,2012arXiv1206.5364N,Mironov:2013xva,Mironov:2021sfo,Koroteev:2019gqi}. In the multiplicative Hitchin system associated to $T^{5d}_{A_r}$,  the duality exchanges the roles of the variables $e^x$ and $e^p$ in the spectral curve \eqref{determ2} (both are the $\mathbb{C}^\times$-valued). In the 5d gauge theory, the  gauge couplings are exchanged with the masses and Coulomb moduli, and the $\Omega$-background parameters are exchanged\footnote{After Higgsing to 3 dimensions, this is also the mirror symmetry map. In particular, the $A$-twisted half-index we computed in this paper would become $B$-twisted instead.}, as
\begin{align}\label{spectral}
x^{\pm}_{d,a},\; e_{a,I} \; &\longleftrightarrow\;1/(g^{5d}_b)^2 \; ,   \nonumber\\ 
t \; &\longleftrightarrow \; q/t \; .
\end{align}
Our Hitchin system featured $L$ arbitrary punctures, the non-compact D5 branes at points on the cylinder $\cC$. Under certain conditions on $L$ and on the effective Chern-Simons levels of the 5d theory, a spectral dual frame may become manifest, with another Lagrangian.

We saw that the Lagrangian description of $T^{5d}_{A_r}$ is lost in the conformal limit to 4 dimensions, as it becomes strongly coupled, but this is not so for the spectral dual theory, which retains a weakly coupled Lagrangian. One way to identify this Lagrangian in 4d is to conformally map the cylinder $\cC$ to a sphere with 2 generic punctures. The theory we are after should then be spectral dual to a theory with 2 generic punctures (at $\{0\}$ and $\{\infty\}$) and $L$ arbitrary ones on the sphere, consistent with the Higgs field profile \eqref{Higgs}. Such a theory was characterized by Gaiotto \cite{Gaiotto:2009we}, and is given by \emph{another} quiver gauge theory.\\

Here, we will look at a simple example, a 5d quiver gauge theory
\beq
T^{5d}_{A_{r-1}}\; : \;\;G_{5d}=\prod_{a=1}^{r-1} SU(n_a) \; ,  \qquad G^F_{5d}=SU(n^F_1)\times  SU(n^F_{r-1})
\eeq
with $n^F_1 = n^F_{r-1} = n_a = 2$. That is, the gauge group is
$(r-1)$ $SU(2)$ factors, and the flavor group is $SU(2)$ on the first node and $SU(2)$ on the right node. The 5d spectral dual theory is readily identified, say from the brane engineering of the theory in type IIB \cite{Mitev:2014jza}, as a $\widetilde{T^{5d}_{A_{r-1}}}=T^{'5d}_{A_1}$ gauge theory, that is
\beq
T^{'5d}_{A_1}\; : \;\;G'_{5d}= SU(r) \; ,  \qquad G^{'F}_{5d}=SU(2r)
\eeq
In the conformal limit to 4 dimensions, $T^{5d}_{A_{r-1}}$ becomes strongly coupled, but $T^{'5d}_{A_1}$ doesn't; instead, it becomes the same gauge theory, but in the 4-dimensional sense:
\beq
T^{'4d}_{A_1}\; : \;\;G'_{4d}= SU(r) \; ,  \qquad G^{'F}_{4d}=SU(2r)
\eeq
Recently, Nekrasov and Tsymbaliuk studied $T^{'4d}_{A_1}$ in the presence of generic codimension-2 monodromy defects \cite{Nekrasov:2021tik}. These defects are introduced by replacing the support manifold $\mathbb{C}_q\times\mathbb{C}_t$ with the orbifold $(\mathbb{C}_q/\mathbb{Z}_r)\times\mathbb{C}_t$. Then, one can show that the instanton partition function of $T^{'4d}_{A_1}$ in the presence of these defects defects takes the form of the following chiral correlator
\beq\label{NekraTsymb}
{\bf F} = \left\langle v_{\nu'_\infty}, \widetilde{V}_m(z)\, \widetilde{V}_{\widetilde{m}}(1) \; v_{\nu'_0} \right\rangle  \; ,
\eeq
which is a solution to the (rational) Knizhnik-Zamolodchikov equations\footnote{Because the orbifold is along $\mathbb{C}_q$ in our conventions, we exchanged $\epsilon_q \leftrightarrow \epsilon_t$ compared to \cite{Nekrasov:2021tik}.}:
\beq
(\kappa'+r)\, \frac{\partial {\bf F}}{\partial z} = \left(\frac{\Omega_{z0}}{z}+\frac{\Omega_{z1}}{z-1}\right) {\bf F} \; , \qquad \;\; \kappa'=\frac{\epsilon_q}{\epsilon_t}
\eeq
with $\Omega$ as in \eqref{KZdef}. In particular, note from the definition of $\kappa'$ that in this setup, it is S-duality and not TS-duality which is more natural.

Just like in our setup, $v_{\nu'_0}$ and $v_{\nu'_\infty}$ are  highest and lowest weight vectors of irreducible Verma modules, respectively. Meanwhile, $\widetilde{V}_m$ and $\widetilde{V}_{\widetilde{m}}$ are so-called twisted Heisenberg-Weyl modules \cite{Etingof:1993wp}. These modules encode the dependence on the masses, with the twist encoding the dependence on the Coulomb moduli. Let us specialize to $r=2$, meaning  $SU(2)$ with 4 flavors. If we Higgs the theory by freezing the Coulomb moduli to the masses, and further specialize to $r=2$, all modules become Verma modules over $\widehat{A_1}$.  

\begin{figure}[h!]
	\emph{}
	\centering
	\includegraphics[trim={0 0 0 0cm},clip,width=0.99\textwidth]{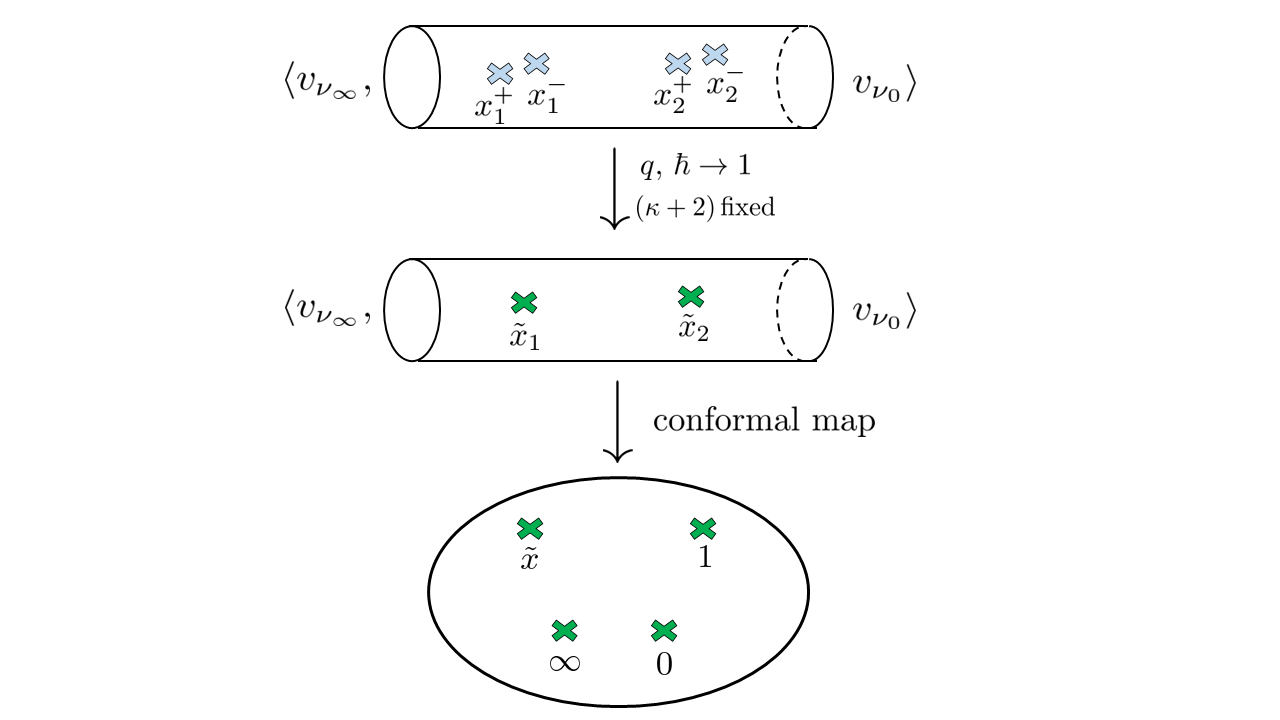}
	\vspace{-1pt}
	\caption{From the insertion of two $q$-primary operators on the cylinder to the 4-point function on the sphere. The first arrow down denotes the CFT limit.} 
	\label{fig:NekTsym}
\end{figure}

Because $SU(2)$ with 4 flavors is known to be self-dual under spectral duality, we should expect this description to agree with ours in that case. And indeed, for $r=2$, and in the conformal limit, we are also describing a 4-point function on the sphere with vertex operator insertions labeled by irreducible Verma modules over $\widehat{A_1}$. The rational limit of our trigonometric KZ equations is 
\beq
(\kappa+2)\, \frac{\partial {\bf F}}{\partial \tilde{x}} = \left(\frac{\Omega_{x0}}{\tilde{x}}+\frac{\Omega_{x1}}{\tilde{x}-1}\right) {\bf F} \; , \qquad \;\; \kappa=\frac{\epsilon_{q}}{\epsilon_\hbar} \; . 
\eeq
Under the spectral duality map \eqref{spectral}, this is precisely the Nekrasov-Tsymbaliuk solution. It would be important to generalize the analysis to all $r\geq 2$.


\subsubsection{Instanton/Vortex counting and a theorem of Atiyah}

A key part of this paper was the use of gauge-vortex duality as a means to relate instanton counting in the 5d theory $T^{5d}_{\fg_\text{o}}$ to vortex counting in the 3d theory $T^{3d}_{\fg_\text{o}}$. 

Mathematically, this is highly reminiscent of a theorem of Atiyah, formulated in the 80's \cite{Atiyah:1984tk}: given a classical group $G$, and a positive integer $k$, the moduli space of $k$ $G$-instantons on $\mathbb{C}^2$ (suitably regularized, and modulo based gauge transformations) is diffeomorphic to the moduli space of (based) degree $k$ holomorphic maps  ${\cal{M}}(\mathbb{P}^1\rightarrow \Omega G)$ to the (based) loop group of $G$. For the instantons, based gauge transformations are those which are trivial at $\{\infty\}\in\mathbb{C}^2$, and based holomorphic maps send $\{\infty\}\in\mathbb{P}^1$ to the identity in $\Omega G$. The based loop group $\Omega G$ is the group of maps $S^1 \rightarrow G$ which send constant maps to the identity in $G$.

When $\fg = A_{r-1}$, this theorem was used to identify the ground states of the $(2,0)$ little string theory, defined through discrete light-cone quantization (DLCQ) \cite{Ashwinkumar:2015sps}. That is, the little string is compactified on a light-like circle $S^1$ with $k$ units of discrete momentum along it, obtained as a limit of compactification on a small space-like circle boosted by a large amount \cite{Seiberg:1997ad}. The DLCQ of the $A_{r-1}$ little string theory with $k$ units of momentum is a 2d $\cN=(4,4)$ sigma model whose target space is the moduli space of $k$ $SU(r)$ instantons on $\mathbb{C}^2$ \cite{Aharony:1997th}. The ground states of this sigma model are $L^2$-cohomology classes of this instanton moduli space\footnote{$L^2$-cohomology arises here because the moduli space is not compact.}, which by Atiyah's theorem are $L^2$-cohomology classes of ${\cal{M}}(\mathbb{P}^1\rightarrow \Omega SU(r))$. By studying the $A$-twisted topological sigma model on $\Omega SU(r)$, it is therefore possible to identify the ground states of the $(2,0)$ $A_{r-1}$ little string theory; the analysis of \cite{Ashwinkumar:2015sps} identifies these states as modules over an affine Lie algebra $\widehat{\text{su}(r)}$, in agreement with a general result of Braverman and Finkelberg \cite{Braverman:2007dvq}.\\

In the context of our paper, it would be important to generalize these results by first considering a K-theoretical count of the maps instead of a cohomological one, and further introducing an $\Omega$-background. One should expect the affine Lie algebra $\widehat{\text{su}(r)}$ to be replaced by the Drinfeld quantization of its universal enveloping algebra, or rather a quantum toroidal algebra\footnote{Indeed, there is growing evidence that for the little string on $S^1$ and in the $\Omega$-background, the BPS strings wrapped on $S^1$, as studied by \cite{Ashwinkumar:2015sps}, should be represented  by the generators of the quantum toroidal algebra  $U_{q,t}(\Hat{\Hat{\fgl_1}})$, and the various D-branes, such as the ones studied in our paper, should be represented by specific modules \cite{Awata:2011ce,Mironov:2016yue,Zenkevich:2020ufs,2022arXiv221214808Z,Bayindirli:2023byn,Kimura:2023bxy}. These matters will also be explored elsewhere \cite{NathanH2}.} \cite{2009arXiv0904.1679F,2009arXiv0905.2555S}. If this is true, then our results would imply a spectral dual version of Atiyah's theorem, at least for $\fg=A_{r-1}$ and in the quantized K-theoretical setting: our starting point was also the $A_{r-1}$ little string, but for us ``$r$" labeled the rank of a \emph{quiver} instead of the rank of a gauge group. The (quantum) loop algebra symmetry we identified in the codimension-2 theory was likewise a symmetry of a quiver. We leave this investigation to future work.

\vspace{8mm}

\section{Case study: tamely ramified $A_1$ Langlands correspondence}
\label{sec:example}

Throughout this section, the Lie algebra will be $\fg=A_1$.\\

\subsection{The $U_\hbar(\widehat{A_1})$-algebra}

The quantum affine algebra $U_\hbar(\widehat{A_1})$ has a presentation in terms of Chevalley generators, which following standard terminology goes by the name of the Drinfeld-Jimbo realization \cite{Jimbo:1985zk,Jimbo:1985vd,Drinfeld1985HopfAA,Drinfeld:1986in}. It consists of generators $e_0^{\pm}, e_1^{\pm}, K_0^{\pm 1}, K_1^{\pm 1}$, and relations\footnote{We choose to match the notations of \cite{Frenkel:1998} rather than the cited references, meaning our quantization parameter is defined as $\hbar_{\text{here}} = \hbar^2_{\text{there}} = q/t$.}:
\begin{align}
&K_0 \, K_1 = K_1 \, K_0 \; ,\label{A1algebra}\nonumber\\
&K_i \, K_i^{-1} = 1 = K_i^{-1}\, K_i \; , \qquad (i=0,1)\nonumber\\
&K_i \, e_i^{\pm} = \hbar^{\pm 1}\, e_i^{\pm}\, K_i \; , \qquad (i=0,1)\nonumber\\
&K_i \, e_j^{\pm} = \hbar^{\mp 1}\, e_j^{\pm}\, K_i \; , \qquad (i\neq j)\nonumber\\
&e_i^{+} \, e_i^{-} - e_i^{-} \, e_i^{+} = \frac{K_i - K_i^{-1}}{\hbar^{1/2} - \hbar^{-1/2}} \; , \qquad (i=0,1)\nonumber\\
&e_0^{\pm} \, e_1^{\mp} = e_1^{\mp} \, e_0^{\pm}  \; ,\nonumber\\
&(e_i^{\pm})^3 e_j^{\pm} - \frac{\hbar^{3/2} - \hbar^{-3/2}}{\hbar^{1/2} - \hbar^{-1/2}} \, (e_i^{\pm})^2 e_j^{\pm} e_i^{\pm} + \frac{\hbar^{3/2} - \hbar^{-3/2}}{\hbar^{1/2} - \hbar^{-1/2}}  \,  e_i^{\pm} e_j^{\pm} (e_i^{\pm})^2 - e_j^{\pm} (e_i^{\pm})^3 =0 \; . \qquad (i\neq j)
\end{align}
The constraints on the last line are a quantized version of the familiar $\widehat{A_1}$ Serre relations. The Hopf algebra \footnote{For simplicity's sake, we skip writing down the action of the counit and antipode on the generators.} nature of $U_\hbar(\widehat{A_1})$ guarantees the existence of a coproduct $\Delta: U_\hbar(\widehat{A_1})\rightarrow U_\hbar(\widehat{A_1})\otimes U_\hbar(\widehat{A_1})$:
\begin{align}
&\Delta(e_i^+) = e_i^+ \otimes K_i + 1 \otimes e_i^+ \; , \qquad (i=0,1)\label{coproduct}\nonumber\\
&\Delta(e_i^-) = e_i^- \otimes 1 + K_i^{-1} \otimes e_i^- \; , \qquad (i=0,1)\nonumber\\
&\Delta(K_i^{\pm 1}) = K_i^{\pm 1} \otimes K_i^{\pm 1} \; . \qquad (i=0,1)
\end{align}
Consider the subalgebra $U_\hbar(A_1)$ of $U_\hbar(\widehat{A_1})$ generated by $e^{\pm}=e^{\pm}_1$ and $K^{\pm 1}=K^{\pm 1}_1$. For all $x\in\mathbb{C}^\times$, evaluation representations are defined through the (surjective) algebra homomorphism\footnote{We called this homomorphism $\text{ev}\circ D_x$ in the main text, where $D_x$ denoted a shift automorphism.} $\text{ev}_x : U_\hbar(\widehat{A_1})\rightarrow U_\hbar(A_1)$, as
\begin{align}
&\text{ev}_x(e^{\pm}_0) = x^{\pm 1}\, \hbar^{\mp 1/2} \, e^{\mp} \; , \qquad\qquad \text{ev}_x(e^{\pm}_1)= e^{\pm} \; , \label{evalmor}\nonumber\\
&\text{ev}_x(K^{\pm 1}_0) = K^{\mp 1} \; , \qquad\qquad \text{ev}_x(K^{\pm 1}_1)= K^{\pm 1}  \; .
\end{align}
Then, let ${\cV}_{i}$ be a highest weight representation of $U_\hbar(A_1)$, with highest weight vector $v_i$ and corresponding weight $\mu_i$, and let ${\cV}_{i}(x)$ be the corresponding evaluation representation in $U_\hbar(\widehat{A_1})$, obtained after applying the homomorphism $\text{ev}_x$. The trigonometric quantum $R$-matrix of $U_\hbar(\widehat{A_1})$ is an endomorphism 
$R_{\cV_i\, \cV_j}(x): \text{End}(\cV_i(x)\otimes\cV_j(1))$, defined so that it satisfies
\beq
\Delta'(y) \, R_{\cV_i\, \cV_j}(x) = R_{\cV_i\, \cV_j}(x)\, \Delta(y) \, , \qquad y\in U_\hbar(\widehat{A_1}) \; ,
\eeq
with $\Delta' = \sigma \circ \Delta$, and $\sigma(y\otimes z) = z \otimes y$.

The $R$-matrix is normalized so that
\beq
R_{\cV_i\, \cV_j}(x) \, v_i \otimes v_j = v_i \otimes v_j \; .
\eeq  
It is sufficient to understand the action of the $R$-matrix on the two-dimensional space with basis $\{e^- v_i \otimes v_j \; , \;  v_i \otimes e^- v_j\}$. One computes
\begin{align}
&R_{\cV_i\, \cV_j}(x) (e^- v_i \otimes v_j) = \frac{x\, \hbar^{m_j}-\hbar^{m_i}}{x-\hbar^{m_i + m_j}}\, e^- v_i \otimes v_j +  \frac{1- \hbar^{2m_j}}{x-\hbar^{m_i + m_j}}\, v_i \otimes e^- v_j \; , \label{Rmatrixaction}\nonumber\\
&R_{\cV_i\, \cV_j}(x) (v_i \otimes e^- v_j) = \frac{x\, (1- \hbar^{2m_j})}{x-\hbar^{m_i + m_j}}\, e^- v_i \otimes v_j +  \frac{x\, \hbar^{m_i}-\hbar^{m_j}}{x-\hbar^{m_i + m_j}}\, v_i \otimes e^- v_j  \; ,
\end{align}
Above, we have defined complex parameters
\beq\label{midef}
m_i = \frac{\langle  \nu_i,e^+ \rangle}{2} \; ,
\eeq
where $e^+=\alpha$ is the positive simple root of $A_1$. 
Integral solutions to the $U_\hbar(\widehat{A_1})$ qKZ equations \eqref{qKZ} have been constructed by Matsuo \cite{cmp/1104252137,cmp/1104254019}.
Such solutions are valued in an $L$-dimensional space of weight
\beq
\nu_0-\nu_\infty = \sum_{d=1}^{L}\nu_d - N\, e^+ \; .
\eeq
All the features of tame ramification will already be manifest for $N=1$, so we henceforth specialize to that case in order not to overburden the presentation.
Then, for all $i=1,\ldots,L$, define the integrals
\beq\label{integrandqKZ}
{\bf F}_i(\vec{x}) = \oint_\Gamma \frac{dy}{y} \, y^\eta \, K_i(y,\vec{x})\, \prod_{d=1}^{L} \frac{(y/\tilde{x}_d; q)_\infty}{(q^{\sigma_d}\,y/\tilde{x}_d; q)_\infty} \; ,
\eeq
where the complex numbers $\sigma_d$ and $\eta$ are determined in terms of the highest weights $\nu_d$ and $\nu_0$ as
\beq
q^{\sigma_d} = \hbar^{\langle \nu_d, e^+\rangle}\; , \qquad\; q^{\eta} = \hbar^{-\langle \nu_0 , e^+\rangle}
\eeq
and
\beq\label{Ki}
K_i(y,\vec{x})=\frac{1}{1-y/\tilde{x}_i} \prod_{d=1}^{i-1}\frac{1-q^{\sigma_d}\, y/\tilde{x}_d}{1-y/\tilde{x}_d}\; .
\eeq
We renormalize the $F_i(\vec{x})$ as
\beq
\varphi_i(\vec{x}) = q^{(\sigma_{i+1}+\ldots+\sigma_{L})/2}\; x^{\sigma_1}_1 \ldots x^{\sigma_{L}}_{L}\, F_i(q^{\sigma_1/2}\, x_1,\ldots, q^{\sigma_{L}/2}\, x_{L}) \; .
\eeq
Then Matsuo showed that 
\beq\label{qKZMatsuosol}
\Psi(\vec{x}) = \sum_{i=1}^{L} \varphi_i(x_1,\ldots,x_{L})\, v_1\otimes\ldots\otimes e^- v_i \otimes\ldots\otimes v_{L}
\eeq
solves the $U_\hbar(\widehat{A_1})$ qKZ equations.
More precisely, this is a solution to the qKZ equations for any choice of contour for which the integral remains invariant under $q$-shift. In particular, the contours are such that $\int \frac{dy}{y} \, I(q\, y) = \int \frac{dy}{y} \, I(y)$, with $I(y)$ the integrand in \eqref{integrandqKZ}.\\

To make the appearance of K-theoretic stable envelopes explicit, it is useful to rescale the equivariant parameters $\tilde{x}_d$ by $q^{\sigma_d}$ and write 
\beq\label{integrandqKZ2}
{\bf F}_i(\vec{x}) = \oint_\Gamma \frac{dy}{y} \, y^\eta \, K_i(y,\vec{x})\, \prod_{d=1}^{L} \frac{(q^{-\sigma_d}\,y/\tilde{x}_d; q)_\infty}{(y/\tilde{x}_d; q)_\infty} \; ,
\eeq
with
\beq\label{Ki2}
K_i(y,\vec{x})=\frac{1}{1-q^{-\sigma_i}\,y/\tilde{x}_i} \prod_{d=1}^{i-1}\frac{1- y/\tilde{x}_d}{1-q^{-\sigma_d}\,y/\tilde{x}_d}\; .
\eeq
But this is equal to 
\beq\label{integrandqKZ3}
{\bf F}_i(\vec{x}) = \oint_\Gamma \frac{dy}{y} \, y^\eta \, \text{Stab}^K_{i}(y,\vec{x})\, \prod_{d=1}^{L} \frac{(q^{1-\sigma_d}\,y/\tilde{x}_d ; q)_\infty}{(y/\tilde{x}_d; q)_\infty} \; ,
\eeq
where
\beq\label{Ki3}
\text{Stab}^K_{i}(y,\vec{x})= \prod_{d=1}^{i-1}(1- y/\tilde{x}_d)\; \prod_{d=i+1}^{L} (1-q^{-\sigma_d}\,y/\tilde{x}_d)
\eeq
is the stable envelope for the $i$-th fixed point.\\

\subsection{The $\cW_{q,t}(A_1)$-algebra}

The $q$-conformal blocks of $\cW_{q,t}(A_1)$ are the magnetic blocks, which are components of the Vertex function 
\beq\label{correlatordefWA1}
\bold{V}=\left\langle v_{\mu_\infty},  {Q}^\vee \prod_{d=1}^{L} \cV_{\{\lambda\}_d}(\tilde{x}_d)\; v_{\mu_0} \right\rangle  \, .
\eeq
The screening charge ${Q}^\vee$ is the $(-1)$-st Fourier coefficient of the screening current,
\begin{align}
&{Q}^\vee =\int dy\, S^\vee(y) \; , \label{screeningdefA1}\nonumber\\
& S^\vee(y) = y^{-\alpha[0]}\,: \exp\left(\sum_{k\neq 0}{ \alpha[k] \over q^{k\, \over 2} - q^{-\,{k \,  \over 2}}} \, y^k\right):  \; .
\end{align}
The modes $\alpha[k]$ with $k\in\mathbb{Z}$ obey the deformed Heisenberg algebra relations
\begin{align}\label{commutatorgeneratorsA1}
[\alpha[k], \alpha[n]] = {1\over k} (q^{k\over 2} - q^{-{k\over 2}})(t^{{k\over 2} }-t^{-{k\over 2} })(q^{k\over 2}\,t^{-k\over 2} + q^{-k\over 2}\,t^{k\over 2} ) \delta_{k, -n} \; .
\end{align}
The vector $v_{\mu_0}$ defines a Fock space representation as:
\begin{align}
\alpha[0]\, v_{\mu_0} &= \langle\mu_0, \alpha\rangle \, v_{\mu_0} \label{eigenvalueA1here}\nonumber\\
\alpha[k]\, v_{\mu_0} &= 0\, , \qquad\qquad\;\; \mbox{for} \; k>0\; .
\end{align}
Our first task is to construct the deformed primary vertex operators $V_{\{{\lambda}\}_d}$. According to our proposal, each such operator, say the $d$-th one among $L$ of them, is defined via a set of $J_d$ weights $\{{\lambda}\}_d=\{\lambda_{d,1},\ldots,\lambda_{d,J_d}\}$ chosen among the fundamental representations of $U_\hbar(\widehat{^L \fg})$, with a constraint on the degree of their respective Drinfeld polynomials.

Explicitly, consider the Cartan eigenspace decomposition of the fundamental representation $\widehat{V}$ in $U_\hbar(\widehat{A_1})$. This representation is two-dimensional, and admits the following linear space decomposition as a $U_\hbar({A_1})$-module:
\beq\label{A1eigenspace}
\widehat{V} = V_Y  \oplus  V_{Y^{-1}} \; .
\eeq
Consider the two weight spaces $V_{\lambda_1}=V_Y$ and $V_{\lambda_2}=V_{Y^{-1}}$ in $U_\hbar(\widehat{A_1})$. As vector spaces, $V_{Y}=\mathbb{C} v_{Y}$ with $v_{Y}$ the highest weight vector of the fundamental representation, while $V_{Y^{-1}}=\mathbb{C} v_{Y^{-1}}$, with $v_{Y^{-1}}$ the lowest weight vector of the fundamental representation. Let us derive the Drinfeld polynomials for $V_{\lambda_1}$ and $V_{\lambda_2}$ explicitly.


\vspace{8mm}

\noindent
------ \emph{FIRST METHOD} ------\\

First, we will work straight from the representation theory of $Rep(U_{\hbar}(\widehat{A_1}))$. Instead of the Drinfeld-Jimbo presentation we reviewed above \eqref{A1algebra}, we will need instead to consider Drinfeld's so-called second realization of the quantum affine algebra \cite{Drinfeld:1987sy}. It consists of generators $x^{\pm}_k, H_r, K^{\pm 1}$, where $k\in\mathbb{Z}$ and $r\in\mathbb{Z}^*$, and $c^{\pm 1}$ in the center, along with the relations:
\begin{align}
& K \, K^{-1} = K^{-1} \, K = 1 \; , \qquad c \, c^{-1} = c^{-1} \, c = 1 \; ,\label{A1algebranew}\nonumber\\
& [K,H_r]=0 \; , \qquad\qquad\qquad\;\,  K\, X^{\pm}_k = \hbar^{\pm 1}\, X^{\pm}_k\, K \; ,\nonumber\\
& [H_r, X^{\pm}_k] = \pm\frac{1}{r}\,\frac{\hbar^{r}-\hbar^{-r}}{\hbar^{1/2}-\hbar^{-1/2}}\, c^{\mp|r|}\, X^{\pm}_{r+k}  \; , \nonumber\\
& [H_r, H_{r'}] = \delta_{r,-r'}\frac{1}{r}\,\frac{\hbar^{r}-\hbar^{-r}}{\hbar^{1/2}-\hbar^{-1/2}}\, \frac{c^{r}-c^{-r}}{\hbar^{1/2}-\hbar^{-1/2}} \; , \nonumber\\
& X^{\pm}_{\kappa+1}\, X^{\pm}_{k'} - \hbar^{\pm 1}\, X^{\pm}_{k'}\, X^{\pm}_{\kappa+1} = \hbar^{\pm 1}\, X^{\pm}_{k}\, X^{\pm}_{k'+1}  - X^{\pm}_{k'+1}\, X^{\pm}_k \; , \nonumber\\
&[X^+_k, X^-_{k'}] = \frac{1}{\hbar^{1/2}-\hbar^{-1/2}}(c^{k-k'}\psi^+_{\kappa+k'} -c^{k'-k} \psi^-_{\kappa+k'})  \; .
\end{align}
with 
\begin{align}
&\sum_{k=0}^\infty \psi^{\pm}_{\pm k}\, z^{\pm k} = K^{\pm} \exp\left(\pm(\hbar-\hbar^{-1})\sum_{s=1}^\infty H_{\pm s}\, z^{\pm s}\right) \; ,\label{A1algebranew2}
\end{align}
and $\psi^+_k=0=\phi^-_{-k}$ for $k<0$. 
In this formalism, a highest weight vector $v$ is defined so as to satisfy\footnote{From the action of $c$ on $v$ and the fourth line in  \eqref{A1algebranew}, it follows that $[H_r, H_{r'}]=0$ for highest weight modules, showing the existence of a commutative ``Cartan subalgebra" for those modules, like in the finite case $Rep(U_{\hbar}({A_1}))$. When one further fixes the sign ambiguity by imposing that $c$ act as the identity on $v$, the corresponding representation is said to be of type 1. All the representations we consider here are of this type.} 
\begin{align}
&\psi^+_k\cdot v=\lambda^+_k\, v\;, \; \qquad k\geq 0, \;\;\lambda^+_k\in\mathbb{C}\; ,\nonumber\\
&\psi^-_k\cdot v=\lambda^-_k\, v\;, \; \qquad k\leq 0, \;\;\lambda^-_k\in\mathbb{C}\; ,\nonumber\\
&X^{+}_k\cdot v = 0\;, \;\;\;\;\;\,\qquad k\in\mathbb{Z}\; ,\nonumber\\
&(c\cdot v= v ) 
\end{align} 
Our task is to identify the weights  $\lambda^+_k$ explicitly. The weights $\lambda^-_k$ will be redundant in this discussion, so we ignore them from now on.
The strategy is to invoke the evaluation representations from $Rep(U_{\hbar}(\widehat{A_1}))$ to $Rep(U_{\hbar}({A_1}))$, and then pull back the eigenvalue content of the latter to obtain the weights $\lambda^+_k$. Explicitly, in Drinfeld's second realization, the surjective homomorphism $\text{ev}_x : U_\hbar(\widehat{A_1})\rightarrow U_\hbar(A_1)$ reads
\begin{align}
&\widetilde{\text{ev}}_x(c^{\pm 1}) = 1 \; , \qquad\qquad\qquad\qquad \;\;\;\;\; \widetilde{\text{ev}}_x(K)= K \; , \label{evalmor2}\nonumber\\
&\widetilde{\text{ev}}_x(x^{+}_k) = \hbar^{-k/2}\, x^k\,  K^k\, e^+ \; , \qquad\qquad   \widetilde{\text{ev}}_x(x^{-}_k) = \hbar^{-k/2}\, x^k\,   e^-\, K^k  \; ,
\end{align}
for $x\in\mathbb{C}^\times$. Then, consider a basis $(v_{Y}, v_{Y^{-1}})$ of the fundamental representation $V$ of $U_\hbar(A_1)$, on which the generators act as\footnote{This is simply the quantum analog of the familiar action of the Pauli matrices $\sigma_z$, $\sigma_+$ and $\sigma_-$ on a qubit in the fundamental representation of $A_1$.}:
\begin{align}
& K\cdot v_{Y} = \hbar^{1/2}\, v_{Y} \; , \qquad\;\;\;\; K\cdot v_{Y^{-1}} = \hbar^{-1/2} v_{Y^{-1}}\; , \label{finiteaction}\nonumber\\
&e^+\cdot v_{Y^{-1}}  = v_{Y} \; , \qquad\qquad   e^+\cdot v_{Y}  = 0 \; ,\nonumber\\
&e^-\cdot v_{Y}  = v_{Y^{-1}} \; , \qquad\qquad   e^-\cdot v_{Y^{-1}}  = 0 \; .
\end{align}
As a vector space, this fundamental representation $V$ coincides with the fundamental representation  $\widehat{V} = V_{Y}  \oplus  V_{Y^{-1}}$ in the quantum affine case: $V_{Y}=\mathbb{C} v_{Y}$ with $v_{Y}$ the highest weight vector of the fundamental representation, while $V_{Y^{-1}}=\mathbb{C} v_{Y^{-1}}$, with $v_{Y^{-1}}$ the lowest weight vector of the fundamental representation. Making use of the evaluation homomorphism $\text{ev}_x$, we derive the action of the quantum affine raising and lowering generators on the basis vectors:
\begin{align}
&X^+_k\cdot v_{Y^{-1}} = x^k \, v_{Y} \; , \qquad\qquad   X^+_k\cdot v_{Y}  = 0 \; ,\label{finiteaction2}\nonumber\\
&X^-_k\cdot v_{Y}  = x^k \, v_{Y^{-1}} \; , \qquad\qquad   X^-_k\cdot v_{Y^{-1}}  = 0 \; .
\end{align}
Using the commutation relation in the last line of \eqref{A1algebranew}, we easily deduce the eigenvalues $\lambda^+_k$:
\begin{align}
&\psi^+_0 \cdot v_{Y} = \hbar^{1/2} \, v_{Y} \; , \qquad\qquad\qquad\;\;\,   \;\;\psi^+_0 \cdot v_{Y^{-1}} = \hbar^{-1/2} \, v_{Y} \; ,\label{finiteaction3}\nonumber\\
&\psi^+_k \cdot v_{Y} = x^k \, (\hbar^{1/2}-\hbar^{-1/2}) \, v_{Y} \; , \qquad   \psi^+_k \cdot v_{Y^{-1}} = x^k \, (\hbar^{-1/2}-\hbar^{1/2}) \, v_{Y^{-1}}\; , \;\;\;\; k>0 \; .
\end{align}
Following Chari and Pressley \cite{cmp/1104248585}, the generating series of eigenvalues is a ratio of polynomials. For the vector space $V_{Y}$, one finds\footnote{The general statement is as follows: for any weight space $V_{\mu}$ present in the decomposition of a $U_{\hbar}(\widehat{\fg})$ finite-dimensional representation $\widehat{V} = \bigoplus_{\mu} V_{\mu}$,  there exist polynomials ${\cA}^{+}_{\mu}(z),\;\; {\cA}^{-}_{\mu}(z)\in\mathbb{C}[z]$, of constant term ${\cA}^{+}_{\mu}(0)=1={\cA}^{-}_{\mu}(0)$, which satisfy
	\beq
	\sum_{k=0}^\infty \lambda^+_{k,\mu} \, z^k = \hbar^{(\text{deg}({\cA}^{+}_{\mu})-\text{deg}({\cA}^{-}_{\mu}))/2}\, \frac{{\cA}^{+}_{\mu}(\hbar^{-1/2}z)}{{\cA}^{+}_{\mu}(\hbar^{1/2}z)}\,\frac{{\cA}^{-}_{\mu}(\hbar^{1/2}z)}{{\cA}^{-}_{\mu}(\hbar^{-1/2}\,z)}\; .
	\eeq
}
\beq\label{highestpoly}
\sum_{k=0}^\infty \lambda^+_{k,Y} \, z^k = \hbar^{1/2} +  (\hbar^{1/2}-\hbar^{-1/2}) \sum_{k=1}^\infty x^k \, z^k = \hbar^{1/2}\, \frac{1-\hbar^{-1}\, x\, z}{1- x\, z} \equiv \hbar^{1/2}\, \frac{{\cA}^{+}_{Y}(\hbar^{-1/2}z)}{{\cA}^{+}_{Y}(\hbar^{1/2}z)}\; ,
\eeq
where in the last step we defined the polynomial ${\cA}^{+}_{Y}(z) = (1- x\, \hbar^{-1/2} z)$. 
Similarly for the vector space $V_{Y^{-1}}$, one finds
\beq\label{lowestpoly}
\sum_{k=0}^\infty \lambda^+_{k,Y^{-1}} \, z^k = \hbar^{-1/2}\, \frac{1-\hbar\, x\, z}{1- x\, z} \equiv \hbar^{-1/2}\, \frac{{\cA}^{-}_{Y^{-1}}(\hbar^{1/2} \,z)}{{\cA}^{-}_{Y^{-1}}(\hbar^{-1/2}\,z)}\; ,
\eeq
where ${\cA}^{-}_{Y^{-1}}(z) = (1- x\, \hbar^{1/2} z)$. 
Each weight space is in bijection with its set of polynomials, so we are done.  We normalize the highest weight polynomial to be 
\beq
{\cA}^{+}_{Y}(z)|_{x=\hbar^{1/2}} = (1- z) 
\eeq
so that
\beq
{\cA}^{-}_{Y^{-1}}(z)|_{x=\hbar^{1/2}} = (1- \hbar\, z) \; .
\eeq
We call these normalized polynomials the Drinfeld polynomials.\\
 
Our definition of a deformed $\cW_{q,t}(A_1)$ primary operator, say the $d$-th one, is 
\beq\label{operator1A1again}
\cV_{\{{\lambda}\}_d}(\tilde{x}_d) =\; :\prod\limits_{s=1}^{J_d=2}\prod_{i=1}^{\deg(\cA^{+}_{\lambda_{d,s}})}\Lambda\left(\hbar^{-\mathfrak{a}^{+}_{\lambda_{d,s},i}}\, x_{d,s}\right) \prod_{j=1}^{\deg(\cA^{-}_{\lambda_{d,s}})}\Lambda^{-1}\left(\hbar^{-\mathfrak{a}^{-}_{\lambda_{d,s},i}}\, x_{d,s}\right): \; ,
\eeq
with the arguments of the operators
\beq\label{coweightvertexdefA1}
\Lambda^{\pm 1}(x) =\; : \exp\left(\pm \, \sum_{k\neq 0}{w[k] \over (q^{k\over 2} - q^{-\,{k \over 2}})(t^{k\over 2} - t^{-\,{k \over 2}})} \, x^k\right):  \; , 
\eeq
given by the Drinfeld roots. We find at once $\mathfrak{a}^{+}_{Y,1}=0$ and $\mathfrak{a}^{-}_{Y^{-1},1}=1$ as the only nonzero integers in ${\cA}^{+}_{Y}(z) = (1- \hbar^{\mathfrak{a}^{+}_{Y,1}} z)$ and   ${\cA}^{-}_{Y^{-1}}(z) = (1- \hbar^{\mathfrak{a}^{-}_{Y^{-1},1}} z)$, respectively. In the end, we propose the following deformed primary operator expression:
\beq\label{operator1A1againlast}
\cV_{\{{\lambda}\}_d}(\tilde{x}_d) =\; :\Lambda\left(x_{d,1}\right) \; \left[\Lambda\left(\hbar^{-1}\, x_{d,2}\right)\right]^{-1}: \; .
\eeq
\begin{remark}
Since $A_1$ is simply-laced, the Drinfeld polynomials can simply be read off the Frenkel-Reshetikhin $\hbar$-character map  \cite{Frenkel:qch}:
\begin{align}
\chi_\hbar : Rep(U_{\hbar}(\widehat{A_1})) &\rightarrow \mathbb{Z}[Y^{\pm 1}(z)]\label{qcharacterA1}\nonumber\\
\widehat{V}\;\;\; &\mapsto t^{\widehat{V}}(z)=Y(z)+ Y^{-1}(\hbar^{-1} z)\, .
\end{align}
The only nontrivial polynomials are ${\cA}^{+}_{Y}(z)  = (1 - z)$, which accounts for the first term $Y(z)$, and ${\cA}^{-}_{Y^{-1}}(z) = (1- \hbar\, z)$, which accounts for the second term  $Y^{-1}(\hbar^{-1} z)$.\\
\end{remark}

%
%
%

\noindent
------ \emph{SECOND METHOD:} ------\\

We construct the generating currents of $\cW_{q,t}(\fg)$ as the commutant of the screening charges, and then invoke the $(q,t)$-character map \eqref{qtcharacter} to make contact with the representation ring of the quantum affine algebra. 
The algebra $\cW_{q,t}(A_1)$ has a single generator, the deformed Virasoro stress tensor, constructed as the commutant of the $A_1$ screening charge \eqref{screeningdefA1} (and its dual). A computation shows that this commutant has the following form:
\beq\label{examplestressA1}
T^{\widehat{V}}(z) = \cY(z) + \left[\cY(\sqrt{t^2/q^2}\, z)\right]^{-1} \; .
\eeq
Here, $\cY$ is the $\fg=A_1$ version of the vertex operators \eqref{YoperatorToda}. 
But  $T^{\widehat{V}}$ can also be understood as a $(q,t)$-character homomorphism $Rep(U_{q}(\widehat{A_1})) \rightarrow \cW_{q,t}(A_1)$, a $t$-refinement of Frenkel and Reshetikhin's $q$-character for the fundamental representation $\widehat{V}$ of $U_q(\widehat{A_1})$. Note the argument of the  $\cY$ operators above does not depend on $q$ and $t$ separately, but only on the ratio $q/t=\hbar$ (as is the case for all simply-laced $\fg_\text{o}$). One reinterprets the $(q,t)$-character as a map $Rep(U_{\hbar}(\widehat{A_1})) \rightarrow \cW_{q,t}(A_1)$, where the current $T^{\widehat{V}}(x)$ stands for the image of the eigenspace decomposition $\widehat{V} = V_{Y}  \oplus  V_{Y^{-1}}$ in $Rep(U_{\hbar}(\widehat{A_1}))$, and the weight spaces $V_{Y}$ and $V_{Y^{-1}}$ have respective Drinfeld polynomials ${\cA}^{+}_{Y}(z) = (1- z)$ and ${\cA}^{-}_{Y^{-1}}(z) = (1- \sqrt{q^2/t^2}\, z)$.

By our proposal, this data uniquely characterizes a deformed $\cW_{q,t}(A_1)$ primary operator:
\begin{align}\label{opA1}
\cV_{\{{\lambda}\}_d}(\tilde{x}_d) =\; :\Lambda\left(x_{d,1}\right) \; \left[\Lambda\left(\sqrt{t^2/q^2}\, x_{d,2}\right)\right]^{-1}: \; .
\end{align}
After rewriting the nontrivial Drinfeld root as $\sqrt{t^2/q^2}=\hbar^{-1}$, this agrees nicely with the operator \eqref{operator1A1againlast} from the previous construction.\\

The deformed primary has a center of mass position $\tilde{x}_d$ on $\cC=\mathbb{C}^\times$. Each individual vertex operator $\Lambda^\pm$ is itself located at a position $x_{d,j}$, which is $q$-shifted relative to $\tilde{x}_d$ by some amount $\sigma_{d,j}\in\mathbb{C}$:
\beq\label{compositionA1}
x_{d,j} \equiv \tilde{x}_d \, q^{\sigma_{d,j}} \; ,\qquad\;\; j=1, 2 \; .
\eeq
Correspondingly, we define the ``momentum" $\fg$-coweight of the $d$-th deformed primary $\cV_{\{{\lambda}\}_d}$ as 
\beq\label{commomentumA1}
\widetilde{\sigma}_{d} \equiv \sigma_{d,1}\, \underline{\lambda_1} + \sigma_{d,2}\, \underline{\lambda_2}  \; , \qquad\text{with}\; \; \underline{\lambda_1}=[\phantom{-}1] \; ,\;\;\; \underline{\lambda_2}=[-1] \; .
\eeq
In the rest of this section, we will also make repeated use of the notation
\beq\label{relativemomA1}
\sigma_d \equiv \langle \widetilde{\sigma}_{d},e^+\rangle = \sigma_{d,1}-\sigma_{d,2} \; ,
\eeq
Inside the chiral correlator \eqref{correlatordefWA1}, we insert $L$ such primaries, which are simply $L$ copies of the above deformed primary,
\beq
\prod_{d=1}^L \cV_{\{{\lambda}\}_d}(\tilde{x}_d)\; ,\qquad \; d=1,\ldots,L \;. 
\eeq
We assume that the positions $\tilde{x}_d$ on $\cC$ are distinct and generic. 
Each of the $L$ operators $\cV_{\{\lambda\}_d}(\tilde{x}_d)$ has a well-defined conformal limit $t=q^{\beta}$, $q\rightarrow 1$, where it becomes the Virasoro primary vertex operator $e^{\langle \widetilde{\sigma}_d , \phi(x)\rangle}$, with $\phi(x)$ the corresponding free boson on $\cC$.\\

The vertex function $\bold{V}$ evaluates to
\beq\label{correlatordefWA1againwow}
\left\langle v_{\mu_\infty},  {Q}^\vee \;\prod_{d=1}^{L} \cV_{\{{\lambda}\}_d}(\tilde{x}_d) \; v_{\mu_0} \right\rangle = \oint_{\Gamma} \frac{dy}{y}\, y^{-\langle  \mu_0, \alpha  \rangle} \, \prod_{d=1}^{L}\frac{\left(\sqrt{q/t} \; y/x_{d,1}\, ; q\right)_\infty}{\left(\sqrt{t/q} \; y/x_{d,2}\, ; q\right)_\infty} \; ,
\eeq
in agreement with the general formula \eqref{vertexnew}.
After rescaling the equivariant parameters $x_{d,s}$ by appropriate factors of $q$ and $t$ and switching to the center of mass coordinates \eqref{compositionA1} and \eqref{relativemomA1}, this chiral correlator is conveniently written as 
\beq\label{correlatordefWA1again2}
\oint_{\Gamma} \frac{dy}{y}\, y^{-\langle  \mu_0, \alpha  \rangle} \, \prod_{d=1}^{L} \frac{(q^{1-\sigma_d}\,y/\tilde{x}_d; q)_\infty}{(y/\tilde{x}_d; q)_\infty} \; .
\eeq
Here, $\Gamma$ is a sum of contours enclosing all possible poles. For $|q|<1$, we choose a contour $C_k$ which encloses ``$\infty$"  and the infinite number of poles at 
\beq
\label{contourA1}
C_k \; : \qquad y=q^{-n}\, \tilde{x}_k\, , \qquad\; n=0,1,2,\ldots
\eeq
This translates to choosing a particular component $\bold{V}_k$ of the vertex function $\bold{V}$. The integral is readily evaluated by residues:
\beq
\label{vertexA1}
\bold{V}_k = (\tilde{x}_k)^\eta \frac{(q^{1-\sigma_k}; q)_\infty}{(q; q)_\infty}\prod_{d\neq k}^L \frac{(q^{1-\sigma_d} \tilde{x}_k/\tilde{x}_d; q)_\infty}{(\tilde{x}_k/\tilde{x}_d; q)_\infty}\sum_{r\geq 0}\left(\frac{z}{\prod_i q^{\sigma_d-1}}\right)^r\prod_{d=1}^L \frac{(q^{\sigma_d} \tilde{x}_d/\tilde{x}_k; q)_r}{(q\, \tilde{x}_d/\tilde{x}_k; q)_r}
\eeq
We have introduced the fugacity $z=q^{\langle  \mu_0, \alpha  \rangle}$, and the identification $\eta=-\langle  \mu_0, \alpha  \rangle$. We also made use of the finite $q$-Pochhammer symbol notation $(x;q)_r=\prod_{k=0}^{r-1}(1-x\, q^k)$, with $(x;q)_0=1$.\\  

Up to normalization, this is a $q$-hypergeometric function:
\beq
\label{vertexA1hyper}
\bold{V}_k = (\tilde{x}_k)^\eta \frac{(q^{1-\sigma_k}; q)_\infty}{(q; q)_\infty}\prod_{i\neq k} \frac{(q^{1-\sigma_i} \tilde{x}_k/\tilde{x}_i; q)_\infty}{(\tilde{x}_k/\tilde{x}_i; q)_\infty}\, \pFq{n}{n}{q^{\sigma_i} \tilde{x}_k/\tilde{x}_i}{q\,\tilde{x}_k/\tilde{x}_i}{\frac{z}{\prod_i q^{\sigma_i-1}}} \; ,
\eeq
with 
\beq
\label{hypergeo}
\pFq{n}{n}{a_i}{b_i}{z}= \sum_{r\geq 0}z^r\prod_{i=1}^n \frac{(a_i; q)_r}{(b_i; q)_r}   \; .
\eeq
The vertex function $\bold{V}$, as well as the qKZ integral solutions \eqref{integrandqKZ}, are analytic in the chamber
\beq\label{chamberz}
\fC_C = \{|z|<1\} \; ,
\eeq  
but they are not jointly analytic in any $\{\tilde{x}_d\}$-chamber. For instance, looking at the components of the vertex function, it is clear from our expressions that $\bold{V}_1$ is analytic when $|x_1|<|x_2|$, while $\bold{V}_2$  is instead analytic when $|x_2|<|x_1|$ (note that both components are $z$-analytic). 
Since bona fide $q$-conformal blocks of $U_\hbar(\widehat{A_1})$ are correlation functions of chiral operators on $\cC$, they must be analytic in \emph{some} $\{\tilde{x}_d\}$-chamber. To remedy this, we first fix a chamber
\beq\label{chamber}
\fC_H=\{|x_1|<|x_2|<\ldots<|x_L|\} \; ,
\eeq
and then perform a linear change of basis between $z$-solutions and $\{\tilde{x}_d\}$-solutions to obtain a new vertex function analytic in chamber $\fC_H$: 
\beq\label{vertexellstab}
\bold{V}_{\fC_H}= \oint_{C} \frac{dy}{y}\, y^{\eta} \, \prod_{d=1}^{L} \frac{(q^{1-\sigma_d}\,y/\tilde{x}_d; q)_\infty}{(y/\tilde{x}_d; q)_\infty}\;  \fB_{\fC_H} \; .
\eeq
The change of basis is implemented using a matrix $\fB_{\fC_H}$ of $q$-periodic pseudo-constants: 
\beq\label{changeofmatrix}
\fB_{\fC_H,k}(y,z,\{\tilde{x}_d\})=U_{\fC_H,k}(z,\{\tilde{x}_d\})\;\frac{\text{Stab}^{Ell}_{\fC_H,k}(y,z,\{\tilde{x}_d\})}{\displaystyle  \prod_{d=1}^L\Theta\left(q^{\sigma_d}\tilde{x}_d/y\right)}\; \text{e}(y,z)^{-1} \, ,
\eeq
where $\text{Stab}^{Ell}_{\fC_H,k}$ is the elliptic stable envelope  for the $k$-th fixed point of a quiver variety we will spell out in the next section. 
\beq\label{ellstab}
\text{Stab}^{Ell}_{\fC_H,k}(y,z,\{\tilde{x}_d\})=\frac{\displaystyle \prod_{d<k}\Theta\left(\tilde{x}_d/y\right)\Theta\left(q^{\sum_{d=1}^k \sigma_d}\tilde{x}_k/y z\right)\prod_{d>k }\Theta\left(q^{\sigma_d}\tilde{x}_d/y\right)}{\displaystyle \Theta\left(q^{\sum_{d=1}^k \sigma_d}/z\right)}
\eeq
The additional factors present in $\fB_{\fC_H,k}$ ensure that the matrix is indeed $q$-periodic in the variables $y, z, x_1, x_2, \ldots, x_L$; we have an overall normalization by the theta function $\Theta\left(q^{\sum_{d=1}^k \sigma_d}/z\right)$, as well as
\beq\label{efunc}
\text{e}(y,z)^{-1}=\exp\left[\frac{\ln(y)\ln(z)}{\ln(q)}\right] 
\eeq
and
\beq\label{Ufunc}
U_{\fC_H,k}(z,\{\tilde{x}_d\})= \exp\left[\frac{\ln(\tilde{x}_k)\ln(q^{\sum_{d=1}^k \sigma_d}/z)-\sum_{d\leq k}\ln(\tilde{x}_d)\ln(q^{\sigma_d})}{\ln(q)}\right] \; .
\eeq
The presence of the factor $\fB_{\fC_H}$ inside the integrand of $\bold{V}_{\fC_H}$ \eqref{vertexellstab} results in new poles, which come in two series: they are either located at 
\beq\label{poles1}
y=q^{-n}\, \tilde{x}_d\, , \qquad\; n=0,1,2,\ldots, \qquad d\geq k \; ,
\eeq
in which case they accumulate at $y=\infty$ (we are assuming throughout that $|q|<1$), or they are located at
\beq\label{poles2}
y=q^{n}\, \tilde{x}_d\, q^{\sigma_d}\, , \qquad\; n=0,1,2,\ldots, \qquad d\leq k \; ,
\eeq
in which case they accumulate at $y=0$. We define the contour $C$ in \eqref{vertexellstab} to enclose the poles \eqref{poles1}, while the poles \eqref{poles2} near 0 are kept outside.\\ 

From the definition of the matrix 
\beq\label{polesub}
\fB^{k'}_{\fC_H,k}=\fB_{\fC_H,k}(x_{k'}) \; ,
\eeq 
it follows at once that 
\beq\label{polesubtriang}
\fB^{k'}_{\fC_H,k}=0 \; \qquad \text{if} \; k>k' \; ,
\eeq 
meaning $\fB^{k'}_{\fC_H,k}$ is triangular and the elliptic stable envelope for $\fC_H$ has adequate support. 

The regularity of the integral $\bold{V}_{\fC_H}$ depends on whether or not the poles of the integrand pinch the contour; the poles  \eqref{poles1} and \eqref{poles2} on opposite sides of the contour coalesce if and only if
\beq\label{pinchcontour}
\frac{\tilde{x}_j}{\tilde{x}_i} = q^{n+\sigma_i} \, , \qquad\; n=0,1,2,\ldots, \qquad j\geq k\geq i \; .
\eeq 
But the chamber $\fC_H$ was defined such that $|\tilde{x}_j/\tilde{x}_i| > 1$ for all $j>i$, ensuring no singularities. In particular, $\bold{V}_{\fC_H}$ is holomorphic in a neighborhood of 0, as long as we stay in the chamber $\fC_H$. Then, the change of basis reads
\beq\label{changebasis}
\bold{V}_{\fC_H,k} = \sum_{k'=1}^L \bold{V}_\fC\,\fB^{k'}_{\fC_H,k} \; .
\eeq 
As an example, let us specialize to $L=2$, the insertion of two $A_1$ punctures, meaning two $q$-primary operators of $\cW_{q,t}(A_1)$, labeled respectively by the momenta $\sigma_1$ and $\sigma_2$. In that case, the vertex function $\bold{V}$ has two components, 
\begin{align}
\nonumber\label{V12z}
&\bold{V}_{1} = (x_1)^\eta \frac{(q^{1-\sigma_1})_\infty}{(q)_\infty}\, \frac{(q^{1-\sigma_2} x_1/x_2)_\infty}{(x_1/x_2)_\infty}\; \pFq{2}{2}{q^{\sigma_1}, \;\;q^{\sigma_2} x_2/x_1}{q,\qquad q\, x_2/x_1}{\frac{z}{q^{\sigma_1+\sigma_2-2}}} \; ,\\
&\bold{V}_{2} = (x_2)^\eta \frac{(q^{1-\sigma_2})_\infty}{(q)_\infty}\, \frac{(q^{1-\sigma_1} x_2/x_1)_\infty}{(x_2/x_1)_\infty}\; \pFq{2}{2}{q^{\sigma_2}, \;\;q^{\sigma_1} x_1/x_2}{q,\qquad q\, x_1/x_2}{\frac{z}{q^{\sigma_1+\sigma_2-2}}} \; .
\end{align}
Note that both components are analytic in $|z|<1$. Using Heine's transformation for $q$-hypergeometric series, it is easy to see that $\bold{V}_{1}$ and $\bold{V}_{2}$ cannot both be analytic in any given $\{\tilde{x}_i\}$-chamber \emph{simultaneously}:
\begin{align}
\nonumber\label{V12zHeine}
&\bold{V}_{1} = (x_1)^{\eta'} \frac{(q^{1-\sigma_1})_\infty}{(q)_\infty}\, \frac{(q^{2-\sigma_2} z)_\infty\Theta(q^{\sigma_2}x_2/x_1)}{(q^{2-\sigma_1-\sigma_2} z)_\infty\Theta(x_1/x_2)}\; \pFq{2}{2}{q^{1-\sigma_2}, \;\;q^{2-\sigma_1-\sigma_2}\, z}{q,\qquad q^{2-\sigma_2}\,z}{q^{\sigma_2}\frac{x_2}{x_1}} \; ,\\
&\bold{V}_{2} =(x_2)^{\eta'} \frac{(q^{1-\sigma_2})_\infty}{(q)_\infty}\, \frac{(q^{2-\sigma_1} z)_\infty\Theta(q^{\sigma_1}x_1/x_2)}{(q^{2-\sigma_1-\sigma_2} z)_\infty\Theta(x_2/x_1)}\; \pFq{2}{2}{q^{1-\sigma_1}, \;\;q^{2-\sigma_1-\sigma_2}\, z}{q,\qquad q^{2-\sigma_1}\,z}{q^{\sigma_1}\frac{x_1}{x_2}} \; .
\end{align}
In order to make the components of $\bold{V}$ analytic in the chamber $\fC_H = \{|x_1|<|x_2|\}$, we apply the linear change of basis \eqref{changebasis}, with the matrix
\renewcommand{\arraystretch}{2.5}
\beq\label{Bmatrix2by2}
\fB^{k'}_{\fC_H,k}=U_{\fC_H,k}\;
\begin{bmatrix} 
\dfrac{1}{\Theta(q^{\sigma_1})} & 0 \\
\dfrac{\Theta(q^{\sigma_1} x_1/x_2\, z)}{\Theta(q^{\sigma_1}\, z)\Theta(q^{\sigma_1} x_1/x_2)} & \dfrac{\Theta(x_1/x_2)}{\Theta(q^{\sigma_2})\Theta(q^{\sigma_1} x_1/x_2)} \\
\end{bmatrix}
\; \text{e}^{-1} \, .
\eeq
Above, both $\text{e}^{-1}$ and $U_{\fC_H,k}$  are 2x2 diagonal matrices with entries \eqref{efunc} and \eqref{Ufunc}. The new vertex function $\bold{V}_\fC$ has scalar components $\bold{V}_{\fC_H,1}$ and $\bold{V}_{\fC_H,2}$. The second component $\bold{V}_{\fC_H,2}$ is simply a rescaling of $\bold{V}_{2}$, since $\fB^{k'}_{\fC_H,k}$ is lower-triangular. The computation of the first component $\bold{V}_{\fC_H,1}$ is more involved: at first sight, it is a sum of two $q$-hypergeometric functions, but making use of the three-term identity
\begin{align}\nonumber\label{qhyperidentity}
\pFq{2}{2}{a, b}{q, c}{f} &= \;\frac{(b)_\infty(c/a)_\infty\Theta(a f)}{(c)_\infty(b/a)_\infty\Theta(f)}\; \pFq{2}{2}{a,\; qa/c}{q,\;  q a/b}{\frac{q c}{a b f}}\\
& \qquad\;\;+\frac{(a)_\infty(c/b)_\infty\Theta(b f)}{(c)_\infty(a/b)_\infty\Theta(f)}\; \pFq{2}{2}{b,\; qb/c}{q,\;  q b/a}{\frac{q c}{a b f}} 
\end{align}
with $a=q^{1-\sigma_2}$, $b =q^{2-\sigma_1-\sigma_2}\, z$, $c=q^{2-\sigma_2}\, z$, $f=q^{\sigma_2}\, x_2/x_1$, a cancellation guarantees that only one $q$-hypergeometric series survives. All in all, we find:
\begin{align}
\nonumber\label{V12x}
&\bold{V}_{\fC_H,1} = (x_1)^{\eta'} \frac{(q^{1-\sigma_1})_\infty}{(q)_\infty\,\Theta(q^{\sigma_1})}\, \frac{(q\,z)_\infty}{(q^{1-\sigma_1} z)_\infty}\; \pFq{2}{2}{q^{1-\sigma_2}, \qquad\;\; 1/z}{\;\;\;q,\qquad q^{1-\sigma_1}/z}{\;q^{\sigma_1}\frac{x_1}{x_2}} \; ,\\
&\bold{V}_{\fC_H,2} =(x_2)^{\eta'} \frac{(q^{1-\sigma_2})_\infty}{(q)_\infty\, \Theta(q^{\sigma_2})}\, \frac{(q^{2-\sigma_1} z)_\infty}{(q^{2-\sigma_1-\sigma_2} z)_\infty}\;\pFq{2}{2}{q^{1-\sigma_1}, \;\;q^{2-\sigma_1-\sigma_2}\, z}{q,\qquad q^{2-\sigma_1}\,z}{\;q^{\sigma_1}\frac{x_1}{x_2}} \; .
\end{align}
In this new basis, the vertex function $\bold{V}_{\fC}$ is manifestly analytic in chamber $\fC_H =\{ |x_1|<|x_2|\}$.\\

Likewise, if we consider the electric blocks \eqref{integrandqKZ3} analytic in chamber $\fC_C$, then the actual $q$-conformal blocks of $U_\hbar(\widehat{A_1})$ are obtained after a change basis to $x$-solutions, using once again the matrix \eqref{changeofmatrix}: 
\beq\label{qKZellstab}
{\bf F}_{\fC_H,k,i}(\{\tilde{x}_d\})= \oint_{\Gamma} \frac{dy}{y}\, y^{\eta} \,\text{Stab}^K_i(y,\{\tilde{x}_d\}) \prod_{d=1}^{L} \frac{(q^{1-\sigma_d}\,y/\tilde{x}_d; q)_\infty}{(y/\tilde{x}_d; q)_\infty}\;  \fB_{\fC_H,k}(y,z,\{\tilde{x}_d\}) \; ,
\eeq
for all $k=1,\ldots,n$.\\

Finally, the explicit relation between electric and magnetic blocks, whether as $z$- or $x$-solutions, is established via the existence of a pairing with a certain covector $W$. For example, in the case of $z$-solutions, the pairing reads
\beq
\sum_{d=1}^L W_d\, {\bf F}^d_{k} = \bold{V}_{k} \; .
\eeq 
This is a particular case of the general formula \eqref{covector}.
We say that ${\bf F}_i(\{\tilde{x}_d\})$ is a (vector) vertex function with descendant insertions at 0. In particular, a trivial insertion at 0 gives a way to identify the covector $W$ explicitly:
\beq
\sum_{d=1}^L W_d\, \text{Stab}^K_{d}(y,\vec{x}) = 1 \; .
\eeq 
with $\text{Stab}^K_{d}(y,\vec{x})$ the K-theoretic stable envelope in $\eqref{Ki3}$. The fact that the stable basis is upper triangular ensures that all coefficients $W_d$  can be solved for recursively, precisely like in the unramified case.\\

\subsection{The 3d gauge theory}

%

Our goal is to show that all of the previous $q$-conformal blocks are realized as the half-indices or partition functions of a certain 3d $\cN=2$ gauge theory, for different choices of 3d manifold and boundary conditions.\\

The Drinfeld gauge theory $T^{3d}_{A_1}$ in question has a single gauge group $U(N)$ which we specialize to be abelian, $N=1$\footnote{The generalization from the gauge group $U(1)$ to the maximal torus of $U(N)$ requires $N$ integration variables as described in the main text, see Section \ref{ssec:3dindex}. Most notably, the 3d $\cN=4$ vector multiplet would then contain additional factors  $\frac{1}{N!}\prod_{1\leq j\neq i\leq N} \frac{\left(y_{i}/y_{j};q\right)_{\infty}}{\left(t\, y_{i}/y_{j};q\right)_{\infty}}$ which are responsible for new poles in the integrand. The poles are classified by Young tableaux with at most $N$ rows \cite{Aganagic:2013tta}. In the case at hand, for the gauge group $U(1)$, summing over Young tableaux with at most 1 row is just a sum over all positive integers.}, and flavor symmetry $F_{H}\times U(1)_t$. The bare Chern-Simons level is set to 0. According to Proposition \ref{prop3d}, $F_{H}$ is uniquely defined via the $\cW_{q,t}(A_1)$ $q$-primary operator content: here, we picked the weight $\lambda_1\in V_{Y}$ and the weight $\lambda_2\in V_{Y^{-1}}$ a total of $L$ times each, so we have a flavor group $F_H= U(L)\times U(L)$.
%
%
%
%
%
\begin{figure}[h!]
	\emph{}
	\centering
	\includegraphics[trim={0 0 0 1cm},clip,width=0.9\textwidth]{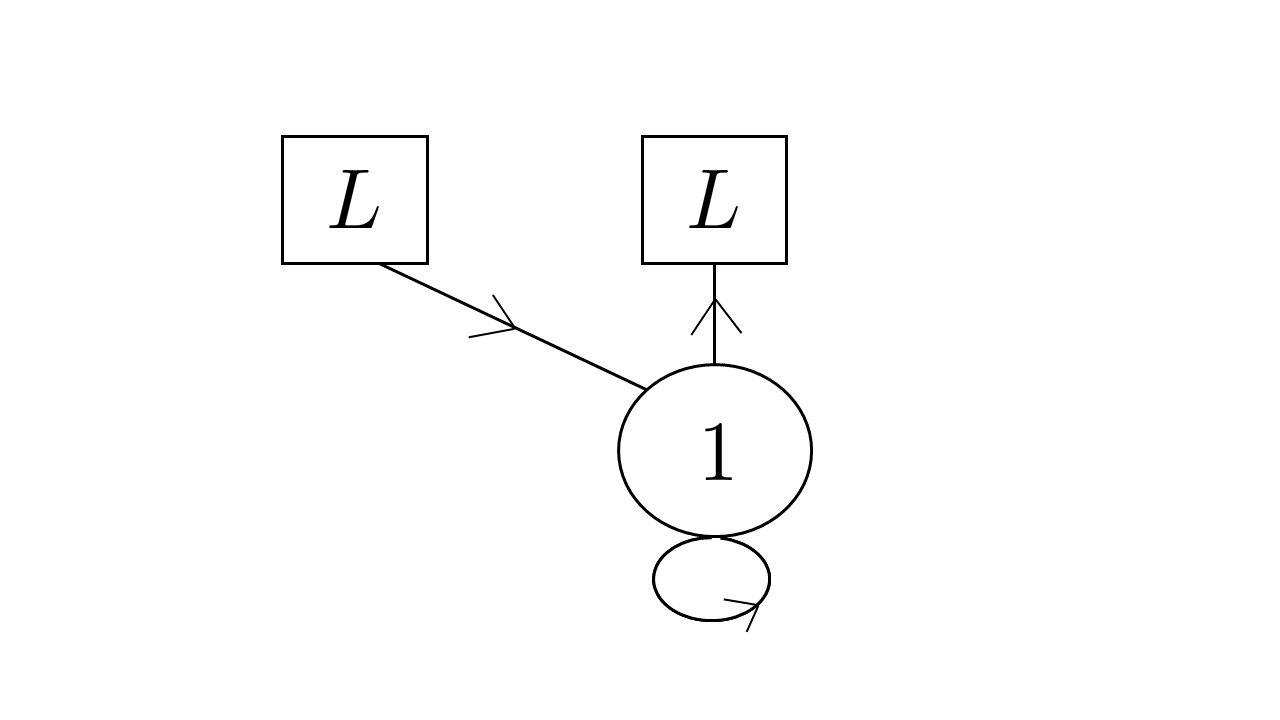}
	\vspace{-10pt}
	\caption{The 3d $\cN=2$ Drinfeld gauge theory $T^{3d}_{A_1}$ in the abelian case $N=1$.} 
	\label{fig:A13d}
\end{figure}
The flavor symmetry maximal tori act by Hamiltonian isometries on the chiral multiplets:
\beq
T_H:\;(\Phi^-_k, \Phi^+_k)\; \rightarrow (e^{-i\theta_k}\Phi^-_k, e^{+i\theta_k}\Phi^+_k)\; , \qquad T_t:\;(\Phi^-_k, \Phi^+_k)\; \rightarrow (e^{+i\theta_k}\Phi^-_k, e^{+i\theta_k}\Phi^+_k)\; .
\eeq
The superpotential is
\beq
W=\sum_{d=1}^L \left(\sigma + m_{d,1} +\frac{\epsilon_t}{2}\right)|\Phi^+_d|^2 + \sum_{d=1}^L \left(-\sigma - m_{d,2} +\frac{\epsilon_t}{2}\right)|\Phi^-_d|^2 - \xi\, \sigma \; ,
\eeq
where $\sigma$ is the real scalar in the vector multiplet. From it, we deduce the real moment map equation
\beq\label{vacvev}
\sum_{d=1}^L \left(|\Phi^+_d|^2 - |\Phi^-_d|^2\right) = \xi\; .
\eeq
For $k\in\{1,\ldots,L\}$, a Higgs vacuum arises when either $\Phi^+_k$ or $\Phi^-_k$ is given a nonzero vev.  When $\xi >0$, the vev is $|\Phi^+_k|^2 = \xi$, with $\sigma =- m_{k,1} -\frac{\epsilon_t}{2}$. When $\xi <0$, the vev is $|\Phi^-_k|^2 = -\xi$, with $\sigma = - m_{k,2} +\frac{\epsilon_t}{2}$. We work in the chamber $\fC_C= \{\xi>0\}$ throughout.


Let us first analyze the magnetic blocks of $\cW_{q,t}(A_1)$ analytic in chamber $\fC_C$, and correspondingly the gauge theory $T^{3d}_{A_1}$ defined on $M_3=S^1_{\cC'}\times D^2$.
Following Section \ref{sec:3dgauge}, the integral form of the partition function takes the form
\beq\label{vec3dpartA1}
{\mathcal Z}(T^{3d}_{{A_1},{\bf N}})=\oint_{C}  \frac{dy}{y}\,y^{\frac{\ln(z)}{\ln(q)}} \prod_{d=1}^{L}\frac{\left(\sqrt{q/t} \; y/x_{d,1}\, ; q\right)_\infty}{\left(\sqrt{t/q} \; y/x_{d,2}\, ; q\right)_\infty}  \; .
\eeq
The power of $y$ containing the complexified F.I. parameter $z$ is the contribution of the mixed topological/gauge anomaly; namely, the boundary anomaly polynomial is $2 \,f\, f_z$, where $f$ is the curvature of the corresponding dynamical $U(1)$ gauge symmetry at the boundary, and $f_z$ is the curvature of the topological symmetry $T_C=U(1)$.


The ratio of $q$-Pochhammer symbols in the integrand is the contribution of the chiral matter, as dictated by the Drinfeld polynomial data. In particular, the R-charge  and $U(1)_t$ charge of the chiral multiplets are fixed by the Drinfeld roots of the polynomials associated to the weights $\{\lambda\}_d$. 

The denominator stands for the fundamental chiral multiplets with boundary conditions N, whose scalars and their derivatives have charge $(n,-1,1/2)$ under $U(1)_J\times U(1)_R\times U(1)_t$. The numerator stands for the anti-fundamental chiral multiplets with boundary conditions D, whose fermions and their derivatives have charge $(n,1,1/2)$ under $U(1)_J\times U(1)_R\times U(1)_t$. The ratio of $q$-Pochhammer factors matches the product of two-points
\beq
\prod_{d=1}^{L}\left\langle S^\vee(y)\, \cV_{\{\lambda\}_d}(\tilde{x}_d) \right\rangle
\eeq
where the $L$ primary operators were defined in \eqref{opA1}. The notation is as before, with each mass $x_{d,s}\in T_H$ redefined with respect to a center of mass $\tilde{x}_d$ as 
\beq\label{massesencoreA1}
x_{d,s} \equiv \tilde{x}_d\, q^{\sigma_{d,s}} \; ,\qquad\;\; s=1, 2 \; .
\eeq
The contour is chosen to enclose the infinite number of $q$-shifted poles between $\infty$ and $x_{k,2}$ for some $k\in\{1,\ldots,L\}$. Then, in center of mass coordinates, and after rescaling,
\beq\label{vec3dpartA1again}
{\mathcal Z}(T^{3d}_{{A_1},{\bf N}})=\oint_{C_k}  \frac{dy}{y}\,y^{\frac{\ln(z)}{\ln(q)}} \prod_{d=1}^{L}\frac{(q^{1-\sigma_d}\,  y/\tilde{x}_d\, ; q)_\infty}{\left(y/\tilde{x}_d\, ; q\right)_\infty}  \; ,
\eeq
with $\sigma_d \equiv \sigma_{d,1}-\sigma_{d,2}$ for all $d$. The half-index ${\mathcal Z}(T^{3d}_{{A_1},{\bf N}})$ and the integral form of the magnetic block clearly match, since they have the same integrand and contour.\\

In order to exhibit the analytic properties of the magnetic block in the F.I. parameter $z$, we can simply evaluate the integral by residue. The result is a $q$-hypergeometric function which takes the form of a power series expansion in $z$ with positive radius of convergence around 0, as in the $q$-conformal block computation.\\

In the gauge theory $T^{3d}_{A_1}$, we can obtain the series form of the $z$-solution from the onset by imposing  Exceptional Dirichlet boundary conditions ${\bf D_{EX}}$. 
We start by imposing usual Dirichlet boundary conditions ${\bf D}$ on the $U(1)$ gauge field. Recall that unlike {\bf N}, imposing  ${\bf D}$ breaks the $U(1)$ bulk gauge symmetry to a global symmetry at the boundary, which we denote as  $U(1)_{\partial}$, with corresponding fugacity $u$ and  generator $\ft_\partial$. The index reads
\beq
\label{3dhalfindexmoreflow1A1}
{\mathcal Z}(T^{3d}_{{A_1},{\bf D}})  = {\rm Tr}\left[(-1)^F\, q^{J+\frac{V}{2}}\, t^{\frac{A-V}{2}} \, u^{\ft_\partial} \, z^{\ft_C}\, x^{\ft_H}  \right]\;\; .
\eeq

The index takes the form of a sum over all abelian magnetic fluxes $m$ through the hemisphere $D^2$,
\beq
\label{3dhalfindexmoreflow2A1}
{\mathcal Z}(T^{3d}_{{A_1},{\bf D}}) = \frac{1}{(q; q)_\infty} \sum_{m\in\mathbb{Z}} \; \prod_{d=1}^L \frac{(q^{1+m} \sqrt{t}\,u\, x_{d,2}; q)_\infty}{(q^{m}\sqrt{t}\, u\, x_{d,1}; q)_\infty}\;z^m\, q^{m} \; .
\eeq
$m$ is an integer since it is an element of the cocharacter lattice $\Lambda_{cochar}=\text{Hom}(U(1),U(1))=\mathbb{Z}$, the space of maps to the $U(1)$ gauge group. 

The exceptional Dirichlet boundary condition is implemented by generating a flow along the boundary to a new boundary condition. This is done by giving a vev $c\neq 0$ to, say, the $k$-th fundamental chiral field $\Phi^+_{k,|\partial}= c$. This chiral field has charge $(1,1,1)$ under $U(1)_V\times U(1)_{\partial}\times U(1)_{\mathcal H}$, where $U(1)_{\mathcal H}$ is part of the maximal torus of the relevant $U(L)$.  In the index, the weight of the corresponding field is set to 1: following our choice of vacuum in \eqref{vacvev}, this is  $u\, x_{k,1}\, t^{1/2} =1$, meaning we simply substitute $u = t^{-1/2}\, x_{k,1}^{-1}$ in the entire summand \eqref{3dhalfindexmoreflow2A1}.  We change to center of  mass notation, $x_{d,j} \equiv \tilde{x}_d\, q^{\sigma_{d,j}}$  ($j=1,2$), and once again set  $\sigma_d \equiv \sigma_{d,1}-\sigma_{d,2}$ for all $d$. After rescaling $x_{d,1}$ and $x_{d,2}$ by the appropriate factors of $q^{\sigma_{d,j}}$, we find in these new center of mass coordinates
\beq
\label{3dhalfindexmoreflow3A1}
{\mathcal Z}(T^{3d}_{{A_1},{\bf D_{EX,k}}}) = \frac{1}{(q; q)_\infty} \sum_{m\geq 0}\; \prod_{d=1}^L \frac{(q^{1+m} \, \tilde{x}_d/\tilde{x}_{k}; q)_\infty}{(q^{\sigma_d+m}\, \tilde{x}_d/\tilde{x}_k; q)_\infty}\; z^m\, q^{m(1-\sigma_d)} \; .
\eeq
Note the summation index $m$ now runs  over the positive integers only, since the numerator is null otherwise. The factors of $z^m$ and $q^{m(1-\sigma_d)}$ are interpreted as a mixed $U(1)_{\partial}-U(1)_C$ anomaly and a mixed $U(1)_{\partial}-U(1)_{\mathcal H}$ anomaly, respectively. 
The index is holomorphic in a punctured neighborhood of $0_{\fC_C}$, the origin of the F.I. chamber $\fC_C=\{|z|<1\}$. 

Up to normalization by the $z$-independent part, we recognize the $k$-th component of the vector vertex function $\bold{V}$ introduced earlier. Namely, 
\beq
\label{vertexA1comp}
\bold{V}_k = \left[(\tilde{x}_k)^\eta\, \frac{\Theta(q^{\sigma_k}; q)}{(q; q)_\infty}\prod_{d\neq k}^L \frac{\Theta(q^{1-\sigma_k} \tilde{x}_k/\tilde{x}_d; q)}{\Theta(\tilde{x}_k/\tilde{x}_d; q)}\right]{\mathcal Z}(T^{3d}_{{A_1},{\bf D_{EX,k}}})
\eeq
with $\bold{V}_k$ as in \eqref{vertexA1}:
\beq
\label{vertexA1comp2}
\bold{V}_k  = (\tilde{x}_k)^\eta \frac{(q^{1-\sigma_k}; q)_\infty}{(q; q)_\infty}\prod_{d\neq k}^L \frac{(q^{1-\sigma_d} \tilde{x}_k/\tilde{x}_d; q)_\infty}{(\tilde{x}_k/\tilde{x}_d; q)_\infty}\sum_{m\geq 0}\left(\frac{z}{\prod_d q^{\sigma_d-1}}\right)^m\prod_{d=1}^L \frac{(q^{\sigma_d} \tilde{x}_d/\tilde{x}_k; q)_m}{(q\, \tilde{x}_d/\tilde{x}_k; q)_m} \; .
\eeq
This is proved by simply invoking the definition of the finite $q$-Pochhammer symbol
\beq
(q^{m} \, x ; q)_\infty = \frac{(x; q)_\infty}{(x; q)_m}   \; \; , \qquad m\in \mathbb{Z}^+\; ,
\eeq
as well as the theta function $\Theta_{q}(x)= (x \,;q)_\infty\,(q/x \,; q)_\infty$. Besides the bulk $U(1)$ contribution $(q; q)_\infty$, the prefactor in brackets represents the contributions of various additional degrees of freedom at the boundary: the $(L-1)$ theta functions on the denominator are the contributions of $(L-1)$ 2d $\cN=(0,2)$ chiral multiplets living on the boundary. We also find $L$  2d $\cN=(0,2)$ Fermi multiplets on the boundary, which make up the numerator theta functions.

The ratio of theta functions in the product is the $\cN=(0,2)$ version of a similar contribution found in the 2d $\cN=(2,2)$ context, where enhanced supersymmetry would require a fixed value $q^{\sigma_d}= \hbar$ for all $d=1,\ldots,L$; there, this ratio represents the contribution of $(L-1)$ boundary 2d $\cN=(2,2)$ chiral multiplets, which are used to ``flip" the polarization of hypermultiplets with ${\bf D_{EX,k}}$ boundary conditions. In our 3d $\cN=2$ theory, we initially chose a polarization $(+,+,\ldots,+)$, as defined in Section \ref{ssec:excep}, but the vertex function $\bold{V}_k$ as we have introduced it is more naturally presented in the polarization $(-,\ldots,-,+,-,\ldots,-)$, where all ``$+$" except for the $k$-th one are flipped to ``$-$", hence the ratio of theta functions appearing in \eqref{vertexA1comp}.

%
We fix the mass parameter chamber as
\beq
\fC_H=\{|\tilde{x}_{1}|<|\tilde{x}_{2}|<\ldots<|\tilde{x}_{L}|\}\, ,
\eeq
and further choose the parameters $\sigma_{d,j}$ such that
\beq
1<|q^{\sigma_{d,1}}|<|q^{\sigma_{d,2}}|
\eeq
for all $d=1,\ldots,L$, with the understanding that $|q|<1$.
The $x$-solution is by definition holomorphic in a punctured neighborhood of $0_{\fC_H}$, the origin of the chamber $\fC_H$. It is obtained by imposing enriched Neumann boundary conditions $\bf {N}_{EN,\fC_H}$ in the theory $T^{3d}_{A_1}$. In what follows, we simply write ${\bf {N}_{EN}}$ and keep the dependence on the chamber $\fC_H$ implicit.

The support of ${\bf {N}_{EN}}$ consists of attracting submanifolds to the vacuum $\{\bf A\}$, defined via the gradient flow from other vacua $\{\bf B\}$. Crucially, we should include all submanifolds where the image of the moment map is strictly greater than the image of the moment map for our vacuum ${\bf{A}}$.  In this simple example, it is enough to proceed in two steps: first, we multiply the integrand of the $\bf {N}$ half-index \eqref{vec3dpartA1again} by 
\beq\label{step1A1}
\frac{\displaystyle \prod_{d<k}\Theta\left(\tilde{x}_d/y\right)\prod_{d>k}\Theta(q^{\sigma_d}\tilde{x}_d/y)}{\displaystyle  \prod_{d=1}^L\Theta(q^{\sigma_d}\tilde{x}_d/y)} \; .
\eeq
This insertion has precisely the desired effect of flipping chiral multiplets from  N to D and vice-versa, such that the support is on attracting manifolds only. The new pole structure is
\begin{align}\label{poles1again}
&y=q^{-n}\, \tilde{x}_d\, , \;\;\;\;\,\qquad\; n=0,1,2,\ldots, \qquad d\geq k \; ,\nonumber\\
&y=q^{n}\, \tilde{x}_d\, q^{\sigma_d}\, , \qquad\; n=0,1,2,\ldots, \qquad d\leq k \; ,
\end{align}
and the contour $C$ is deformed so as to enclose only the poles from the first line, which accumulate at $y=\infty$. 
Second, the remaining insertions are determined from anomaly cancellation, for all $k=1,\ldots,L$. In the end, we recover the expression
\beq\label{vec3dpartA1againwow}
{\mathcal Z}(T^{3d}_{{A_1},{\bf N_{EN}}_{,\fC_H}})=\oint_{C}  \frac{dy}{y}\,y^{\frac{\ln(z)}{\ln(q)}} \prod_{d=1}^{L}\frac{(q^{1-\sigma_d}\,  y/\tilde{x}_d\, ; q)_\infty}{\left(y/\tilde{x}_d\, ; q\right)_\infty}  \; \fB_\fC 
\eeq 
for the magnetic $x$-solution, where the pole-subtraction matrix $\fB_\fC$ is as before,
\beq\label{changeofmatrixwow}
\fB_{\fC_H,k}(y,z,\{\tilde{x}_d\})= U_{\fC_H,k}(z,\{\tilde{x}_d\})\;\frac{\text{Stab}^{Ell}_{\fC_H,k}(y,z,\{\tilde{x}_d\})}{\displaystyle  \prod_{d=1}^L\Theta\left(q^{\sigma_d}\tilde{x}_d/y\right)}\; \text{e}(y,z)^{-1} \, .
\eeq 
The factor $\text{Stab}^{Ell}_{\fC_H,k}(y,z,\{\tilde{x}_d\})$ was written in \eqref{ellstab}; it includes the numerator in \eqref{step1A1} and various other contributions of 2d $\cN=(0,2)$ Fermi and chiral multiplets on the boundary, which are needed to cancel anomalies, including the bulk mixed topological/gauge anomaly $\text{e}(y,z)$ in \eqref{efunc}. There will be leftover global symmetry mixed anomalies which further need to be canceled: these are all encoded in the exponential factor $U_{\fC_H,k}(z,\{\tilde{x}_d\})$ as written in \eqref{Ufunc}.

As a concrete example, consider the case $L=2$, meaning two surface defect insertions at center of mass position $\tilde{x}_1$ and $\tilde{x}_2$ on $\cC$ (with chamber $\fC_H =\{|\tilde{x}_1|<|\tilde{x}_2|\}$), and the reference vacuum labeled by $k=2$. Our choice of chamber $\fC_H$ instructs us to introduce two boundary chiral multiplets and a Fermi multiplet, which contribute to the index as
\beq\label{step1A1L2}
\frac{\Theta(\tilde{x}_1/y)}{\Theta(q^{\sigma_1}\tilde{x}_1/y)\,\Theta(q^{\sigma_2}\tilde{x}_2/y)} 
\eeq
inside the integrand. This will flip the two 3d chiral multiplets from having D to N boundary conditions, and likewise it will flip one chiral multiplet from having  ${\bf N}$ to  ${\bf D}$ boundary conditions, providing the adequate support on the attracting submanifold under gradient flow. Indeed, note that evaluation at $y= x_1$ makes the numerator vanish, so the $k'=1$ vacuum ${\bf{B}}$ is not supported on ${\bf {N}_{EN}}$, in accordance with the definition of the chamber  $\fC_H$. The boundary anomalies are canceled by computing the anomaly polynomial and canceling the quadratic terms. We find that additional Fermi and chiral multiplets are required, with contribution
\beq\label{step2A1L2}
\frac{\Theta(q^{\sigma_1+\sigma_2}x_2/z\,y)}{\Theta(q^{\sigma_1+\sigma_2}/z)} \; ,
\eeq
where $z$ is the F.I. parameter. The leftover anomalies are canceled by \eqref{efunc} and \eqref{Ufunc}, for $L=2$ and $k=2$.\\

%
%
%

The electric solutions are $U_\hbar(\widehat{A_1})$-algebra $q$-conformal blocks, which solve the  $A_1$-type qKZ equations. To engineer them from the 3d gauge theory $T^{3d}_{A_1}$, we simply change the support manifold from  $M_3= S^1_{\cC'}\times \mathbb{C}_q$ to $M^\times_3= S^1_{\cC'}\times \mathbb{C}^\times_q$, with the origin removed. The index is as before, but now includes 3d/1d contributions of 1/2-BPS  Wilson loop operators on $S^1_{\cC'}\times \{0\}$ coupled to the 3d gauge theory:
\beq\label{Ki3more}
\text{Stab}^K_{i}(y,\vec{x})= \prod_{d=1}^{i-1}(1- y/\tilde{x}_d)\; \prod_{d=i+1}^{L} (1-q^{-\sigma_d}\,y/\tilde{x}_d)
\eeq
Namely, the loops are supported on the K-theoretic stable envelopes.\\

\subsection{The 5d gauge theory}

Consider a 5d $\cN=1$ gauge theory $T^{5d}_{\fg_\text{o}}$ with gauge group
\beq
G_{5d}=U(n) \qquad \text{with}\;\; n=L \; ,
\eeq 
and flavor group
\beq
G^F_{5d} = U(n^F)  \qquad \text{with}\;\; n^F = 2 \,L \; ,
\eeq 
with the $U(1)$ centers understood as nondynamical. The matter content consists of $2\, L$ fundamental hypermultiplets in representation  $(n,\overline{n^F})$ of $U(n)\times U(n^F)$. There is a remaining global symmetry, the topological symmetry
\beq
G^{top}_{5d}= U(1)^{top} \; ,
\eeq  
whose associated charge is the instanton number $k$. The mass for this topological symmetry is the instanton counting parameter $z$, related to the 5d gauge coupling as $z=e^{-8\pi^2\,R_{\cC'}/g^2_{5d}}$. The bare Chern-Simons level is set to 0.
The 5d gauge theory $T^{5d}_{\fg_{\text{o}}}$ should be understood as a relevant deformation of a fixed point in the UV, with the deformation scale set by the gauge coupling\footnote{This is a particular case of a more general statement \cite{Intriligator:1997pq}: 5d $SU(n)$ gauge theory with $n^F$ fundamental hypermultiplets has a UV fixed point if $n^F + 2 \,|k_{CS, bare}| \leq 2\, n$, with $k_{CS, bare}$ an integer whenever $n^F$ is even and  $k_{CS, bare}$ half an odd integer whenever $n^F$ is odd. In our case, we have $n_F=2\,L$, $n=L$, and $k_{CS, bare}=0$, which saturates the inequality.}.


\begin{figure}[h!]
	\emph{}
	\centering
	\includegraphics[trim={0 0 0 1cm},clip,width=0.9\textwidth]{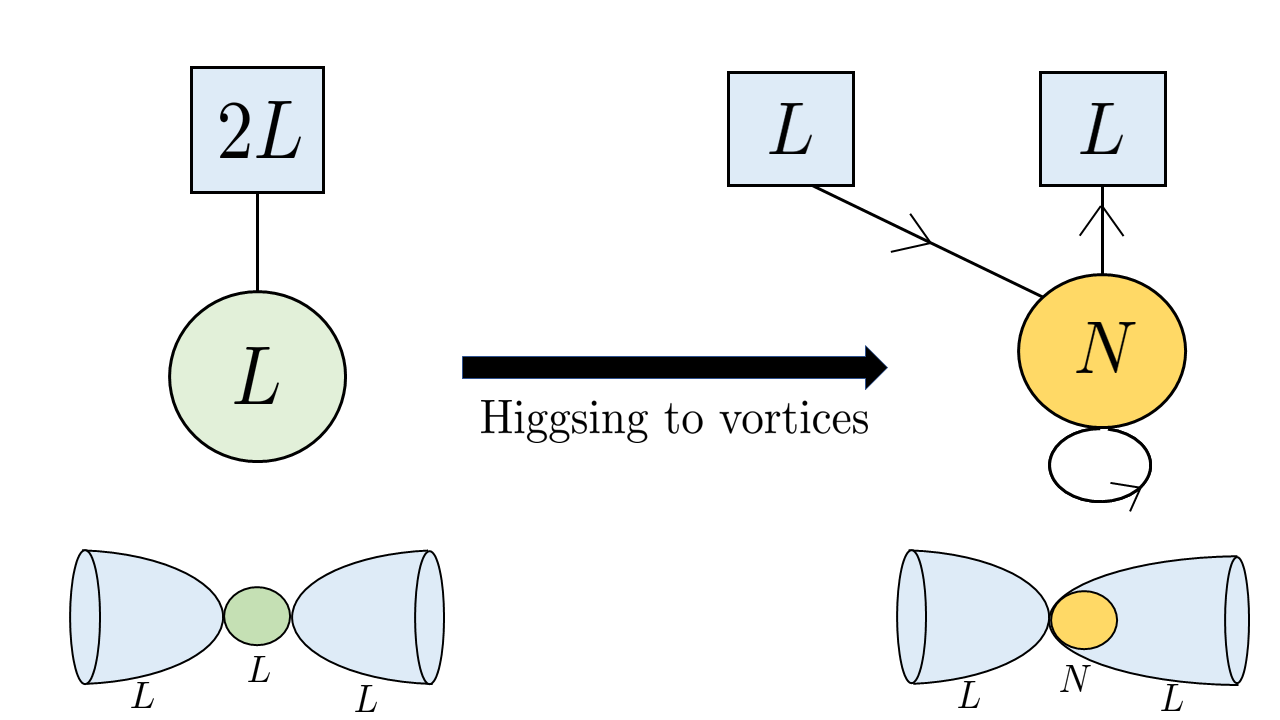}
	\vspace{-10pt}
	\caption{The 5d $\cN=1$ gauge theory and its Higgsing to the Drinfeld gauge theory $T^{3d}_{A_1}$. Here, we focus on the moduli space of charge $N=1$ vortices. The string theory realization is shown at the bottom, with branes wrapping a resolved $A_1$ singularity. The green and blue colors stand for D5 branes, while the yellow color stands for D3 branes.} 
	\label{fig:5dto3d}
\end{figure}

We take $T^{5d}_{\fg_\text{o}}$ to be supported on $M_5= S^1_{\cC'} \times \mathbb{C}_q \times \mathbb{C}_t$, 
or on $M^\times_5= S^1_{\cC'} \times \mathbb{C}^\times_q \times \mathbb{C}_t$ with the origin removed, depending on whether we want to make contact with $\cW_{q,t}(A_1)$ or $U_\hbar(\widehat{A_1})$ $q$-conformal blocks, respectively.\\

First consider the manifold $M_5$. The instanton partition function of $T^{5d}_{A_1}$ is computed as the Witten index of a 1d $\cN=4$ gauged quantum mechanics on $S^1_{\cC'}$, follwoing \eqref{5dhalfindexmore}.
We fix the F.I. parameter of the quantum mechanics to lie in the positive chamber $\zeta>0$. Because $n^F = 2\, n$ and the 5d bare Chern-Simons level is 0, the Coulomb branch continuum contributes additional spurious UV states to the index, which will factorize as
\beq
\label{extraA1}
{\mathcal Z}_{inst} = Z_{inst}\cdot {\mathcal Z}_{extra}
\eeq
with ${\mathcal Z}_{extra}\neq 1$. In the rest of this section, we will only make statements about ${\mathcal Z}_{inst}$, so we won't need to identify ${\mathcal Z}_{extra}$ explicitly. The details of the factorization can be found in \cite{Bergman:2013ala,Bao:2013pwa,Hayashi:2013qwa,Taki:2013vka}.

The quantum mechanical index is a sum over all instanton numbers,
\beq
{\mathcal Z}_{inst} =\sum_{k=0}^{\infty}  {\mathcal Z}_{k}\;  \frac{z^{k}}{k!} \; .
\eeq
where $k!$ is the order of the Weyl group for $U(k)$. 
The quantum mechanical index is evaluated in the F.I. chamber $\zeta>0$ with coupling $g_{QM}\rightarrow 0$, in two steps: first, we integrate over non-zero modes via Gaussian integrals over massive fluctuations, keeping the zero modes fixed. Second, we integrate over these zero modes, which are the scalars $\varphi$ inside the 1d $\cN=2$ vector multiplet of the $U(k)$ quantum mechanics (and the gaugini fermion zero modes). The scalars $\varphi$ have $k$ eigenvalues $I=1,\ldots,k$, and get complexified as $\phi_{I}=\varphi_{I}+ i \, A^{\theta}_{I}$, with $A^{\theta}$ the holonomy around $S^1_{\cC'}$. Explicitly, we obtain \cite{Moore:1997dj,Losev:1997hx,Moore:1998et,Nekrasov:2002qd,Hwang:2014uwa}:
\begin{align}
\label{5dintegralA1}
{\mathcal Z}_{k}  =\oint  \prod_{I=1}^{k}  \left[\frac{d\phi_{I}}{2\pi i}\right]{\mathcal Z}^{vec}\cdot {\mathcal Z}^{fund} \; ,
\end{align}
with
\begin{align}
\label{5dintegralA1integrand}
& {\mathcal Z}^{vec} =\prod_{\substack{I\neq J\\ I,J=1}}^k\sh(\phi_I-\phi_J)\prod_{I, J=1}^k\frac{\sh(\phi_I-\phi_J+\E_q+\E_t)}{\sh(\phi_I-\phi_J+\E_q)\sh(\phi_I-\phi_J+\E_t)}\nonumber\\
&\qquad\times\prod_{I=1}^{k} \prod_{i=1}^{L} \frac{1}{\sh\left(\phi_I-a_i+(\E_q+\E_t)/2 \right)\sh\left(\phi_I-a_i-(\E_q+\E_t)/2 \right)}\; , \\
&Z^{fund} =\prod_{I=1}^{k} \prod_{s=1}^{2 L} \sh\left(\phi_I-m_s\right)\; .\nonumber
\end{align}
We used the notation $\sh(x)\equiv 2\sinh(R_{\cC'}\, x/2)$. The factors $Z^{vec}$ and $Z^{fund}$ are the 1-loop determinants for the quantum mechanical modes of the vector multiplet and fundamental hypermultiplets of the 5d gauge theory, respectively. 

The equivariant parameters appearing in the argument of the $\sh(x)$ function are related to the K-theoretic fugacities we have been using in this paper, as
\begin{align}\label{fugacities5d}
q=e^{R_{\cC'}\,\epsilon_q},\qquad t=e^{-R_{\cC'}\,\epsilon_t},\qquad e_i=e^{-R_{\cC'}\,a_i},\qquad f_s=e^{-R_{\cC'}\,m_s}\; ,
\end{align}
where nonzero holonomies are understood.

In the F.I. chamber $\zeta>0$, the contours are chosen to enclose a subset of the poles in ${\mathcal Z}^{vec}$. The poles that end up contributing to the residue integral are famously classified by $n$-colored Young tableaux $\{\overrightarrow{\boldsymbol{\mu}}\}=\{\boldsymbol{\mu}_1, \boldsymbol{\mu}_2, \ldots, \boldsymbol{\mu}_n\}$ \cite{Nekrasov:2002qd,Nekrasov:2003rj}:
\beq
\label{A1youngtuples}
\phi_I=a_i +\frac{(\epsilon_q+\epsilon_t)}{2} - s_1\, \epsilon_1 - s_2\, \epsilon_2 , \text{with}\; (s_q, s_t)\in \boldsymbol{\mu}_i \; .
\eeq
Summing over residues, the instanton partition function is the sum
\beq\label{bulk5dA1}
{\mathcal Z}_{inst}(T^{5d}_{A_1}) =  \sum_{\{\overrightarrow{\boldsymbol{\mu}}\}}   z^{\sum_{i=1}^{n}{\left|\boldsymbol{\mu}_{i}\right|}}\, {\boldsymbol Z}^{5d,vec} \cdot {\boldsymbol Z}^{5d,fund}\cdot {\boldsymbol Z}^{5d,CS} \, ,
\eeq
with
\begin{align}
&{\boldsymbol Z}^{5d,vec} = \prod_{i,j=1}^{L}\left[\cN_{\boldsymbol{\mu}_i\boldsymbol{\mu}_j}\left(\frac{e_{i}}{e_{j}}\, ;q\right)\right]^{-1}\label{5dbulkvecA1}\\
&{\boldsymbol Z}^{5d,fund} = \prod_{d=1}^{2L} \prod_{i=1}^{L} \cN_{\boldsymbol{\emptyset}\, \boldsymbol{\mu}_i}\left(\sqrt{\frac{q}{t}}\, \frac{f_{a,s}}{e_{a,i}}\, ; q\right) \label{5dbulkmatterA1}\\
&{\boldsymbol Z}^{5d,CS} = \prod\limits_{i=1}^{L} \left((-1)^{|\boldsymbol{\mu}_i|} q^{\Arrowvert \boldsymbol{\mu}_i\Arrowvert^{2}/2}t^{-\Arrowvert \boldsymbol{\mu}_i^{t}\Arrowvert^{2}/2}\right)^{L} \label{5dbulkCSA1}
\end{align}
and 
\beq\label{nekrasovNA1}
\cN_{\boldsymbol{\mu}_i\boldsymbol{\mu}_j}(Q \,\; q) = \prod\limits_{k,s = 1}^{\infty} 
\frac{\big( Q \, q^{\boldsymbol{\mu}_{i,k}-\boldsymbol{\mu}_{j,s}} \,t^{s - k + 1}\big)_{\infty}}{\big( Q\,  q^{\boldsymbol{\mu}_{i,k}-\boldsymbol{\mu}_{j,s}}\, t^{s - k}\big)_{\infty}} \,
\frac{\big( Q\,  t^{s - k}\big)_{\infty}}{\big( Q\,  t^{s - k + 1}\big)_{\infty}} \; .
\eeq
The notation $\boldsymbol{\mu}_{i,k}$ stands for the length of the $k$-th row in the partition $\boldsymbol{\mu}_{i}$.\\

Gauge-vortex duality to 3 dimensions is achieved by freezing each of the $L$ Coulomb moduli $e_i$ to a distinct mass among the $2\, L$ ones $f_s$.
After turning on vortices of charge $N_{i}$, the Coulomb parameters get shifted as:
\beq\label{5drootHiggsvortexA1}
e_{i} = f_{2\,i} \,  t^{N_{i}} \, \sqrt{t/q}  \;, \qquad\; i=1,\ldots,L \; .
\eeq
This truncates each partition $\boldsymbol{\mu}_{i}$: if there are more than $N_i$ rows, the entire summand of the partition function vanishes. 
In the gauge-vortex dictionary, the vortex charges $N_{i}$ are identified with the rank of the 3d gauge group, so let us turn on a single unit of vortex charge $N=1$ on the Higgs branch of $T^{5d}_{A_1}$. At the level of the instanton partition function, this means we freeze 
\beq\label{5drootHiggsvortexA1ex1}
e_{k} = x_{k,2}^{-1} \,  t \, \sqrt{t/q}  \;, \qquad\; \text{for some}\; k\in\{1,\ldots,L\} \; 
\eeq
and
\beq\label{5drootHiggsvortexA1ex2}
e_{i} = x_{i,2}^{-1} \, \sqrt{t/q}  \;, \qquad\; \text{for all}\; i\neq k \; .
\eeq
(We implicitly relabeled all 5d hypermultiplet masses as 3d chiral multiplet masses, $f_{a,s}\rightarrow x^{-1}_{d,s}$, see footnote \ref{footnotewow}).
Specializing \eqref{bulk5dA1} as such, we recover our previous expression \eqref{3dhalfindexmoreflow3A1} for the 3d half-index with exceptional boundary conditions for a $U(1)$ gauge group and $U(L)\times U(L)$ flavor group:
\begin{align}\label{partequalityA1}
{\mathcal Z}(T^{3d}_{A_1,\bf D_{EX}}) =c_{3d}\cdot  {\mathcal Z}_{inst}(T^{5d}_{A_1})_{e_{i}= x^{-1}_{i,2}\,t^{N_{a,i}}}\, .
\end{align}
This is the UV version of the IR statement proved in \cite{Aganagic:2013tta}, where the Higgsed 5d instanton partition function was identified as a 3d holomorphic block on $S^1_{\cC'}\times\mathbb{C}_q$.\\

To recover the electric blocks, one works on $M^{\times}_5$ instead, and simply inserts the Fermi contributions of Wilson loops directly into the quantum mechanics integrand \eqref{5dintegralA1}. The contour is not modified by this insertion, and the electric form of the magnetic block follows after the same Higgsing we performed.

\section*{Acknowledgments}
We are grateful to Mina Aganagic, Mykola Dedushenko, Gurbir Dhillon, Ahsan Khan, Spencer Tamagni, Oleksandr Tsymbaliuk and Edward Witten for valuable discussions at various stages of this project. NH acknowledges support from NSF Grant PHY-2207584 and the Sivian Fund at the Institute for Advanced Studies.

\vspace{8mm}
 
\appendix
\section{Basics of Lie theory and Langlands duality}
\label{sec:appendixlie}

Let us briefly review some Lie theory terminology: Let $G$ be a simple Lie group\footnote{The discussion also applies to a compact Lie group with minor modifications.} with Lie algebra $\fg$ of rank $r$. We denote the maximal torus of $G$ as $T$, with Lie algebra $\fh$, also called the Cartan subalgebra of $\fg$.  

One defines a symmetric bilinear form $K:\fg \times \fg \rightarrow \mathbb{C}$  as\footnote{For convenience, We have chosen a normalization with a factor of half the dual Coxeter number in front of the trace, $1/2\, h^\vee$. For reference, the explicit value of the dual Coxeter number of $\fg$ is $h^\vee_{A_r}=r+1$, $h^\vee_{B_r}=2r-1$, $h^\vee_{C_r}=r+1$, $h^\vee_{D_r}=2r-2$, $h^\vee_{E_6}=12$, $h^\vee_{E_7}=18$, $h^\vee_{E_8}=30$, $h^\vee_{F_4}=9$, and $h^\vee_{G_2}=4$.}
\beq
K(X,Y)=\frac{1}{2\,h^\vee}\text{Tr}\left(\text{ad}X\,\text{ad}Y\right)
\eeq
called the Killing form on $\fg$. The role of this form is to establish an isomorphism between the Cartan subalgebra $\fh$ and its dual $\fh^\vee$. Namely, for $a=1,\ldots,r$ and  ${\bold h}_a\in\fh$, the form $K({\bold h}_a,\cdot)$ maps every element of $\fh$ to a scalar. Then, for every $\beta\in\fh^\vee$, there exists ${\bold h}_\beta\in\fh$ such that $\beta({\bold h}_a)= K({\bold h}_a,{\bold h}_\beta)$. Equipped with this isomorphism, it is possible to write a positive-definite scalar product on the dual space as
\beq\label{scalarprod}
\langle \beta, \gamma\rangle = K(h_\beta,h_\gamma) \; .
\eeq
For an arbitrary representation of $\fg$, there always exists a basis $\{|\lambda\rangle\}$ such that 
\beq
{\bold h}_a |\lambda\rangle =\lambda_a |\lambda\rangle\; , \;\qquad a=1,\ldots,r \; .
\eeq 
The eigenvalues $\lambda_a$ make up a vector $\lambda=(\lambda_1,\ldots,\lambda_r)$ called a weight. Weights live in the dual space $\fh^\vee$, meaning $\lambda({\bold h}_a)=\lambda_a$. As such, the scalar product between weights is fixed by the Killing form, and we denote it as \eqref{scalarprod}. Weights in the adjoint representation of $\fg$ are called roots. In particular, roots are also valued in the space $\fh^\vee$, so we can take \eqref{scalarprod} to define the scalar product on the root space. The dimension of the root space is the dimension of $\fg$ minus its rank $r$, which is larger than $r$. This means that roots are linearly dependent, so we fix a basis $\{\alpha_1,\ldots,\alpha_r\}$ in $\fh^\vee$. A given root $\alpha$ can therefore be be written in that basis as $\alpha=\sum_{a=1}^r u_a\, \alpha_a$, and we define an ordering by declaring $\alpha$ positive if the first nonzero entry in the vector $(u_1,\ldots,u_r)$ is positive. We denote the set of positive roots as $\Delta_+$. A root $\alpha_a$ is called simple root when it cannot be written as the sum of two elements of $\Delta_+$. We denote the set of simple roots as $\Delta$. Since there are precisely $r$ such simple roots, we conveniently choose the basis $\{\alpha_1,\ldots,\alpha_r\}$ to be exclusively made out of the simple roots.  

The Weyl group $W$ of $\fg$ is generated by the reflections with respect to the simple roots. Explicitly, a reflection with respect to a simple root $\alpha_b$  is the map 
\beq\label{Weyl}
\alpha_a \mapsto \alpha_a - \alpha_b \frac{2\,\langle \alpha_a, \alpha_b\rangle}{\langle \alpha_b, \alpha_b\rangle} \; .
\eeq
In Physics, the Lie group $G$ is called the gauge group, and the choice of a Cartan subalgebra $\fh$ in $\fg$ specifies a Higgsing, or breaking, of $G$ to $T\times W$  at a generic point on the moduli space, where $T \cong U(1)^r$ denotes a maximal torus, and $W$ denotes the Weyl group of $\fg$.

The classification of simple Lie algebras guarantees that when $\fg$ is simply-laced, meaning that when $\fg=A_r, D_r, E_6, E_7, E_8$, all simple roots have the same length squared. Meanwhile, when $\fg$ is non simply-laced, meaning when $\fg=B_r, C_r, F_4, G_2$, simple roots are divided into two sets $\Delta_l$ and $\Delta_s$, whose elements are respectively called long roots and short roots. The long roots all have the same length squared, and the short roots likewise all have the same length squared. The ratio of the length squared of long roots over short roots is an integer $\fn_\fg$ called the lacing number of $\fg$. When $\fg$ is simply-laced, long and short roots are one and the same, so $\fn_\fg=1$. In the non simply-laced case, one finds $\fn_\fg=2$ for $\fg=B_r, C_r, F_4$, while $\fn_{G_2}=3$.

Additionally, one defines a set of $r$ simple coroots of $\fg$,
\beq 
\alpha_a^{\vee}=2\alpha_a/\langle\alpha_a,\alpha_a\rangle \; .
\eeq 
Note from the denominator that this definition of coroots depends on the normalization of the scalar product of roots in $\fg$. In order to give a definition of coroots which is intrinsically independent of such a normalization, we should understand the coroots of $\fg$ as elements of the Cartan subalgebra $\fh$, since the roots of $\fg$ were elements of the dual space $\fh^\vee$. 
The existence of coroots gives a definition of the Langlands dual Lie group $^L G$, with corresponding Lie algebra $^L\fg$, such that the roots of $^L\fg$ are the coroots of $\fg$. It implies that $^L\fg=\fg$ whenever $\fg$ is simply-laced. Moreover, $^L B_r=C_r$, $^L C_r=B_r$, $^L F_4=\widetilde{F_4}$ and  $^L G_2=\widetilde{G_2}$, by which we mean that $B_r$ and $C_r$ get mapped to each other under Langlands duality, while $F_4$ and $G_2$ get mapped to themselves, but in a nontrivial way, since their long and short roots get swapped under duality (hence the use of a different notation $\widetilde{F_4}$ and $\widetilde{G_2}$). Note that $^L(^L \fg)=\fg$ for all simple $\fg$.

The Weyl group $^L W$ of $^L\fg$ is generated by the reflections with respect to its simple roots, which are the simple coroots of $\fg$. Explicitly, a reflection with respect to a simple coroot $\alpha^\vee_b$  is the map 
\beq\label{Weyldual}
\alpha_a \mapsto \alpha_a - \alpha^\vee_b \frac{2\,\langle \alpha_a, \alpha^\vee_b\rangle}{\langle \alpha^\vee_b, \alpha^\vee_b\rangle} \; .
\eeq
It follows from this action and the definition of a coroot that the Weyl groups of $\fg$ and $^L\fg$ coincide:  $^L W =  W$.\\

The Cartan matrix of $\fg$ has size $r\times r$, with entries 
\beq
C_{ab}=\langle\alpha_a,\alpha_b^{\vee}\rangle \; .
\eeq
This definition directly implies that the Cartan matrix of $^L\fg$  is the transpose of the Cartan matrix of $\fg$.

The fundamental weights of $\fg$ are a distinguished set of $r$ weights, defined as dual to the simple coroots,  
\beq
\langle \lambda_a, \alpha_b^{\vee}\rangle=\delta_{ab} \; .
\eeq  
The expansion coefficients of a weight in the fundamental weight basis are called Dynkin labels. In particular, the Dynkin labels of a weight in a finite-dimensional irreducible representation of $\fg$ are always integers. Finally, one defines $r$ fundamental coweights of $\fg$ as dual to the simple roots: 
\beq
\langle \lambda_a^{\vee}, \alpha_b\rangle = \delta_{ab} \; .
\eeq 
From the Langlands dual point of view, the fundamental weights of $^L\fg$ are the fundamental coweights of $\fg$.\\

The simple roots span a rank $r$ lattice $\Lambda_{rt}$ called the root lattice of $\fg$. Physically, $\Lambda_{rt}$ is the lattice of electric charges of massive gauge bosons in 4d Yang-Mills. There exists a dual rank $r$ lattice $\Lambda_{cowt}$ spanned by the fundamental coweights, called the coweight lattice of $\fg$. This is the lattice of magnetic charges of a physical state, as prescribed by the Dirac quantization condition.
Meanwhile, the simple coroots span a rank $r$ lattice $\Lambda_{cort}$ called the coroot lattice of $\fg$. $\Lambda_{cort}$ is the lattice of magnetic charges of elementary BPS monopoles. The dual rank $r$ lattice $\Lambda_{wt}$ spanned by the fundamental weights is called the weight lattice of $\fg$. This is the lattice of electric charges of a physical state.
From the definition of the lattices, it is clear that $\Lambda_{rt}\subset \Lambda_{wt}$, and that $\Lambda_{cort}\subset \Lambda_{cowt}$.\\

We now fix the normalization of the scalar product $\langle , \rangle$ on $\fg$ for the entire paper. Whenever $\fg$ is simply-laced, the simple roots $\alpha_a$ (which are also the simple coroots) will all have length squared $2$, meaning $\langle \alpha_a, \alpha_a\rangle=2$ for all $a=1,\ldots,r$.

Whenever $\fg$ is non simply-laced, we choose to give the long roots $\alpha_{l,a}$ a length squared $2$, and therefore the short roots $\alpha_{s,a}$ a length squared  $2/\fn_{\fg}$. It follows that the short coroots $\alpha^\vee_{s,a}$ coincide with the long roots, while the long coroots are $\fn_{\fg}$ times the short roots:
\begin{align}
&\alpha^\vee_{s,a} = \alpha_{l,a} \; , \;\;\;\;\,\ \qquad a=1,\ldots,|\Delta_l| \nonumber\label{rootlengths}\\
&\alpha^\vee_{l,a} = \fn_{\fg}\cdot\alpha_{s,a} \; , \qquad a=1,\ldots,|\Delta_s|
\end{align}
For reference, our conventions agree with the ones used in \cite{Argyres:2006qr}, which are different from the ones used in \cite{Gukov:2006jk}\footnote{In \cite{Gukov:2006jk}, Gukov and Witten normalize their inner products such that short roots have length squared 2, meaning the long roots will have length $2\,\fn_{\fg}$. Their short and long coroots are defined as
	\begin{align}
	&\alpha^\vee_{s,a} = \alpha_{l,a}/\fn_{\fg} \; , \;\qquad a=1,\ldots,|\Delta_l| \nonumber\label{rootlengthsGW}\\
	&\alpha^\vee_{l,a} = \alpha_{s,a} \; ,\;\;\;\;\;\; \qquad a=1,\ldots,|\Delta_s| \; .
	\end{align}}. Our conventions were chosen to match the existing literature on deformed $\cW_{q,t}(\fg)$-algebras of non simply-laced type \cite{Frenkel:1998}, and their recent string theoretic realizations \cite{Aganagic:2017smx,Haouzi:2017vec,Haouzi:2019jzk}.\\

The fundamental weights $w_a$ of $\fg$ are defined as dual to the simple coroots,  $\langle w_a, \alpha_b^{\vee} \rangle =\delta_{ab}$. Just like roots, arbitrary weights live in the dual space $\fh^\vee$, so the Killing form also fixes the scalar product between weights. The expansion coefficients of a weight in the fundamental weight basis are called Dynkin labels. In particular, the Dynkin labels of a weight in a finite-dimensional irreducible representation of $\fg$ are always integers. 
The fundamental coweights $w^\vee_a$ of $\fg$ are likewise defined as dual to the simple roots, $\langle \lambda_a^{\vee}, \alpha_b\rangle = \delta_{ab}$. 

Langlands duality exchanges the Lie algebras $\fg$ and $^L\fg$ by taking the transpose of the Cartan matrix $C_{ab}$. It follows that the simple roots and fundamental weights of $\fg$ are mapped to the simple coroots and fundamental coweights of $^L\fg$, respectively. The only simple Lie algebras which are exchanged under Langlands duality are $B_r$ and $C_r$. The other simple Lie algebras are mapped to themselves, although nontrivially for $G_2$ and $F_4$, since long and short roots are swapped in $^LG_2$ and $^LF_4$.

For completeness, we mention here the existence of intermediate lattices: let $\Lambda_{char}$ be a character lattice of $\fg$, which is any lattice lying between the root and weight lattices, 
\beq
\Lambda_{rt}\subset\Lambda_{char}\subset\Lambda_{wt} \; .
\eeq
The lattice $\Lambda_{char}$ is also called the group of characters, which is the space of maps
\beq
\Lambda_{char} = \text{Hom}(T,U(1)) 
\eeq
from the maximal torus $T$ of $G$ to $U(1)$. The choice of such a lattice is an important ingredient in the classification of compact simple Lie groups, as $G$ is uniquely determined from $\fg$ and the choice of a lattice $\Lambda_{char}$.

The lattice $\Lambda_{char}$ admits a dual $\Lambda_{cochar}$, called the cocharacter lattice of $\fg$, which lies between the coroot and coweight lattices, 
\beq
\Lambda_{cort}\subset\Lambda_{cochar}\subset\Lambda_{cowt} \; .
\eeq
The lattice $\Lambda_{cochar}$ is the space of maps 
\beq
\Lambda_{cochar} = \text{Hom}(^L T,U(1)) 
\eeq
from the maximal torus $^L T$ of $^L G$ to $U(1)$. Equivalently, it is the space of maps $\text{Hom}(U(1),T)$.\\

The lattices $\Lambda_{char}$ and $\Lambda_{cochar}$ also play a role in probing the physics of extended operators in 4d, most notably using Wilson or 't Hooft line operators \cite{Kapustin:2005py,Kapustin:2006pk}, or the surface operators we study here \cite{Gukov:2006jk,Gukov:2008sn}. A sharper statement, then, is that the lattice of electric charges of $G$ is really $\Lambda_{char}$, while the lattice of magnetic charges is $\Lambda_{cochar}$.  From the point of view of the gauge groups, this amounts to specifying the global form of $G$ and $^L G$, respectively. Because the quantum $q$-Langlands program is formulated at the level of the Lie algebras $\fg$ and $^L \fg$, the global form of the  gauge groups is ignored in the correspondence.\\

When we discuss 4d $\cN=4$ Super Yang-Mills of type $G$ and its surface defects, the group $G$ will be simply-connected in this paper. This is because we will want $\text{T}:\tau\mapsto\tau+1$ to be a symmetry of the gauge coupling  $\tau= \frac{\theta}{2\pi} + i\, \frac{4\pi}{g^2_{4d}}$, and is argued as follows: the symmetry reflects the invariance of the theory under a shift of the theta angle $\theta$  by $2\pi$, where $\theta$ appears in the Yang-Mills action through an instanton term:
\beq
-i\,\theta\, k \; , \;\;\; \text{where} \;\;k=\frac{-1}{8\pi^2} \int_{M_4} \text{Tr}F\wedge F \; .
\eeq
The trace $\text{Tr}$ is a negative-definite quadratic form on $\fg$, normalized so that if $\alpha^\vee_s$ is a short coroot, then $\text{Tr}(\alpha^\vee_s\cdot\alpha^\vee_s)=-2$. With this normalization\footnote{A different normalization of the trace $\text{Tr}$ would rescale the definition of the instanton number by a factor of a root length, which we decide against for clarity of presentation. We therefore adopt the same normalization on the short coroot length.}, and assuming $k$ is an integer (the instanton number), the action $\text{T}:\theta \mapsto \theta + 2\pi$, or equivalently $\text{T}:\tau\mapsto\tau+1$, is a symmetry only if $G$ is simply-connected\footnote{The 4-manifold $M_4$ should also be closed. For us, the manifold $M_4$ we typically encounter is $\mathbb{C}^2$, and therefore not closed; this causes a non-compactness of the instanton moduli space in the IR, which we cure via an $\Omega$-background $\mathbb{C}_{q}\times\mathbb{C}_t$ \cite{Nekrasov:2002qd,Losev:2003py,Nekrasov:2003rj}.}.  If $G$ is not simply-connected, the symmetry is no longer T, but instead $\text{T}^p:\tau\mapsto\tau+p$, for some integer $p$. The quantum $q$-Langlands correspondence (with or without ramification) we study is well-defined at least for $p=1$, hence the requirement that $G$ be simply-connected. In that case, the character lattice will coincide with the weight lattice:
\beq
\Lambda_{char}=\Lambda_{wt} \; .
\eeq
The simple-connectedness of $G$ further implies that the Langlands dual group $^L G$ will be of adjoint type, meaning that its root and weight lattices coincide. Equivalently, this sets the cocharacter lattice of $G$ to coincide with its coroot lattice:
\beq
\Lambda_{cochar}=\Lambda_{cort} \; .
\eeq

\vspace{8mm}

\section{A quick guide to the 3d $\cN=2$ half-index}
\label{sec:appendixindex}

In an effort to keep our presentation self-contained, we summarize here the various building blocks that make up the 3d $\cN=2$ half-index \eqref{3dhalfindex}:
\beq
\label{3dhalfindexagain}
{\mathcal Z}(T^{3d}_{\fg_\text{o}})  = {\rm Tr}\left[(-1)^F\, q^{J+\frac{R}{2}} \; {\bf{x}}^{\Pi} \right]\;\; .
\eeq
We use the following notations: 
\beq
(x \,; q)_\infty = \prod_{k=0}^\infty \left(1-q^k x\right) \; ,
\eeq
denotes the $q$-Pochhammer symbol, and 
\beq
\Theta(x\, ; q)= (x \,;q)_\infty\,(q/x \,; q)_\infty \; . 
\eeq
the genus 1 odd theta function.\\

A 3d chiral multiplet $\Phi$ satisfies chirality constraints with respect to differential operators in superspace: $\overline{D}_+\Phi=0=\overline{D}_-\Phi$, with off-shell superfield content
\beq
\Phi=\phi+\theta^+\psi_++\theta^-\psi_-+\theta^+\theta^-\, F+\ldots \; .
\eeq 
The fields $\phi$ and $\psi_\alpha$ are respectively a complex boson and a complex fermion, the field $F$ is an auxiliary complex boson set to 0 by the equations of motion. We denoted by ``$\ldots$" additional terms which will not contribute to the index. The 3d chiral multiplet decomposes on the boundary into a (0,2) chiral multiplet,
\beq
\Phi_{|\partial}=\phi+\theta^+\psi_+-i\theta^+\overline{\theta}^+\partial_+\phi \; ,
\eeq
which satisfies $\overline{D}_+\Phi_{|\partial}=0$, and a (0,2) Fermi multiplet,
\beq
\Psi_{|\partial}={\psi}_- + \theta^+\, G - i\theta^+\overline{\theta}^+\partial_+{\psi}_- \; ,
\eeq
which satisfies $\overline{D}_+\Psi_{|\partial}=0$. On-shell, the auxiliary field $G$ is equal to $\partial_{\bot}\phi$, and $\Psi_{|\partial}=\overline{D}_+\Phi|_{\theta^-=\overline{\theta}^-}=0$.\\

The bulk action for the 3d chiral multiplet couples to a  boundary Fermi multiplet via a J-term, as $\Psi_{|\partial} \, \partial_\bot \Phi_{|\partial}$. In components, this means the action contains the term ${\psi}_- \, \partial_\bot \psi_+$. In order to avoid boundary terms in the equations of motion, one should therefore either set the fermion $\psi_- = 0$ or the fermion $\psi_+ = 0$ at the boundary. Following the terminology of \cite{Dimofte:2017tpi}, we call the supersymmetric completion of these boundary conditions Neumann (N) and Dirichlet (D), respectively. Then, a boundary condition N sets to zero the entire Fermi multiplet $\Psi_{|\partial}$; in particular, $\psi_- = 0$ and $\partial_\bot \phi=0$ at the boundary. Meanwhile, a  boundary condition D sets to zero the entire chiral multiplet $\Phi_{|\partial}=0$; in particular, $\psi_+ = 0$ and $\phi=0$ at the boundary.

It is also possible to consider a ``deformed" Dirichlet boundary condition  $\text{D}_c$, by setting $\Phi_{|\partial}=c$, with $c\in\mathbb{C}$ a constant chiral multiplet at the boundary; in particular, $\psi_+ = 0$ and $\phi=c$ at the boundary. This boundary condition will break $U(1)_R$ unless $\phi$ has R-charge 0. In isolation, the R-charge of such a chiral field is too low for $\text{D}_c$  to flow to a superconformal boundary condition in the IR; however, such a flow is possible if the field is part of a full fledged gauge theory (with the gauge group partially broken at the boundary because of  $\text{D}_c$). This is the typical situation we encounter in this paper: some of our chiral multiplets have  $\text{D}_c$  boundary conditions, but they are always part of a gauge theory $T^{3d}_{\fg_\text{o}}$ whose IR boundary conditions are superconformal.\\

We denote by $U(1)_J$ the action rotating the boundary plane $\mathbb{R}^{1,1}$.
Suppose then that the 3d chiral has charge $+1$ under a flavor symmetry $U(1)_x$, with associated fugacity $x$, and charge $\rho$ under the R-symmetry $U(1)_R$. In the decomposition into (0,2) chiral and Fermi multiplets, the operators in $\overline{Q}_+$-cohomology counted by the index are $\phi$ (the top component of the chiral multiplet) and its derivatives $\partial^n_z\phi$, as well as  ${\psi}_-$ (the top component of the Fermi multiplet) and its derivatives $\partial^n_z{\psi}_-$, and products thereof. Under $U(1)_x\times U(1)_R \times U(1)_{J}$, the operators $\partial^n_z\phi$ carry charge $(1,\rho,n)$, while  $\partial^n_z{\psi}_-$ carry charge $(-1,1-\rho,n+\frac{1}{2})$. 

When the 3d chiral multiplet is given Neumann boundary conditions N, it follows that the operators  $\partial^n_z{\psi}_-$ are set to zero, so only the bosons  $\partial^n_z\phi$ contribute to the index:
\beq\label{chiral1}
\prod_{n=0}^\infty\left(1-q^{n+\frac{\rho}{2}}\, x\right)^{-1}= \left(x \, q^{\frac{\rho}{2}} ; q\right)^{-1}_\infty \; .
\eeq
Meanwhile, when the 3d chiral multiplet is given Dirichlet boundary conditions D, the operators $\partial^n_z\phi$ are set to zero, so only the fermions $\partial^n_z{\psi}_-$  contribute to the index:
\beq\label{chiral2}
\prod_{n=0}^\infty\left(1-q^{n+(1-\frac{\rho}{2})}/x\right) = \left(q^{1-\frac{\rho}{2}}/x ; q\right)_\infty \; . 
\eeq
For the 3d chiral multiplet $\Phi$ with deformed Dirichlet boundary conditions $\text{D}_c$, the fermions $\partial^n_z{\psi}_-$ likewise contribute to the index, but now carry charge $(0,1,n+\frac{1}{2})$ under $U(1)_x\times U(1)_R \times U(1)_{J}$. Indeed, $U(1)_x$ is broken at the boundary, and $U(1)_R$ is unbroken only when $\Phi$ has R-charge 0. 
At the level of the index, one can simply specialize the fugacity $x\rightarrow 1$ in the {D} boundary condition. For a single chiral multiplet, this gives a contribution
\beq
\left(q \; ; q\right)_\infty \; .
\eeq
When the chiral multiplet is part of a gauge theory $T^{3d}_{\fg_\text{o}}$, one should first impose {D} and then specialize the corresponding fugacity $x\rightarrow 1$ inside the entire index.\\

A 3d $\cN=2$ gauge multiplet can be written as a gauge invariant linear multiplet \cite{Aharony:1997bx}:
\beq
\Sigma=\fa-\theta\overline{\lambda}+\overline{\theta}\lambda+i\overline{\theta}\theta \fD + \overline{\theta}\sigma^{ij}\theta F_{ij}+\ldots \; ,
\eeq 
which satisfies ${D}_\alpha{D}^\alpha \Sigma_{3d} = 0 = \overline{D}^\alpha\overline{D}_\alpha \Sigma_{3d}$. In the multiplet, one finds the field strength $F_{ij}$ for a  connection $A_i$, a real scalar $\fa$ and  off-shell auxiliary field $\fD$, and a complex fermion $\lambda_\alpha$. The 3d linear multiplet $\Sigma$ decomposes on the boundary into a (0,2) chiral multiplet,
\beq
S_{|\partial}=(\fa+i A_{\bot}) -2\,\theta^+\overline{\lambda}_+\ldots \; ,
\eeq
which satisfies $\overline{D}_+S_{|\partial}=0$, and a (0,2) gauge multiplet, or alternatively a chiral field-strength superfield
\beq\label{fieldstrengthgauge}
\Upsilon_{|\partial}=\lambda_-+\theta^+(F_{01}+i\, \fD') +\ldots \; ,
\eeq
which satisfies $\overline{D}_+\Upsilon_{|\partial}=0$,  with $\fD'$ an auxiliary field equal to $\fD-\partial_{\bot}\fa$  on-shell.\\

The Neumann boundary condition on $\Sigma$ sets $F_{\bot\mu|\partial}=0$, and in particular leaves an unbroken gauge symmetry on the 2d boundary. In contrast, the Dirichlet boundary condition on $\Sigma$ sets $A_{\mu|\partial}=0$, which implies the local gauge symmetry on the 2d boundary is broken to a global symmetry. The supersymmetric completion of these boundary conditions will also be denoted as Neumann ({\bf N}) and Dirichlet (${\bf D}$).

The boundary condition ${\bf N}$ preserves a subset of fields present in the $\Upsilon_{|\partial}$ multiplet only. Let us first ignore the surviving gauge invariance at the boundary, and treat the symmetry there as global. The operators in $\overline{Q}_+$-cohomology counted by the index are the gaugino $\lambda_-$, its covariant derivatives $D^n_z\lambda_-$, and products thereof. If we denote the gauge group by $G_{3d}$, then under $G_{3d}\times U(1)_R \times U(1)_{J}$, the operators $D^n_z\lambda_-$ carry charge $(\text{adj},1,n+\frac{1}{2})$. Factorizing the adjoint contribution as the contribution of $\text{rank}(G_{3d})$ null weights and the roots, the index is
\beq\label{gaugeneum}
\left(q \; ; q\right)^{\text{rank}(G_{3d})}_\infty\prod_{\alpha_a\in\text{roots}\,[G_{3d}]}\left(q \,y_{\alpha_a} \; ; q\right)_\infty \; ,
\eeq
with $y_{\alpha_a}$ valued in the maximal torus of $G_{3d}$. Imposing gauge invariance is done by integrating over these  eigenvalues and inserting the Vandermonde factor $\frac{1}{W_{G_{3d}}}(1-y_{\alpha_a})$:
\beq
\frac{\left(q \; ; q\right)^{\text{rank}(G_{3d})}_\infty}{W_{G_{3d}}}\oint\frac{dy}{y}\prod_{\alpha_a\in\text{roots}\,[G_{3d}]}\left(y_{\alpha_a} \; ; q\right)_\infty \; ,
\eeq
with $W(G_{3d})$ the Weyl group of $G_{3d}$.
In our work, $G_{3d}$ is always a product of unitary gauge groups. For instance, for $G_{3d}=U(N)$, the above reads
\beq
\left(q \; ; q\right)^N_\infty\oint\prod_{i=1}^N\frac{dy_i}{y_i}\prod_{j\neq i}^N\left( \,y_i/y_j \; ; q\right)_\infty \; .
\eeq
The boundary condition ${\bf D}$ preserves a subset of fields in the $S_{|\partial}$ multiplet instead. The operators in $\overline{Q}_+$-cohomology counted by the index are $\fa+i A_{\bot}$ and its derivatives, or more precisely  $D_z\fa+i F_{z\bot}$ and its derivatives; the reasoning here is that even though the gauge symmetry is broken at the boundary to a global symmetry $G_{|\partial}$, there is a residual gauge symmetry along the $x_{\bot}$ coordinate, under which $\fa+i A_{\bot}$ is not invariant, but the field strength  $D_z\fa+i F_{z\bot}$ is. The operators $D^{n+1}_z\fa+i D^{n}_z F_{z\bot}$ carry charge $(\text{adj},0,n+\frac{1}{2})$ under $G_{|\partial}\times U(1)_R \times U(1)_{J}$. Factorizing the adjoint contribution as the contribution of $\text{rank}(G_{3d})$ null weights and the roots, the index is
\beq
\left(q \; ; q\right)^{-\text{rank}(G_{|\partial})}_\infty\prod_{\alpha_a\in\text{roots}\,[G_{|\partial}]}\left(q \,y_{\alpha_a} \; ; q\right)^{-1}_\infty \; ,
\eeq
with $y_{\alpha_a}$ valued in the maximal torus of the boundary global symmetry $G_{|\partial}$.

Crucially, there are non-perturbative contributions to this  ${\bf D}$  half-index coming from BPS monopole operators, which should be counted as they are in $\overline{Q}_+$-cohomology. A monopole operator in the 3d bulk of a $\cN=2$ theory is a disorder operator given by a singular solution to
\beq
F = \star D \fa\, , \qquad\; D \star\fa = 0 \; .
\eeq
In the abelian case $G_{3d} = U(1)$, the basic solution is a Dirac monopole $\fa = m/2\,r$, with $r$ is the radial distance from the singularity and $m$ is an integer, due to the quantization of flux through a 2-sphere centered at the origin:
\beq
\oint_{S^2}\frac{F}{2\pi}=m\in\mathbb{Z} \; .
\eeq
Likewise, boundary monopoles contribute to the index. This time, a monopole operator at the origin should be surrounded by a hemisphere $D^2$, with boundary $S^1_{D^2}$. Crucially, $F_{z\bot}$ and $\fa$ are left unconstrained at the boundary, so the presence of such a monopole is compatible with the Dirichlet boundary condition  ${\bf D}$, which we impose to trivialize the $U(1)$-bundle at the boundary $S^1_{D^2}$. The curvature integral now takes the form
\beq
\frac{1}{2\pi}\oint_{D^2}F +\frac{1}{2\pi}\oint_{S^1_{D^2}} A = m \in\mathbb{Z} \; .
\eeq
But $\oint_{S^1_{D^2}} A=0$ from the boundary condition $A_{\mu|\partial}=0$, so the monopole charge $m$ is nothing but the quantized flux through the hemisphere:
\beq
\frac{1}{2\pi}\oint_{D^2}F = m\in\mathbb{Z} \; .
\eeq
More generally, when $G_{3d}$ is nonabelian, we embed the Dirac monopole into $G_{3d}$, so by definition $m$ is an element of the cocharacter lattice  $m\in\Lambda_{cochar}=\text{Hom}(U(1),T_{G_{3d}})$, the space of maps to the maximal torus $T_{G_{3d}}$.
The index takes the form of a sum over all abelian magnetic fluxes $m$ on the hemisphere $D^2$:
\beq
\left(q \; ; q\right)^{-\text{rank}(G_{|\partial})}_\infty\sum_{m\in\Lambda_{cochar}}\;\prod_{\alpha_a\in\text{roots}\,[G_{|\partial}]}\left(q^{1+m\cdot\alpha_a} \,y_{\alpha_a} \; ; q\right)^{-1}_\infty \;\ldots\;,
\eeq
where the boundary fugacity $y_{\alpha_a}$ is now shifted by $q^m$, reflecting the fact that the corresponding state acquires spin $m$ due to the monopole. The dots ``$\ldots$" stand for the contribution of the remaining matter content of the theory, as well as boundary mixed 't Hooft anomalies.\\

We now write the 2d degrees of freedom living exclusively on the boundary.
Namely, consider a 2d $(2,0)$ chiral multiplet $\Phi_{|\partial}={\phi} +\ldots$ with charge $+1$ under a flavor symmetry $U(1)_x$ and charge $\rho$ under the R-symmetry $U(1)_R$. The operators in $\overline{Q}_+$-cohomology counted by the index are $\phi$, $\partial_z\overline{\phi}$, their derivatives, and products thereof. Under $U(1)_x\times U(1)_R \times U(1)_{J}$, the operators $\partial^n_z\phi$ carry charge $(1,\rho,n)$, while  $\partial^{n+1}_z\overline{\phi}$ carry charge $(-1,-\rho,n+1)$.  
Their contribution to the index is
\beq\label{chiral2d}
\prod_{n=0}^\infty \left(1-q^{n+\frac{\rho}{2}}\, x\; ; q\right)_{\infty}^{-1} \left(1-q^{n+1-\frac{\rho}{2}}\, x^{-1}\; ; q\right)_{\infty}^{-1}=\frac{1}{\Theta\left(x \, q^{\frac{\rho}{2}} ; q\right)} \; .
\eeq
Note that the index counts more operators compared to the 3d chiral multiplet with Neumann boundary conditions \eqref{chiral1}. Indeed, by the supersymmetry transformation $\overline{Q}_+ \overline{\phi} \sim \overline{\psi}_+$, the operators $\partial^{n+1}_z\overline{\phi}$ in the 2d chiral multiplet are $\overline{Q}_+$-closed after imposing the Dirac equation $\partial_z \overline{\psi}_+=0$, so they are counted by the index. This is no longer true for the 3d chiral, as $\partial_z \overline{\psi}_+ \neq 0$.
	
	Similarly, consider a 2d $(2,0)$ Fermi multiplet $\Psi_{|\partial}={\psi}_- +\ldots$  with charge $-1$ under a flavor symmetry $U(1)_x$ and charge $1-\rho$ under the R-symmetry $U(1)_R$. The operators in $\overline{Q}_+$-cohomology counted by the index are  $\partial^{n}_z{\psi}_-$, $\partial^{n}_z\overline{\psi}_-$, their derivatives, and products thereof.
	Under $U(1)_x\times U(1)_R \times U(1)_{J}$, the operators $\partial^{n}_z{\psi}_-$ carry charge  $(-1,1-\rho,n+\frac{1}{2})$, while  $\partial^{n}_z\overline{\psi}_-$ carry charge $(1,\rho-1,n+\frac{1}{2})$.  
	Their contribution to the index is
	\beq\label{fermi2d}
	\prod_{n=0}^\infty \left(1-q^{n+1-\frac{\rho}{2}}\, x^{-1}\; ;q\right)_{\infty}^{-1} \left(1-q^{n+\frac{\rho}{2}}\, x \; ; q\right)_{\infty}^{-1} =\frac{1}{\Theta\left(x \, q^{\frac{\rho}{2}} ; q\right)} \; .
	\eeq
	
	Finally, consider a 2d $(2,0)$ gauge multiplet with gauge group $G$, which we wrote before in terms of a  chiral field-strength superfield $\Upsilon$ \eqref{fieldstrengthgauge}. The R-charge of the multiplet is uniquely fixed to be 1, since the gauge fields must have R-charge 0. The operators in $\overline{Q}_+$-cohomology counted by the index are the gauginos $\lambda_-$, $D^n_z\overline{\lambda}_-$, their covariant derivatives, and products thereof. Note that when we discussed the 3d gauge multiplet with ${\bf N}$ boundary conditions \eqref{gaugeneum}, $D^n_z\overline{\lambda}_-$ was not $\overline{Q}_+$-closed, by the 3d equations of motion.
	Under $G\times U(1)_R \times U(1)_{J}$, the operators $D^n_z{\lambda}_-$ carry charge  $(\text{adj},1,n+\frac{1}{2})$, while  $D^{n+1}_z\overline{\lambda}_-$ carry charge $(\text{adj},-1,n+\frac{3}{2})$.  
	
	If $G$ were a global symmetry then the index would be
	\beq
	\left(q \; ; q\right)^{2\,\text{rank}(G)}_\infty\prod_{\alpha_a\in\text{roots}\,[G]}\left(q \,y_{\alpha_a} \; ; q\right)_\infty\left(q \,y^{-1}_{\alpha_a} \; ; q\right)_\infty \; ,
	\eeq
	with $y_{\alpha_a}$ valued in the maximal torus of $G$. In this expression, we factorized (twice) the adjoint contribution as that of $\text{rank}(G)$ null weights and the roots. Imposing gauge invariance is done by integrating over the $y_{\alpha_a}$  eigenvalues and inserting the Vandermonde factor $\frac{1}{W_{G}}(1-y_{\alpha_a})$:
	\beq
	\frac{\left(q \; ; q\right)^{2\, \text{rank}(G)}_\infty}{W_{G}}\oint\frac{dy}{y}\prod_{\alpha_a\in\text{roots}\,[G]}\Theta\left(y_{\alpha_a}  ; q\right) \; .
	\eeq

\bibliographystyle{kp}
\begin{spacing}{0.70}
{\small \bibliography{summaryqq}}
\end{spacing}
\end{document}